\documentclass[%
  reprint,amsmath,amssymb,aps,rmp]{revtex4-2}

\usepackage{url}
\usepackage[dvipsnames]{xcolor}
\usepackage{color,soul}

\usepackage[colorlinks=true]{hyperref}
\hypersetup{allcolors = {black},}

\setstcolor{red}
\sethlcolor{yellow}
\usepackage{soul}

\usepackage{natbib}
\usepackage{graphicx}
\usepackage{dcolumn}
\usepackage{bm}
\usepackage{booktabs}
\usepackage{float}

\begin{document}
\setstcolor{TealBlue}

\preprint{APS/123-QED}

\title{Sliding and Pinning in Structurally Lubric 2D Material Interfaces}

\author{Jin Wang}
\affiliation{International School for Advanced Studies (SISSA), Via Bonomea 265, 34136 Trieste, Italy}
 
\author{Ali Khosravi}%
\affiliation{International School for Advanced Studies (SISSA), Via Bonomea 265, 34136 Trieste, Italy}
\affiliation{International Centre for Theoretical Physics (ICTP), Strada Costiera 11,34151 Trieste,Italy}%

\author{Andrea Vanossi}
\affiliation{CNR-IOM, Consiglio Nazionale delle Ricerche - Istituto Officina dei Materiali, c/o SISSA, Via Bonomea 265, 34136 Trieste, Italy}%
\affiliation{International School for Advanced Studies (SISSA), Via Bonomea 265, 34136 Trieste, Italy}%

\author{Erio Tosatti}
  \email{tosatti@sissa.it}
\affiliation{International School for Advanced Studies (SISSA), Via Bonomea 265, 34136 Trieste, Italy}
\affiliation{International Centre for Theoretical Physics (ICTP), Strada Costiera 11,34151 Trieste,Italy}
\affiliation{CNR-IOM, Consiglio Nazionale delle Ricerche - Istituto Officina dei Materiali, c/o SISSA, Via Bonomea 265, 34136 Trieste, Italy}


\begin{abstract}
A plethora of two-dimensional (2D) materials entered the physics and engineering scene in the last two decades. Their robust, membrane-like sheet permit – mostly require -- deposition, giving rise to solid-solid dry interfaces whose bodily mobility, pinning, and general tribological properties under shear stress are currently being understood and controlled, experimentally and theoretically. In this Colloquium we use simulation case studies of twisted graphene system as a prototype workhorse tool to demonstrate and discuss the general picture of 2D material interface sliding. First, we highlight the crucial mechanical difference, often overlooked, between small and large incommensurabilities, corresponding e.g., to small and large twist angles in graphene interfaces. In both cases, focusing on flat, structurally lubric, “superlubric” geometries, we elucidate and review the generally separate scaling with area of static friction in pinned states and of kinetic friction during  sliding, tangled as they are with the effects of velocity, temperature, load, and defects. Including the role of island boundaries and of elasticity, and corroborating when possible the existing case-by-case results in literature beyond graphene, the overall picture proposed is meant for general 2D material interfaces, that are of importance for the physics and technology of existing and future bilayer and multilayer systems. 
\end{abstract}

\maketitle

\tableofcontents

\section{Introduction}

\begin{figure}[ht!]
\centering
\includegraphics[width=\linewidth]{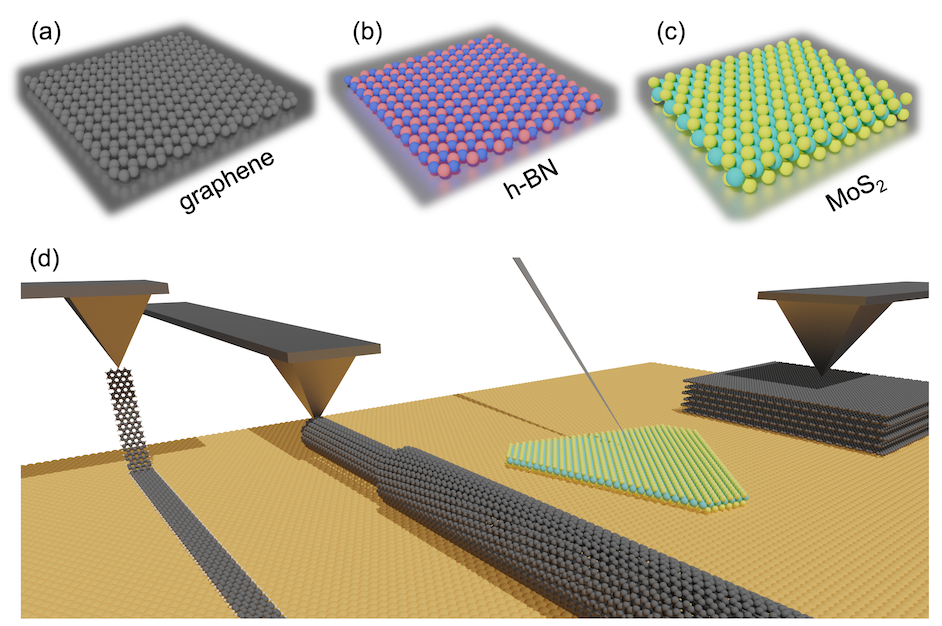}
\caption{(a-c). Some relevant 2D materials; (d) common 
experimental setups involving their interfaces.}
\label{fig:1}
\end{figure}

A great variety -- qualitatively pictured in Fig.~\ref{fig:1}(a-c) -- of strong, graphene-like 2D materials \cite{Geim.natmater.2007,Geim.nature.2013,Neto2016}, increasingly pervades materials science, physics and technology.
That includes graphene bilayers and transition metal dichalcogenides 
that are especially important for new phenomena and applications \cite{Geim.nature.2013, Neto2016, Rao.AngChemInter.2009, Butler.acsnano.2013, Yankowitz.NatRevPhys.2019, Wang.royalSCh.2017, Liu.NatureReviewsMaterials.2016,Pham.ChemRev.2022}.

Besides the electronic properties that attract most of the attention \cite{Cao.nature.2018-1,Cao.nature.2018-2,Kerelsky.nature.2019,Yankowitz.science.2019,Stepanov.nature.2020,Park.nature.2021}, their mechanical, tribological and rheological characterization must in parallel be physically understood and controlled \cite{Zhang.MaterToday.2019} 
for many reasons.
A first, practical one is that mutual sliding of two juxtaposed  2D material layers is known to be quite easy, owing to their weak interlayer van der Waals interaction.
Beyond that,
mutual incommensurability of 2D materials interfaces can lead to ultra-low sliding friction, a property known as ``superlubricity'' \cite{Hirano.prb.1990,Dienwiebel.prl.2004}.
That property makes it naturally attractive for energy saving and potentially for lifetime increase.
Manipulations involving these interfaces also occur in several micro/nanoscale experiments (see Fig.~\ref{fig:1}d).
Good lubricity makes them promising for the application in micro/nano electromechanical systems, e.g., micro/nano-generators and nano-oscillators \cite{Huang.nanoenergy.2020,Huang.nc.2021,Wu.commumater.2021,Zheng.prl.2002}.
Another one, more substantial, is that our current understanding of ``superlubricity" is still too vague. 
On the whole, the friction's connection to incommensurability and rotation is as we shall see a source of surprises.\\

\begin{figure}[ht!]
\centering
\includegraphics[width=\linewidth]{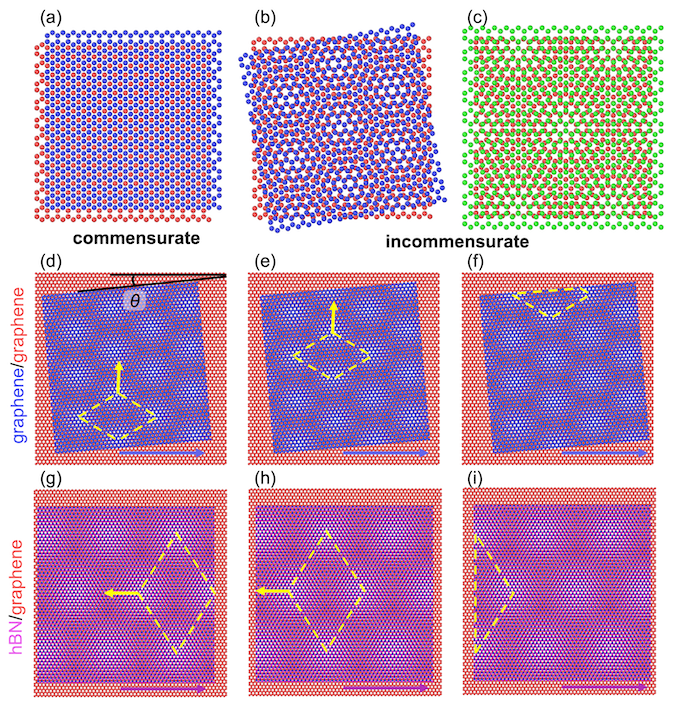}
\caption{Sketch of 2D materials interfaces: (a) commensurate;
(b) “homo” incommensurate with twist angle $\theta$.
(c) “hetero” incommensurate with lattice mismatch $ a \neq b$ and zero twist;
(d-f) graphene-graphene moiré cell (yellow), shown for a small twist $\theta$, obliquely translating upon horizontal interface sliding (arrow); (g-i) same for hBN/graphene, for $\theta$ =0.
The angle $\beta$ of relative motion,  Eq.~\eqref{Eq:1.2}, is discussed in text.
}
\label{fig:2}
\end{figure}

The general incommensurate interface is composed of two monolayers that are either identical but 
rotated by a ``twist'' angle $\theta$  (homo case) that are different, with an inherent lattice mismatch (hetero case).
It is characterized by the so-called moir\'e pattern (see Fig.~\ref{fig:2}a-c), an almost-periodic superlattice resulting from the beating of the two 2D lattices.
The length $\lambda$ of the moir\'e superlattice 
between layers with triangular or hexagonal symmetry is determined by the lattice constants of the two contacting layers $a$, $b$, and the twist angle $\theta$, 
\begin{equation}
    \lambda(\theta)=\frac{a b}{\sqrt{a^2+b^2-2ab \cos(\theta)}}
    \label{Eq:1}
\end{equation}
For homostructures with $a=b$, the moir\'e size $\lambda (\theta )=a/\sqrt{2-2 \cos{\theta}}$ scales as $\theta^{-1}$.
When one of the layers (the ``slider") is sheared relative to the other,
the moiré pattern drifts at an angle $\beta$ relative to the sliding direction, \begin{equation}
    \beta (\theta) = \arccos{\frac{b -a\cos(\theta)}{\sqrt{a^2+b^2-2ab \cos(\theta)}}}
    \label{Eq:1.2}
\end{equation}
As Eq.~\eqref{Eq:1.2} shows, the moir\'e pattern generally drifts askew of the sliding direction.
The ratio of the sliding velocity of the moiré $v_\mathrm{M}$ to the slider's $v$ is $\lambda/a$ \cite{Hermann.jpcm.2012, Wang.tobe.2023}.
Larger than one, this ratio diverges when $a=b$ and $\theta \to $ 0.
As sketched in Fig.~\ref{fig:2} (d-f), for graphene/graphene homostructure, thus with $a=b$ and twist angle $\theta$, then $\beta=90^{\circ}-\theta/2$, i.e. the moir\'e direction is nearly perpendicular to the sliding direction of the slider at small twists.
For a hetero bilayer such as hBN/graphene  with mismatched lattices
$a/b \approx 1.018$, at zero twist angle (Fig.~\ref{fig:2} g-i), $\beta=180^{\circ}$. \\

A host of questions arise about the facile sliding of these incommensurate 2D crystalline interfaces,
or alternatively on their mechanical pinning against shear.
The exceptional in-plane robustness coupled with great out-of-plane membrane-like flexibility are brand new elements. Another one is the unusually important adhesion between the very flat layers. These elements make 2D materials interfaces different from other frictional systems \cite{Vanossi.nc.2020} as 3D solids \cite{Bowden.2001, Liu.surfcoat.1996, Eriksson.wear.2002, Zhao.acsnano.2021}, adsorbed layers and clusters \cite{He.Science.1999,Jacqueline.AdvPhys.2012, Varini.nanoscale.2015, Pierno.NatNano.2015}, colloid monolayers \cite{Bohlein.natmat.2012, Vanossi.natmat.2012, Mandelli.prl.2015, Brazda.prx.2018, Cao.NaturePhysics.2019}, etc.
For nano to mesoscale, the size and temperature dependence of static friction of 2D material contacts, as well as their comparison with that of (velocity-dependent) kinetic friction generally differs from classic macroscopic laws in a way that is currently addressed case by case.
The widespread concept in the sliding of crystalline interfaces is structural lubricity between incommensurate faces is
\cite{Aubry.physD.1983, Peyrard.Aubry.1983, Sacco.PhysRevLett.1978, Muser.EPL.2004},
generally believed to imply superlubricity \cite{Sokoloff.PhysRevB.1990, Shinjo.surfsci.1993, Dienwiebel.prl.2004, Vanossi.RevModPhys.2013, Baykara.APR.2018,  Martin.PhysicsToday.2018}.
It should be noted that this term is nowadays used with different meanings in physics and in engineering.\\

Standard superlubricity \cite{Peyrard.Aubry.1983},
which we adopt here as synonymous to \textbf{\emph{structural superlubricity}} \cite{Muser.EPL.2004}, can be 
defined for infinite defect-free systems as an {\it unpinned}, free sliding state.
That is a state where the static friction $F_\mathrm{s}$ – the smallest force needed to initiate the sliding motion – is mathematically zero even down to $T=0$.
That also implies the ability to slide, albeit with infinitesimal velocity, under an infinitesimal applied force.
In the real world of course everything has finite size, even without defects. The edges destroy the slider's perfect translational invariance, inevitably causing static friction.
However, if the slider's bulk is structurally superlubric and defect-free, $F_\mathrm{s}$ scales as $A^{\alpha}$ with $\alpha \leq 1$. For $\alpha <1$ we shall refer to this real-world cousin of structural superlubricity as \textbf{\emph{structural lubricity}}.
Finally, engineers and practitioners, who handle real materials with systematic defects, finite temperature, etc., sometimes call
superlubricity for any system with low sliding friction and friction coefficient ($\mu <10^{-2}$)
\cite{Martin.PhysRevB.1993, Baykara.APR.2018, Hod.nature.2018, Martin.PhysicsToday.2018}.
To avoid confusion, which this superposition of names sometimes generates, we will use the further term \textbf{\emph{engineering superlubricity}} when needed, for the latter.
Despite the celebrated early detection of superlubricity
\cite{Hirano.PhysRevLett.1997, Liu.prl.2012} in structurally lubric twisted graphene flakes and graphite interfaces \cite{Dienwiebel.prl.2004, Liu.prl.2012, Lee.science.2010,Qu.prl.2020,Koren.science.2015,Urbakh.natnano.2013,Zhang.natnano.2013, Filleter.PhysRevLett.2009,Kawai.Science.2016, Berman.Science.2015} and theoretical work \cite{van.Jphys.2012,Consoli.PhysRevLett.2000,deWijn.prb.2012,Leven.jpcl.2013,Hod.chemphyschem.2013,Mandelli.PhysRevLett.2019},
the community remains in need of a broadly applicable roadmap applicable to the sliding of strong 2D layered materials. We list some of the questions that seem currently open and/or debatable.\\

To what extent does a large size twisted bilayer, incommensurate and therefore candidate to structural superlubricity, or at least to structural lubricity, realize free sliding?
What is the origin of friction and how are static and kinetic friction related at 2D materials interfaces?
What kind of area, temperature, velocity and load dependence of interface sliding friction should one generally anticipate?
Specifically, different experiments report different scaling of friction with area $A^{\alpha}$, with exponent $\alpha$ ranging from 0 to 1. What is the origin of this dispersion?
What role does the extreme 3D anisotropy of these poorly extensible, yet very flexible and easily corrugated membranes play?
Why is it that structurally lubric 2D material experimental sliders generally exhibit a logarithmic velocity dependence (the earmark of stick-slip) \cite{Gnecco.PRL.2000, Riedo.prl.2003, Muser.PhysRevB.2011}, instead of a linear one (the earmark of smooth sliding)?
Can temperature  (or load) bring about a change between high and low friction states?\cite{Krylov.physstatussolid.2014, Pellegrini.prb.2019}
And what are numerically the actual friction coefficients of the 2D materials interfaces?
Is the differential friction coefficients, generally used for 2D materials, really adequate?\\


This Colloquium is motivated by questions like those posed above. Answers are currently scarce, despite extensive investigations of a variety of 2D materials and models.
We shall proceed through  examination of literature abundantly augmented by our own simulations.
Ultimately, we aim to provide a more comprehensive framework for nanofriction of structurally lubric 2D material interfaces.
The workhorse helping us throughout our discussion will mostly be a graphene/graphene twisted interface.
Convenient as it is for a direct exemplification through molecular dynamics simulations, that choice also describes encapsulated bilayers, systems of current interest for their own sake \cite{Andrei.natmater.2020,Torma.natrevphys.2022,Pham.ChemRev.2022,Mogera.carbon.2020,Nimbalkar.nanomicro.2020}
Beyond the specific example, the scope of our approach is broader.
Results should be applicable for generic twisted, structurally lubric, incommensurate contacts among 2D materials \cite{Geim.nature.2013, Neto2016}, which we consider our broader target.

The Colloquium will unroll as follows. Prior to actual sliding, we first examine in Sect.~II the equilibrium interface geometry.
That will highlight the existence of qualitatively different configurations as function of decreasing twist angle: large, intermediate, and small.  We then introduce in Sect.~III the setup of non-equilibrium molecular dynamics (NEMD) simulations of a twisted graphene interface, modeling  the structure and sliding with a state-of-the-art force field.
Simulations which, as said, are a tool to embody our questions and to propose answers along with discussion and review.
Energy barriers and related static friction considerations, together with simulated frictional force traces, that represent  the  background data for subsequent analysis, are presented in Sect.~IV.
Sect.~V is devoted to the generally distinct – yet sometimes coincident – area dependencies of static and kinetic friction in different regimes and geometries.
Temperature dependence, where viscous friction characteristic of truly superlubric free sliding, is contrasted in Sect. VI with the so-called ``thermolubric” evolution, where thermal barrier crossing
gradually smears out stick-slip friction. A similar contrast between free sliding and stick-slip regimes shows up in the velocity dependence of kinetic friction, which correspondingly varies from linear to logarithmic.
The load dependence of 2D materials friction of Sect.~VII highlights the concept and nature of the friction coefficient, for which we provide a tentative table. Friction coefficient data of 2D materials are a rarity
in literature, where only differential friction coefficients usually appear. That point is corrected by considering adhesion, unusually important in 2D materials contacts.
We also clarify that the load dependence of friction in structurally lubric 2D interfaces arises insofar as pressure modifies the barriers created by edges and defects.
The effect of interfacial elasticity on static and kinetic friction -- directly related to the size of the system, and particularly important at the mesoscale, is discussed in Sect.~VIII.
The variety of defects that mix up and modify the frictional properties of clean perfect interfaces is the subject of Sect.~IX, where we limit ourselves to the main ones.
A discussion and outlook conclude the paper in Sect.~X. In the spirit of Colloquia -- a Rev. Mod. Phys. format invented by Ugo Fano, who was a mentor for one of us -- we will try to debate the questions lined up at the outset, as opposed to offering a scholarly review of the countless contributions in the field. For that, we duly apologize in advance to all the authors whose work we were not able to cite explicitly.\\

\section{Static structure of 2D twisted interfaces}


\subsection{Three regimes}
Prior to directly addressing tribology, with resulting shear stress, pinning and sliding, it is necessary to understand the static structure of a generically incommensurate 2D material crystalline interface,
where the moir\'e patterns do not match the crystal lattices.
First of all, the two layers are not rigid, and deform statically in order adapt to each other, bringing the total energy to a minimum.
This deformation, which we will detail, is important in all our discussions. Its nature is a mere structural relaxation, rather than a ``reconstruction" as it is sometimes dubbed \cite{Yoo.NatMater.2019}, since it changes neither the moir\'e periodicity nor any other symmetry.
The deformation mostly consists of out-of-plane $z$-corrugations of the two layers, very stiff in plane but easy to bend. It is unsymmetrical in a hetero interface, and antisymmetrical across the bilayer's mid-plane in a homo interface at rest.  

\begin{figure}[ht!]
\centering
\includegraphics[width=\linewidth]{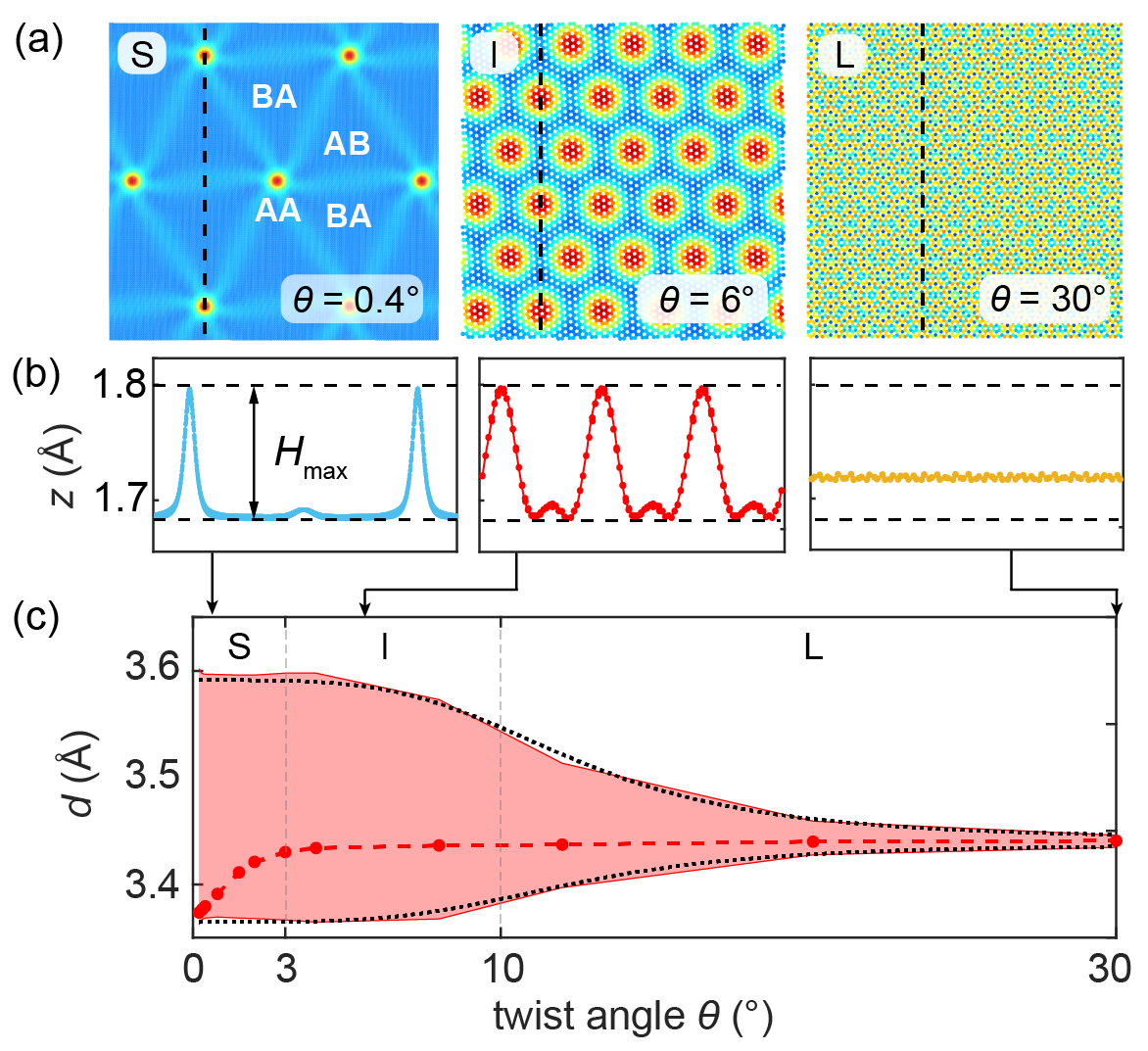}
\caption{
(a) Out-of-plane $z$-distance (corrugation), of a single TBG layer from the mid-interface plane,
fully relaxed at $T=0$~K for S, I, and L regimes. Colormap: red (blue) highlights large (small) corrugation.
(b) corrugation profile along the dashed lines in (a).
(c) Average (red dashed line), minimum and maximum (pink-shaded region) relaxed interlayer spacing as a function of twist angle $\theta$. Dotted lines mark the analytical prediction of Eq.~\eqref{Eq:2}.Data obtained as described in Section III.}
\label{fig:3}
\end{figure}

\begin{figure}[ht!]
\centering
\includegraphics[width=\linewidth]{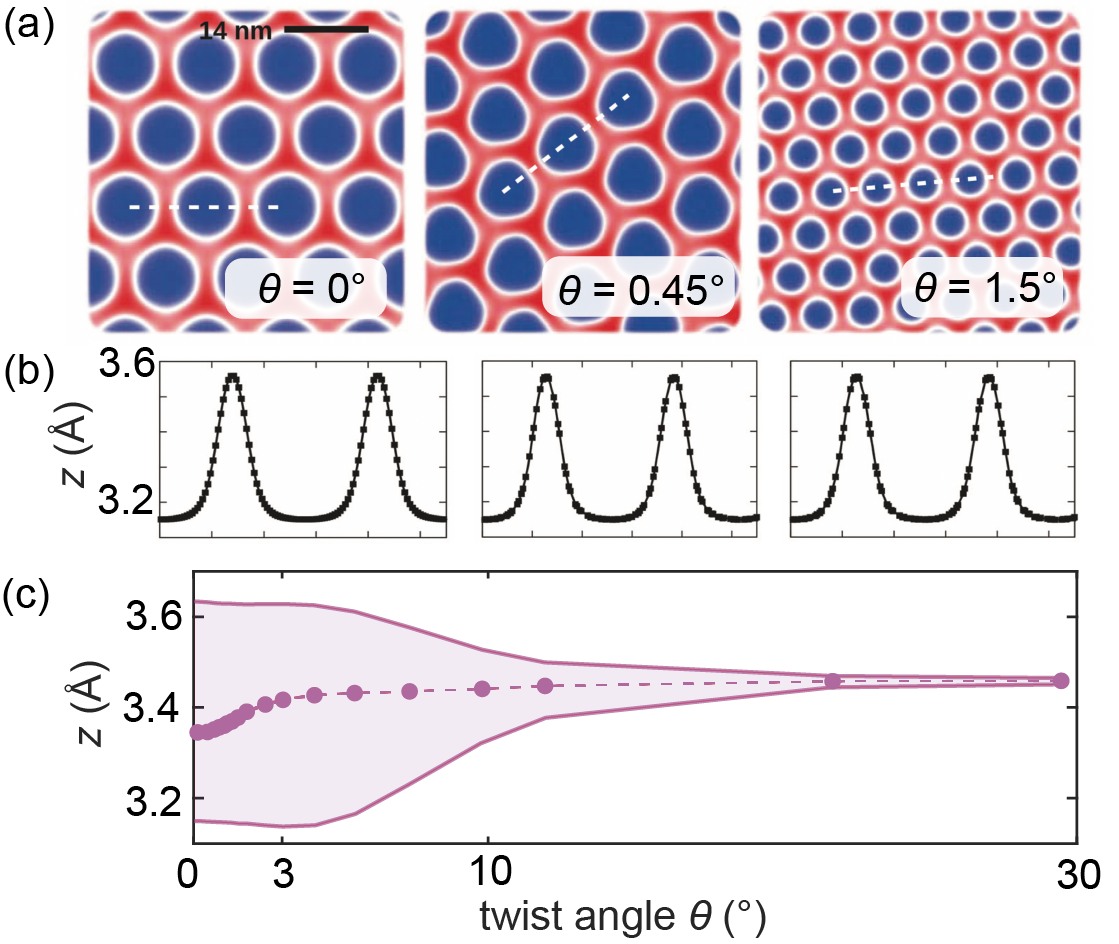}
\caption{Out-of-plane moir\'e corrugation for graphene/hBN simulated heterostructures at various twist angles $\theta=0^\circ$, $0.45^\circ$ and $1.5^\circ$. (a) The moir\'e pattern and (b) the out-of-plane corrugation magnitude along the white dashed traces in panel (a). (c) Maximum, minimum and average value of the $z$ coordinate of carbon atoms in upper graphene layer over the flat hBN substrate as a function of misalignment angle \cite{Guerra.Nanoscale.2017}.}
\label{fig:4}
\end{figure}

We employ here the twisted bilayer graphene (TBG), simulated as will be detailed in Section III, as the simplest showcase of the relaxed structure, see Fig.~\ref{fig:3}.
Generally, there are different regimes to the structure of a 2D material interface.
In a TBG they are realized as a function of the twist angle $\theta$: large (L), intermediate (I),  and small (S), described in Fig.~\ref{fig:3}(a,b).
In the first regime L, with twist $\theta$ ranging from 30$^\circ$ down to about 10$^\circ$, the moiré pattern with small lateral size $\lambda (\theta)$ is accompanied by an out-of-plane corrugation $z(x, y)$.
Here, $z$ is sinusoidal in shape and of small magnitude relative to the interlayer spacing,  while  moderately growing for decreasing $\theta$.
In the opposite small twist regime S below 3$^\circ$, the corrugation pattern is totally different from sinusoidal.
It consists of a sequence of narrowly peaked, well-spaced misfit dislocations, or discommensurations \cite{McMillan.PRB.1976} of constant $z$-magnitude, separating nearly AB and BA commensurate domains.
The twist regime I, roughly from 10$^\circ$ to about 3$^\circ$ in TBG, is intermediate, the corrugation growth less pronounced, and the shape deviating from sinusoidal. 
These different regimes and crossovers were encountered in many previous studies and simulations \cite{Kim.PRL.2012, Uchida.prb.2014, Woods.NatPhys.2014, Jain.2dmater.2016, Zhang.Mech.Phys.Solids.2018, Maity.prr.2020,Jonathan.pnas.2013,Yoo.NatMater.2019,Weston.NatNano.2020,Kazmierczak.NatMater.2021,Zhang.NatMater.2022}, where some aspects were described.
These twist dependent structural differences also imply different frictional properties, an aspect seldom addressed that will make an important point of our discussion. As prototypes of the two extreme limits L and S, we will use $\theta=30^\circ$ (ignoring its special “quasi-incommensurate” geometry \cite{Stampfli.HPA.1986, Koren.prb.quasi.2016}), and $\theta=2^\circ$ respectively.
Smaller twist angles with moir\'e size growing as $\theta^{-1}$ for $a=b$ (Eq.~\ref{Eq:1}) are computationally more expensive and therefore harder to explore.
This problem does not arise in heterostructures, where $a \neq b$ and by Eq.~\eqref{Eq:1} the moir\'e size $\lambda$ remains finite at $\theta=0$.
That is illustrated for hBN/graphene in Fig.~\ref{fig:4}(a).

\begin{figure*}[ht!]
\centering
\includegraphics[width=0.96\linewidth]{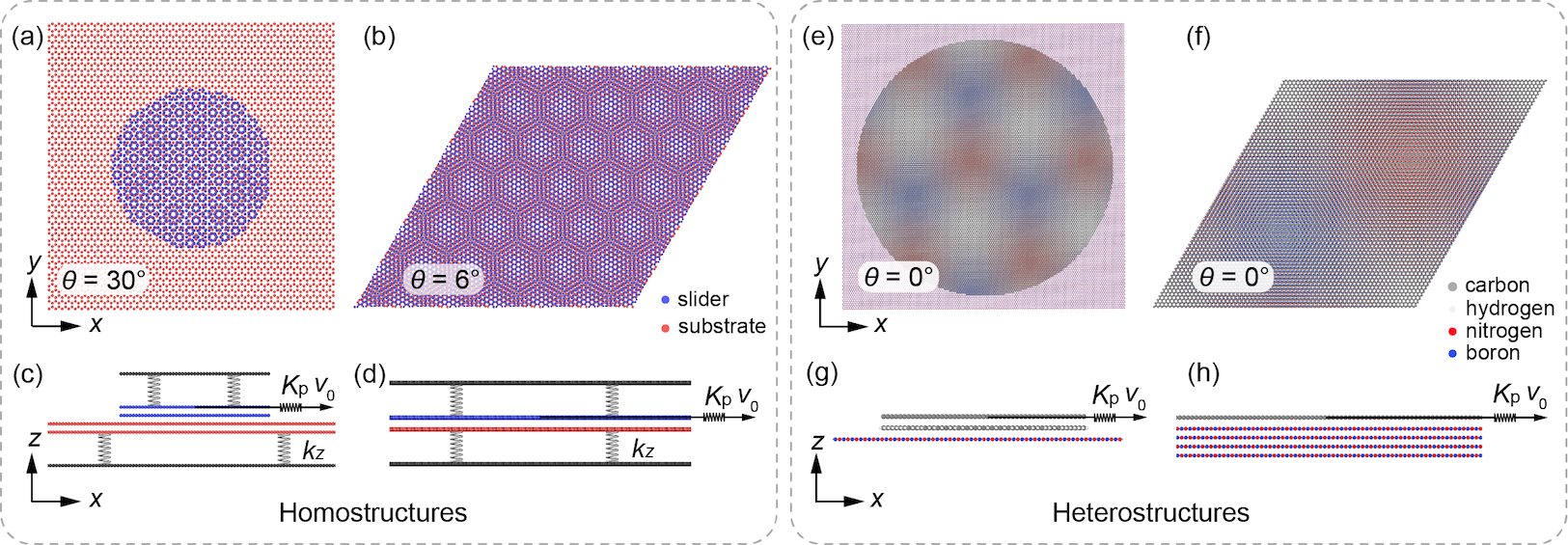}
\caption{Sketched ``top views" of prototypical nanofriction setups. (a) Circular island, OBC bilayer interface; (b) Boundary-free PBC bilayer interface.  Note the different moiré patterns of 30$^\circ$ and 6$^\circ$ twist angles respectively.
(c-d) Side view of two configurations.
Slider and substrate layers are blue and red; the encapsulating stage is black, its vertical grip mimicked by springs $k_z$.
The slider's  center-of-mass is pulled with velocity $v_0$, through a spring $K_\mathrm{p}$ representing the overall effective stiffness of the pulling system.
(e-h) Sliding configurations for graphene/hBN heterostructures in literature \cite{Song.NatureMaterials.2018,Wang.Nano.Letters.2019,Mandelli.PhysRevLett.2019}}
\label{fig:5}
\end{figure*}

\subsection{Understanding the relaxed structural regimes}

Good analytical understanding for the upper and lower envelopes of the pink zone in Fig.~\ref{fig:3}(c),
$d_\mathrm{upper}=d_0+4H/3$ and $d_\mathrm{lower}=d_0-2H/3$, is given by the (monolayer) out-of-plane corrugation magnitude
\begin{equation}
    H(\theta)=\frac{\alpha U_0 \lambda^4(\theta)}{\frac{32 \pi^4 a^2 D d_0}{3\sqrt{3}(1-\nu^2)}
    +\frac{k_z d_0}{2}
    +\frac{144\varepsilon \lambda^4(\theta)}{d_0}}
    \label{Eq:2}
\end{equation}
where $\lambda(\theta)$ is the moir\'e size at twist angle $\theta$,
Eq.~\eqref{Eq:1}, $U_0$ is the energy barrier against sliding at the interlayer equilibrium distance $d=d_0$, $\alpha$ describes the decay rate of $U_0(d)$, $\varepsilon$ is the adhesive energy per atom in the perfect AB stacked bilayer, $D$ and $\nu$ are the bending stiffness and Poisson's ratio of monolayer graphene.
Moreover, $k_z$ is an effective interatomic stiffness, qualitatively representing the rigidity of the semi-infinite bulk supporting the bilayer interface.
Lennard-Jones parameters $\varepsilon$ and $d_0$ generically describe the $z$-dependent interlayer adhesion energy, assumed to be weakly $(x,y)$ sinusoidally dependent \cite{Wang.tobe.2023}.
Its transparent contents is that the out-of-plane corrugation originates from the in-plane energy barrier $U_0$, hindered both by bending stiffness $D$ and by adhesion $\varepsilon$.
Because Eq.~\eqref{Eq:2} assumes a weak out-of-plane corrugation, it applies only to L and part of I regimes, where the local interlayer spacing is still above its minimum $d_\mathrm{AB}$ and below its maximum $d_\mathrm{AA}$ of bulk graphite.
When TBG is ``encapsulated" between supports (sketched in next Section) that regime is realized for $\theta \geq 5^\circ$.
For smaller twist angles regime S, $\min(z)$ approaches $z_\mathrm{AB}$ and $\max(z)$ approaches $z_\mathrm{AA} $ so that $H_\mathrm{max}=z_\mathrm{AA}-z_\mathrm{AB}$ as shown in Fig.~\ref{fig:3}(b).
The range of interlayer spacing is shown as a pink zone in Fig.~\ref{fig:3}(c). In the graphene bilayer, the maximum width reached by the pink zone, $\sim 0.2$ \AA, corresponds to the excess interlayer distance of the AA meta-stable stacking relative to that in the stable AB stacking. The globally flat bilayer is symmetrically moir\'e corrugated with magnitude 2$H$, where $H$ describes the single layer corrugation.
In heterobilayers the corrugation is asymmetrical, but the overall behaviour under twist is similar.
For example, in flexible-graphene/rigid-hBN of Fig.~\ref{fig:4}(a,b), the variable twist angle gives rise to a similar broadening of the interlayer distance (Fig.~\ref{fig:4}c).

\section{Twisted graphene interface simulations - a demonstration tool}

Here we describe the simulated systems which we use for demonstration and discussion.
They are two representative model setups, sketched in Fig.~\ref{fig:5}, of 2D
graphitic homo interfaces,
as well as graphene/hBN hetero interfaces.
For all simulations, the box size is relaxed at the beginning to set the in-plane stresses $p_{xx}$ and $p_{yy}$ to zero. All MD simulations used the LAMMPS code \cite{Plimpton.jcp.1995,Thompson.compphyscomm.2022}.
The interlayer and intralayer interaction of graphene layers are described by registry-dependent interlayer potential (ILP) \cite{Leven.jctc.2016, Ouyang.nanolett.2018} and the second-generation reactive empirical bond order potential (REBO) \cite{Stuart.jcp.2000,Brenner.jpcm.2002} respectively.
Optimized for high pressure conditions, \cite{Ouyang.jctc.2020} the former is meant to reproduce the sliding energy barrier and binding energy of 2D materials, the latter describes well the mechanical behavior of graphene \cite{Rowe.prb.2018}.
A Langevin thermostat at temperature $T$ is applied to the lowest red-colored layer in each of the configurations of Fig.~\ref{fig:5}.

{\itshape {\bfseries Open boundary conditions (OBC)-- a sliding island}}. 
The first setup (Fig.~\ref{fig:5}a,c) is, idealizing some experimental systems: a finite size graphene flake in the form of a circular island (blue), sliding over an infinite, PBC graphene substrate (red).
For simulations, the model is composed of 4 layers of 2D material.
The two lower layers represent the “substrate”, the two upper layers the “slider”.
This OBC setup allows an arbitrary twist angle and arbitrary incommensurability of in-plane lattice spacing for hetero-interfaces. The island edge of course breaks translational invariance.
An issue is the spontaneous torque, generally nonzero and pushing the
twist angle $\theta$ towards energetically favorable alignments, such as $\theta$ =0.
To prevent that in simulations, the overall twist $\theta$ is kept strictly constant, by setting the overall angular momentum to zero at all times.
In a homo interface, such as a TBG, the torque gets negative and huge at smaller twist angles where the island tends to rotate towards $\theta =0$.
While for a mesoscopic or macroscopic island this negative torque may not really succeed to produce a rotation, in microscopic islands it surely can, and that poses an obstacle to the study of twist angle dependence of all frictional properties. This spontaneous rotation occurs not just in homojunctions \cite{Filippov.prl.2008}, but in heterojunctions too \cite{Wang.prl.2016}.\\

{\itshape {\bfseries Periodic boundary conditions (PBC) - a perfect infinite interface}}.
The second model pictured in Fig.~\ref{fig:5}(b,d), is an infinite supported graphene/graphene interface with fully periodic boundary conditions (PBC).
A discrete series of twist angles ranging from 0.1$^\circ$ to 30$^\circ$, are constructed by means of artificially commensurate periodic supercells \cite{Trambly.nanolett.2010}. There are no edges or defects, making the model suitable for the study of delicate structural superlubricity questions at small twist angles.
The residual commensurability energy barrier is irrelevant at sufficiently large system size. A more serious drawback is in this case that the cell size depends by necessity upon the twist angle (Eq.~\ref{Eq:1}), which complicates comparisons between different twists.
At small twist angles, the homo moir\'e and thus the simulation size gets really large. Thus in PBC we limit the number of 2D layers, still supported by the $k_z$ vertical springs,
to two - no extra “encapsulating” layers.\\

\begin{figure*}[ht!]
\centering
\includegraphics[width=\linewidth]{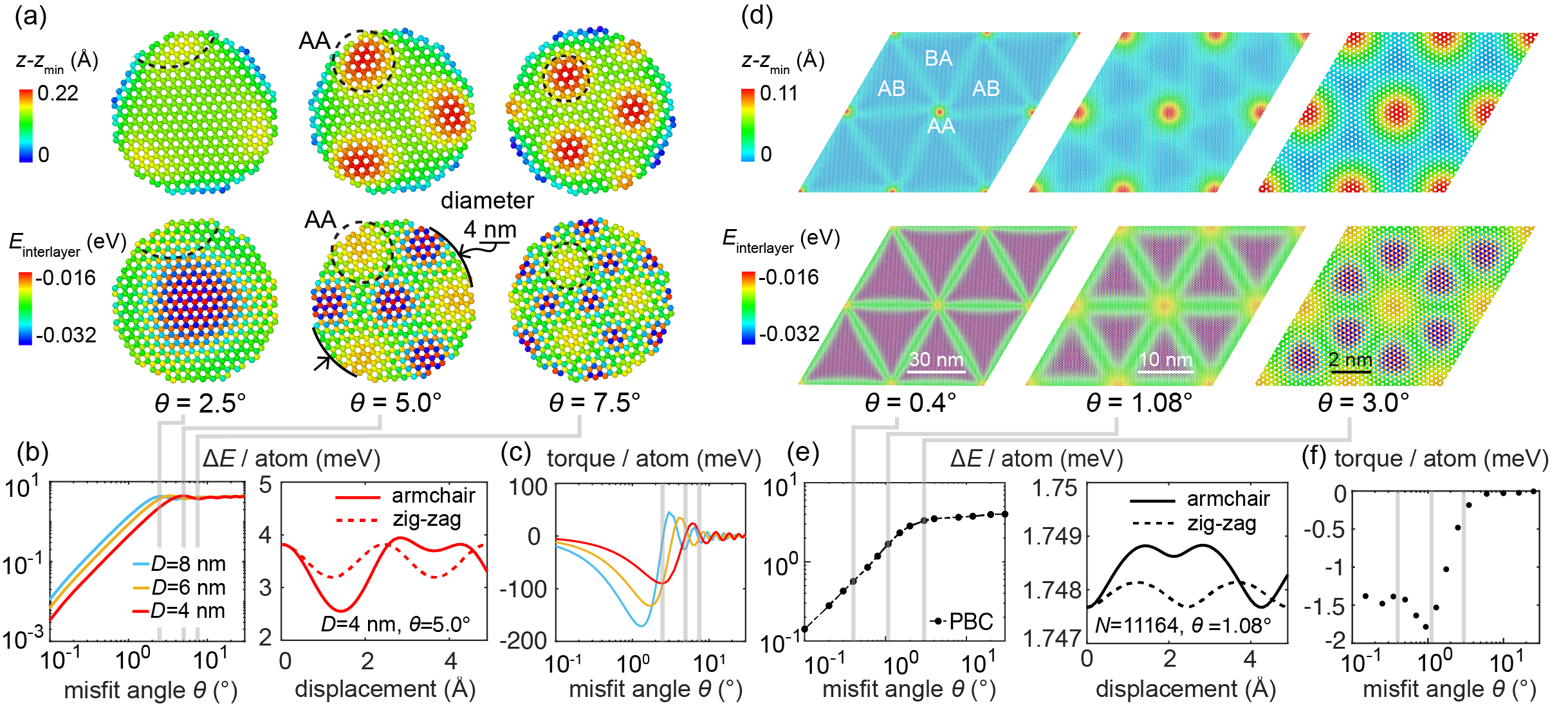}
\caption{Moir\'e corrugation patterns, energy  and rotational torque of the OBC (island) and PBC (infinite) graphene/graphene interfaces.
(a) $z$-coordinates (upper panel) and per-atom interlayer interaction (lower panel) for circular islands with diameter $D=4 nm$ and twist angle $\theta=2.5^\circ$, $5.0^\circ$ and $7.5^\circ$.
(b) Per-atom interlayer interaction energy difference relative to perfect AB stacking for islands with different diameters as a function of twist angle (left), and for one moving island as a function of displacement along substrate armchair (vertical) and zig-zag (horizontal) directions,
with periodicity $\sqrt{3}a=4.26~\mathrm{\AA}$ and $a=2.46~\mathrm{\AA}$ respectively).
(c) per-atom rotational torque of graphene islands with different radii. Three red vertical lines in (b) and (e) correspond to the twist angles shown in (a): the rotationally most unstable angle and two locally stable angles (zero torque).
(d) same plots as (a) for PBC infinite interfaces with twist angles $\theta=0.4^\circ$, $1.08^\circ$ and $3.0^\circ$.
(e) same as (b) for PBC. The three gray vertical lines in (f) and (h) correspond to the twist angles shown in (e).
By comparing  OBC island with PBC energy against slider position note the direction-dependent energy barriers. Despite apparent similarity, the large island barriers are real, whereas the small PBC barriers are an artifact due to the finite cell.
Also the OBC torque oscillations are real features of the island edge interfering with the moir\'{e}. The torque minimum near $1^{\circ}$ signals a change of misfit regime from independent to dependent on elasticity (see text).}
\label{fig:6}
\end{figure*}

{\itshape {\bfseries Twist dependent energy and torque}}.
Structural optimization of the above models provides qualitative and quantitative information reflecting the different regimes. They are shown in Fig.6 for a graphene-graphene interface.
The main one is the interface energy/atom relative to that of the commensurate Bernal AB stacking,
$\Delta E=[E_\mathrm{inter}(\theta )-E_\mathrm{inter}(0)]/N$, where $N$ is the total number of atoms in the upper (or lower) layer at the interface.
Another is the torque, the $z$-component of
$\boldsymbol{Q}=\sum_{i}^N \boldsymbol{r}_i\times\boldsymbol{F}_i$,
where $\boldsymbol{r}_i$ and $\boldsymbol{F}_i$ are the position (relative to the slider's center of mass) and force of $i$-th slider atom.
From energy, the torque magnitude is obtained
$|\boldsymbol{Q}|=  - \mathrm{d}\Delta E/\mathrm{d}\theta$,
extracted respectively from island and from PBC simulations (Fig.~\ref{fig:6}).
With the present force field the interface interaction energy (adhesive, thus negative) evolves from constant and -44.7 meV/atom in the large incommensurability regime L, to $\sim$ -48.7 meV/atom at $\theta =0$ in full commensurate AB stacking, thus with $\Delta E/E_\mathrm{inter}(0)\sim 9\%$.
The interaction energy, and also the average interlayer spacing (red dashed line of Fig.~\ref{fig:3}c) actually appears to indicate just two regimes, i.e. (L+I) and S. Evidence of a separate intermediate regime I between L and S is provided by the onset of a 
torque pushing monotonically towards $\theta=0$.
That coincides with the beginning of regime S, where upon decreasing twist there is a levelling off of the corrugation magnitude $H_\mathrm{max}= z_\mathrm{AA}-z_\mathrm{AB}$.\\

{\itshape {\bfseries Extracting static and kinetic friction}}. 
All layers (four in the island case of Fig.~\ref{fig:5}c, two in the infinite case of Fig.~\ref{fig:5}d) are mechanically flexible, with atoms in the upper layer of the slider and the lower layer of the substrate connected to rigid planes (black in Fig.~\ref{fig:5}c) through $z$ harmonic springs $k_z$=2.7 N/m,
the elastic modulus of graphite
\footnote{This is an extreme value, mimicking a very rigid encasing. An alternative value for a twisted graphite interface could be \cite{Wang.tobe.2023}  $\sim 0.1~\mathrm{N/m}$, i.e., much smaller. The actual magnitude of $k_z$ has little effect on the moir\'e corrugation $H$.}
\cite{Liu.prb.2012, Wang.acsami.2019}.
The center of mass of the slider’s upper layer is connected to a virtual stage through the pulling spring with stiffness $K_\mathrm{p}$, an effective lateral stiffness of the pulling system.
In NEMD simulations, the stage moves along $x$-direction with a constant velocity $v_0$.

The static friction $F_\mathrm{s}$, the smallest force required to initiate sliding, is calculated by quasi-static protocols \cite{Bonelli.epj.2009, Mandelli.scirep.2017}.
Simulations with $\theta$ smaller than $1^\circ$ contain one moir\'e, while systems with larger twist angle afford multiple moirés (Fig.~\ref{fig:5}b).
The kinetic friction $F_\mathrm{k}$ is the average force necessary to maintain steady-state sliding with average velocity $v$. It characterizes the average energy dissipation during sliding. It is calculated as $F_\mathrm{k}=\langle K_\mathrm{p}\left(v_0t-X_\mathrm{com}\right) \rangle$, where $X_\mathrm{com}$ is the center of mass position of the slider and $\langle \dots \rangle$ denotes long time average over a steady sliding state.
To address correlations between friction and contact area $A$, temperature $T$, sliding velocity $v$, and normal load $F_\mathrm{N}$, we control these variables and adjust the circular slider diameter $D$, the stage velocity $v_0$, and the thermostat $T$. Load is controlled by keeping the rigid bottom plane fixed and positioning the top rigid plane at variable $z$.
Compared with the method of directly applying a uniform normal force to each slider atom, this method 
is more realistic, allowing the slider atoms to adaptively experience a nonuniform normal force.

\section{Simulated 2D sliding friction}

The model setups just described provide typical energy and friction data that can be used for the discussion of sliding in 2D materials.

\subsection{Static friction and energy barriers}

The total potential energy, sum of intralayer and interlayer contributions, depends on the relative lateral coordinate of the two facing layers. The resulting periodic moir\'{e} corrugation is depicted in Fig.~\ref{fig:6}, panels (a) and (d)
for the island (OBC) and infinite bilayer (PBC) cases respectively.
The AA regions stand out for their higher interlayer distance and weaker interlayer attraction,
the AB and BA regions for the opposite. The AB/BA dislocation network joining AA nodes is visible in the small twist angle cases of Fig.~\ref{fig:6}(d).
The twist concentrates to a local rotation angle of AA nodes in regime S, visible in Fig.~\ref{fig:6}(d), as also noted earlier \cite{Zhang.Mech.Phys.Solids.2018, Angeli.prb.2018}.
The average interlayer energy of Fig.~\ref{fig:6}(b) evolves from constant in L-regime incommensurability $\theta > 10^\circ$, dropping towards the commensurate value with the asymptotic power law
$\Delta E \propto \theta$ in the S-regime $\theta \ll 3^\circ$. For PBC structural superlubric case (Fig.~\ref{fig:6}e), the prefactor is $1.43~\mathrm{meV}$.
At the same time a torque arises in the regime pushing towards $\theta =0$, see panels (c) and (f). The energy evolution upon static sliding is instructive.
To begin with, total energy in the perfectly incommensurate infinite bilayer should be strictly constant, independent of relative position of the two layers. For a finite size slider however the energy depends on position, with barriers that depend on size and sliding direction. That is exemplified on the right panel of Fig.~\ref{fig:6}(b) for a circular island, where the energy oscillations correspond to moir\'{e} cells entering/exiting the perimetral edge.
For curiosity, we also show the energy barrier of the PBC case of Fig.~\ref{fig:6}(e), completely artificial, and utterly negligible.

\begin{figure}[ht!]
\centering
\includegraphics[width=\linewidth]{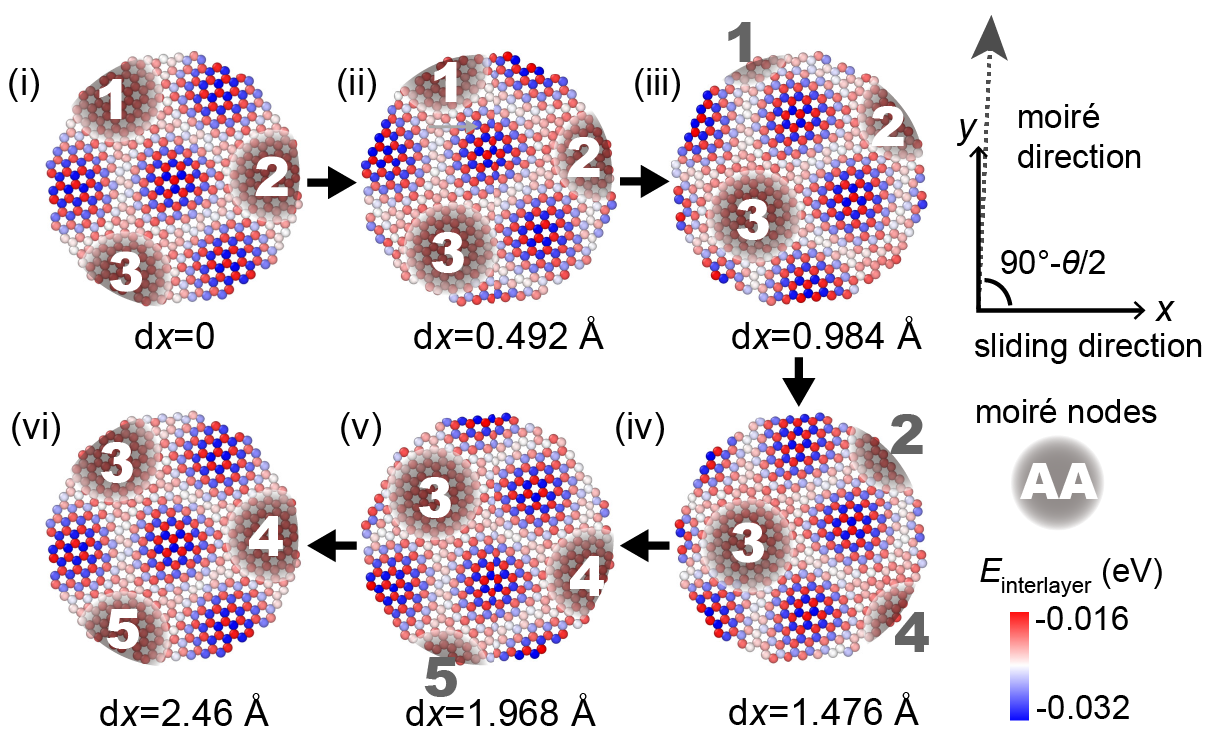}
\caption{
Evolution of interlayer energy map during sliding by $dx$ along $x$ of a circular island of diameter 4 nm and twist angle $\theta=5^\circ$.
Three AA nodes near the edge in panel (i) are highlighted and tracked as they move. When they cross the edges the interlayer energy develops a barrier.}
\label{fig:7}
\end{figure}

\begin{figure*}[ht!]
\centering
\includegraphics[width=\linewidth]{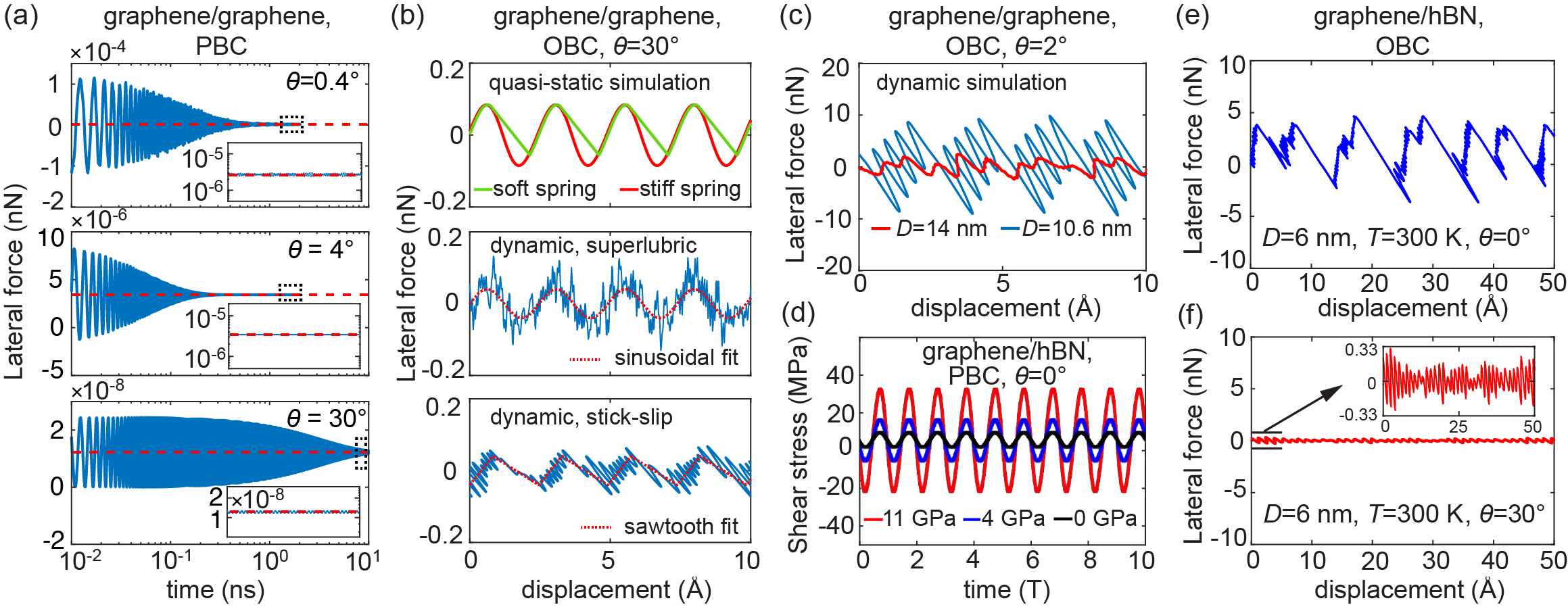}
\caption{
Exemplifying kinetic friction traces from our simulations (a-c) and published simulation studies (d-f). The underlying static friction $F_\mathrm{s}$
is zero in PBC signaling essentially perfect structural superlubricity, but nonzero in OBC, where pinning is due to the island edges.
Kinetic friction $F_\mathrm{k}$ is the average lateral force.
Friction force of PBC systems is normalized per slider atom, and can be converted to shear stress $\tau =F/A_\mathrm{C}$ by dividing the per-atom area of carbon atom, $A_\mathrm{C}=\sqrt{3}a_\mathrm{Gr}^2/4$.
(a) Frictional transient and convergence of PBC graphene/graphene contacts evolving in time ($v_0=10$ m/s, $T$=0 K).
The pulling stiffness $K_\mathrm{p}$ is adjusted case by case in order to keep the underdamped inertial oscillation $\omega = \sqrt{K_\mathrm{p}/M}$ with $M=Nm_\mathrm{C}$ the slider mass within reasonable range.
The mean friction (red dashed line, magnified in insets), here proportional to velocity, increases as the twist angle decreases, reflecting the severe mobility drop from L, to I, to S regimes (slider atom number $N$ is 11644, 6542 and 40838 for $\theta$ =  $30^\circ$, $4^\circ$ and $0.4^\circ$ respectively).
(b) Friction force of OBC graphene/graphene island of diameter $D=4 $ nm with twist angle $\theta=30^\circ$ as a function of sliding distance.
Upper panel: adiabatic protocol, with stiff and soft pulling springs $K_\mathrm{p}$=100 N/m and 1 N/m respectively.
Lower panels: dynamic sliding simulations ($v_0$=10 m/s, $T$=298 K) and the same spring stiffness, stiff and soft, respectively. Note the evolution from smooth sliding to stick-slip upon softening. (c) OBC friction force at small twist angle $2^\circ$.
Blue and red represent island diameters $D$=10.6 and $D$=14 nm respectively. Note the evolution from stick-slip at small size, where edge pinning dominates, to larger size, where the edge effect disappears.
(d) Time evolution of shear stress (per-area friction force) in aligned graphene/hBN heterojunctions under different normal pressures \cite{Mandelli.PhysRevLett.2019}. Time unit $\sim$ 25 ps.The incommensurability of the two lattices and the absence of edges in PBC predicts smooth sliding in this case too.
(e, f) Friction force of graphene/hBN island contacts (OBC) for $0^\circ$ and $30^\circ$ twist angles \cite{Wang.Nano.Letters.2019}. Here the evolution from stick-slip (e) to smooth sliding (f) is associated with a dramatic shrinking of the moir\'{e} cell size with increasing twist.
}
\label{fig:8}
\end{figure*}

To clarify the origin of the frictional barrier in OBC systems,
Fig.~\ref{fig:7} shows the evolution of the moir\'e pattern (via energy distribution) as an island slides.
Two aspects emerge. 
First, as the island slides along $(1,0)$ with velocity $v$, the moir\'e pattern moves along $(\sin{\frac{\theta}{2}},\cos{\frac{\theta}{2}})$ with velocity $v_\mathrm{M} (\theta)=\lambda (\theta) v/a$.
Second, the manner in which this moving moir\'{e} hits the island edge is responsible for the sliding energy barrier (middle panel of Fig.~\ref{fig:6}b).
Every time one or more of the AA nodes in the moir\'{e} cross inwards or outwards the edge (Fig.~\ref{fig:7}), total energy undergoes a local maximum, because local AA stacking costs about 15 meV/atom more than AB stacking \cite{Wen.prb.2018}.

The reluctance of the island to admit or expel AA moir\'e nodes
generates the barriers that cause the static friction against the island motion.
The evolution of this edge-induced barrier with island radius, its direct connection with uncompensated moir\'e cells, and the approximate description of its oscillatory and  progressively decreasing impact on friction, have been repeatedly emphasized in different contexts \cite{Varini.nanoscale.2015,Koren.prb.moire.2016,Cao.prx.2022}.
Theoretical modeling also led to express these friction oscillations with the radius of a circular island in terms of sinusoidal fitting for kinetic friction \cite{Wang.acsami.2019,Wang.Nano.Letters.2019} and of Bessel functions for static friction of rigid colloidal flakes \cite{Cao.prx.2022}.

\subsection{Kinetic friction: surfing moir\'e}

Typical room temperature simulated force traces -- whose steady state average yields the kinetic friction $F_\mathrm{k}$ -- are shown in Fig.~\ref{fig:8}(a-c) for PBC and OBC graphene contacts.
In addition to homojunctions, force traces for PBC and OBC graphene/hBN heterojunctions from existing literature \cite{Mandelli.PhysRevLett.2019,Wang.Nano.Letters.2019}
are reproduced in Fig.~\ref{fig:8}(d-f).

The PBC traces show a weak oscillation, and a sharply increasing mean value of friction as the twist angle decreases across from L regime at $\theta = 30^\circ$, to I and S, eventually close to $\theta = 0$.
The rise of friction reflects the decrease of interface mobility upon
rise of the moir\'e size $\lambda$ (Eq.~\ref{Eq:1}). That reduces the density of AB/BA dislocations
and changes their appearance, from diffuse to localized as shown in Fig.~\ref{fig:3}(a) and Fig.~\ref{fig:6}(d).
The twist angle decrease also entails the rise of moir\'e velocity,
$|\textbf{\textit{v}}_\mathrm{M}|=v_0 \lambda /a$,
along a direction increasingly close to orthogonal to the sliding force (Eq.~\ref{Eq:1.2}).

The dragging spring force constant is adjusted in simulations to scale as the number of slider atom $N$ as $K_\mathrm{p}=N k_i$, where $k_i=0.1$ N/m, which conveniently
keeps their frequency roughly constant.
The transient inertial oscillation is a mere damping and parameter-dependent artifact, of little further significance. The relevant information is the final parameter-independent steady-state friction, see Fig.~\ref{fig:8}(a).

\begin{figure}[ht!]
\centering
\includegraphics[width=\linewidth]{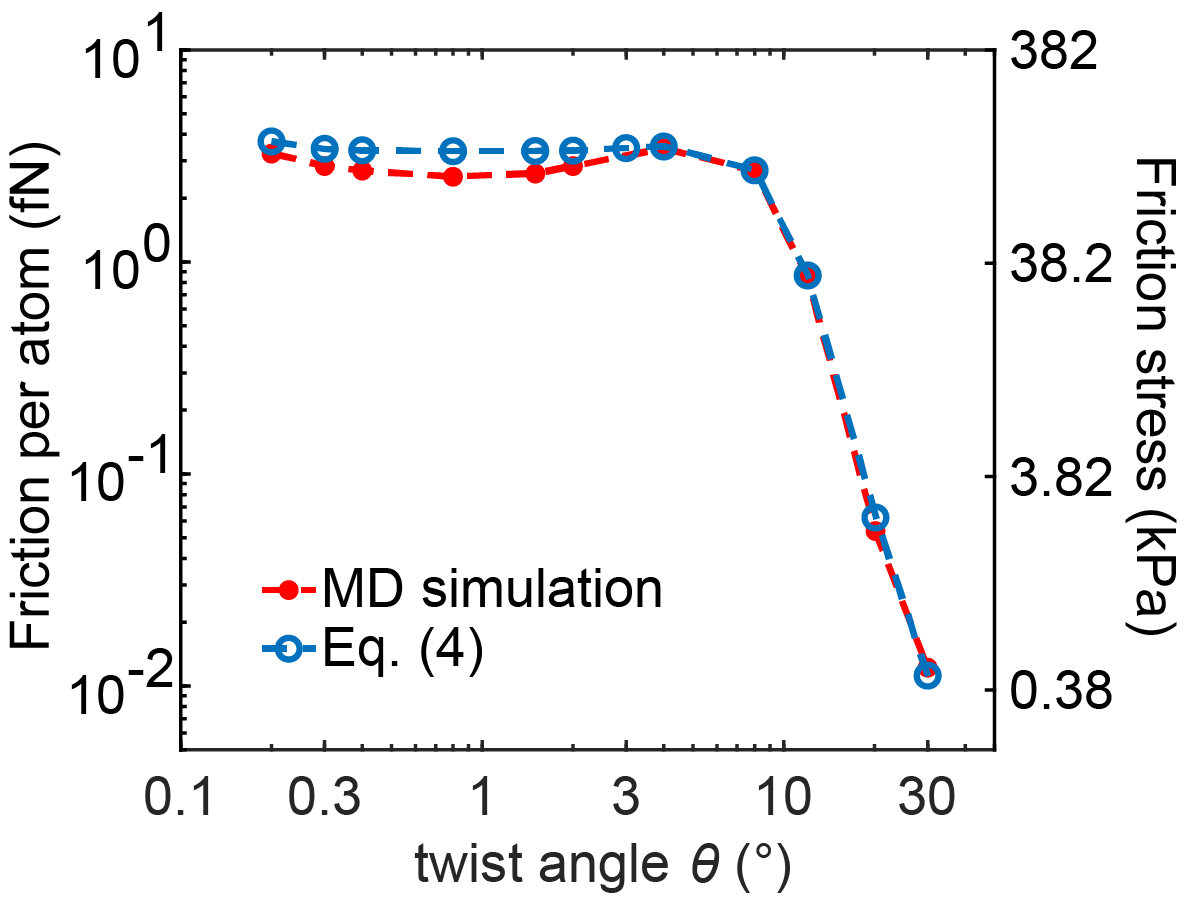}
\caption{Dependence of per-atom kinetic friction on the twist angle, where red dots and blue circles represent simulation results and theoretical results estimated from Eq.~\eqref{Eq:3}. Parameters used here are $\zeta=2~\mathrm{ps}^{-1}$, $v_0=10$~m/s.}
\label{fig:9}
\end{figure}

Based on the reasonable assumption that dissipation comes mostly from out-of-plane motion of atoms, the kinetic friction per atom of PBC
superlubric sliding at $T=0$ K is analytically derived as \cite{Wang.tobe.2023}:
\begin{equation}
    F_\mathrm{k} =c m \zeta v_0 \left[\frac{H(\theta)}{a} \right]^2
    \label{Eq:3}
\end{equation}
where $H$ is the out-of-plane corrugation, $m$ is the mass per atom, $v_0$ is the sliding velocity, $a$ is the lattice spacing of the substrate, and $c$ is a geometrical prefactor depends on the twist angle and sliding direction, in our case $c=\frac{32}{81}\pi^2$.
The phenomenological damping $\zeta$, a simulation parameter mimicking phonon dissipation in the supporting layers \cite{benassi.prb.2010}, is here empirically set in the $\mathrm{ps}^{-1}$ range.
Substituting $H$ of (Eq.~\ref{Eq:2}), good agreement between this theoretically obtained friction and simulation results is obtained, see
Fig.~\ref{fig:9}.
The linear size and velocity dependence of kinetic friction suggested by Eq.~\eqref{Eq:3} also agrees with simulations \cite{Wang.tobe.2023}. That indicates that the kinetic friction is utterly negligible once velocity is scaled down from the m/s scale (typical of MD simulations) to experimental scales of $\mu$m/s or nm/s.\\


Kinetic friction is also controlled, unlike static friction,  by the stiffness of the dragging spring, $K_\mathrm{p}$. For the OBC island with $\theta=30^\circ$,
the lateral forces for stiff and soft springs show sinusoidal and sawtooth patterns respectively (Fig.~\ref{fig:8}b), corresponding to structurally lubric and stick-slip regimes.
As in a Prandtl-Tomlinson model, the regime is determined by competition between the energy barrier $U_0$ (here introduced by edges) and the effective lateral stiffness $K_\mathrm{p}$ \cite{Vanossi.RevModPhys.2013}.
When the dimensionless parameter
\begin{equation}
    \eta= \frac{2\pi^2 U_0}{K_\mathrm{p} a^2}
    \label{Eq:4}
\end{equation}
is larger than 1, the edge barrier dominates and there is stick-slip. This is considered to be the case for most of existing experiments, as will be detailed in the subsequent Sections. In the opposite case the spring stiffness dominates and there is smooth sliding, with kinetic friction almost the same as that of the superlubric edge-free PBC geometry \cite{Wang.Nano.Letters.2019}.

\section{Area dependence}

\begin{figure*}[ht!]
\centering
\includegraphics[width=\linewidth]{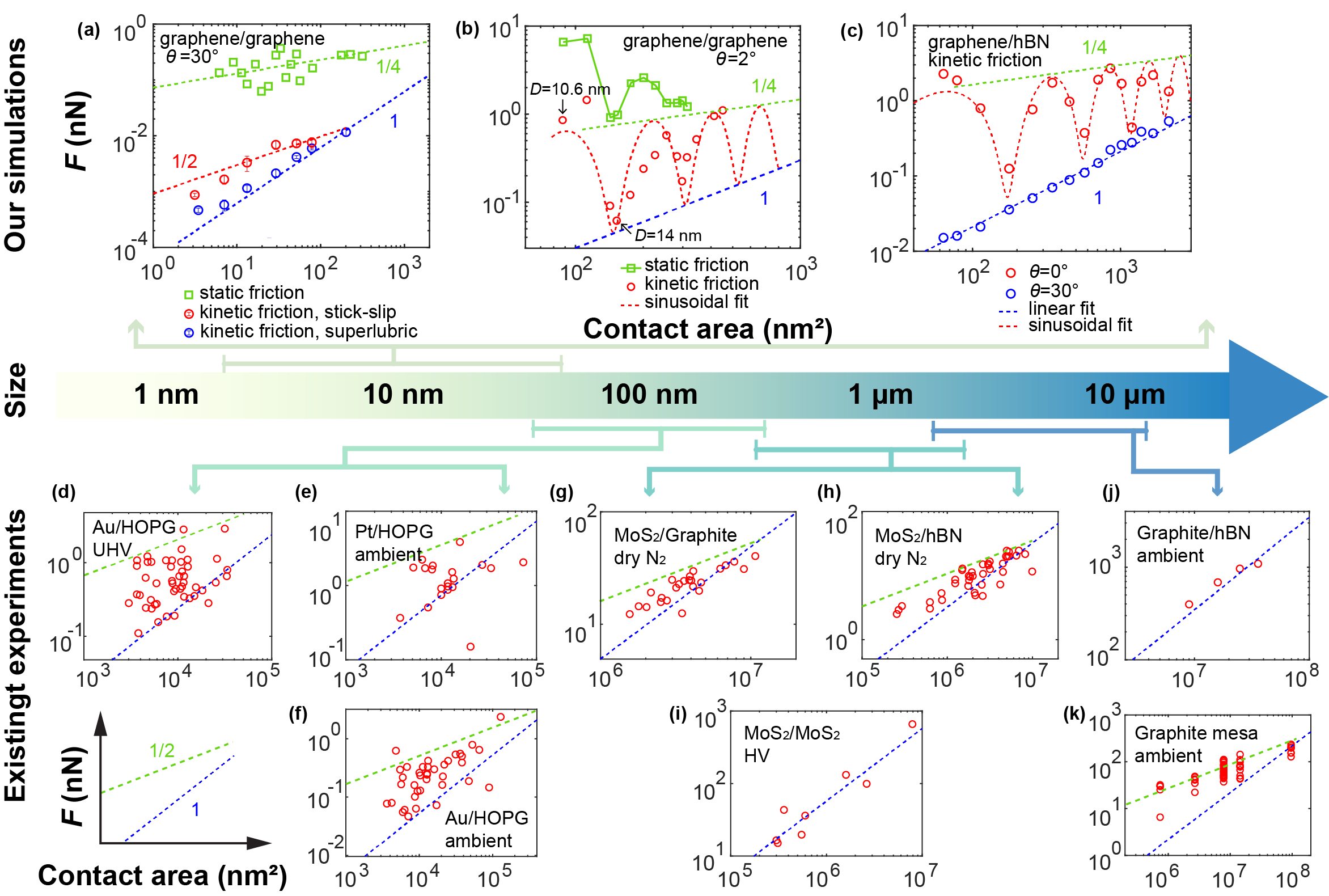}
\caption{Area dependence of nano and mesoscale kinetic and static friction  of 2D materials from simulations (upper panel) and experimental kinetic friction data (lower panel) across the scale.
(a) Graphene/graphene contacts with large twist angle $\theta=30^\circ$ ($v_0$=30 m/s, $T$=300 K). The green and red data points represent the OBC circular island static friction and kinetic friction respectively, the blue data points the superlubric regime kinetic friction (the static friction is zero). The green, blue and red dashed lines are power-law fits $F \sim A^{\alpha}$, with exponent 0.25, 0.5 and 1.
(b) Graphene/graphene contacts with small twist angle $2^\circ$ ($v_0$=10 m/s, $T$=300 K). The green and red data points represent the static and kinetic friction of OBC islands, and the red dotted curve is the sinusoidal fitting whose oscillations reflect successive moir\'e entering/exiting the island \cite{Wang.Nano.Letters.2019}.
Two typical cases are labeled, with diameters $10.6$ nm and $14$ nm, respectively at local friction maximum and minimum.
(c) Graphene/hBN contacts with twist angle $\theta=30^\circ$ and $\theta=0^\circ$ ($v_0$=10 m/s, $T$=300 K). The red ($\theta=0^\circ$) and blue ($\theta=30^\circ$) data points represent the kinetic friction for OBC graphene island (Fig.~\ref{fig:5}e).
Figure replotted with data from \cite{Wang.Nano.Letters.2019}.
By comparing (b) and (c) with (a), note that the lowest friction values, where edge does not contribute to friction, scale linearly with area, indicating structural superlubricity at all special island diameters of the sequence $D_n=\sqrt{3}\lambda (n/2+1/8)$, with integer $n\geq 2$.
The largest friction values, edge-related, scale instead sublinearly indicating proportionality to static friction (red dashed upper envelopes).
The scaling exponent 1/4 is special for the circular shape. Most other shapes, possessing straight edge portions, scale with exponent 1/2.
(d-k) Collection of 2D material kinetic friction data. Red circles represent data extracted and replotted from original papers \cite{Dietzel.prl.2013,Cihan.nc.2016, Ozogul.apl.2017,Li.Adv.Mat.2017,Wang.Nano.Letters.2019, Qu.prl.2020,Liao.natmater.2022}. 
Blue dashed lines underline linear scaling of the lowest friction points, analogous to simulations.}
\label{fig:10}
\end{figure*}

We are now set to discuss the friction dependence upon all the various parameters. Here we begin with the contact area $A$-dependence of static and kinetic friction in a structurally lubric 2D interface.
Kinetic friction is well defined for any system or condition; static friction is zero or undefined for a nanoscale system at finite temperature $T$.
Thermal fluctuations will always, after a sufficient time $t_D(A,T)$, overcome the pinning energy barriers and allow diffusion.
That is a condition, sometimes called thermolubricity \cite{Krylov.physstatussolid.2014, Pellegrini.prb.2019}, to be discussed later, that permits smooth sliding of a nanosystem no matter how weak the applied force.
Considering that $t_D$ grows with mass, and therefore with area, we shall in the present discussion of 2D interfaces, generally assume $t_D$ to be large enough so that static friction could be defined and measured if present. Thermolubric conditions are more difficult to reach for large sliders whose low-temperature friction is stick-slip.
Force traces for different boundary conditions and different systems including our prototype graphene/graphene simulated sliding, are presented in
Fig.~\ref{fig:10}(a-c)
for both the static and kinetic friction of $\theta =2^\circ $ and $\theta =30^\circ $ systems.

First, consider static friction $F_\mathrm{s}$, originating from energy barriers that impede free sliding.
In principle, in a simulated slider the energy barrier
is always nonzero, both in the island OBC and in the defect-free PBC case. 
Yet, in PBC the energy barrier is small and nonphysical, caused by the artificial high order commensurability introduced by the finite simulation cell. This barrier magnitude drops very fast with size, much faster than that of an island. As exemplified in 1D \cite{Theodorou.PRB.1978}, if the simulation cell size
is (in this case in the direction of sliding) $p$ times larger that of the moiré cell, of order $\lambda$, then the PBC barrier is qualitatively expected drop like $\exp(-p)$.
That is to say in our case like $\exp(-N^\beta)$, with $0<\beta <1/2$. This ready 
disappearance of size effects is of course the standard reason why PBC are generally chosen to approach infinite defect-free systems. In defect-free PBC simulations of 2D incommensurate materials the static friction is generally undetectable for most sizes -- a circumstance that allows structural superlubric conditions to be explored, at least theoretically.

Static friction of incommensurate interfaces is associated with defects that break the perfect translational invariance of potential energy (barring here the possibility of Aubry pinning \cite{Peyrard.Aubry.1983}). 
Let the defects occupy a fraction $f(A)=\delta A/A$ of the total area $A$. Unless their number is itself proportional to $A$, $f$ will tend to zero in the large size limit, generally as $f\sim A^{\alpha-1}$ with $0<\alpha <1$. The energy barrier and the corresponding static friction, entirely due to defects in a structurally lubric interface, should thus be at most proportional to $f(A)$. 
For an island of $N$ atoms where $A=\rho_\mathrm{2D} N$ and $\rho_\mathrm{2D}$ is the 2D atom density, the edge is the omnipresent minimal defect. The energy barrier and corresponding static friction can in that case vary with increasing size at most as $N_\mathrm{edge}/N\sim N^{\alpha-1}$ with $\alpha = 1/2$. However, not all edge regions are equally efficient in producing barriers.
The simulated OBC static friction of $\theta =30^\circ $ (green squares in Fig.~\ref{fig:10}a) has an area dependence $F_\mathrm{s}\propto A^{1/4}$, showing that the circular island edge, smooth as it is, is a relatively mildly pinning defect.
The simulated sublinear static scaling results are compatible with existing studies \cite{Koren.prb.moire.2016, Sharp.prb.2016, Dietzel.Nanotechnology.2018}.

The simulated area dependence of kinetic friction, shown in Fig.~\ref{fig:10}(a-c), is generally different from static friction.
First of all, in the large area limit $A\to \infty$, $F_\mathrm{k}$ is proportional to $A$ itself for the defect-free, superlubric PBC case.
That friction however is tiny and would hardly be measurable at the very low experimental velocities (Eq.~\ref{Eq:3}). The OBC case is different. The edges give rise to a large kinetic friction, sublinear with area.
Because of sublinearity, at sufficiently large area and/or large speed and large pulling stiffness the edge friction should become irrelevant, with a crossover to linear growth with area, as will be exemplified later, when discussing velocity dependence. That said, it is useful to note that edge and defect friction appears to be crucial and in fact dominant in most experimental data.\\

The linear growth with area of defect-free, large size, kinetic friction compared with the sublinear of static friction is understood by considering that the frictional power $P_\mathrm{ins}=\frac{1}{\tau}\int_{0}^{\tau}{Fv}\mathrm{d}t$ must be dissipated roughly as
\begin{equation}
P_\mathrm{diss}=\sum\limits_{\alpha}^{x,y,z}\zeta_{\alpha}\sum\limits_{i=1}^N m_i \langle v_{i,\alpha}^2 \rangle
    \label{Eq:5}
\end{equation}
where $\zeta_{\alpha}$ is the damping coefficient along $\alpha$ direction ($\alpha=x, y, z$), $m_i$ is the atomic mass and $v_{i,\alpha}$ is the velocity $\alpha$-component of interface atom $i$. The main source of atom velocity -- a vector quantity whose most important dissipative component in 2D materials is out-of-plane \cite{Song.NatureMaterials.2018,Mandelli.PhysRevLett.2019} -- is the sliding-induced in-plane motion of the moir\'{e} pattern. Moir\'e solitons act here as quasi-particles moving in a viscous medium, dissipating energy proportionally to their velocity, which is in turn proportional to the sliding velocity. The  moir\'{e} pattern (unlike the island edges) occupies a fixed fraction of the total interface area $A$. Therefore the (defect-free) superlubric friction caused by moir\'{e} motion grows proportional to area $A$.

The above conclusion, namely that the area dependence of structurally lubric 2D kinetic friction is different from that of static friction, is valid for sufficiently large areas. 
In island sliders of small diameter, or other realistically
defected systems where near-defect atoms are an important fraction, pinning is strong and both static and kinetic friction grow sublinearly with area with the same power law, being indeed related and proportional.
The island's edge-related energy barrier, demonstrated earlier,
causes the stick-slip sliding at low velocity and for sufficiently soft pulling spring stiffness $K_\mathrm{p}$.

In stick-slip sliding, the (single contact) force trace is sawtooth-like (Fig.~\ref{fig:8}b).
When the damping is sufficient, the average kinetic friction (the force-displacement area) is ruled by static friction (the sawtooth's maxima), and no longer by the moir\'e viscous translation, as in the superlubric case.
As a result, both static friction and the corresponding stick-slip kinetic friction grow sublinearly with area. In Fig.~\ref{fig:10}(a), that regime is seen in the initial part of the island kinetic friction, where the edge related stick-slip dominates.
Such low friction 2D  interfaces are quite commonly encountered \cite{Qu.prl.2020,Dietzel.prl.2013,Cihan.nc.2016,Ozogul.apl.2017,Kawai.Science.2016,Ouyang.nanolett.2018,Mandelli.prm.2018,Muser.PRL.2001,deWijn.prb.2012}. Even if stick-slip is invisible in force traces, generally owing to cancellation among multiple contact regions, it must be present when the velocity dependence is less than linear.
At the same time, the sub-linear dependence of friction force on the contact area provides the signature of underlying structural superlubricity of the defect-free portions. In that sense it can be used instead of the more general definition of superlubricity that friction coefficients is lower than 0.01
\cite{Baykara.APR.2018,Martin.PhysicsToday.2018}. \\


\textbf{\emph{Static friction.}}
Direct static friction experimental values are rarely published because of poor reproducible statistics. Operatively, one could extract them as the maxima of frictional traces, in cases where sliding is impeded by a single obstacle or barrier, and stick-slip sawtooths are evident. In most cases however, experimentally sliding interface sizes are large, with multi-barriers, or asperities acting at the same time.
Their interference, and the meso/macroscopic nature and stiffness of the sliding equipment, lead in most cases to the cancellation of stick-slip oscillations and sawtooths in the traces, which therefore appear smooth, even if noisy. (In principle, a Fourier transform of frictional noise could in the future be very instructive in that respect).
As will be clarified in the next section, the unmistakable earmark of stick-slip kinetics, implying in turn a nonzero static friction, is anyway signaled by velocity dependence, only weak or nil when friction is stick-slip.

\begin{figure*}[ht!]
\centering
\includegraphics[width=0.92\linewidth]{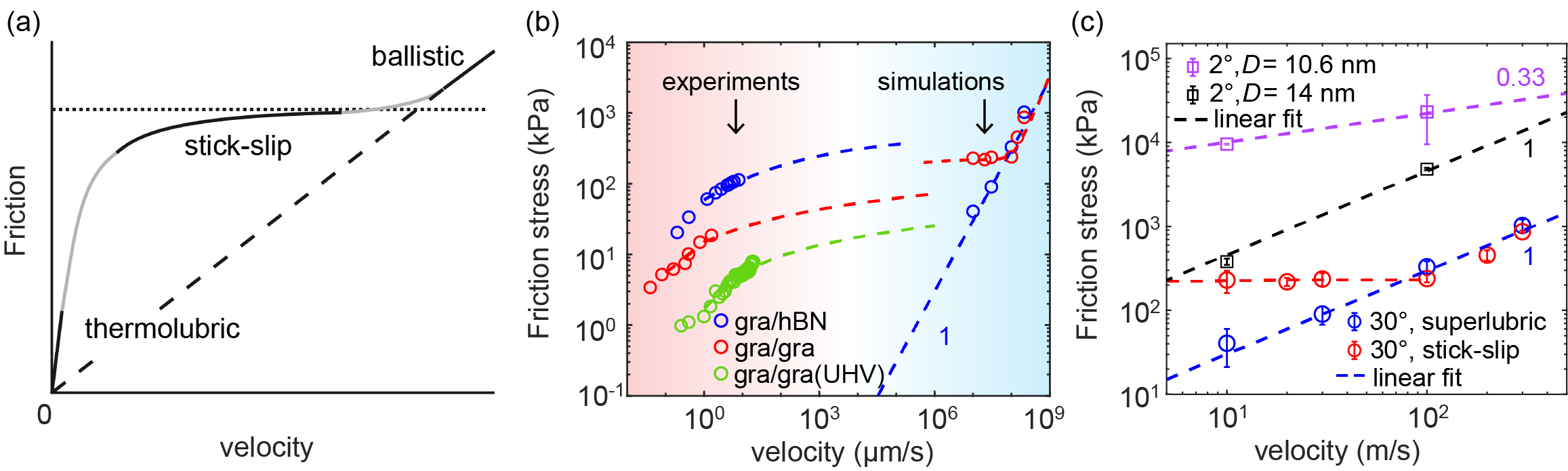}
\caption{
Velocity dependence of kinetic friction of structurally lubric contacts.
(a) Three friction regimes of a pinned nanoslider as a function of velocity: thermolubric, stick-slip, and ballistic; sketch reproduced from \cite{Krylov.physstatussolid.2014}.
(b) A direction comparison between the experimental results (left pink region) and $\theta=30^\circ$ simulation results (right blue region). For experiments, blue, red and green circles are kinetic friction data \cite{Song.NatureMaterials.2018, Wang.epl.2019,Liu.App.Mat.2020}. Colored dashed lines are logarithmic fits.
For MD simulations, the data and the color are consistent with (c).
The blue dashed line indicates linear scaling.
Note the huge velocity difference between simulations and experiments, whereby high velocity ballistic regime is rarely reached in experiments, and no low velocity thermolubric regime is reached in simulations.
(c) Simulated kinetic friction in superlubric (blue circles, $K_\mathrm{p}=100$ N/m) and stick-slip (red circles, $K_\mathrm{p}=1$ N/m) sliding for large twist angle $\theta=30^\circ$.
For the larger stiffness the island friction is viscous in the whole range.
Purple and black squares:  kinetic friction for small twist angle $\theta=2^\circ$ and two different island sizes. All lines are power-law fits.
Note that the velocity scaling is size-dependent, with the larger island close to linear, the smaller one sublinear.
}
\label{fig:11}
\end{figure*}


\textbf{\emph{Kinetic friction.}}
Fig.~\ref{fig:10}(d-f) shows friction results for the sliding of gold and platinum nanoparticles on HOPG. There is a scatter of data, many with a substantially sublinear area dependence. Current understanding is that those growing sublinearly with $\alpha \sim 1/2$ must do so owing to defects whose density does not grow with area, most likely edge regions \cite{Dietzel.acsnao.2017}.
Despite their apparently random spread, the log-log plot of data reveals that they all fall between two lines, one of slope $\alpha=1$ on the lowest friction side, the other with sublinear slopes $\alpha \leq 1/2$ as mentioned above. 
The data scatter between the two lines can be given a suggestive interpretation, considering the size and shape of different sliding islands. Theory shows, see Fig.~\ref{fig:10}(b-c) that circular islands of increasing radius should yield friction oscillating regularly between two straight lines, also with different slopes, 1 and 1/4.
Moreover, the shape dependent exponent 1/4 appears to be an extreme value for the circular shape, above which it is easy, but below which it is seems difficult to go. 
It seems reasonable that every new experimental data point corresponds to a random sequence of different size and shape of the sliding islands, therefore falling between the two lines.
If that was the case, one should actually expect aliasing, that is measuring a regularly oscillating variable at random points. Past works that have discussed this problem include \cite{Dietzel.prl.2013} and \cite{Wang.Nano.Letters.2019}.
More insights could be obtained if/when friction force traces became available for sufficiently small size interfaces.

Another point for further inquiry is the apparently erratic value of the critical size where area scaling of the kinetic friction crosses over from sublinear to linear. 
For graphene/hBN, experiments \cite{Liao.natmater.2022} show that crossover around  $10^4$ $\mathrm{nm}^2$.
This is consistent with the linear scaling of graphite/hBN at $10^7$ $\mathrm{nm}^2$ (Fig.~\ref{fig:10}j).
For $\mathrm{MoS}_2$/hBN or $\mathrm{MoS}_2$/graphite, the size scaling remains sublinear up to $10^6$ $\mathrm{nm}^2$ (Fig.~\ref{fig:10}g, h).
For graphite island/graphite contact, where edge friction dominates \cite{Qu.prl.2020}, the sublinear scaling even remains up to $10^8$ $\mathrm{nm}^2$ (Fig.~\ref{fig:10}k).
There is a single-asperity theoretical estimate of the connection between critical velocity and area for that crossover.
It is obtained by equating the slider's washboard frequency $v/a$ to the characteristic oscillation frequency of the system inside the energy well, $\frac{1}{2\pi}\sqrt{(K_\mathrm{p}+ 2\pi^2 U_0/a^2)/M}$.
The resulting connection between stick-slip -- ballistic crossover velocity and area is
$v_\mathrm{c}=\frac{1}{2\pi}\sqrt{[K_\mathrm{p}a^2 (1 +\eta)]/\rho_\mathrm{2D} A}$,
where the total mass $M = \rho_\mathrm{2D} A$ is proportional to area $A$.
Here $\eta$ is the Prandtl-Tomlinson mechanical instability parameter (Eq.~\ref{Eq:4}), whereby stick-slip will only occur for $\eta > 1$ \cite{Vanossi.RevModPhys.2013}.\\

\section{Velocity and temperature}

After area, we discuss the dependence of friction on other parameters, starting with a closely related pair of variables, velocity and temperature.
Before specializing to 2D material interfaces, it is useful to review the current generic understanding of nanoscale friction.

Stick-slip generally implies a weak logarithmic velocity dependence, roughly $F_\mathrm{k} \propto (\ln {v})^{\gamma}$ with $2/3 < \gamma \leq 1$ \cite{Muser.PhysRevB.2011,Gnecco.PRL.2000,Sang.prl.2001,Riedo.prl.2003,Reimann.prl.2004,Dong.TribLett.2011,Dudko.chemphyslett.2002,Krylov.pre.2005}.
The well-known qualitative reason for this weak dependence is that increasing velocity just increases the frequency of stick-slips but basically not the average frictional force.
On the other hand, perfect structurally superlubric sliding is expected on general grounds to yield a linear dependence $F_\mathrm{k} \propto v$.
As mentioned earlier, the qualitative picture in this case is that the misfit dislocations that form the moir\'e pattern move during sliding as fast dissipative ``Stokes quasi-particles", thus exciting phonons in the medium and giving rise to smooth, viscous friction, proportional to velocity.

Viscous friction is actually more general than that, and applies to pinned interfaces too, in two extreme limits.
Physical reasoning and model studies show that stick-slip sliding, universal for pinned interfaces driven by sufficiently weak springs, only survives in a limited -- even if very large -- velocity and temperature window.
As sketched in Fig.~\ref{fig:11}(a), drawn after Ref. \cite{Krylov.physstatussolid.2014}, stick-slip turns viscous when velocity is either high or low enough.
At high velocity, typically when the kinetic energy exceeds the largest energy barrier energy $\frac{1}{2}M v^2 > U_0$, barriers lose their grip, and friction turns ``ballistic" \cite{Guerra.NatMater.2010}, a regime where friction rises linearly with velocity.
Conversely, at the low velocity limit and nonzero temperature, a nanoslider has ample time to thermally diffuse back and forth across barriers. In that condition Einstein's viscous drift regime applies -- a regime also referred to as ``thermolubric" \cite{Krylov.physstatussolid.2014,Pellegrini.prb.2019}.
In the very vast velocity interval between these two limits (thermolubric and ballistic), friction is stick-slip like, with logarithmic velocity dependence pervasively seen in 2D materials sliding data, as in Fig.~\ref{fig:11}(b).\\

\begin{figure}[ht!]
\centering
\includegraphics[width=\linewidth]{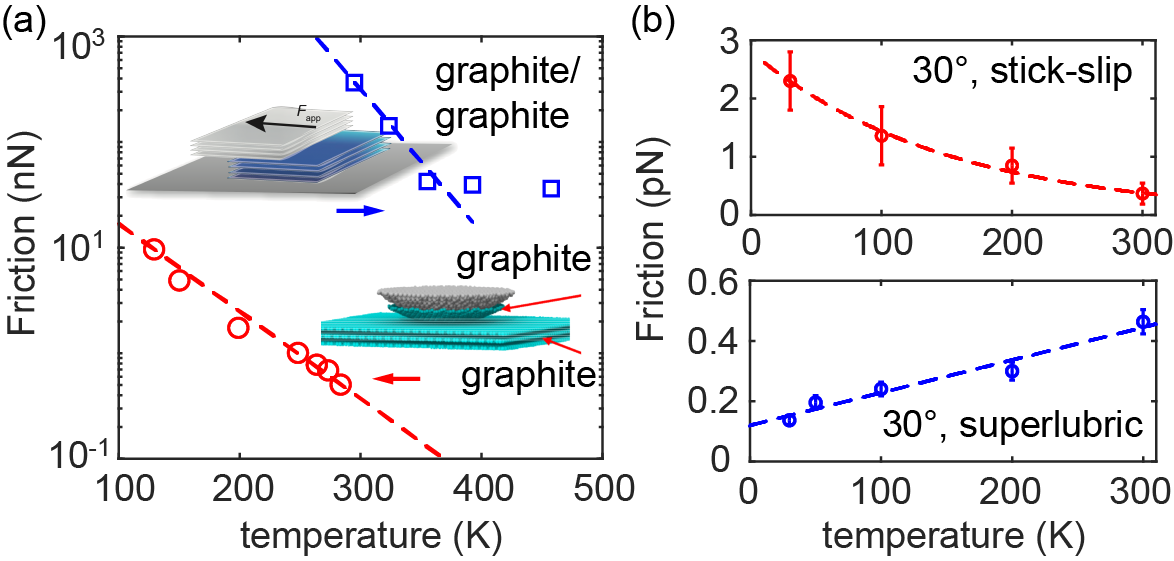}
\caption{
Temperature dependence of graphene/graphene kinetic friction.
(a) Experimental results. Red data for nano-sized ultra-high vacuum sliding \cite{Liu.App.Mat.2020};
blue data from micro-sized, ambient conditions sliding \cite{Wang.epl.2019}.
Dashed lines are exponential fits, reflecting activated barrier crossing. The high temperature saturation of micro-size data is an artifact due the 5nN sensitivity limit of the force sensor.
(b) Simulated kinetic friction of a graphene island (fig.~\ref{fig:5}a) with $D=2$~nm and $\theta =30^\circ$.
Red: stick-slip regime ($K_\mathrm{p}=1$ N/m, Fig.~\ref{fig:8}b);
blue: superlubric regime ($K_\mathrm{p}=100$ N/m, Fig.~\ref{fig:8}b), mimicking the structurally superlubric case.
Dashed lines are exponential and linear fits respectively.
}
\label{fig:12}
\end{figure}

Temperature dependence will also differ in these three regimes. Medium and low speed stick-slip friction should gently drop upon increasing temperature by terms roughly like $-\ln{T}$ \cite{Sang.prl.2001,Dudko.chemphyslett.2002,Krylov.pre.2005} as thermal fluctuations help overcoming barriers.
The negative temperature dependence shown in Fig.~\ref{fig:12}(a) is compatible with the the logarithmic velocity dependence of stick-slip \cite{Liu.App.Mat.2020,Wang.epl.2019}.
Conversely, ultra-low speed thermolubric and viscous friction drops much faster than logarithmic when temperature rises. That is because in this regime, the externally forced drift is a weak perturbation of thermal random walk, so that the frictional damping $\zeta$ obeys Einstein's relation $\zeta = k_\mathrm{B} T /D$. Here  the linear $T$ growth of $\zeta$ is overcome by the much stronger Arrhenius-like exponential growth of diffusion coefficient  $D \propto \exp{(-W/k_\mathrm{B} T)}$ with $W$ some typical barrier energy.
On the other hand, high speed ballistic friction, also viscous with velocity, is predicted to grow linearly with temperature. In that regime, as exemplified {\it e.g.,} in ref \cite{Guerra.NatMater.2010}, phonon scattering of a fast slider is enhanced by the growing dynamic corrugation of the interface at higher temperature. 
After this preamble, we can look once again to the frictional simulation data of a prototype 2D graphene interface.

In the OBC islands with large twist angle and soft driving spring, stick-slip is responsible for the weak velocity dependence (red curve in Fig.~\ref{fig:11}c) and negatively correlated temperature dependence of friction (red curve in Fig.~\ref{fig:12}b).
The same slider pulled with a harder spring switches over to linear velocity and temperature scaling frictional regime.
In the latter case, we still expect a crossover to stick-slip
at much lower velocities, but that only for sufficiently large slider sizes.
The situation changes for islands at small twist angle, where the moiré size becomes large, and the uncompensated moiré at the edges contributes additional sliding barrier.
Thus, the roughly constant stick-slip friction at large twists (red in Fig.~\ref{fig:11}c) is replaced by the size-dependent velocity scaling: linear for $D_n=\sqrt{3}\lambda (n/2+1/8)$ cases ($D_2 \approx 14$ nm) and sublinear for others (Fig.~\ref{fig:11}c).

Experimentally, the limited but representative 2D material velocity dependent results we found (green for nano-sized ultrahigh vacuum (UHV) environment,
red and blue for micro-sized ambient environment, Fig.~\ref{fig:11}b) display a clear logarithmic velocity dependence.
That confirms that stick-slip friction must be present at the nanoscale level \cite{Buzio.carbon.2021,Buzio.Langmuir.2022}, even when not visible in the overall friction traces.
These systems, \cite{Song.NatureMaterials.2018,Wang.epl.2019,Liu.App.Mat.2020,Li.nanoscale.2020} while not structurally {\it superlubric}, may still be structurally lubric, their friction controlled by pinning from defects or edges, not from the slider's interior.
For practitioners nevertheless, these systems are very often engineering superlubric because of the small friction coefficient, as will be detailed in next section.

An interesting digression before closing the discussion of area and velocity dependence of friction in structurally lubric interfaces, concerns the sliding of coaxial carbon nanotubes (CNT), and of graphene nanoribbons (GNR) on metal surfaces.
In multi-walled CNT of different chirality friction is weakly dependent on length \cite{Zhang.natnano.2013,Nigues.natmater.2014}.
That shows that the CNT friction originates from the edge -- just two ends, while the interior is structurally superlubric and frictionless. That very same observation applies to GNR sliding on metal surfaces \cite{Benassi.scirep.2015, Gigli.2dmater.2017,Gigli.nanoscale.2018}.

\section{Load}

\subsection{Ordinary and differential friction coefficients}
The load dependence of sliding friction is historically represented by the friction coefficient $\mu=F_f/F_\mathrm{N}$, where $F_f$ and $F_\mathrm{N}$ are the friction force and normal load respectively. 
The empirical proportionality between friction and load, dating back to Leonardo da Vinci and Amontons hundreds of years ago \cite{Dowson.1979, Amontons.1699}, makes the friction coefficient a widely used phenomenological parameter in macroscopic tribology.
In the sliding of rough interfaces, the increase of friction with load is ordinarily attributed to the increase of contact points (the real contact area) and to the enhanced role of interface corrugation \cite{Bowden.2001}. In nanoscale friction, the variation with load is instead system dependent and generally different from linear.\\

\begin{figure}[ht!]
\centering
\includegraphics[width=\linewidth]{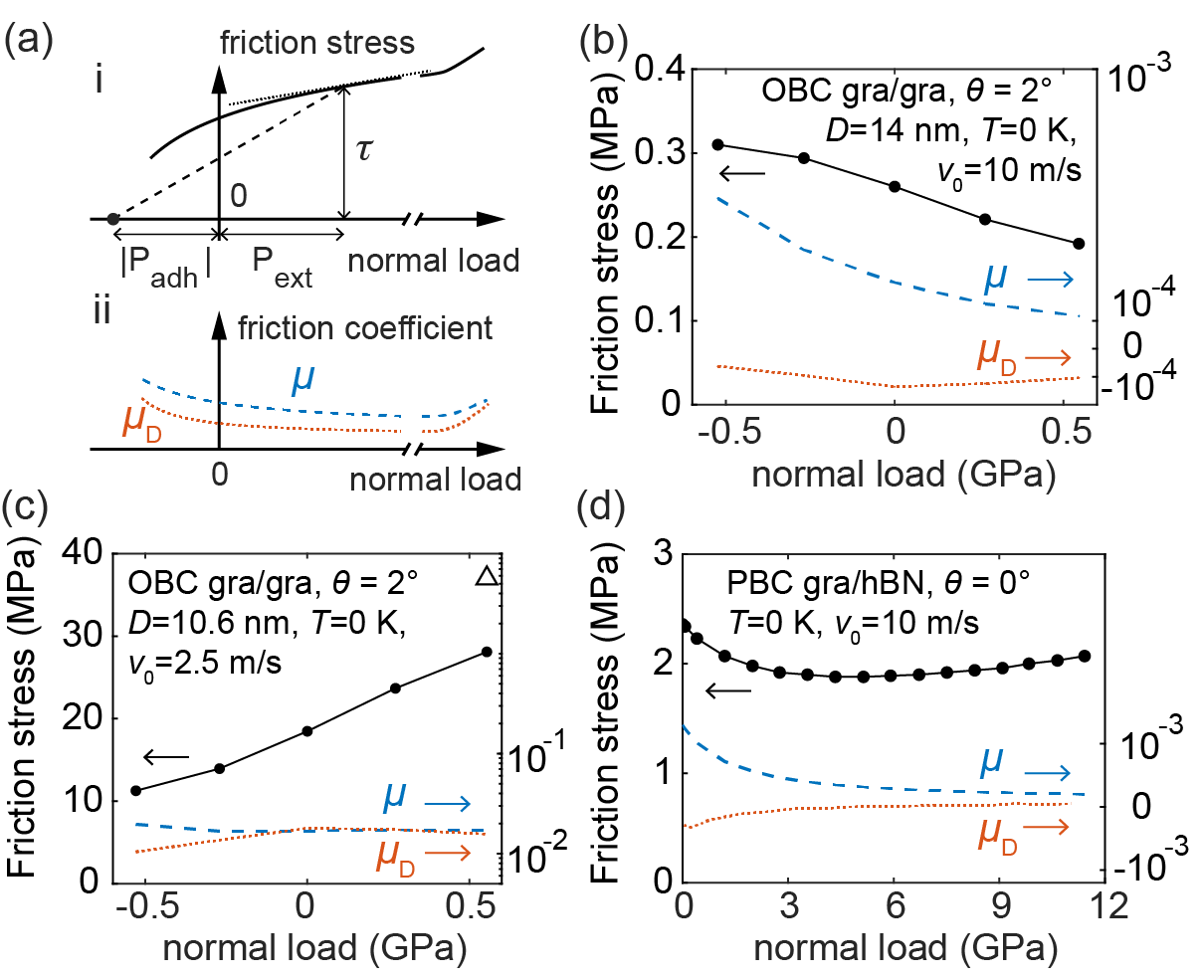}
\caption{
Load dependence of 2D materials sliding friction.
(a) Schematic diagram contrasting the friction coefficient $\mu$ and the differential friction coefficient $\mu_\mathrm{D}$. The slope of dashed and dotted line represents $\mu$ and $\mu_\mathrm{D}$ respectively.
(b-c) Simulated load dependence of kinetic friction in $2^\circ$-twisted islands with different sizes: (b) $D=14$ nm representative of the superlubric regime and (c) $D=10.6$ nm representative of  stick-slip regime, see Fig  10 b.
Inset triangle $\Delta$ in the upper right corner, shows a mere 30\% increase in stick-slip friction force, when $v_0$ is increased 5000\% under 0.5GPa load
(d) Simulated load dependence of large speed kinetic friction in infinite aligned graphene/hBN heterostructures (an ideal structurally superlubric case). Figure adapted from \cite{Mandelli.PhysRevLett.2019}. Blue dashed and red dotted lines represent $\mu$ and $\mu_\mathrm{D}$, respectively.}
\label{fig:13}
\end{figure}

In 2D materials interfaces, with wide atomically smooth terraces in flat contact \cite{Song.NatureMaterials.2018, Peng.nsr.2021}, the load dependence of friction also differs from linear.
To begin with, friction is not zero even at zero external load.
Physical adhesion effects, due to van der Waals attractions and/or electrostatic interactions, cancel out in rough interfaces,
but they do not in 2D material interfaces, owing to their large terraces and limited roughness \cite{Luan.Nature.2005,Erdemir.Superlubricity.2007}.
The result may be seen as an adhesion pressure $P_\mathrm{adh} \sim -A^{-1} \mathrm{d}E_\mathrm{adh}/\mathrm{d}z|_{z=z_0}$ where $z_0$ is the nominal interlayer detachment distance. That is the distance where $\mathrm{d}^2E_\mathrm{adh}/\mathrm{d}z^2 =0$
where the friction force would vanish. At zero external load the kinetic frictional stress of real 2D materials interfaces is generally nonzero as sketched in Fig.~\ref{fig:13}(a), even at extremely low velocity.
If the flat contact area is structurally lubric, then its interior contribution is negligible at such low velocities, and the zero-load friction is entirely due to stick-slip of edges and defects.
The sketch on Fig.~\ref{fig:13}(a) shows how the large adhesion in a 2D material interface influences the friction-load curve.
The early onset of edge and defect friction already at large negative load is the reason why friction remains nearly constant upon weakly increasing positive load.
Conversely, above certain very large loads,
both static and kinetic friction may again increase \cite{van.prb.2013}, reflecting the increased grip associated with distortions at the edge/defects.

The nonzero friction at zero load obviously renders the naively defined friction coefficient $\tilde{\mu} = \tau/P_\mathrm{ext} $, where $\tau = F/A$ is the frictional shear stress, and $P_\mathrm{ext} = F_\mathrm{N}/A$ the external normal pressure, divergent at zero load, and therefore useless. To avoid that problem, practitioners 
often adopt the differential form $\mu_\mathrm{D}=\mathrm{d}\tau/\mathrm{d}P_\mathrm{ext}$ \cite{Gao.jpcb.2004, Hod.nature.2018}.
Yet, the differential friction coefficient $\mu_\mathrm{D}$ 
does not measure the energy dissipation, which is just what the friction coefficient was meant to do. A better definition of friction coefficient $\mu$ is obtained considering adhesion \cite{Derjaguin.ZPhys.1934}, in the form
\begin{equation}
    \mu=\frac{\tau}{P_\mathrm{ext}+|P_\mathrm{adh}|}
    \label{Eq:6}
\end{equation}
where $P_\mathrm{adh}$ is the adhesion pressure, defined above, a constant for each given 2D interface. 
The differential friction coefficient can be expressed as
\begin{equation}
    \mu_\mathrm{D}=\frac{\partial \tau}{\partial P_\mathrm{ext}}=\mu+\frac{\partial \mu}{\partial P_\mathrm{ext}} (P_\mathrm{ext}+|P_\mathrm{adh}|)
    \label{Eq:7}
\end{equation}
We illustrate the behaviour of $\mu$ and $\mu_D$ versus load in  the schematic of Fig.~\ref{fig:13}(a).
For $P_\mathrm{ext}\to -|P_\mathrm{adh}|$, the detachment negative pressure, the effective total load vanishes, and $\mu_\mathrm{D}$ coincides with $\mu$. In this very hypothetical limit however neither is well defined, for the numerator would vanish too. Moreover even before reaching the nominal detachment distance, non-uniform detachment would nucleate, and flatness of the interface would be lost.
For all other loads $P_\mathrm{ext}>-|P_\mathrm{adh}|$, $\mu$, geometrically corresponding to the slope of the dashed line in  Fig.~\ref{fig:13}(a), makes very good sense, and there are all reasons to use it.

First, $\mu$ is generally larger than $\mu_\mathrm{D}$.
However, since $|P_\mathrm{adh}| \sim$ 1 GPa  is generally much larger than the experimental $P_\mathrm{ext}$, $\mu_\mathrm{D}$ and $\mu$ should have similar orders of magnitude.  The fact that $\mu_\mathrm{D}$ is straightforward to both measure and simulate 
``justifies" historically, if not physically, its adoption to qualify the load behaviour of structurally lubric systems \cite{Hod.nature.2018}.
Currently in fact, all 2D material interfaces with a very small $\mu_\mathrm{D}$ are generically dubbed superlubric.
Experiments and simulations in 2D materials-based homo- and hetero-interfaces (Table I \& II), report extremely low $\mu_\mathrm{D}$ values \cite{Liu.acsnano.2018, Li.Adv.Science.2018,Liu.App.Mat.2020,Liu.Nat.Comm.2017,Deng.nanoscale.2018}, down to $10^{-5}$ for structurally lubric twisted graphite/graphite contacts and graphite/MoS$_2$ contacts \cite{Liu.App.Mat.2020} -- but generally do not provide $\mu$.

Second, $\mu_\mathrm{D}$ does not measure frictional dissipation -- it is not even strictly positive -- whereas $\mu$, strictly positive, does. A case in point where that issue strikes comes just when $\mu_\mathrm{D} < 0$.
A negative $\mu_\mathrm{D}$ (see Fig.~\ref{fig:13}d), generally  due to load-induced ``ironing" of corrugations has been found in simulations, either of palladium/graphite systems \cite{Sun.jpcl.2018}, or of systems with grain boundaries \cite{Gao.Nat.Comm.2021} (with buckling corrugations), and twisted graphene/graphene contacts (with moir\'e corrugations). \cite{Wang.frontchem.2021, Mandelli.PhysRevLett.2019}. 
Early experiments in graphite contacts \cite{Vu.prb.2016, Deng.nanoscale.2018} also probably involved negative $\mu_\mathrm{D}$, but did not make it emerge clearly
because the fitting of $\mu_\mathrm{D}$ had a large error.
Despite its popularity, also mentioned in connection with the bearing capacity of structural superlubricity under load, \footnote{A series of studies are pushing the upper limit of external load $P_\mathrm{ext}$ to the GPa range \cite{Liu.Nat.Comm.2017, Li.Adv.Science.2018}. There is, of course, a practical upper limit to the bearing capacity. In graphite for example, at high normal loads (pressure in the order of 10 GPa) \cite{Wendy.science.2003, Martins.nc.2017}, bonds appear to begin forming between layers, causing the demise of structural lubricity.}
negative $\mu_\mathrm{D}$ has not yet been systematically confirmed in structurally lubric experiments.

Our own take on this theme is that the pursuit of a negative $\mu_\mathrm{D}$ is not in itself of real significance.
It will at best signal efficient load-induced ``ironing" of edge or defect corrugations and barriers, and suggest nonmonotonicity of frictional stress and of friction coefficient versus load, but not more than that.
It may even be dangerous, since the real friction coefficient can nonetheless be large -- and unknown. Moreover, as we shall show later, load can in fact play the opposite effect, increasing the edge barrier rather than ironing it away, in which case $\mu_\mathrm{D}$ remains simply positive.

How large or small is the actual friction coefficient $\mu$ for 2D structurally lubric interfaces? At present, the straight answer is that we do not know.
To obtain a simple order of magnitude, we insert in Eq.~\eqref{Eq:6} an estimate of $P_\mathrm{adh}$ as suggested from experiment \cite{Li.NatNanotech.2019} and theory \cite{Ouyang.nanolett.2018}, to be around 1 GPa in the case of graphene. This crude evaluation of $\mu$ for various measured interfaces is given in table I.  The magnitude of $\mu$ turns out to be generally of order $10^{-4}$, sometimes higher, and occasionally as low as $10^{-6}$ for microscale graphite/graphite \cite{Peng.pnas.2020} and $\mathrm{MoS}_2$/graphite contacts \cite{Liao.natmater.2022}.
For comparison, the ice surface, the slipperiest surface most familiar in everyday life, has a friction coefficient in the order of $10^{-2}$ \cite{Kietzig.jap.2010}, much larger.
Summing up, in order to provide evidence of the low dissipative frictional character,
it seems mandatory to go back using $\mu$ rather than $\mu_\mathrm{D}$ for the load dependence of friction in 2D material structural lubricity.

\begin{table*}[t]
\caption{\label{tab:table1} Kinetic friction coefficient for structurally lubric homostructures from existing studies.}
\begin{ruledtabular}
\begin{tabular}{cccccc}
Homostructures&Environment & $\mu^*$ & $\mu_\mathrm{D}$ &$P_\mathrm{ext}^{**}$& Ref.\\
\hline
Graphene/graphene & MD, OBC & $4.8\times 10^{-5}$ & $-3.5\times10^{-4} \sim 5.6\times10^{-5}$ & 4 GPa & \cite{Wang.frontchem.2021} \\
Graphite/graphite & ambient & $4.0\times10^{-6}$ & $-4 \times 10^{-4}\sim 10^{-4}$ & 1.67 MPa & \cite{Vu.prb.2016} \\
Graphene/graphite & ambient & $4.7\times10^{-3}$ & $3 \times 10^{-3}$ & $\sim$1 GPa & \cite{Liu.Nat.Comm.2017} \\
MoS$_2$/MoS$_2$ & HV & $1.3\times10^{-4}$ & \slash  & \slash &  \cite{Li.Adv.Mat.2017} \\
Graphite/graphite & ambient & $1.4\times 10^{-5}$ & $-3.3\times10^{-4}\sim 1.6\times10^{-4}$ & 7.78 MPa & \cite{Deng.nanoscale.2018}\\
Graphite/graphite & UHV & $1.3\times 10^{-4}$ & $4\times10^{-5}$ & $\sim 40$ MPa & \cite{Liu.App.Mat.2020} \\
MoS$_2$/MoS$_2$ & UHV  & \slash & $2.5\times10^{-4}$ & $\sim 25$ MPa & \cite{Liu.App.Mat.2020} \\
Graphite/graphite & ambient & $<3.6\times10^{-6}$ & \slash & 28 MPa & \cite{Peng.pnas.2020} \\
\end{tabular}
\end{ruledtabular}
\raggedright
\end{table*}

\begin{table*}[t]
\caption{\label{tab:table2} Kinetic friction coefficient for structurally lubric heterostructures from existing studies. }

\begin{ruledtabular}
\begin{tabular}{cccccc}
Heterostructures & Environment & $\mu^*$ & $\mu_\mathrm{D}$ &$P_\mathrm{ext}^{**}$ & Ref.\\
\hline
Graphene/hBN & MD, PBC & $1.6\times 10^{-4}$ & $-2.5\times10^{-4} \sim 5 \times 10^{-5}$ & 12 GPa & \cite{Mandelli.PhysRevLett.2019}\\
Graphene/hBN & ambient & $2.7\times10^{-3}$ & $2.5\times10^{-3}$ & $\sim$1 GPa & \cite{Liu.Nat.Comm.2017} \\
Graphite/hBN & ambient & $2.2\times 10^{-5}$ & $\leq1.4\times10^{-4}$ & 11.1 MPa & \cite{Song.NatureMaterials.2018}\\
Graphite/hBN & UHV & $1.2\times 10^{-4}$ & $4\times10^{-5}$  & $\sim 40$ MPa & \cite{Liu.App.Mat.2020}\\
Graphite/MoS$_2$ & UHV & $1.5\times 10^{-4}$ & $1.3\times10^{-4}$ & $\sim 40$ MPa & \cite{Liu.App.Mat.2020}\\
Au/graphite & ambient & $1.2\times 10^{-3}$ & $1\times10^{-3}$ & 34.9 MPa & \cite{Li.carbon.2020}\\
MoS$_2$/graphite & dry N$_2$ & $2.6\times10^{-6}$ & $\ll10^{-3}$ & $\sim 1$ MPa & \cite{Liao.natmater.2022}\\
MoS$_2$/hBN & dry N$_2$ & $2.3\times10^{-6}$ & $\ll10^{-3}$ & $\sim 1$ MPa & \cite{Liao.natmater.2022} \\
\end{tabular}
\end{ruledtabular}
\raggedright
*  Friction coefficient $\mu$ at zero load estimated from Eq.~\eqref{Eq:6}, with $P_\mathrm{ext}$ =0 and $|P_\mathrm{adh}| \approx 1$ GPa. This quantity, and not $\mu_\mathrm{D}$ should be used to calculate dissipation. \\
** $P_\mathrm{ext}$ is the external normal load per unit area.
\end{table*}

\subsection{Friction coefficient of 2D material interfaces}

It is instructive at this point to illustrate the load dependence of 2D material friction by showing actual values from existing studies, in comparison with token simulation results.
Tables I and II summarize some existing results, mostly experimental.
Table III shows graphene/graphene simulated friction coefficients, for large and small twist angles typified by 30 and 2 degrees. We restrict to OBC island friction, while the PBC results are omitted as unrealistic because simulation velocities are many orders of magnitude too large when compared with standard nanofriction experiments, where $v \sim 10^{-6} $ m/s or lower.
The friction coefficient is estimated through Eq.~\eqref{Eq:6} in all cases where it is missing in the original data.

In the experimental data the nature of kinetic friction can generally be argued to be stick-slip on account of the weak logarithmic velocity dependence, while the interface twist angle is generally unknown. The simulation results offer the advantage of exploring a broader regime of parameters, and the twist angle dependence.

On the whole, experimental and simulated friction coefficients appear quite comparable, even though as explained below that need not be significant. Experimental values confirm the well-known engineering superlubric, and structurally lubric, quality of 2D interfaces. Usefully, simulated results show that the different behaviour of friction and differential friction coefficients, $\mu$ and $\mu_\mathrm{D}$.
The first positive and quite stable, the second of variable sign, more volatile and less physically meaningful, although often of similar absolute magnitude, as expected because of the large value of $|P_\mathrm{adh}|$.
Where they differ, $\mu$ is often not as low as $\mu_\mathrm{D}$ might have suggested. We stress again that, while  $\mu_\mathrm{D} <0 $ merely indicates an  effective pressure ``ironing out" of edge and/or defect barriers, only $\mu$ can gauge how important that effect is in terms of frictional energy dissipation.

{\it Large Twist Angles}.
Looking at $\theta = 30^\circ$ in Table III, several features stand out.
The first is that the static friction coefficient is much larger than kinetic friction.
The second is that the absolute value of the (kinetic) friction coefficient $\mu$ is extremely low, of order $10^{-5}$. Both features are compatible with a structurally lubric friction.
On the other hand, friction is not completely structurally superlubric owing to the nonzero static friction coefficient, of order $10^{-3}$ for that case.
The third point is that the differential friction coefficient is negative. The negative sign indicates that the  barrier-generating defects -- here the island edges -- are being effectively ``ironed out" by load. The ironing effect, here very effective for kinetic friction, is seen in this case to affect much less the static friction.

While the above observations are quite instructive about friction of strongly incommensurate interfaces epitomized by graphene/graphene at large twist angle, the quantitative comparison with experimental friction coefficients of Table I and II demands very critical attention.
On the whole in fact, large twist angle theoretical friction coefficients show similar orders of magnitude as experiments, but that in itself is not significant. The reason is that as discussed earlier, smooth sliding implies proportionality of shear stress $F/A$ to $v_0$, while edge-dominated friction stress demands proportionality to $A^{\alpha - 1}$ with $\alpha < 1/2$.
Extrapolation from the simulation velocity of 10 m/s to realistically small velocities and  from the small island area of just 8 nm diameter to realistically larger areas leads to theoretical friction coefficients many orders of magnitude below the experimental ones.
That disagreement indicates in our view that in most experiments there must be large defect-related barriers, possibly acting at different spatial points, that give rise to local stick-slip, thus canceling smooth sliding. A careful study of experimental velocity, twist angle, and area dependence is needed to clear that crucial point.

{\it Small Twist Angles}.
The friction coefficient is quite complex in the small twist S regime where, as shown by Fig.~\ref{fig:10}(b), the frictional behaviour of small islands oscillates between high friction stick-slip maxima and low superlubric-like minima.
Table III exemplifies some of that complexity by comparing the kinetic friction coefficients at velocities between 2.5 and 10 m/s  of $\theta=2^\circ$ islands with two representative small diameters, $D$=10.6 nm (maximum friction, stick-slip) and 14 nm (minimum friction, smooth sliding), chosen as the first members of their families (see Fig.~\ref{fig:10}b) (simulated at $T=0$ for additional clarity).
The radius $D$=10.6 nm and all larger members of its family exhibit a large $\mu$ and a positive $\mu_\mathrm{D}$ owing to strong stick-slip, caused by the edge, creating a barrier against entry/exit of uncompensated moir\'e portions, also illustrated in Fig.~\ref{fig:7} \cite{Koren.prb.moire.2016}. Conversely, Table III and Fig.~\ref{fig:13}(b-c) show much smaller $\mu$ and a negative $\mu_\mathrm{D}$ in the 14 nm diameter where,
as shown by Fig.~\ref{fig:11}(c), there is only a weak edge effect,
and friction at this large speed is viscous. 
That is because no uncompensated moir\'e are attempting to cross the edge.

It is important to remember that these results strongly depend on the island area and velocity.
Based on Fig.~\ref{fig:10}(b), we anticipate that {\it at sufficiently large island sizes} (Eq.~\ref{Eq:10}) stick-slip behaviour with larger $\mu$ and positive $\mu_\mathrm{D}$ becomes universal.
In that limit, the weak friction family will merge with the strong ones.
Moreover, we expect that the merging will also occur at constant island area at low enough velocity. That happens with a crossover from viscous to stick-slip similar to that of Fig.~\ref{fig:11}(b).
This simplified understanding is based on the circular shape of island. In real systems, polygonal tracts of the edge are more likely to take over. That will make the picture more complicated and generic size dependence, thus rather unpredictable.

To become more specific before closing the load dependence, 
we compare stick-slip friction coefficients of the small twist angle simulated graphene interface in Table III with data of Tables I and II. These stick-slip frictions poorly dependent on velocity and elicits complementary comments to the large twist case. 
First, the theoretical values of Table III are orders of magnitude larger than experimental ones. The island sizes are small, but since friction grows sublinearly with area, the friction coefficient will decrease at large areas down to values closer to the experimental range.
Secondly, stick-slip of Table III has a positive differential friction coefficient, showing that, unlike suggestions of PBC simulations (which might in this respect be considered academic) load appears to increase the energy barrier felt by the moir\'e  rather than ironing it away, an observation that may explain why negative $\mu_\mathrm{D}$ is so hard to find in experiments.

\begin{table}[t]
\caption{Friction coefficient from graphene/graphene interface simulations.}
\includegraphics[width=\linewidth]{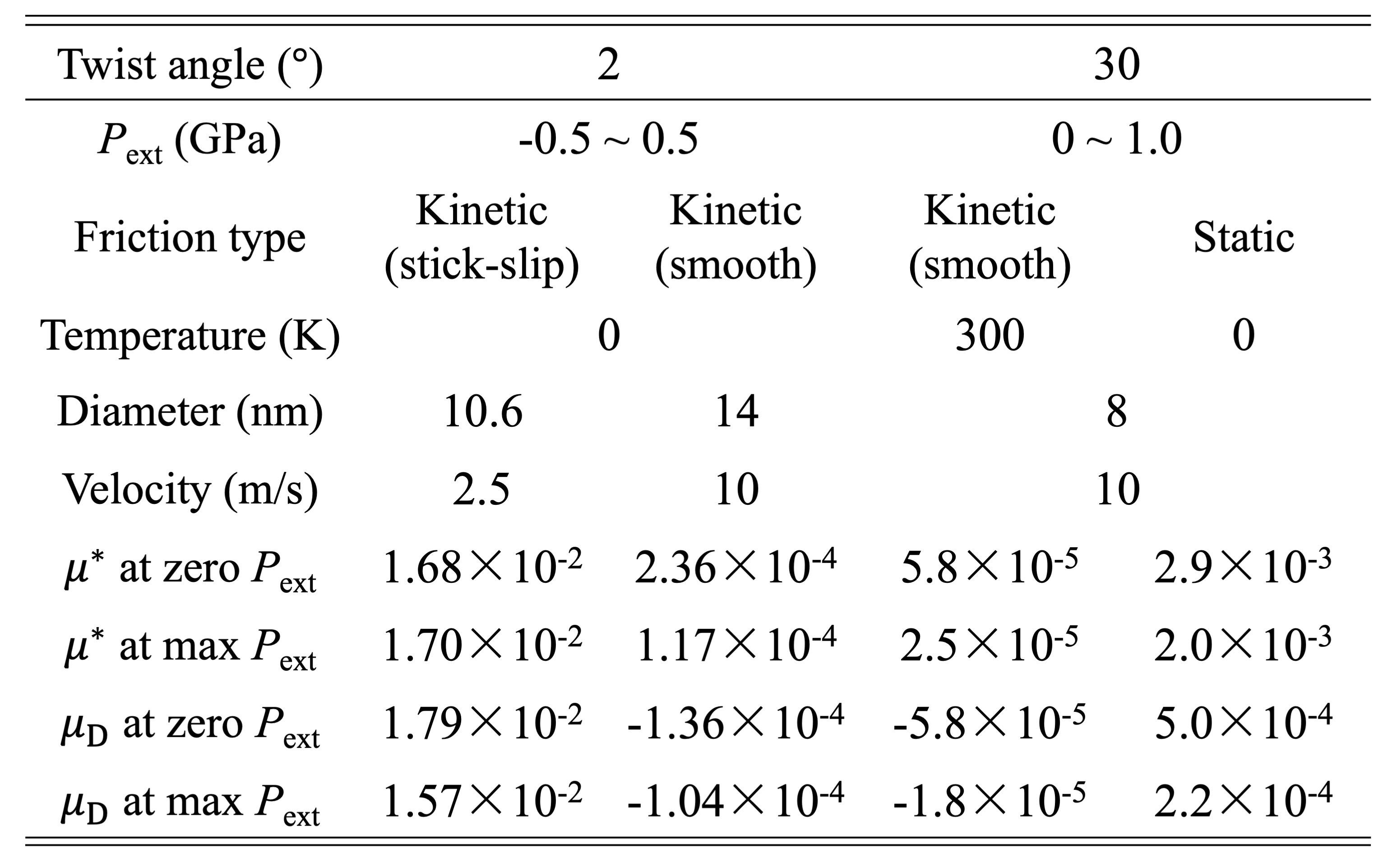}
\raggedright
* The value of friction coefficient $\mu$, according to Eq.~\eqref{Eq:6}, is obtained with $|P_\mathrm{adh}|=1.1$ GPa \cite{Li.NatNanotech.2019}. The twist angle dependence of $P_\mathrm{adh}$ is negligible \cite{Wang.nc.2015}. \end{table}

\section{Elasticity}

Theoretical work has long discussed the structural and tribological effect of elasticity \cite{Hurtado.prsl.1999,Persson.solidstatecommu.1999,Muser.EPL.2004, Ma.prl.2015, Benassi.scirep.2015, Sharp.prb.2016,Mandelli.prm.2018, Feng.friction.2022}.
Elastic effects are dependent upon system dimensionality \cite{Muser.EPL.2004} and size:
important at large slider diameters, and irrelevant for small ones,
where the slider is closer to a rigid flake.
In this Section we discuss what mechanism determines a typical crossover diameter $D$ separating the two regimes, and in what way that might depend on the specific properties of a 2D material interface. Different mechanisms suggest different critical lengths.

\begin{figure}[ht!]
\centering
\includegraphics[width=\linewidth]{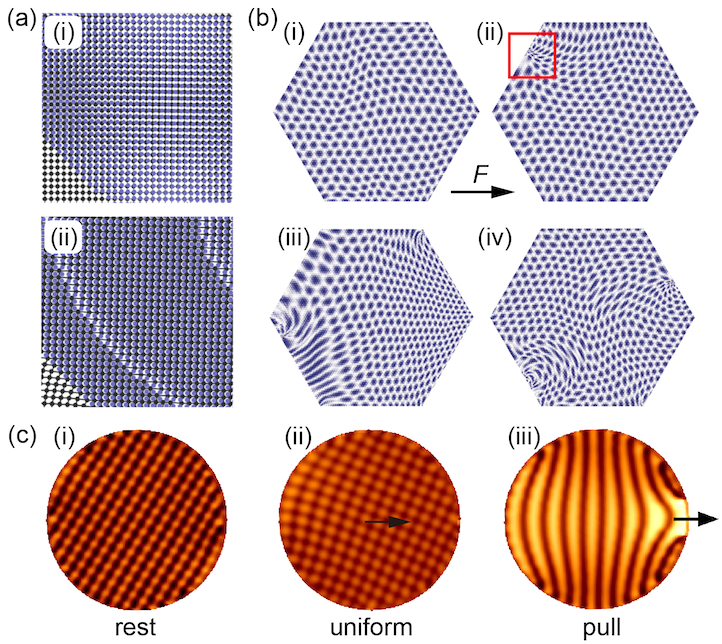}
\caption{(a) Elastic deformation of the flexible substrate (black) in incommensurate contact with a rigid circular slider (blue): (i) and (ii) correspond to stiff ($G/\tau_\mathrm{max}=256$) and soft ($G/\tau_\mathrm{max}=1$) substrate \cite{Sharp.prb.2016}.
(b) Moir\'e evolution of the simulated incommensurate Kr/Pb(111) contacts \cite{Varini.nanoscale.2015}. Colors range from blue when Kr atoms are maximally coincident with Pb atoms, to white when they are minimally coincident. (i) stationary system (dragging force $F= 0$); (ii) system with $F= 2.24$ nN (just above the maximum static friction). Soliton entering from the left edge is highlighted by the red square; (iii) system after sliding of the island center by 1.0 surface lattice spacing; (iv) system when the sliding is 1.5 lattice spacings, and a soliton exits the island on the right hand side.
(c) Strain distribution of a uniformly driven (ii) versus an edge-pulled (iii) 2D harmonic lattice island submitted to a slightly less dense, therefore incommensurate, 2D sinusoidal potential (``Frenkel-Kontorova" type model) compared to its static state (i). Colors range from white for perfect local coincidence, to black for maximal local mismatch. The edge pulling procedure favours the formation of locally strained commensurate regions, enhancing the static friction, and the ensuing stick-slip, compared to the uniformly driven system \cite{Vanossi.nc.2020,Mandelli.prm.2018}. Note also the corresponding moir\'e  switch from 2D to striped. }
\label{fig:14}
\end{figure}

For macroscopic interfaces, generally rough and multi-contact, a well known critical size is the elastic coherence length, or Larkin length, $D_L$ \cite{Larkin.JLTP.1979}, dividing the regimes where different contact points act collectively or act separately. Its theoretical estimate is
$D_L / D \sim \exp {[Yd/(\sigma_y D)]^2}$,
where $D$ is the diameter of the contact,
$Y$ is Young's modulus, $d$ is the average distance between random contacts, and $\sigma_y$ is the yield stress \cite{Persson.solidstatecommu.1999, Vanossi.RevModPhys.2013}.
Originally devised for flux lattice creep and important all the way up to earthquakes, this rough interface scenario is generally far from relevant in real 2D material 
nano and mesoscale interfaces, and we shall not be concerned further with it. 

In mesoscopic, relatively defect-free crystalline contacts, which is the case for pristine
2D material interfaces, the relevant crossover size is the elastic length $D_e$. Generally invoked to describe the competition between intra-slider shear modulus $G$ and the slider-substrate interaction, in turn controlling the shear strength $\tau_\mathrm{max}$ \cite{Muser.EPL.2004, Sharp.prb.2016}, one has
\begin{equation}
   D_e /a \sim G/\tau_\mathrm{max}.  
   \label{Eq:8}
\end{equation}
Once the slider diameter $D$ exceeds $D_e$ -- whose magnitude varies for different systems and driving methods \cite{Mandelli.prm.2018,Benassi.scirep.2015} -- a dislocation nucleates at the edge as exemplified in Fig.~\ref{fig:14}(a,b), thus enhancing static friction \cite{Hurtado.prsl.1999, Varini.nanoscale.2015, Sharp.prb.2016}.
In structurally lubric graphene interfaces where $G\approx 300$ GPa \cite{Liu.Nanolett.2012},
$\tau_\mathrm{max}\approx 100$ MPa \cite{Wang.jmps.2022}, one would estimate $D_e \approx 1 \mu \mathrm{m}$.
Even after the dislocation has nucleated, sublinear static friction controlled by edges must generally persist \cite{Varini.nanoscale.2015}, up to some larger size $D_t$ where a train of dislocations enters the slider's interior.
For $D > D_t$  the static friction will eventually grow proportional to area \cite{Hurtado.prsl.1999, Sharp.prb.2016}.

For very stiff sliders, as 2D materials are, the sublinear size scaling of static friction is in fact known to persist for much larger diameters $D \gg D_e$ \cite{Sharp.prb.2016}. Model estimates of the $D_t$ where structural lubricity fails in 2D material stacks \cite{Ma.prl.2015} are way larger than even the size of experiments.
Among examples, graphite mesas of $\mu \mathrm{m}$-size \cite{Qu.prl.2020} and 1D carbon nanotubes of cm-length \cite{Zhang.natnano.2013}, support a much wider range of structural lubricity than ordinary 3D crystalline interfaces.

Apart from static friction, elasticity influences kinetic friction of 2D materials interfaces, deciding in particular whether it will be smooth sliding or stick-slip.
Two extra factors matter for kinetic friction:
(i) large static friction favors stick-slip, but does not actually determine it;
(ii) the effective lateral stiffness $K_\mathrm{p}$ becomes very important.
With layer elasticity in mind, we therefore look for an upper size $D_c$ above which it will cause stick-slip, even without destroying structural lubricity. A useful route is suggested by the PT model. Consider a 2D material island whose
sliding must negotiate an energy barrier caused by the edge.
Contact elasticity and spring stiffness together determines whether the sliding can be smooth despite the barrier, or whether a mechanical instability will take place and give rise, for sufficiently low velocities, to stick-slip \cite{Vanossi.RevModPhys.2013}.
If the effective lateral stiffness $K_\mathrm{p}$ is soft enough, resulting in
$\eta = 2\pi^{2}U_{0}/K_\mathrm{p} a^2 \gg 1$,
stick-slip will necessarily ensue for all slider diameters.
This regime may indeed be close to actual experiments, where the underlying stick-slip regime is proven by the ubiquitous logarithmic velocity dependence.
On the other hand, if $K_\mathrm{p}$ is hard enough, the sliding can still be either smooth or stick-slip depending whether the diameter of the slider is smaller or larger than a maximal critical $D^\mathrm{max}_c$.

Determined by three factors, the barrier $U_0$, the pulling spring stiffness $K_\mathrm{p}$ and the intra-island stiffness $k_\mathrm{isl}$, a more general $D_c$
is estimated by requiring that the joint spring-island inverse stiffness
$k_\mathrm{isl}^{-1}(D) + K_\mathrm{p}^{-1}$  satisfy the PT-like inequality $\eta \geq 1$ (Eq.~\ref{Eq:4}).
An upper bound for the critical circular slider diameter $D_c$ is obtained in the limit $K_\mathrm{p}^{-1} =0$ where stick-slip occurs if the effective island in plane stiffness $k_\mathrm{isl}$ alone satisfies the inequality
$k_\mathrm{isl}(D) < 2\pi^{2}U_{0}(D)/a^{2}$, where $a$ is the lattice spacing.
For an edge-pinned circular flake with uniformly distributed in-plane dragging force along $x$, we obtain by fitting simulations of a circular island with a pinned circumference a displacement field extremely well represented by the form
\begin{equation}
    u_x (x,y)=\delta_x [1-\frac{4(x^2+y^2)}{D^2}], 
    u_y (x,y)=0
    \label{Eq:9}
\end{equation}
with the corresponding strain field under the assumption of small deformation, $\varepsilon_{xx} (x,y)=-8 \delta_x x/D^2$ and $\varepsilon_{xy} (x,y)=-4\delta_x y/D^2$. By equating strain energy $\frac{1}{2} k_\mathrm{isl} \delta_x^2=\frac{1}{2}\iint_A (Y_\mathrm{2D} \varepsilon_{xx}^2+2 G_\mathrm{2D} \varepsilon_{xy}^2) \,\mathrm{d}x\,\mathrm{d}y$, we get $k_\mathrm{isl} \sim \pi Y_\mathrm{2D}$, where $Y_\mathrm{2D}$ and $G_\mathrm{2D}$ are the (2D) Young's modulus and shear modulus, and $A=\pi D^2/4$ is the contact area of the flake.
We see that the effective stiffness $k_\mathrm{isl}$ of a circular 2D island is, remarkably, independent of diameter. However, due to the emergence of relevant localized elastic distortions, this property will not hold if the slider is pulled or pushed inhomogeneously, e.g., by one side. That situation is illustrated in Fig.~\ref{fig:14}(c) and detailed in the caption, where stick-slip is strongly enhanced \cite{Vanossi.nc.2020}.

What does depend on diameter in all cases is the effective barrier $U_{0}$ which pins the circular island, growing like $U_0 \sim U_a (D/a)^{1/2}$.
Hence $D_c^\mathrm{max}/a = (\frac{Y_\mathrm{2D} a^2}{2\pi U_a})^2$ (obtained for $K_\mathrm{p} = \infty$) is the slider's diameter upper bound above which smooth sliding is impossible, and mechanical instability will arise -- for low velocity and low temperatures.
For a general spring stiffness $K_\mathrm{p}$ we finally obtain,
\begin{equation}
    D_c/a = [\frac{a^2 Y_\mathrm{2D}}{2 \pi U_{a}(1+ \pi K_\mathrm{p}^{-1} Y_\mathrm{2D})}]^2
    \label{Eq:10}
\end{equation}
a clean new result for a uniformly pulled circular slider.

This predicted critical diameter $D_c$ for elasticity-induced stick-slip will get arbitrarily small for a sufficiently soft external spring $K_\mathrm{p}$: but how large will it get in practice for a hypothetically super hard spring $K_\mathrm{p} \to \infty$?
Approximating $Y_\mathrm{2D} a^2 \sim G a^3$ and $U_a \sim \tau_\mathrm{max} a^3$, Eq.~\eqref{Eq:10} could be rewritten for $K_\mathrm{p} = \infty$ as $D_c^\mathrm{max}/a \sim (G/\tau_\mathrm{max})^2$.
Compared with $D_e/a \sim G/\tau_\mathrm{max}$, one sees that in this limit $D_c > D_e$ for structurally lubric 2D materials where $G/\tau_\mathrm{max}\gg1$.
Therefore, if a 2D slider can be pulled with a hard effectively spring $K_p$, and if the edge energy barrier $U_a$ is large enough, there is a range of diameters between $D_c$ and $D_e$ where the sliding remains smooth.
To illustrate the opposite behaviour, for the $30^\circ$ twisted graphene interface of Fig.~\ref{fig:10}(a), the barrier $U_a$ is only about $17$ meV. With $Y_\mathrm{2D}=300$ N/m \cite{Memarian.supermicro.2015}, the maximal critical size $D_c^\mathrm{max}$ is as large as 280 $\mu$m, which exceeds $D_e \approx 1~\mu \mathrm{m}$.

More generally, besides a soft pulling device, the critical size $D_c$ for stick-slip onset will also depend on other variables, including particularly shape, mutual lattice orientation and shearing direction \cite{Wang.Nano.Letters.2019}. 
In graphene interfaces the critical $D_c$ will decreases for decreasing twist angles from regimes L to regimes I and S, where the barrier $U_a$ is larger resulting in the larger friction shown in Fig.~\ref{fig:10}(a, b).

Summing up, for 2D materials, where $G/\tau_\mathrm{max}$ is large, the elastic length $D_e$ of Eq.~\eqref{Eq:8} identifies the size where elasticity permits nucleation of an edge dislocation, raising static friction.
Independently, $D_c$ of Eq.~\eqref{Eq:10} is the elasticity-induced threshold of stick-slip. Structural lubricity persists in both  cases. There must however be an even larger size $D_t$ where dislocations finally invade uniformly the slider,
above which the friction turns proportional to area, marking the failure of structural lubricity.

\section{Defects}

In the preceding Sections, we reviewed and discussed the size, velocity, temperature and load dependence of friction together with the effects from elasticity for structurally lubric 2D material interfaces.
NEMD simulations helped us understand and rationalize physical phenomena and frictional data. Defect-free PBC simulations in particular, are important in order to get the idealized physical picture.  By comparing PBC simulations with finite islands OBC ones, it was invariably noted that many key differences were made by defects, in our case exemplified by edges that mark the boundary of the islands. 

In practice, real 2D material interfaces are more complex than that. Physical and chemical complications arising from defects include besides shapes of sliders \cite{Dietzel.prl.2013, Luo.acsnano.2011,Ozogul.apl.2017}, edge chemisorption, \cite{Gongyang.friction.2020}, corrugation by contaminants at the interfaces, other imperfections such as grain boundaries \cite{Cervenka.prb.2009, Yazyev.prb.2010, Yazyev.natnano.2014} and vacancies \cite{Hashimoto.nature.2004, Gajurel.afm.2017} and more, all playing an important role. Each type of defect introduces its own specific energy barriers against sliding, influencing in turn the frictional behavior of real 2D material interfaces, even when nominally structurally lubric. While a review of these defects would open a vast chapter which it is beyond our scope to explore, we should at least mention some of the most important ones before closing this Colloquium.

\begin{figure}[htb!]
\centering
\includegraphics[width=\linewidth]{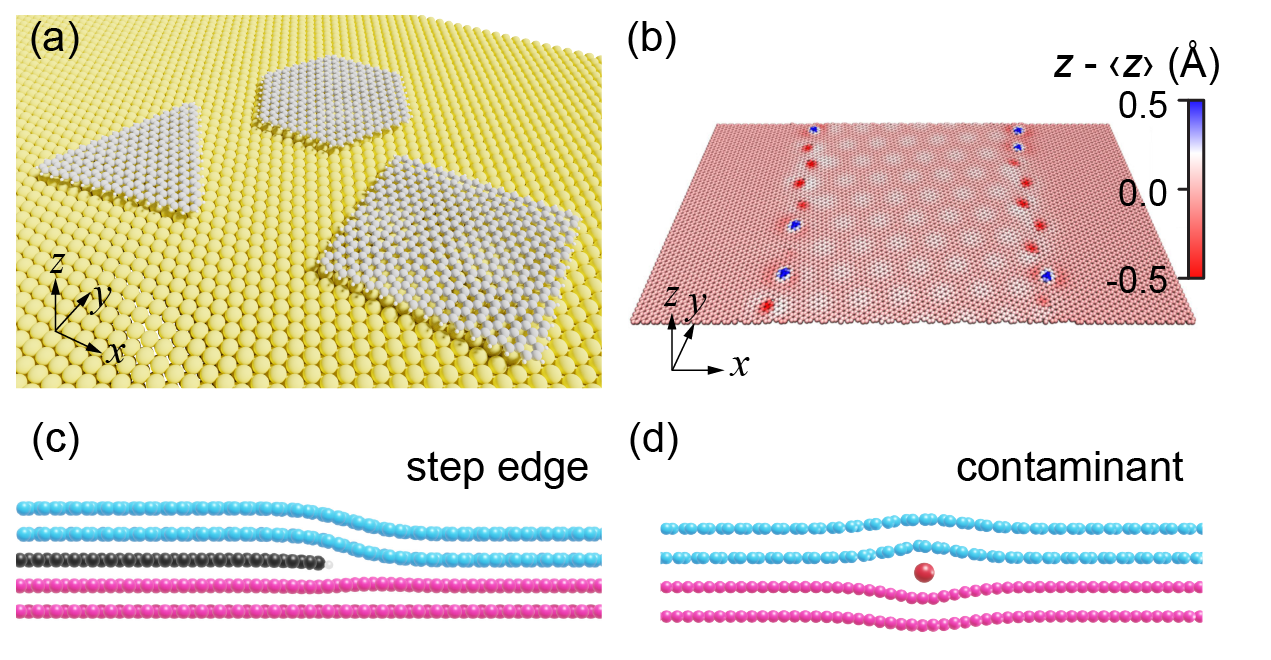}
\caption{Sketch of various defects. (a) Slider with different shapes, i.e., different number of edges and corners. (b) Grain boundaries at the graphene interface \cite{Gao.Nat.Comm.2021}. Color represents the out-of-plane displacement.
(c) Step edge exposed to the sliding interface.  Blue, black, and pink are the slider, step edge, and substrate, respectively. (d) Particle-contaminated graphene interface. Blue, red, and pink are the slider, contaminated particle, and substrate, respectively.}
\label{fig:15}
\end{figure}

\subsection{Shape of sliders} 
As mentioned in the previous section,
circular islands/flakes discussed in our modeling are heavily idealized. Despite of a few circular-shape-based experiments \cite{Koren.science.2015, Li.Adv.Mat.2017,Finney.NatNano.2019,Ribeiro.Science.2018}, most data refer to different shapes.
In particular, straight-edged polygon shapes are common (the straight edges reflecting the robust 2D crystalline structure of the island or flake), as sketched in Fig.~\ref{fig:15}(a). Several theoretical studies discuss this point \cite{deWijn.prb.2012, Varini.nanoscale.2015, Koren.prb.moire.2016,Dietzel.Nanotechnology.2018, Wang.Nano.Letters.2019},
and find different size scalings of static friction for different shapes, with the exponent ranging from 0 to $1/2$ -- this variability reflecting the different mismatch of the edge orientation with the substrate lattice. 
Although the kinetic friction size scaling should eventually tend to linear when the contact area $A\to \infty$, the limit where the sublinear contribution from the edge can be neglected is hard to reach in practice.
Generally, owing also to the relatively high dissipation at edges \cite{Varini.nanoscale.2015,Liao.natmater.2022},
the size and shape dependent sublinear scaling behavior prevails.
An especially large shape contribution to friction may be expected from sharp corners and protrusions, locations which enhanced adhesion may transform into ``pivots" causing islands to rotate. We expect future theoretical and experimental studies should verify these expectations and provide further understanding.

\subsection{Grain boundaries} 
In the real 2D materials samples where macroscale engineering superlubricity is being sought and studied, extended lattice defects such as grain boundaries are inevitable \cite{Yazyev.natnano.2014,Tison.nanolett.2014}, and  understanding their effect on friction is important. An effort was attempted in a series of carefully designed MD simulations \cite{Gao.prb.2021,Gao.Nat.Comm.2021,Gao.prl.2023}.
Shear induced buckling/unbuckling of the grain boundary is shown to affect the kinetic friction (Fig.~\ref{fig:15}b), resulting in a non-monotonic dependence of friction upon load and temperature. Nevertheless, from the perspective of engineering superlubricity, the differential friction coefficient of the grain boundary-containing systems is still below $10^{-3}$ for a linear concentration of grain boundaries of $0.0838~\mathrm{nm}^{-1}$.\\

\subsection{Step edges}
Experiments and simulations find that the ultralow friction of structural lubricity can be destroyed by the presence of steps (Fig.~\ref{fig:15}c) within the sliding interface \cite{Wang.prl.2020}. Simulations further suggest that a large part of the extra friction is due to the violent, large size  out-of-plane displacement of the step-free slider as it surmounts the step edge of the substrate \cite{Wang.prl.2020,Chen.Sci.Adv.2019}.
The pinning caused by the step can change friction from smooth to stick-slip, destroying the main feature of structural lubricity.
Buried steps, on the other hand \cite{Wang.prl.2020} and their milder corrugation propagated at the sliding interface \cite{Peng.nsr.2021} have not been reported to cause effects of comparable significance. A remaining question is at what depth should a step lie for the influence on friction to become irrelevant.

\subsection{Contaminants}
It is generally accepted that structural lubricity requires the interface to be perfectly regular and in particular ultra-clean.
In common frictional setups however, adsorbed gases and contaminants are rife, including measurements in fully ambient conditions \cite{Li.natmater.2013,Li.AccountChemRes.2015,He.Science.1999}.
The presence of contaminants should intuitively lead to an increase of friction \cite{Ouyang.nanoscale.2018} including the loss of superlubricity \cite{Muser.PRL.2001}, an expectation confirmed by recent experiments and simulations \cite{Deng.nanoscale.2018, Cheng.prm.2020, Wang.jmps.2022}.
In our prototypical simulation with $30^\circ$-twisted graphite system ($T=300$ K and $v_0=30$ m/s), the friction stress increases from $\sim 10~\mathrm{kPa}$ to $352~\pm 56~\mathrm{kPa}$ by introducing one contaminate particle with number density $0.0244 \mathrm{nm}^{-2}$ (Fig.~\ref{fig:15}d). The presence of the contaminants result in stick-slip motion with a sublinear velocity scaling: by reducing the velocity to 10 m/s, the friction stress becomes $345 \pm 39 \mathrm{kPa}$ -- almost unchanged.
On the other hand, tribological running in, showing a friction decrease during repeated sliding of graphite contacts, was observed \cite{Deng.nanoscale.2018} and explained by self-cleaning \cite{Liu.nanotech.2011, Ma.prx.2015}.
Contaminant may also be removed by temperature, thereby artificially causing lower friction at higher temperatures \cite{Wang.epl.2019}.
On the other hand, contaminants such as water molecules \cite{Falk.nanolett.2010} or graphitic nanoflakes \cite{Li.Adv.Science.2018},
serve as lubricants, yielding lower friction.

A recent experiment conducted by intentionally introducing airborne contaminants into microscale superlubric graphite contacts
shows that although the friction stress increases (from 10 kPa to 50 kPa) as the concentration of contaminants increase, the key features of engineering superlubricity, i.e., the ultralow friction and differential friction coefficient, are well preserved \cite{Wang.jmps.2022}.
In the case of graphite/hBN heterostructures, data in ambient conditions (where contaminants are expected) observed that the friction force exhibits a $60^\circ$ rotational symmetry with both the twist angle \cite{Song.NatureMaterials.2018} and the sliding angle \cite{Song.prm.2021}, suggesting that the interface retains its bulk-like structure and symmetry. The sliding of gold nanoparticles on HOPG have been studied in both 
UHV \cite{Dietzel.prl.2013} and ambient conditions \cite{Cihan.nc.2016}, yielding friction stresses of the same magnitude and similar scaling effects (Fig.~\ref{fig:10}d,f).
These experiments imply the robustness of structural lubricity, possibly implying low contamination, or else the low impact of contaminants, that could even be swept out when the slider-substrate adhesion is strong.
One should expect future experiments as well as simulations to provide clearer understandings of these phenomena, following well designed surface science protocols, starting with clean interfaces then introducing single well defined contaminants.

\section{Summary}

We wind up the Colloquium by recalling the main outcomes and conclusions about friction of 2D structurally lubric interfaces. Our understanding, based on existing studies, and illustrated by simulations of a specific graphene/graphene example, is proposed as relevant to the broader variety of 2D materials sliding interfaces. \\

{\it Area dependence.}
Structurally {\it super}lubric sliding causes a linear growth of kinetic friction with area -- due to swift moir\'e flight -- along with zero static friction.
In the experimentally more relevant case of structural lubricity, the area dependence of static friction is generally sublinear.
That of kinetic friction too,
for soft effective stiffness, $\eta >1$ (Eq.~\ref{Eq:4}), 
and for contact size below critical size $D_t$.
Above size $D_t$, both static and kinetic friction are proportional to area.
For defect-ridden 2D materials interfaces that are merely engineering superlubric,
the frictional behaviour depends on the nature and distribution of the pinning agents.
A uniform distribution will naturally cause static and kinetic friction to grow proportional to area.
The moir\'e pattern, its size and evolution are also crucial in 2D materials friction.
For example, at small twist angle (S regime) and for
a circular slider shape, the kinetic friction is predicted to oscillate with increasing size.
That reflects the energy barriers caused by the uncompensated moir\'es at the edges
as the  moir\'e nodes cross the slider's perimeter.
These oscillations with size, even if difficult to address experimentally, are both predicted and also observed in colloidal flakes \cite{Cao.prx.2022} and should deserve a specific search in 2D materials.

{\it Velocity and temperature dependence.} 
A linear velocity dependence of kinetic friction is a hallmark of structural superlubricity.
With negligible static friction and weak moir\'e dissipation, structurally superlubric friction falls way below measurable magnitudes. Measurable values would require a sliding velocity of 1 m/s or higher -- many orders of magnitude above current experiments, even if common in simulations.
In experimental structurally lubric sliders, whose static friction is nonzero, the velocity dependence is not linear but logarithmic.
It always implies stick-slip friction, even when interference among pinning spots might have cancelled its straight sawtooth-like evidence in force traces, a cancellation expected for large size sliders \cite{Braun.Tribolett.2012}.
Crossovers from logarithmic to linear regimes are always possible. Stick-slip
can demise in favor of thermolubric sliding \cite{Krylov.physstatussolid.2014} at the lowest speeds and/or highest temperatures, and of ballistic sliding\cite{Guerra.NatMater.2010} at the highest speeds.
Both regimes imply linear velocity dependence, even if
neither is experimentally common in 2D materials friction.
In some simulation studies actually, the high speed ballistic behaviour is sometimes mistaken for structural superlubricity.

{\it Load dependence.}
In genuine structural superlubricity the static friction is zero, independent of load.
Unlike, e.g., colloid monolayers \cite{Brazda.prx.2018}, we note, incidentally, that so far there is no evidence that a load increase could induce an Aubry-type pinning transition in 2D materials sliding.
Interestingly, simulations (high velocity) suggest that load may ``iron" out the moir\'e corrugation, reducing structurally superlubric friction \cite{Mandelli.PhysRevLett.2019}.
The load dependence of experimentally relevant structural or engineering lubric friction on the other hand, depends on the diverse effects that load can have on the edges and other pinning defects.
The effective load determining the true, absolute friction coefficient must include a large adhesive pressure, as large as 1 GPa in graphene type interfaces.
That makes the physically significant friction coefficient of 2D materials interfaces small, but not as small as the commonly used differential friction coefficient. Use of the latter is dubious, because it does not measure frictional dissipation.
It thus seems advisable to resume the use of ordinary friction coefficient,
properly defined following Derjaguin \cite{Derjaguin.ZPhys.1934}.

{\it Elasticity.} The effect of elasticity in large 2D sliding islands demands a fresh approach, owing to the extreme in-plane stiffness of these systems.
For a circular shape with edge pinning, one can identify a novel critical diameter, Eq.~\eqref{Eq:10}, above which structurally lubric sliding will develop a mechanical instability and necessarily enter stick-slip regime.

{\it Defects}. Finally and not surprisingly, realistic defects bring havoc into this picture -- each type opening a special chapter, deserving its own special attention.
The special role of the omnipresent slider's edge has been especially emphasized.  

{\it Terminology}. 
It is suggested to avoid confusing terms sometimes found in literature,
by introducing a distinction between three types of frictional behaviour.
Structural superlubricity is reserved for the essentially academic case of zero static friction. Although only accessible in simulations so far, understanding this regime is important in the physics of 2D materials interfaces \cite{Wang.tobe.2023}.
Structural lubricity designates cases where the interface is incommensurate with a defect-free interior, but where static friction is nonzero owing to pinning by, e.g., the slider's edge.
We then propose that cases where the friction coefficient is just very small (typically lower than $10^{-2}$) be called engineering superlubric.
In this connection, further semantic confusion should be avoided between ordinary and differential friction coefficients, somehow a dangerous current habit.\\

Many key points of course still remain to be addressed.
Multi-scale experiments/simulations are called for to verify the large scale size friction scaling, presently mostly extrapolated. Experiments could and should be designed, for example examining frictional noise, to pinpoint directly the elusive but unquestionable underlying stick-slip friction of engineering superlubric systems.
Experiments could explore the crossover from stick-slip to superlubric which a sliding island may undergo when the driver changes from soft to stiff, provided shape and size is such to permit smooth sliding, such as in Fig.~\ref{fig:8}(b). 
That would enable access to  superlubric-like kinetic friction in a real world finite size system.
The sliding behaviors for small twist systems, where the interfacial structure is different from the large twists, are also interesting and lacking existing studies.
It will also be desirable to extend the experimental velocity ranges to locate the crossover from stick-slip to the high velocity “ballistic" regime.
Attacking these and related problems will not just solidify theoretical bases of “superlubricity", but will also offer keys to its practical application.
Many other points, more speculative but rich of real or potential questions, promising applications, and corresponding urgent problems, deserve to be at least mentioned. Among them: \\

-- Friction of freestanding or quasi-freestanding 2D materials interfaces \cite{Butz.Nature.2014,Riedl.prl.2009}. Despite the strong restriction from the substrate, the out-of-plane deformation of adsorbed 2D materials dominates the friction of structurally lubric systems. For freestanding systems, new phenomena brought about by out-of-plane flexural deformability are worth exploring \cite{Wang.buckling.2023}.\\

-- Variety of 2D materials with structurally lubric behavior \cite{Gao.acsami.2021}. Is graphene the ideal material for structural lubricity? Are there many more natural or engineered functional materials with better properties, such as larger intralayer to interlayer interaction ratios?\\

-- Electronic friction in twisted bilayers,  friction under external field (thermal, electric, magnetic...) \cite{Belviso.prb.2020,Wang.friction.2015,Song.nanolett.2022}.\\

-- Rotational friction is also interesting but less explored \cite{Koren.science.2015, Yang.acsami.2021, Cao.prx.2022, Finney.NatNano.2019,Ribeiro.Science.2018}.\\

-- Potential applications:
Structural lubricity is promising in the application of nano-(micro-)electromechanical systems, including nano-generators, nano-oscillators, nano-tip of hard drives and nano-robotics, etc. \cite{Huang.nc.2021,Huang.nanoenergy.2020, Zheng.prl.2002, Wu.commumater.2021}.
The dry interfaces with ultra-low friction and high current stability \cite{Lang.carbon.2021}
can play important roles in precision instruments such as aerospace slip rings.
Beyond these, the large number of combinations of homo/hetero-interfaces composed of diverse 2D materials give the interface various physical properties, and thus wider applications.\\

We close in the hope to have brought at least some light and discussion into the questions that animate the pinning and sliding of 2D “superlubric” material interfaces, important for its  physics, as well as for current and future technologies.

\begin{acknowledgments}
We express our gratitude to
A. Silva, N. Manini, X. Gao, E. Meyer, M. Kisiel, R. Pawlak, and M. Urbakh
for collaboration and helpful discussions.
This work is carried out under ERC ULTRADISS Contract No. 834402. Support by the Italian Ministry of University and Research through PRIN UTFROM N. 20178PZCB5 is also acknowledged. J.W. acknowledges the computing resources support from National Supercomputer Center in Tianjin.
\end{acknowledgments}

\newpage

\bibliography{ref}

\begin{thebibliography}{200}%
\makeatletter
\providecommand \@ifxundefined [1]{%
 \@ifx{#1\undefined}
}%
\providecommand \@ifnum [1]{%
 \ifnum #1\expandafter \@firstoftwo
 \else \expandafter \@secondoftwo
 \fi
}%
\providecommand \@ifx [1]{%
 \ifx #1\expandafter \@firstoftwo
 \else \expandafter \@secondoftwo
 \fi
}%
\providecommand \natexlab [1]{#1}%
\providecommand \enquote  [1]{``#1''}%
\providecommand \bibnamefont  [1]{#1}%
\providecommand \bibfnamefont [1]{#1}%
\providecommand \citenamefont [1]{#1}%
\providecommand \href@noop [0]{\@secondoftwo}%
\providecommand \href [0]{\begingroup \@sanitize@url \@href}%
\providecommand \@href[1]{\@@startlink{#1}\@@href}%
\providecommand \@@href[1]{\endgroup#1\@@endlink}%
\providecommand \@sanitize@url [0]{\catcode `\\12\catcode `\$12\catcode
  `\&12\catcode `\#12\catcode `\^12\catcode `\_12\catcode `\%12\relax}%
\providecommand \@@startlink[1]{}%
\providecommand \@@endlink[0]{}%
\providecommand \url  [0]{\begingroup\@sanitize@url \@url }%
\providecommand \@url [1]{\endgroup\@href {#1}{\urlprefix }}%
\providecommand \urlprefix  [0]{URL }%
\providecommand \Eprint [0]{\href }%
\providecommand \doibase [0]{https://doi.org/}%
\providecommand \selectlanguage [0]{\@gobble}%
\providecommand \bibinfo  [0]{\@secondoftwo}%
\providecommand \bibfield  [0]{\@secondoftwo}%
\providecommand \translation [1]{[#1]}%
\providecommand \BibitemOpen [0]{}%
\providecommand \bibitemStop [0]{}%
\providecommand \bibitemNoStop [0]{.\EOS\space}%
\providecommand \EOS [0]{\spacefactor3000\relax}%
\providecommand \BibitemShut  [1]{\csname bibitem#1\endcsname}%
\let\auto@bib@innerbib\@empty
\bibitem [{\citenamefont {Alden}\ \emph {et~al.}(2013)\citenamefont {Alden},
  \citenamefont {Tsen}, \citenamefont {Huang}, \citenamefont {Hovden},
  \citenamefont {Brown}, \citenamefont {Park}, \citenamefont {Muller},\ and\
  \citenamefont {McEuen}}]{Jonathan.pnas.2013}%
  \BibitemOpen
  \bibfield  {author} {\bibinfo {author} {\bibnamefont {Alden}, \bibfnamefont
  {Jonathan~S}}, \bibinfo {author} {\bibfnamefont {Adam~W.}\ \bibnamefont
  {Tsen}}, \bibinfo {author} {\bibfnamefont {Pinshane~Y.}\ \bibnamefont
  {Huang}}, \bibinfo {author} {\bibfnamefont {Robert}\ \bibnamefont {Hovden}},
  \bibinfo {author} {\bibfnamefont {Lola}\ \bibnamefont {Brown}}, \bibinfo
  {author} {\bibfnamefont {Jiwoong}\ \bibnamefont {Park}}, \bibinfo {author}
  {\bibfnamefont {David~A.}\ \bibnamefont {Muller}}, and\ \bibinfo {author}
  {\bibfnamefont {Paul~L.}\ \bibnamefont {McEuen}}} (\bibinfo {year} {2013}),\
  \bibfield  {title} {\enquote {\bibinfo {title} {Strain solitons and
  topological defects in bilayer graphene},}\ }\href
  {https://doi.org/10.1073/pnas.1309394110} {\bibfield  {journal} {\bibinfo
  {journal} {Proceedings of the National Academy of Sciences}\ }\textbf
  {\bibinfo {volume} {110}}~(\bibinfo {number} {28}),\ \bibinfo {pages}
  {11256--11260}}\BibitemShut {NoStop}%
\bibitem [{\citenamefont {Amontons}(1699)}]{Amontons.1699}%
  \BibitemOpen
  \bibfield  {author} {\bibinfo {author} {\bibnamefont {Amontons},
  \bibfnamefont {Guillaume}}} (\bibinfo {year} {1699}),\ \bibfield  {title}
  {\enquote {\bibinfo {title} {De la resistance cause'e dans les machines},}\
  }\href@noop {} {\bibinfo  {journal} {Memoires De L'Acedemie Royale A}\ ,\
  \bibinfo {pages} {257--282}}\BibitemShut {NoStop}%
\bibitem [{\citenamefont {Andrei}\ and\ \citenamefont
  {MacDonald}(2020)}]{Andrei.natmater.2020}%
  \BibitemOpen
\bibfield  {journal} {  }\bibfield  {author} {\bibinfo {author} {\bibnamefont
  {Andrei}, \bibfnamefont {Eva~Y}}, and\ \bibinfo {author} {\bibfnamefont
  {Allan~H.}\ \bibnamefont {MacDonald}}} (\bibinfo {year} {2020}),\ \bibfield
  {title} {\enquote {\bibinfo {title} {Graphene bilayers with a twist},}\
  }\href {https://doi.org/10.1038/s41563-020-00840-0} {\bibfield  {journal}
  {\bibinfo  {journal} {Nature Materials}\ }\textbf {\bibinfo {volume}
  {19}}~(\bibinfo {number} {12}),\ \bibinfo {pages} {1265--1275}}\BibitemShut
  {NoStop}%
\bibitem [{\citenamefont {Angeli}\ \emph {et~al.}(2018)\citenamefont {Angeli},
  \citenamefont {Mandelli}, \citenamefont {Valli}, \citenamefont {Amaricci},
  \citenamefont {Capone}, \citenamefont {Tosatti},\ and\ \citenamefont
  {Fabrizio}}]{Angeli.prb.2018}%
  \BibitemOpen
  \bibfield  {author} {\bibinfo {author} {\bibnamefont {Angeli}, \bibfnamefont
  {M}}, \bibinfo {author} {\bibfnamefont {D.}~\bibnamefont {Mandelli}},
  \bibinfo {author} {\bibfnamefont {A.}~\bibnamefont {Valli}}, \bibinfo
  {author} {\bibfnamefont {A.}~\bibnamefont {Amaricci}}, \bibinfo {author}
  {\bibfnamefont {M.}~\bibnamefont {Capone}}, \bibinfo {author} {\bibfnamefont
  {E.}~\bibnamefont {Tosatti}}, and\ \bibinfo {author} {\bibfnamefont
  {M.}~\bibnamefont {Fabrizio}}} (\bibinfo {year} {2018}),\ \bibfield  {title}
  {\enquote {\bibinfo {title} {Emergent ${D}_{6}$ symmetry in fully relaxed
  magic-angle twisted bilayer graphene},}\ }\href
  {https://doi.org/10.1103/PhysRevB.98.235137} {\bibfield  {journal} {\bibinfo
  {journal} {Phys. Rev. B}\ }\textbf {\bibinfo {volume} {98}},\ \bibinfo
  {pages} {235137}}\BibitemShut {NoStop}%
\bibitem [{\citenamefont {Aubry}(1983)}]{Aubry.physD.1983}%
  \BibitemOpen
  \bibfield  {author} {\bibinfo {author} {\bibnamefont {Aubry}, \bibfnamefont
  {Serge}}} (\bibinfo {year} {1983}),\ \bibfield  {title} {\enquote {\bibinfo
  {title} {The twist map, the extended frenkel-kontorova model and the devil's
  staircase},}\ }\href
  {https://doi.org/https://doi.org/10.1016/0167-2789(83)90129-X} {\bibfield
  {journal} {\bibinfo  {journal} {Physica D: Nonlinear Phenomena}\ }\textbf
  {\bibinfo {volume} {7}}~(\bibinfo {number} {1}),\ \bibinfo {pages}
  {240--258}}\BibitemShut {NoStop}%
\bibitem [{\citenamefont {Baykara}\ \emph {et~al.}(2018)\citenamefont
  {Baykara}, \citenamefont {Vazirisereshk},\ and\ \citenamefont
  {Martini}}]{Baykara.APR.2018}%
  \BibitemOpen
  \bibfield  {author} {\bibinfo {author} {\bibnamefont {Baykara}, \bibfnamefont
  {Mehmet~Z}}, \bibinfo {author} {\bibfnamefont {Mohammad~R.}\ \bibnamefont
  {Vazirisereshk}}, and\ \bibinfo {author} {\bibfnamefont {Ashlie}\
  \bibnamefont {Martini}}} (\bibinfo {year} {2018}),\ \bibfield  {title}
  {\enquote {\bibinfo {title} {Emerging superlubricity: A review of the state
  of the art and perspectives on future research},}\ }\href
  {https://doi.org/10.1063/1.5051445} {\bibfield  {journal} {\bibinfo
  {journal} {Applied Physics Reviews}\ }\textbf {\bibinfo {volume}
  {5}}~(\bibinfo {number} {4}),\ \bibinfo {pages} {041102}}\BibitemShut
  {NoStop}%
\bibitem [{\citenamefont {Belviso}\ \emph {et~al.}(2020)\citenamefont
  {Belviso}, \citenamefont {Cammarata}, \citenamefont {Missaoui},\ and\
  \citenamefont {Polcar}}]{Belviso.prb.2020}%
  \BibitemOpen
  \bibfield  {author} {\bibinfo {author} {\bibnamefont {Belviso}, \bibfnamefont
  {Florian}}, \bibinfo {author} {\bibfnamefont {Antonio}\ \bibnamefont
  {Cammarata}}, \bibinfo {author} {\bibfnamefont {Jamil}\ \bibnamefont
  {Missaoui}}, and\ \bibinfo {author} {\bibfnamefont {Tomas}\ \bibnamefont
  {Polcar}}} (\bibinfo {year} {2020}),\ \bibfield  {title} {\enquote {\bibinfo
  {title} {Effect of electric fields in low-dimensional materials:
  Nanofrictional response as a case study},}\ }\href
  {https://doi.org/10.1103/PhysRevB.102.155433} {\bibfield  {journal} {\bibinfo
   {journal} {Phys. Rev. B}\ }\textbf {\bibinfo {volume} {102}},\ \bibinfo
  {pages} {155433}}\BibitemShut {NoStop}%
\bibitem [{\citenamefont {Benassi}\ \emph {et~al.}(2015)\citenamefont
  {Benassi}, \citenamefont {Ma}, \citenamefont {Urbakh},\ and\ \citenamefont
  {Vanossi}}]{Benassi.scirep.2015}%
  \BibitemOpen
  \bibfield  {author} {\bibinfo {author} {\bibnamefont {Benassi}, \bibfnamefont
  {A}}, \bibinfo {author} {\bibfnamefont {Ming}\ \bibnamefont {Ma}}, \bibinfo
  {author} {\bibfnamefont {M.}~\bibnamefont {Urbakh}}, and\ \bibinfo {author}
  {\bibfnamefont {A.}~\bibnamefont {Vanossi}}} (\bibinfo {year} {2015}),\
  \bibfield  {title} {\enquote {\bibinfo {title} {The breakdown of
  superlubricity by driving-induced commensurate dislocations},}\ }\href
  {https://doi.org/10.1038/srep16134} {\bibfield  {journal} {\bibinfo
  {journal} {Scientific Reports}\ }\textbf {\bibinfo {volume} {5}}~(\bibinfo
  {number} {1}),\ \bibinfo {pages} {16134}}\BibitemShut {NoStop}%
\bibitem [{\citenamefont {Benassi}\ \emph {et~al.}(2010)\citenamefont
  {Benassi}, \citenamefont {Vanossi}, \citenamefont {Santoro},\ and\
  \citenamefont {Tosatti}}]{benassi.prb.2010}%
  \BibitemOpen
  \bibfield  {author} {\bibinfo {author} {\bibnamefont {Benassi}, \bibfnamefont
  {A}}, \bibinfo {author} {\bibfnamefont {A.}~\bibnamefont {Vanossi}}, \bibinfo
  {author} {\bibfnamefont {G.~E.}\ \bibnamefont {Santoro}}, and\ \bibinfo
  {author} {\bibfnamefont {E.}~\bibnamefont {Tosatti}}} (\bibinfo {year}
  {2010}),\ \bibfield  {title} {\enquote {\bibinfo {title} {Parameter-free
  dissipation in simulated sliding friction},}\ }\href
  {https://doi.org/10.1103/PhysRevB.82.081401} {\bibfield  {journal} {\bibinfo
  {journal} {Phys. Rev. B}\ }\textbf {\bibinfo {volume} {82}},\ \bibinfo
  {pages} {081401}}\BibitemShut {NoStop}%
\bibitem [{\citenamefont {Berman}\ \emph {et~al.}(2015)\citenamefont {Berman},
  \citenamefont {Deshmukh}, \citenamefont {Sankaranarayanan}, \citenamefont
  {Erdemir},\ and\ \citenamefont {Sumant}}]{Berman.Science.2015}%
  \BibitemOpen
  \bibfield  {author} {\bibinfo {author} {\bibnamefont {Berman}, \bibfnamefont
  {Diana}}, \bibinfo {author} {\bibfnamefont {Sanket~A.}\ \bibnamefont
  {Deshmukh}}, \bibinfo {author} {\bibfnamefont {Subramanian K. R.~S.}\
  \bibnamefont {Sankaranarayanan}}, \bibinfo {author} {\bibfnamefont {Ali}\
  \bibnamefont {Erdemir}}, and\ \bibinfo {author} {\bibfnamefont {Anirudha~V.}\
  \bibnamefont {Sumant}}} (\bibinfo {year} {2015}),\ \bibfield  {title}
  {\enquote {\bibinfo {title} {Macroscale superlubricity enabled by graphene
  nanoscroll formation},}\ }\href {https://doi.org/10.1126/science.1262024}
  {\bibfield  {journal} {\bibinfo  {journal} {Science}\ }\textbf {\bibinfo
  {volume} {348}}~(\bibinfo {number} {6239}),\ \bibinfo {pages}
  {1118--1122}}\BibitemShut {NoStop}%
\bibitem [{\citenamefont {Bohlein}\ \emph {et~al.}(2012)\citenamefont
  {Bohlein}, \citenamefont {Mikhael},\ and\ \citenamefont
  {Bechinger}}]{Bohlein.natmat.2012}%
  \BibitemOpen
  \bibfield  {author} {\bibinfo {author} {\bibnamefont {Bohlein}, \bibfnamefont
  {Thomas}}, \bibinfo {author} {\bibfnamefont {Jules}\ \bibnamefont {Mikhael}},
  and\ \bibinfo {author} {\bibfnamefont {Clemens}\ \bibnamefont {Bechinger}}}
  (\bibinfo {year} {2012}),\ \bibfield  {title} {\enquote {\bibinfo {title}
  {Observation of kinks and antikinks in colloidal monolayers driven across
  ordered surfaces},}\ }\href {https://doi.org/10.1038/nmat3204} {\bibfield
  {journal} {\bibinfo  {journal} {Nature Materials}\ }\textbf {\bibinfo
  {volume} {11}}~(\bibinfo {number} {2}),\ \bibinfo {pages}
  {126--130}}\BibitemShut {NoStop}%
\bibitem [{\citenamefont {Bonelli}\ \emph {et~al.}(2009)\citenamefont
  {Bonelli}, \citenamefont {Manini}, \citenamefont {Cadelano},\ and\
  \citenamefont {Colombo}}]{Bonelli.epj.2009}%
  \BibitemOpen
  \bibfield  {author} {\bibinfo {author} {\bibnamefont {Bonelli}, \bibfnamefont
  {F}}, \bibinfo {author} {\bibfnamefont {N.}~\bibnamefont {Manini}}, \bibinfo
  {author} {\bibfnamefont {E.}~\bibnamefont {Cadelano}}, and\ \bibinfo {author}
  {\bibfnamefont {L.}~\bibnamefont {Colombo}}} (\bibinfo {year} {2009}),\
  \bibfield  {title} {\enquote {\bibinfo {title} {Atomistic simulations of the
  sliding friction of graphene flakes},}\ }\href
  {https://doi.org/10.1140/epjb/e2009-00239-7} {\bibfield  {journal} {\bibinfo
  {journal} {The European Physical Journal B}\ }\textbf {\bibinfo {volume}
  {70}}~(\bibinfo {number} {4}),\ \bibinfo {pages} {449--459}}\BibitemShut
  {NoStop}%
\bibitem [{\citenamefont {Bowden}\ and\ \citenamefont
  {Tabor}(2001)}]{Bowden.2001}%
  \BibitemOpen
  \bibfield  {author} {\bibinfo {author} {\bibnamefont {Bowden}, \bibfnamefont
  {Frank~Philip}}, and\ \bibinfo {author} {\bibfnamefont {David}\ \bibnamefont
  {Tabor}}} (\bibinfo {year} {2001}),\ \href@noop {} {\emph {\bibinfo {title}
  {The friction and lubrication of solids}}},\ Vol.~\bibinfo {volume} {1}\
  (\bibinfo  {publisher} {Oxford university press})\BibitemShut {NoStop}%
\bibitem [{\citenamefont {Braun}\ \emph {et~al.}(2012)\citenamefont {Braun},
  \citenamefont {Peyrard}, \citenamefont {Stryzheus},\ and\ \citenamefont
  {Tosatti}}]{Braun.Tribolett.2012}%
  \BibitemOpen
  \bibfield  {author} {\bibinfo {author} {\bibnamefont {Braun}, \bibfnamefont
  {O~M}}, \bibinfo {author} {\bibfnamefont {Michel}\ \bibnamefont {Peyrard}},
  \bibinfo {author} {\bibfnamefont {D.~V.}\ \bibnamefont {Stryzheus}}, and\
  \bibinfo {author} {\bibfnamefont {Erio}\ \bibnamefont {Tosatti}}} (\bibinfo
  {year} {2012}),\ \bibfield  {title} {\enquote {\bibinfo {title} {Collective
  effects at frictional interfaces},}\ }\href
  {https://doi.org/10.1007/s11249-012-9913-z} {\bibfield  {journal} {\bibinfo
  {journal} {Tribology Letters}\ }\textbf {\bibinfo {volume} {48}}~(\bibinfo
  {number} {1}),\ \bibinfo {pages} {11--25}}\BibitemShut {NoStop}%
\bibitem [{\citenamefont {Brazda}\ \emph {et~al.}(2018)\citenamefont {Brazda},
  \citenamefont {Silva}, \citenamefont {Manini}, \citenamefont {Vanossi},
  \citenamefont {Guerra}, \citenamefont {Tosatti},\ and\ \citenamefont
  {Bechinger}}]{Brazda.prx.2018}%
  \BibitemOpen
  \bibfield  {author} {\bibinfo {author} {\bibnamefont {Brazda}, \bibfnamefont
  {T}}, \bibinfo {author} {\bibfnamefont {A.}~\bibnamefont {Silva}}, \bibinfo
  {author} {\bibfnamefont {N.}~\bibnamefont {Manini}}, \bibinfo {author}
  {\bibfnamefont {A.}~\bibnamefont {Vanossi}}, \bibinfo {author} {\bibfnamefont
  {R.}~\bibnamefont {Guerra}}, \bibinfo {author} {\bibfnamefont
  {E.}~\bibnamefont {Tosatti}}, and\ \bibinfo {author} {\bibfnamefont
  {C.}~\bibnamefont {Bechinger}}} (\bibinfo {year} {2018}),\ \bibfield  {title}
  {\enquote {\bibinfo {title} {Experimental observation of the aubry transition
  in two-dimensional colloidal monolayers},}\ }\href
  {https://doi.org/10.1103/PhysRevX.8.011050} {\bibfield  {journal} {\bibinfo
  {journal} {Phys. Rev. X}\ }\textbf {\bibinfo {volume} {8}},\ \bibinfo {pages}
  {011050}}\BibitemShut {NoStop}%
\bibitem [{\citenamefont {Brenner}\ \emph {et~al.}(2002)\citenamefont
  {Brenner}, \citenamefont {Shenderova}, \citenamefont {Harrison},
  \citenamefont {Stuart}, \citenamefont {Ni},\ and\ \citenamefont
  {Sinnott}}]{Brenner.jpcm.2002}%
  \BibitemOpen
  \bibfield  {author} {\bibinfo {author} {\bibnamefont {Brenner}, \bibfnamefont
  {Donald~W}}, \bibinfo {author} {\bibfnamefont {Olga~A}\ \bibnamefont
  {Shenderova}}, \bibinfo {author} {\bibfnamefont {Judith~A}\ \bibnamefont
  {Harrison}}, \bibinfo {author} {\bibfnamefont {Steven~J}\ \bibnamefont
  {Stuart}}, \bibinfo {author} {\bibfnamefont {Boris}\ \bibnamefont {Ni}}, and\
  \bibinfo {author} {\bibfnamefont {Susan~B}\ \bibnamefont {Sinnott}}}
  (\bibinfo {year} {2002}),\ \bibfield  {title} {\enquote {\bibinfo {title} {A
  second-generation reactive empirical bond order ({REBO}) potential energy
  expression for hydrocarbons},}\ }\href
  {https://doi.org/10.1088/0953-8984/14/4/312} {\bibfield  {journal} {\bibinfo
  {journal} {Journal of Physics: Condensed Matter}\ }\textbf {\bibinfo {volume}
  {14}}~(\bibinfo {number} {4}),\ \bibinfo {pages} {783--802}}\BibitemShut
  {NoStop}%
\bibitem [{\citenamefont {Butler}\ \emph {et~al.}(2013)\citenamefont {Butler},
  \citenamefont {Hollen}, \citenamefont {Cao}, \citenamefont {Cui},
  \citenamefont {Gupta}, \citenamefont {Gutiérrez}, \citenamefont {Heinz},
  \citenamefont {Hong}, \citenamefont {Huang}, \citenamefont {Ismach},
  \citenamefont {Johnston-Halperin}, \citenamefont {Kuno}, \citenamefont
  {Plashnitsa}, \citenamefont {Robinson}, \citenamefont {Ruoff}, \citenamefont
  {Salahuddin}, \citenamefont {Shan}, \citenamefont {Shi}, \citenamefont
  {Spencer}, \citenamefont {Terrones}, \citenamefont {Windl},\ and\
  \citenamefont {Goldberger}}]{Butler.acsnano.2013}%
  \BibitemOpen
  \bibfield  {author} {\bibinfo {author} {\bibnamefont {Butler}, \bibfnamefont
  {Sheneve~Z}}, \bibinfo {author} {\bibfnamefont {Shawna~M.}\ \bibnamefont
  {Hollen}}, \bibinfo {author} {\bibfnamefont {Linyou}\ \bibnamefont {Cao}},
  \bibinfo {author} {\bibfnamefont {Yi}~\bibnamefont {Cui}}, \bibinfo {author}
  {\bibfnamefont {Jay~A.}\ \bibnamefont {Gupta}}, \bibinfo {author}
  {\bibfnamefont {Humberto~R.}\ \bibnamefont {Gutiérrez}}, \bibinfo {author}
  {\bibfnamefont {Tony~F.}\ \bibnamefont {Heinz}}, \bibinfo {author}
  {\bibfnamefont {Seung~Sae}\ \bibnamefont {Hong}}, \bibinfo {author}
  {\bibfnamefont {Jiaxing}\ \bibnamefont {Huang}}, \bibinfo {author}
  {\bibfnamefont {Ariel~F.}\ \bibnamefont {Ismach}}, \bibinfo {author}
  {\bibfnamefont {Ezekiel}\ \bibnamefont {Johnston-Halperin}}, \bibinfo
  {author} {\bibfnamefont {Masaru}\ \bibnamefont {Kuno}}, \bibinfo {author}
  {\bibfnamefont {Vladimir~V.}\ \bibnamefont {Plashnitsa}}, \bibinfo {author}
  {\bibfnamefont {Richard~D.}\ \bibnamefont {Robinson}}, \bibinfo {author}
  {\bibfnamefont {Rodney~S.}\ \bibnamefont {Ruoff}}, \bibinfo {author}
  {\bibfnamefont {Sayeef}\ \bibnamefont {Salahuddin}}, \bibinfo {author}
  {\bibfnamefont {Jie}\ \bibnamefont {Shan}}, \bibinfo {author} {\bibfnamefont
  {Li}~\bibnamefont {Shi}}, \bibinfo {author} {\bibfnamefont {Michael~G.}\
  \bibnamefont {Spencer}}, \bibinfo {author} {\bibfnamefont {Mauricio}\
  \bibnamefont {Terrones}}, \bibinfo {author} {\bibfnamefont {Wolfgang}\
  \bibnamefont {Windl}}, and\ \bibinfo {author} {\bibfnamefont {Joshua~E.}\
  \bibnamefont {Goldberger}}} (\bibinfo {year} {2013}),\ \bibfield  {title}
  {\enquote {\bibinfo {title} {Progress, challenges, and opportunities in
  two-dimensional materials beyond graphene},}\ }\href
  {https://doi.org/10.1021/nn400280c} {\bibfield  {journal} {\bibinfo
  {journal} {ACS Nano}\ }\textbf {\bibinfo {volume} {7}}~(\bibinfo {number}
  {4}),\ \bibinfo {pages} {2898--2926}},\ \bibinfo {note} {pMID:
  23464873}\BibitemShut {NoStop}%
\bibitem [{\citenamefont {Butz}\ \emph {et~al.}(2014)\citenamefont {Butz},
  \citenamefont {Dolle}, \citenamefont {Niekiel}, \citenamefont {Weber},
  \citenamefont {Waldmann}, \citenamefont {Weber}, \citenamefont {Meyer},\ and\
  \citenamefont {Spiecker}}]{Butz.Nature.2014}%
  \BibitemOpen
  \bibfield  {author} {\bibinfo {author} {\bibnamefont {Butz}, \bibfnamefont
  {Benjamin}}, \bibinfo {author} {\bibfnamefont {Christian}\ \bibnamefont
  {Dolle}}, \bibinfo {author} {\bibfnamefont {Florian}\ \bibnamefont
  {Niekiel}}, \bibinfo {author} {\bibfnamefont {Konstantin}\ \bibnamefont
  {Weber}}, \bibinfo {author} {\bibfnamefont {Daniel}\ \bibnamefont
  {Waldmann}}, \bibinfo {author} {\bibfnamefont {Heiko~B.}\ \bibnamefont
  {Weber}}, \bibinfo {author} {\bibfnamefont {Bernd}\ \bibnamefont {Meyer}},
  and\ \bibinfo {author} {\bibfnamefont {Erdmann}\ \bibnamefont {Spiecker}}}
  (\bibinfo {year} {2014}),\ \bibfield  {title} {\enquote {\bibinfo {title}
  {Dislocations in bilayer graphene},}\ }\href
  {https://doi.org/10.1038/nature12780} {\bibfield  {journal} {\bibinfo
  {journal} {Nature}\ }\textbf {\bibinfo {volume} {505}}~(\bibinfo {number}
  {7484}),\ \bibinfo {pages} {533--537}}\BibitemShut {NoStop}%
\bibitem [{\citenamefont {Buzio}\ \emph {et~al.}(2021)\citenamefont {Buzio},
  \citenamefont {Gerbi}, \citenamefont {Bernini}, \citenamefont {Repetto},\
  and\ \citenamefont {Vanossi}}]{Buzio.carbon.2021}%
  \BibitemOpen
  \bibfield  {author} {\bibinfo {author} {\bibnamefont {Buzio}, \bibfnamefont
  {Renato}}, \bibinfo {author} {\bibfnamefont {Andrea}\ \bibnamefont {Gerbi}},
  \bibinfo {author} {\bibfnamefont {Cristina}\ \bibnamefont {Bernini}},
  \bibinfo {author} {\bibfnamefont {Luca}\ \bibnamefont {Repetto}}, and\
  \bibinfo {author} {\bibfnamefont {Andrea}\ \bibnamefont {Vanossi}}} (\bibinfo
  {year} {2021}),\ \bibfield  {title} {\enquote {\bibinfo {title} {Graphite
  superlubricity enabled by triboinduced nanocontacts},}\ }\href
  {https://doi.org/https://doi.org/10.1016/j.carbon.2021.08.071} {\bibfield
  {journal} {\bibinfo  {journal} {Carbon}\ }\textbf {\bibinfo {volume} {184}},\
  \bibinfo {pages} {875--890}}\BibitemShut {NoStop}%
\bibitem [{\citenamefont {Buzio}\ \emph {et~al.}(2022)\citenamefont {Buzio},
  \citenamefont {Gerbi}, \citenamefont {Bernini}, \citenamefont {Repetto},\
  and\ \citenamefont {Vanossi}}]{Buzio.Langmuir.2022}%
  \BibitemOpen
  \bibfield  {author} {\bibinfo {author} {\bibnamefont {Buzio}, \bibfnamefont
  {Renato}}, \bibinfo {author} {\bibfnamefont {Andrea}\ \bibnamefont {Gerbi}},
  \bibinfo {author} {\bibfnamefont {Cristina}\ \bibnamefont {Bernini}},
  \bibinfo {author} {\bibfnamefont {Luca}\ \bibnamefont {Repetto}}, and\
  \bibinfo {author} {\bibfnamefont {Andrea}\ \bibnamefont {Vanossi}}} (\bibinfo
  {year} {2022}),\ \bibfield  {title} {\enquote {\bibinfo {title} {Sliding
  friction and superlubricity of colloidal afm probes coated by tribo-induced
  graphitic transfer layers},}\ }\href
  {https://doi.org/10.1021/acs.langmuir.2c02030} {\bibfield  {journal}
  {\bibinfo  {journal} {Langmuir}\ }\textbf {\bibinfo {volume}
  {null}}~(\bibinfo {number} {null}),\ \bibinfo {pages} {null}}\BibitemShut
  {NoStop}%
\bibitem [{\citenamefont {Cao}\ \emph {et~al.}(2019)\citenamefont {Cao},
  \citenamefont {Panizon}, \citenamefont {Vanossi}, \citenamefont {Manini},\
  and\ \citenamefont {Bechinger}}]{Cao.NaturePhysics.2019}%
  \BibitemOpen
  \bibfield  {author} {\bibinfo {author} {\bibnamefont {Cao}, \bibfnamefont
  {Xin}}, \bibinfo {author} {\bibfnamefont {Emanuele}\ \bibnamefont {Panizon}},
  \bibinfo {author} {\bibfnamefont {Andrea}\ \bibnamefont {Vanossi}}, \bibinfo
  {author} {\bibfnamefont {Nicola}\ \bibnamefont {Manini}}, and\ \bibinfo
  {author} {\bibfnamefont {Clemens}\ \bibnamefont {Bechinger}}} (\bibinfo
  {year} {2019}),\ \bibfield  {title} {\enquote {\bibinfo {title}
  {Orientational and directional locking of colloidal clusters driven across
  periodic surfaces},}\ }\href {https://doi.org/10.1038/s41567-019-0515-7}
  {\bibfield  {journal} {\bibinfo  {journal} {Nature Physics}\ }\textbf
  {\bibinfo {volume} {15}}~(\bibinfo {number} {8}),\ \bibinfo {pages}
  {776--780}}\BibitemShut {NoStop}%
\bibitem [{\citenamefont {Cao}\ \emph {et~al.}(2022)\citenamefont {Cao},
  \citenamefont {Silva}, \citenamefont {Panizon}, \citenamefont {Vanossi},
  \citenamefont {Manini}, \citenamefont {Tosatti},\ and\ \citenamefont
  {Bechinger}}]{Cao.prx.2022}%
  \BibitemOpen
  \bibfield  {author} {\bibinfo {author} {\bibnamefont {Cao}, \bibfnamefont
  {Xin}}, \bibinfo {author} {\bibfnamefont {Andrea}\ \bibnamefont {Silva}},
  \bibinfo {author} {\bibfnamefont {Emanuele}\ \bibnamefont {Panizon}},
  \bibinfo {author} {\bibfnamefont {Andrea}\ \bibnamefont {Vanossi}}, \bibinfo
  {author} {\bibfnamefont {Nicola}\ \bibnamefont {Manini}}, \bibinfo {author}
  {\bibfnamefont {Erio}\ \bibnamefont {Tosatti}}, and\ \bibinfo {author}
  {\bibfnamefont {Clemens}\ \bibnamefont {Bechinger}}} (\bibinfo {year}
  {2022}),\ \bibfield  {title} {\enquote {\bibinfo {title} {Moir\'e-pattern
  evolution couples rotational and translational friction at crystalline
  interfaces},}\ }\href {https://doi.org/10.1103/PhysRevX.12.021059} {\bibfield
   {journal} {\bibinfo  {journal} {Phys. Rev. X}\ }\textbf {\bibinfo {volume}
  {12}},\ \bibinfo {pages} {021059}}\BibitemShut {NoStop}%
\bibitem [{\citenamefont {Cao}\ \emph {et~al.}(2018{\natexlab{a}})\citenamefont
  {Cao}, \citenamefont {Fatemi}, \citenamefont {Demir}, \citenamefont {Fang},
  \citenamefont {Tomarken}, \citenamefont {Luo}, \citenamefont
  {Sanchez-Yamagishi}, \citenamefont {Watanabe}, \citenamefont {Taniguchi},
  \citenamefont {Kaxiras}, \citenamefont {Ashoori},\ and\ \citenamefont
  {Jarillo-Herrero}}]{Cao.nature.2018-1}%
  \BibitemOpen
  \bibfield  {author} {\bibinfo {author} {\bibnamefont {Cao}, \bibfnamefont
  {Yuan}}, \bibinfo {author} {\bibfnamefont {Valla}\ \bibnamefont {Fatemi}},
  \bibinfo {author} {\bibfnamefont {Ahmet}\ \bibnamefont {Demir}}, \bibinfo
  {author} {\bibfnamefont {Shiang}\ \bibnamefont {Fang}}, \bibinfo {author}
  {\bibfnamefont {Spencer~L.}\ \bibnamefont {Tomarken}}, \bibinfo {author}
  {\bibfnamefont {Jason~Y.}\ \bibnamefont {Luo}}, \bibinfo {author}
  {\bibfnamefont {Javier~D.}\ \bibnamefont {Sanchez-Yamagishi}}, \bibinfo
  {author} {\bibfnamefont {Kenji}\ \bibnamefont {Watanabe}}, \bibinfo {author}
  {\bibfnamefont {Takashi}\ \bibnamefont {Taniguchi}}, \bibinfo {author}
  {\bibfnamefont {Efthimios}\ \bibnamefont {Kaxiras}}, \bibinfo {author}
  {\bibfnamefont {Ray~C.}\ \bibnamefont {Ashoori}}, and\ \bibinfo {author}
  {\bibfnamefont {Pablo}\ \bibnamefont {Jarillo-Herrero}}} (\bibinfo {year}
  {2018}{\natexlab{a}}),\ \bibfield  {title} {\enquote {\bibinfo {title}
  {Correlated insulator behaviour at half-filling in magic-angle graphene
  superlattices},}\ }\href {https://doi.org/10.1038/nature26154} {\bibfield
  {journal} {\bibinfo  {journal} {Nature}\ }\textbf {\bibinfo {volume}
  {556}}~(\bibinfo {number} {7699}),\ \bibinfo {pages} {80--84}}\BibitemShut
  {NoStop}%
\bibitem [{\citenamefont {Cao}\ \emph {et~al.}(2018{\natexlab{b}})\citenamefont
  {Cao}, \citenamefont {Fatemi}, \citenamefont {Fang}, \citenamefont
  {Watanabe}, \citenamefont {Taniguchi}, \citenamefont {Kaxiras},\ and\
  \citenamefont {Jarillo-Herrero}}]{Cao.nature.2018-2}%
  \BibitemOpen
  \bibfield  {author} {\bibinfo {author} {\bibnamefont {Cao}, \bibfnamefont
  {Yuan}}, \bibinfo {author} {\bibfnamefont {Valla}\ \bibnamefont {Fatemi}},
  \bibinfo {author} {\bibfnamefont {Shiang}\ \bibnamefont {Fang}}, \bibinfo
  {author} {\bibfnamefont {Kenji}\ \bibnamefont {Watanabe}}, \bibinfo {author}
  {\bibfnamefont {Takashi}\ \bibnamefont {Taniguchi}}, \bibinfo {author}
  {\bibfnamefont {Efthimios}\ \bibnamefont {Kaxiras}}, and\ \bibinfo {author}
  {\bibfnamefont {Pablo}\ \bibnamefont {Jarillo-Herrero}}} (\bibinfo {year}
  {2018}{\natexlab{b}}),\ \bibfield  {title} {\enquote {\bibinfo {title}
  {Unconventional superconductivity in magic-angle graphene superlattices},}\
  }\href {https://doi.org/10.1038/nature26160} {\bibfield  {journal} {\bibinfo
  {journal} {Nature}\ }\textbf {\bibinfo {volume} {556}}~(\bibinfo {number}
  {7699}),\ \bibinfo {pages} {43--50}}\BibitemShut {NoStop}%
\bibitem [{\citenamefont {Chen}\ \emph {et~al.}(2019)\citenamefont {Chen},
  \citenamefont {Khajeh}, \citenamefont {Martini},\ and\ \citenamefont
  {Kim}}]{Chen.Sci.Adv.2019}%
  \BibitemOpen
  \bibfield  {author} {\bibinfo {author} {\bibnamefont {Chen}, \bibfnamefont
  {Zhe}}, \bibinfo {author} {\bibfnamefont {Arash}\ \bibnamefont {Khajeh}},
  \bibinfo {author} {\bibfnamefont {Ashlie}\ \bibnamefont {Martini}}, and\
  \bibinfo {author} {\bibfnamefont {Seong~H.}\ \bibnamefont {Kim}}} (\bibinfo
  {year} {2019}),\ \bibfield  {title} {\enquote {\bibinfo {title} {Chemical and
  physical origins of friction on surfaces with atomic steps},}\ }\href
  {https://doi.org/10.1126/sciadv.aaw0513} {\bibfield  {journal} {\bibinfo
  {journal} {Science Advances}\ }\textbf {\bibinfo {volume} {5}}~(\bibinfo
  {number} {8}),\ \bibinfo {pages} {eaaw0513}}\BibitemShut {NoStop}%
\bibitem [{\citenamefont {Cheng}\ and\ \citenamefont
  {Ma}(2020)}]{Cheng.prm.2020}%
  \BibitemOpen
  \bibfield  {author} {\bibinfo {author} {\bibnamefont {Cheng}, \bibfnamefont
  {Yao}}, and\ \bibinfo {author} {\bibfnamefont {Ming}\ \bibnamefont {Ma}}}
  (\bibinfo {year} {2020}),\ \bibfield  {title} {\enquote {\bibinfo {title}
  {Understanding the effects of intercalated molecules on structural
  superlubric contacts},}\ }\href
  {https://doi.org/10.1103/PhysRevMaterials.4.113606} {\bibfield  {journal}
  {\bibinfo  {journal} {Phys. Rev. Materials}\ }\textbf {\bibinfo {volume}
  {4}},\ \bibinfo {pages} {113606}}\BibitemShut {NoStop}%
\bibitem [{\citenamefont {Cihan}\ \emph {et~al.}(2016)\citenamefont {Cihan},
  \citenamefont {{\.{I}}pek}, \citenamefont {Durgun},\ and\ \citenamefont
  {Baykara}}]{Cihan.nc.2016}%
  \BibitemOpen
  \bibfield  {author} {\bibinfo {author} {\bibnamefont {Cihan}, \bibfnamefont
  {Ebru}}, \bibinfo {author} {\bibfnamefont {Semran}\ \bibnamefont
  {{\.{I}}pek}}, \bibinfo {author} {\bibfnamefont {Engin}\ \bibnamefont
  {Durgun}}, and\ \bibinfo {author} {\bibfnamefont {Mehmet~Z.}\ \bibnamefont
  {Baykara}}} (\bibinfo {year} {2016}),\ \bibfield  {title} {\enquote {\bibinfo
  {title} {Structural lubricity under ambient conditions},}\ }\href
  {https://doi.org/10.1038/ncomms12055} {\bibfield  {journal} {\bibinfo
  {journal} {Nature Communications}\ }\textbf {\bibinfo {volume} {7}}~(\bibinfo
  {number} {1}),\ \bibinfo {pages} {12055}}\BibitemShut {NoStop}%
\bibitem [{\citenamefont {Consoli}\ \emph {et~al.}(2000)\citenamefont
  {Consoli}, \citenamefont {Knops},\ and\ \citenamefont
  {Fasolino}}]{Consoli.PhysRevLett.2000}%
  \BibitemOpen
  \bibfield  {author} {\bibinfo {author} {\bibnamefont {Consoli}, \bibfnamefont
  {L}}, \bibinfo {author} {\bibfnamefont {H.~J.~F.}\ \bibnamefont {Knops}},
  and\ \bibinfo {author} {\bibfnamefont {A.}~\bibnamefont {Fasolino}}}
  (\bibinfo {year} {2000}),\ \bibfield  {title} {\enquote {\bibinfo {title}
  {Onset of sliding friction in incommensurate systems},}\ }\href
  {https://doi.org/10.1103/PhysRevLett.85.302} {\bibfield  {journal} {\bibinfo
  {journal} {Phys. Rev. Lett.}\ }\textbf {\bibinfo {volume} {85}},\ \bibinfo
  {pages} {302--305}}\BibitemShut {NoStop}%
\bibitem [{\citenamefont {Deng}\ \emph {et~al.}(2018)\citenamefont {Deng},
  \citenamefont {Ma}, \citenamefont {Song}, \citenamefont {He},\ and\
  \citenamefont {Zheng}}]{Deng.nanoscale.2018}%
  \BibitemOpen
  \bibfield  {author} {\bibinfo {author} {\bibnamefont {Deng}, \bibfnamefont
  {He}}, \bibinfo {author} {\bibfnamefont {Ming}\ \bibnamefont {Ma}}, \bibinfo
  {author} {\bibfnamefont {Yiming}\ \bibnamefont {Song}}, \bibinfo {author}
  {\bibfnamefont {Qichang}\ \bibnamefont {He}}, and\ \bibinfo {author}
  {\bibfnamefont {Quanshui}\ \bibnamefont {Zheng}}} (\bibinfo {year} {2018}),\
  \bibfield  {title} {\enquote {\bibinfo {title} {Structural superlubricity in
  graphite flakes assembled under ambient conditions},}\ }\href
  {https://doi.org/10.1039/C7NR09628C} {\bibfield  {journal} {\bibinfo
  {journal} {Nanoscale}\ }\textbf {\bibinfo {volume} {10}},\ \bibinfo {pages}
  {14314--14320}}\BibitemShut {NoStop}%
\bibitem [{\citenamefont {Derjaguin}(1934)}]{Derjaguin.ZPhys.1934}%
  \BibitemOpen
  \bibfield  {author} {\bibinfo {author} {\bibnamefont {Derjaguin},
  \bibfnamefont {B}}} (\bibinfo {year} {1934}),\ \bibfield  {title} {\enquote
  {\bibinfo {title} {Molekulartheorie der {\"a}u{\ss}eren reibung},}\ }\href
  {https://doi.org/10.1007/BF01333114} {\bibfield  {journal} {\bibinfo
  {journal} {Zeitschrift f{\"u}r Physik}\ }\textbf {\bibinfo {volume}
  {88}}~(\bibinfo {number} {9}),\ \bibinfo {pages} {661--675}}\BibitemShut
  {NoStop}%
\bibitem [{\citenamefont {Dienwiebel}\ \emph {et~al.}(2004)\citenamefont
  {Dienwiebel}, \citenamefont {Verhoeven}, \citenamefont {Pradeep},
  \citenamefont {Frenken}, \citenamefont {Heimberg},\ and\ \citenamefont
  {Zandbergen}}]{Dienwiebel.prl.2004}%
  \BibitemOpen
  \bibfield  {author} {\bibinfo {author} {\bibnamefont {Dienwiebel},
  \bibfnamefont {Martin}}, \bibinfo {author} {\bibfnamefont {Gertjan~S.}\
  \bibnamefont {Verhoeven}}, \bibinfo {author} {\bibfnamefont {Namboodiri}\
  \bibnamefont {Pradeep}}, \bibinfo {author} {\bibfnamefont {Joost W.~M.}\
  \bibnamefont {Frenken}}, \bibinfo {author} {\bibfnamefont {Jennifer~A.}\
  \bibnamefont {Heimberg}}, and\ \bibinfo {author} {\bibfnamefont {Henny~W.}\
  \bibnamefont {Zandbergen}}} (\bibinfo {year} {2004}),\ \bibfield  {title}
  {\enquote {\bibinfo {title} {Superlubricity of graphite},}\ }\href
  {https://doi.org/10.1103/PhysRevLett.92.126101} {\bibfield  {journal}
  {\bibinfo  {journal} {Phys. Rev. Lett.}\ }\textbf {\bibinfo {volume} {92}},\
  \bibinfo {pages} {126101}}\BibitemShut {NoStop}%
\bibitem [{\citenamefont {Dietzel}\ \emph {et~al.}(2017)\citenamefont
  {Dietzel}, \citenamefont {Brndiar}, \citenamefont {{\v{S}}tich},\ and\
  \citenamefont {Schirmeisen}}]{Dietzel.acsnao.2017}%
  \BibitemOpen
  \bibfield  {author} {\bibinfo {author} {\bibnamefont {Dietzel}, \bibfnamefont
  {Dirk}}, \bibinfo {author} {\bibfnamefont {J{\'a}n}\ \bibnamefont {Brndiar}},
  \bibinfo {author} {\bibfnamefont {Ivan}\ \bibnamefont {{\v{S}}tich}}, and\
  \bibinfo {author} {\bibfnamefont {Andr{\'e}}\ \bibnamefont {Schirmeisen}}}
  (\bibinfo {year} {2017}),\ \bibfield  {title} {\enquote {\bibinfo {title}
  {Limitations of structural superlubricity: Chemical bonds versus contact
  size},}\ }\href {https://doi.org/10.1021/acsnano.7b02240} {\bibfield
  {journal} {\bibinfo  {journal} {ACS Nano}\ }\textbf {\bibinfo {volume}
  {11}}~(\bibinfo {number} {8}),\ \bibinfo {pages} {7642--7647}}\BibitemShut
  {NoStop}%
\bibitem [{\citenamefont {Dietzel}\ \emph {et~al.}(2013)\citenamefont
  {Dietzel}, \citenamefont {Feldmann}, \citenamefont {Schwarz}, \citenamefont
  {Fuchs},\ and\ \citenamefont {Schirmeisen}}]{Dietzel.prl.2013}%
  \BibitemOpen
  \bibfield  {author} {\bibinfo {author} {\bibnamefont {Dietzel}, \bibfnamefont
  {Dirk}}, \bibinfo {author} {\bibfnamefont {Michael}\ \bibnamefont
  {Feldmann}}, \bibinfo {author} {\bibfnamefont {Udo~D.}\ \bibnamefont
  {Schwarz}}, \bibinfo {author} {\bibfnamefont {Harald}\ \bibnamefont {Fuchs}},
  and\ \bibinfo {author} {\bibfnamefont {Andr\'e}\ \bibnamefont {Schirmeisen}}}
  (\bibinfo {year} {2013}),\ \bibfield  {title} {\enquote {\bibinfo {title}
  {Scaling laws of structural lubricity},}\ }\href
  {https://doi.org/10.1103/PhysRevLett.111.235502} {\bibfield  {journal}
  {\bibinfo  {journal} {Phys. Rev. Lett.}\ }\textbf {\bibinfo {volume} {111}},\
  \bibinfo {pages} {235502}}\BibitemShut {NoStop}%
\bibitem [{\citenamefont {Dietzel}\ \emph {et~al.}(2018)\citenamefont
  {Dietzel}, \citenamefont {de~Wijn}, \citenamefont {Vorholzer},\ and\
  \citenamefont {Schirmeisen}}]{Dietzel.Nanotechnology.2018}%
  \BibitemOpen
  \bibfield  {author} {\bibinfo {author} {\bibnamefont {Dietzel}, \bibfnamefont
  {Dirk}}, \bibinfo {author} {\bibfnamefont {Astrid~S}\ \bibnamefont
  {de~Wijn}}, \bibinfo {author} {\bibfnamefont {Matthias}\ \bibnamefont
  {Vorholzer}}, and\ \bibinfo {author} {\bibfnamefont {Andre}\ \bibnamefont
  {Schirmeisen}}} (\bibinfo {year} {2018}),\ \bibfield  {title} {\enquote
  {\bibinfo {title} {Friction fluctuations of gold nanoparticles in the
  superlubric regime},}\ }\href {https://doi.org/10.1088/1361-6528/aaac21}
  {\bibfield  {journal} {\bibinfo  {journal} {Nanotechnology}\ }\textbf
  {\bibinfo {volume} {29}}~(\bibinfo {number} {15}),\ \bibinfo {pages}
  {155702}}\BibitemShut {NoStop}%
\bibitem [{\citenamefont {Dong}\ \emph {et~al.}(2011)\citenamefont {Dong},
  \citenamefont {Vadakkepatt},\ and\ \citenamefont
  {Martini}}]{Dong.TribLett.2011}%
  \BibitemOpen
  \bibfield  {author} {\bibinfo {author} {\bibnamefont {Dong}, \bibfnamefont
  {Yalin}}, \bibinfo {author} {\bibfnamefont {Ajay}\ \bibnamefont
  {Vadakkepatt}}, and\ \bibinfo {author} {\bibfnamefont {Ashlie}\ \bibnamefont
  {Martini}}} (\bibinfo {year} {2011}),\ \bibfield  {title} {\enquote {\bibinfo
  {title} {Analytical models for atomic friction},}\ }\href
  {https://doi.org/10.1007/s11249-011-9850-2} {\bibfield  {journal} {\bibinfo
  {journal} {Tribology Letters}\ }\textbf {\bibinfo {volume} {44}}~(\bibinfo
  {number} {3}),\ \bibinfo {pages} {367}}\BibitemShut {NoStop}%
\bibitem [{\citenamefont {Dowson}(1979)}]{Dowson.1979}%
  \BibitemOpen
  \bibfield  {author} {\bibinfo {author} {\bibnamefont {Dowson}, \bibfnamefont
  {D}}} (\bibinfo {year} {1979}),\ \href@noop {} {\emph {\bibinfo {title}
  {History of tribology}}}\ (\bibinfo  {publisher} {Longman,
  London})\BibitemShut {NoStop}%
\bibitem [{\citenamefont {Dudko}\ \emph {et~al.}(2002)\citenamefont {Dudko},
  \citenamefont {Filippov}, \citenamefont {Klafter},\ and\ \citenamefont
  {Urbakh}}]{Dudko.chemphyslett.2002}%
  \BibitemOpen
  \bibfield  {author} {\bibinfo {author} {\bibnamefont {Dudko}, \bibfnamefont
  {OK}}, \bibinfo {author} {\bibfnamefont {A.E.}\ \bibnamefont {Filippov}},
  \bibinfo {author} {\bibfnamefont {J.}~\bibnamefont {Klafter}}, and\ \bibinfo
  {author} {\bibfnamefont {M.}~\bibnamefont {Urbakh}}} (\bibinfo {year}
  {2002}),\ \bibfield  {title} {\enquote {\bibinfo {title} {Dynamic force
  spectroscopy: a fokker–planck approach},}\ }\href
  {https://doi.org/https://doi.org/10.1016/S0009-2614(01)01469-5} {\bibfield
  {journal} {\bibinfo  {journal} {Chemical Physics Letters}\ }\textbf {\bibinfo
  {volume} {352}}~(\bibinfo {number} {5}),\ \bibinfo {pages}
  {499--504}}\BibitemShut {NoStop}%
\bibitem [{\citenamefont {van~den Ende}\ \emph {et~al.}(2012)\citenamefont
  {van~den Ende}, \citenamefont {de~Wijn},\ and\ \citenamefont
  {Fasolino}}]{van.Jphys.2012}%
  \BibitemOpen
  \bibfield  {author} {\bibinfo {author} {\bibnamefont {van~den Ende},
  \bibfnamefont {Joost~A}}, \bibinfo {author} {\bibfnamefont {Astrid~S}\
  \bibnamefont {de~Wijn}}, and\ \bibinfo {author} {\bibfnamefont {Annalisa}\
  \bibnamefont {Fasolino}}} (\bibinfo {year} {2012}),\ \bibfield  {title}
  {\enquote {\bibinfo {title} {The effect of temperature and velocity on
  superlubricity},}\ }\href {https://doi.org/10.1088/0953-8984/24/44/445009}
  {\bibfield  {journal} {\bibinfo  {journal} {Journal of Physics: Condensed
  Matter}\ }\textbf {\bibinfo {volume} {24}}~(\bibinfo {number} {44}),\
  \bibinfo {pages} {445009}}\BibitemShut {NoStop}%
\bibitem [{\citenamefont {Erdemir}\ and\ \citenamefont
  {Martin}(2007)}]{Erdemir.Superlubricity.2007}%
  \BibitemOpen
  \bibfield  {author} {\bibinfo {author} {\bibnamefont {Erdemir}, \bibfnamefont
  {Ali}}, and\ \bibinfo {author} {\bibfnamefont {Jean-Michel}\ \bibnamefont
  {Martin}}} (\bibinfo {year} {2007}),\ \href@noop {} {\emph {\bibinfo {title}
  {Superlubricity}}}\ (\bibinfo  {publisher} {Elsevier})\BibitemShut {NoStop}%
\bibitem [{\citenamefont {Eriksson}\ \emph {et~al.}(2002)\citenamefont
  {Eriksson}, \citenamefont {Bergman},\ and\ \citenamefont
  {Jacobson}}]{Eriksson.wear.2002}%
  \BibitemOpen
  \bibfield  {author} {\bibinfo {author} {\bibnamefont {Eriksson},
  \bibfnamefont {Mikael}}, \bibinfo {author} {\bibfnamefont {Filip}\
  \bibnamefont {Bergman}}, and\ \bibinfo {author} {\bibfnamefont {Staffan}\
  \bibnamefont {Jacobson}}} (\bibinfo {year} {2002}),\ \bibfield  {title}
  {\enquote {\bibinfo {title} {On the nature of tribological contact in
  automotive brakes},}\ }\href
  {https://doi.org/https://doi.org/10.1016/S0043-1648(01)00849-3} {\bibfield
  {journal} {\bibinfo  {journal} {Wear}\ }\textbf {\bibinfo {volume}
  {252}}~(\bibinfo {number} {1}),\ \bibinfo {pages} {26--36}}\BibitemShut
  {NoStop}%
\bibitem [{\citenamefont {Falk}\ \emph {et~al.}(2010)\citenamefont {Falk},
  \citenamefont {Sedlmeier}, \citenamefont {Joly}, \citenamefont {Netz},\ and\
  \citenamefont {Bocquet}}]{Falk.nanolett.2010}%
  \BibitemOpen
  \bibfield  {author} {\bibinfo {author} {\bibnamefont {Falk}, \bibfnamefont
  {Kerstin}}, \bibinfo {author} {\bibfnamefont {Felix}\ \bibnamefont
  {Sedlmeier}}, \bibinfo {author} {\bibfnamefont {Laurent}\ \bibnamefont
  {Joly}}, \bibinfo {author} {\bibfnamefont {Roland~R.}\ \bibnamefont {Netz}},
  and\ \bibinfo {author} {\bibfnamefont {Lyd{\'e}ric}\ \bibnamefont {Bocquet}}}
  (\bibinfo {year} {2010}),\ \bibfield  {title} {\enquote {\bibinfo {title}
  {Molecular origin of fast water transport in carbon nanotube membranes:
  Superlubricity versus curvature dependent friction},}\ }\href
  {https://doi.org/10.1021/nl1021046} {\bibfield  {journal} {\bibinfo
  {journal} {Nano Letters}\ }\textbf {\bibinfo {volume} {10}}~(\bibinfo
  {number} {10}),\ \bibinfo {pages} {4067--4073}}\BibitemShut {NoStop}%
\bibitem [{\citenamefont {Feng}\ and\ \citenamefont
  {Xu}(2022)}]{Feng.friction.2022}%
  \BibitemOpen
  \bibfield  {author} {\bibinfo {author} {\bibnamefont {Feng}, \bibfnamefont
  {Shizhe}}, and\ \bibinfo {author} {\bibfnamefont {Zhiping}\ \bibnamefont
  {Xu}}} (\bibinfo {year} {2022}),\ \bibfield  {title} {\enquote {\bibinfo
  {title} {Robustness of structural superlubricity beyond rigid models},}\
  }\href {https://doi.org/10.1007/s40544-021-0548-7} {\bibfield  {journal}
  {\bibinfo  {journal} {Friction}\ }\textbf {\bibinfo {volume} {10}}~(\bibinfo
  {number} {9}),\ \bibinfo {pages} {1382--1392}}\BibitemShut {NoStop}%
\bibitem [{\citenamefont {Filippov}\ \emph {et~al.}(2008)\citenamefont
  {Filippov}, \citenamefont {Dienwiebel}, \citenamefont {Frenken},
  \citenamefont {Klafter},\ and\ \citenamefont {Urbakh}}]{Filippov.prl.2008}%
  \BibitemOpen
  \bibfield  {author} {\bibinfo {author} {\bibnamefont {Filippov},
  \bibfnamefont {Alexander~E}}, \bibinfo {author} {\bibfnamefont {Martin}\
  \bibnamefont {Dienwiebel}}, \bibinfo {author} {\bibfnamefont {Joost W.~M.}\
  \bibnamefont {Frenken}}, \bibinfo {author} {\bibfnamefont {Joseph}\
  \bibnamefont {Klafter}}, and\ \bibinfo {author} {\bibfnamefont {Michael}\
  \bibnamefont {Urbakh}}} (\bibinfo {year} {2008}),\ \bibfield  {title}
  {\enquote {\bibinfo {title} {Torque and twist against superlubricity},}\
  }\href {https://doi.org/10.1103/PhysRevLett.100.046102} {\bibfield  {journal}
  {\bibinfo  {journal} {Phys. Rev. Lett.}\ }\textbf {\bibinfo {volume} {100}},\
  \bibinfo {pages} {046102}}\BibitemShut {NoStop}%
\bibitem [{\citenamefont {Filleter}\ \emph {et~al.}(2009)\citenamefont
  {Filleter}, \citenamefont {McChesney}, \citenamefont {Bostwick},
  \citenamefont {Rotenberg}, \citenamefont {Emtsev}, \citenamefont {Seyller},
  \citenamefont {Horn},\ and\ \citenamefont
  {Bennewitz}}]{Filleter.PhysRevLett.2009}%
  \BibitemOpen
  \bibfield  {author} {\bibinfo {author} {\bibnamefont {Filleter},
  \bibfnamefont {T}}, \bibinfo {author} {\bibfnamefont {J.~L.}\ \bibnamefont
  {McChesney}}, \bibinfo {author} {\bibfnamefont {A.}~\bibnamefont {Bostwick}},
  \bibinfo {author} {\bibfnamefont {E.}~\bibnamefont {Rotenberg}}, \bibinfo
  {author} {\bibfnamefont {K.~V.}\ \bibnamefont {Emtsev}}, \bibinfo {author}
  {\bibfnamefont {Th.}\ \bibnamefont {Seyller}}, \bibinfo {author}
  {\bibfnamefont {K.}~\bibnamefont {Horn}}, and\ \bibinfo {author}
  {\bibfnamefont {R.}~\bibnamefont {Bennewitz}}} (\bibinfo {year} {2009}),\
  \bibfield  {title} {\enquote {\bibinfo {title} {Friction and dissipation in
  epitaxial graphene films},}\ }\href
  {https://doi.org/10.1103/PhysRevLett.102.086102} {\bibfield  {journal}
  {\bibinfo  {journal} {Phys. Rev. Lett.}\ }\textbf {\bibinfo {volume} {102}},\
  \bibinfo {pages} {086102}}\BibitemShut {NoStop}%
\bibitem [{\citenamefont {Finney}\ \emph {et~al.}(2019)\citenamefont {Finney},
  \citenamefont {Yankowitz}, \citenamefont {Muraleetharan}, \citenamefont
  {Watanabe}, \citenamefont {Taniguchi}, \citenamefont {Dean},\ and\
  \citenamefont {Hone}}]{Finney.NatNano.2019}%
  \BibitemOpen
  \bibfield  {author} {\bibinfo {author} {\bibnamefont {Finney}, \bibfnamefont
  {Nathan~R}}, \bibinfo {author} {\bibfnamefont {Matthew}\ \bibnamefont
  {Yankowitz}}, \bibinfo {author} {\bibfnamefont {Lithurshanaa}\ \bibnamefont
  {Muraleetharan}}, \bibinfo {author} {\bibfnamefont {K.}~\bibnamefont
  {Watanabe}}, \bibinfo {author} {\bibfnamefont {T.}~\bibnamefont {Taniguchi}},
  \bibinfo {author} {\bibfnamefont {Cory~R.}\ \bibnamefont {Dean}}, and\
  \bibinfo {author} {\bibfnamefont {James}\ \bibnamefont {Hone}}} (\bibinfo
  {year} {2019}),\ \bibfield  {title} {\enquote {\bibinfo {title} {Tunable
  crystal symmetry in graphene--boron nitride heterostructures with coexisting
  moir{\'e} superlattices},}\ }\href
  {https://doi.org/10.1038/s41565-019-0547-2} {\bibfield  {journal} {\bibinfo
  {journal} {Nature Nanotechnology}\ }\textbf {\bibinfo {volume}
  {14}}~(\bibinfo {number} {11}),\ \bibinfo {pages} {1029--1034}}\BibitemShut
  {NoStop}%
\bibitem [{\citenamefont {Gajurel}\ \emph {et~al.}(2017)\citenamefont
  {Gajurel}, \citenamefont {Kim}, \citenamefont {Wang}, \citenamefont {Dai},
  \citenamefont {Liu},\ and\ \citenamefont {Cen}}]{Gajurel.afm.2017}%
  \BibitemOpen
  \bibfield  {author} {\bibinfo {author} {\bibnamefont {Gajurel}, \bibfnamefont
  {Prakash}}, \bibinfo {author} {\bibfnamefont {Mina}\ \bibnamefont {Kim}},
  \bibinfo {author} {\bibfnamefont {Qiang}\ \bibnamefont {Wang}}, \bibinfo
  {author} {\bibfnamefont {Weitao}\ \bibnamefont {Dai}}, \bibinfo {author}
  {\bibfnamefont {Haitao}\ \bibnamefont {Liu}}, and\ \bibinfo {author}
  {\bibfnamefont {Cheng}\ \bibnamefont {Cen}}} (\bibinfo {year} {2017}),\
  \bibfield  {title} {\enquote {\bibinfo {title} {Vacancy-controlled contact
  friction in graphene},}\ }\href
  {https://doi.org/https://doi.org/10.1002/adfm.201702832} {\bibfield
  {journal} {\bibinfo  {journal} {Advanced Functional Materials}\ }\textbf
  {\bibinfo {volume} {27}}~(\bibinfo {number} {47}),\ \bibinfo {pages}
  {1702832}}\BibitemShut {NoStop}%
\bibitem [{\citenamefont {Gao}\ \emph {et~al.}(2021{\natexlab{a}})\citenamefont
  {Gao}, \citenamefont {Wu}, \citenamefont {Wang}, \citenamefont {Jia},
  \citenamefont {Ouyang},\ and\ \citenamefont {Liu}}]{Gao.acsami.2021}%
  \BibitemOpen
  \bibfield  {author} {\bibinfo {author} {\bibnamefont {Gao}, \bibfnamefont
  {Enlai}}, \bibinfo {author} {\bibfnamefont {Bozhao}\ \bibnamefont {Wu}},
  \bibinfo {author} {\bibfnamefont {Yelingyi}\ \bibnamefont {Wang}}, \bibinfo
  {author} {\bibfnamefont {Xiangzheng}\ \bibnamefont {Jia}}, \bibinfo {author}
  {\bibfnamefont {Wengen}\ \bibnamefont {Ouyang}}, and\ \bibinfo {author}
  {\bibfnamefont {Ze}~\bibnamefont {Liu}}} (\bibinfo {year}
  {2021}{\natexlab{a}}),\ \bibfield  {title} {\enquote {\bibinfo {title}
  {Computational prediction of superlubric layered heterojunctions},}\ }\href
  {https://doi.org/10.1021/acsami.1c04870} {\bibfield  {journal} {\bibinfo
  {journal} {ACS Applied Materials \& Interfaces}\ }\textbf {\bibinfo {volume}
  {13}}~(\bibinfo {number} {28}),\ \bibinfo {pages} {33600--33608}}\BibitemShut
  {NoStop}%
\bibitem [{\citenamefont {Gao}\ \emph {et~al.}(2004)\citenamefont {Gao},
  \citenamefont {Luedtke}, \citenamefont {Gourdon}, \citenamefont {Ruths},
  \citenamefont {Israelachvili},\ and\ \citenamefont
  {Landman}}]{Gao.jpcb.2004}%
  \BibitemOpen
  \bibfield  {author} {\bibinfo {author} {\bibnamefont {Gao}, \bibfnamefont
  {Jianping}}, \bibinfo {author} {\bibfnamefont {W.~D.}\ \bibnamefont
  {Luedtke}}, \bibinfo {author} {\bibfnamefont {D.}~\bibnamefont {Gourdon}},
  \bibinfo {author} {\bibfnamefont {M.}~\bibnamefont {Ruths}}, \bibinfo
  {author} {\bibfnamefont {J.~N.}\ \bibnamefont {Israelachvili}}, and\ \bibinfo
  {author} {\bibfnamefont {Uzi}\ \bibnamefont {Landman}}} (\bibinfo {year}
  {2004}),\ \bibfield  {title} {\enquote {\bibinfo {title} {Frictional forces
  and amontons' law: From the molecular to the macroscopic scale},}\ }\href
  {https://doi.org/10.1021/jp036362l} {\bibfield  {journal} {\bibinfo
  {journal} {The Journal of Physical Chemistry B}\ }\textbf {\bibinfo {volume}
  {108}}~(\bibinfo {number} {11}),\ \bibinfo {pages} {3410--3425}}\BibitemShut
  {NoStop}%
\bibitem [{\citenamefont {Gao}\ \emph {et~al.}(2021{\natexlab{b}})\citenamefont
  {Gao}, \citenamefont {Ouyang}, \citenamefont {Hod},\ and\ \citenamefont
  {Urbakh}}]{Gao.prb.2021}%
  \BibitemOpen
  \bibfield  {author} {\bibinfo {author} {\bibnamefont {Gao}, \bibfnamefont
  {Xiang}}, \bibinfo {author} {\bibfnamefont {Wengen}\ \bibnamefont {Ouyang}},
  \bibinfo {author} {\bibfnamefont {Oded}\ \bibnamefont {Hod}}, and\ \bibinfo
  {author} {\bibfnamefont {Michael}\ \bibnamefont {Urbakh}}} (\bibinfo {year}
  {2021}{\natexlab{b}}),\ \bibfield  {title} {\enquote {\bibinfo {title}
  {Mechanisms of frictional energy dissipation at graphene grain boundaries},}\
  }\href {https://doi.org/10.1103/PhysRevB.103.045418} {\bibfield  {journal}
  {\bibinfo  {journal} {Phys. Rev. B}\ }\textbf {\bibinfo {volume} {103}},\
  \bibinfo {pages} {045418}}\BibitemShut {NoStop}%
\bibitem [{\citenamefont {Gao}\ \emph {et~al.}(2021{\natexlab{c}})\citenamefont
  {Gao}, \citenamefont {Ouyang}, \citenamefont {Urbakh},\ and\ \citenamefont
  {Hod}}]{Gao.Nat.Comm.2021}%
  \BibitemOpen
  \bibfield  {author} {\bibinfo {author} {\bibnamefont {Gao}, \bibfnamefont
  {Xiang}}, \bibinfo {author} {\bibfnamefont {Wengen}\ \bibnamefont {Ouyang}},
  \bibinfo {author} {\bibfnamefont {Michael}\ \bibnamefont {Urbakh}}, and\
  \bibinfo {author} {\bibfnamefont {Oded}\ \bibnamefont {Hod}}} (\bibinfo
  {year} {2021}{\natexlab{c}}),\ \bibfield  {title} {\enquote {\bibinfo {title}
  {Superlubric polycrystalline graphene interfaces},}\ }\href
  {https://doi.org/10.1038/s41467-021-25750-w} {\bibfield  {journal} {\bibinfo
  {journal} {Nature Communications}\ }\textbf {\bibinfo {volume}
  {12}}~(\bibinfo {number} {1}),\ \bibinfo {pages} {5694}}\BibitemShut
  {NoStop}%
\bibitem [{\citenamefont {Gao}\ \emph {et~al.}(2022)\citenamefont {Gao},
  \citenamefont {Urbakh},\ and\ \citenamefont {Hod}}]{Gao.prl.2023}%
  \BibitemOpen
  \bibfield  {author} {\bibinfo {author} {\bibnamefont {Gao}, \bibfnamefont
  {Xiang}}, \bibinfo {author} {\bibfnamefont {Michael}\ \bibnamefont {Urbakh}},
  and\ \bibinfo {author} {\bibfnamefont {Oded}\ \bibnamefont {Hod}}} (\bibinfo
  {year} {2022}),\ \bibfield  {title} {\enquote {\bibinfo {title} {Stick-slip
  dynamics of moir\'e superstructures in polycrystalline 2d material
  interfaces},}\ }\href {https://doi.org/10.1103/PhysRevLett.129.276101}
  {\bibfield  {journal} {\bibinfo  {journal} {Phys. Rev. Lett.}\ }\textbf
  {\bibinfo {volume} {129}},\ \bibinfo {pages} {276101}}\BibitemShut {NoStop}%
\bibitem [{\citenamefont {Geim}\ and\ \citenamefont
  {Grigorieva}(2013)}]{Geim.nature.2013}%
  \BibitemOpen
  \bibfield  {author} {\bibinfo {author} {\bibnamefont {Geim}, \bibfnamefont
  {A~K}}, and\ \bibinfo {author} {\bibfnamefont {I.~V.}\ \bibnamefont
  {Grigorieva}}} (\bibinfo {year} {2013}),\ \bibfield  {title} {\enquote
  {\bibinfo {title} {Van der waals heterostructures},}\ }\href
  {https://doi.org/10.1038/nature12385} {\bibfield  {journal} {\bibinfo
  {journal} {Nature}\ }\textbf {\bibinfo {volume} {499}}~(\bibinfo {number}
  {7459}),\ \bibinfo {pages} {419--425}}\BibitemShut {NoStop}%
\bibitem [{\citenamefont {Geim}\ and\ \citenamefont
  {Novoselov}(2007)}]{Geim.natmater.2007}%
  \BibitemOpen
  \bibfield  {author} {\bibinfo {author} {\bibnamefont {Geim}, \bibfnamefont
  {A~K}}, and\ \bibinfo {author} {\bibfnamefont {K.~S.}\ \bibnamefont
  {Novoselov}}} (\bibinfo {year} {2007}),\ \bibfield  {title} {\enquote
  {\bibinfo {title} {The rise of graphene},}\ }\href
  {https://doi.org/10.1038/nmat1849} {\bibfield  {journal} {\bibinfo  {journal}
  {Nature Materials}\ }\textbf {\bibinfo {volume} {6}}~(\bibinfo {number}
  {3}),\ \bibinfo {pages} {183--191}}\BibitemShut {NoStop}%
\bibitem [{\citenamefont {Gigli}\ \emph {et~al.}(2017)\citenamefont {Gigli},
  \citenamefont {Manini}, \citenamefont {Benassi}, \citenamefont {Tosatti},
  \citenamefont {Vanossi},\ and\ \citenamefont {Guerra}}]{Gigli.2dmater.2017}%
  \BibitemOpen
  \bibfield  {author} {\bibinfo {author} {\bibnamefont {Gigli}, \bibfnamefont
  {L}}, \bibinfo {author} {\bibfnamefont {N}~\bibnamefont {Manini}}, \bibinfo
  {author} {\bibfnamefont {A}~\bibnamefont {Benassi}}, \bibinfo {author}
  {\bibfnamefont {E}~\bibnamefont {Tosatti}}, \bibinfo {author} {\bibfnamefont
  {A}~\bibnamefont {Vanossi}}, and\ \bibinfo {author} {\bibfnamefont
  {R}~\bibnamefont {Guerra}}} (\bibinfo {year} {2017}),\ \bibfield  {title}
  {\enquote {\bibinfo {title} {Graphene nanoribbons on gold: understanding
  superlubricity and edge effects},}\ }\href
  {https://doi.org/10.1088/2053-1583/aa7fdf} {\bibfield  {journal} {\bibinfo
  {journal} {2D Materials}\ }\textbf {\bibinfo {volume} {4}}~(\bibinfo {number}
  {4}),\ \bibinfo {pages} {045003}}\BibitemShut {NoStop}%
\bibitem [{\citenamefont {Gigli}\ \emph {et~al.}(2018)\citenamefont {Gigli},
  \citenamefont {Manini}, \citenamefont {Tosatti}, \citenamefont {Guerra},\
  and\ \citenamefont {Vanossi}}]{Gigli.nanoscale.2018}%
  \BibitemOpen
  \bibfield  {author} {\bibinfo {author} {\bibnamefont {Gigli}, \bibfnamefont
  {L}}, \bibinfo {author} {\bibfnamefont {N.}~\bibnamefont {Manini}}, \bibinfo
  {author} {\bibfnamefont {E.}~\bibnamefont {Tosatti}}, \bibinfo {author}
  {\bibfnamefont {R.}~\bibnamefont {Guerra}}, and\ \bibinfo {author}
  {\bibfnamefont {A.}~\bibnamefont {Vanossi}}} (\bibinfo {year} {2018}),\
  \bibfield  {title} {\enquote {\bibinfo {title} {Lifted graphene nanoribbons
  on gold: from smooth sliding to multiple stick-slip regimes},}\ }\href
  {https://doi.org/10.1039/C7NR07857A} {\bibfield  {journal} {\bibinfo
  {journal} {Nanoscale}\ }\textbf {\bibinfo {volume} {10}},\ \bibinfo {pages}
  {2073--2080}}\BibitemShut {NoStop}%
\bibitem [{\citenamefont {Gnecco}\ \emph {et~al.}(2000)\citenamefont {Gnecco},
  \citenamefont {Bennewitz}, \citenamefont {Gyalog}, \citenamefont {Loppacher},
  \citenamefont {Bammerlin}, \citenamefont {Meyer},\ and\ \citenamefont
  {G\"untherodt}}]{Gnecco.PRL.2000}%
  \BibitemOpen
  \bibfield  {author} {\bibinfo {author} {\bibnamefont {Gnecco}, \bibfnamefont
  {E}}, \bibinfo {author} {\bibfnamefont {R.}~\bibnamefont {Bennewitz}},
  \bibinfo {author} {\bibfnamefont {T.}~\bibnamefont {Gyalog}}, \bibinfo
  {author} {\bibfnamefont {Ch.}\ \bibnamefont {Loppacher}}, \bibinfo {author}
  {\bibfnamefont {M.}~\bibnamefont {Bammerlin}}, \bibinfo {author}
  {\bibfnamefont {E.}~\bibnamefont {Meyer}}, and\ \bibinfo {author}
  {\bibfnamefont {H.-J.}\ \bibnamefont {G\"untherodt}}} (\bibinfo {year}
  {2000}),\ \bibfield  {title} {\enquote {\bibinfo {title} {Velocity dependence
  of atomic friction},}\ }\href {https://doi.org/10.1103/PhysRevLett.84.1172}
  {\bibfield  {journal} {\bibinfo  {journal} {Phys. Rev. Lett.}\ }\textbf
  {\bibinfo {volume} {84}},\ \bibinfo {pages} {1172--1175}}\BibitemShut
  {NoStop}%
\bibitem [{\citenamefont {Gongyang}\ \emph {et~al.}(2020)\citenamefont
  {Gongyang}, \citenamefont {Ouyang}, \citenamefont {Qu}, \citenamefont
  {Urbakh}, \citenamefont {Quan}, \citenamefont {Ma},\ and\ \citenamefont
  {Zheng}}]{Gongyang.friction.2020}%
  \BibitemOpen
  \bibfield  {author} {\bibinfo {author} {\bibnamefont {Gongyang},
  \bibfnamefont {Yujie}}, \bibinfo {author} {\bibfnamefont {Wengen}\
  \bibnamefont {Ouyang}}, \bibinfo {author} {\bibfnamefont {Cangyu}\
  \bibnamefont {Qu}}, \bibinfo {author} {\bibfnamefont {Michael}\ \bibnamefont
  {Urbakh}}, \bibinfo {author} {\bibfnamefont {Baogang}\ \bibnamefont {Quan}},
  \bibinfo {author} {\bibfnamefont {Ming}\ \bibnamefont {Ma}}, and\ \bibinfo
  {author} {\bibfnamefont {Quanshui}\ \bibnamefont {Zheng}}} (\bibinfo {year}
  {2020}),\ \bibfield  {title} {\enquote {\bibinfo {title} {Temperature and
  velocity dependent friction of a microscale graphite-dlc heterostructure},}\
  }\href {https://doi.org/10.1007/s40544-019-0288-0} {\bibfield  {journal}
  {\bibinfo  {journal} {Friction}\ }\textbf {\bibinfo {volume} {8}}~(\bibinfo
  {number} {2}),\ \bibinfo {pages} {462--470}}\BibitemShut {NoStop}%
\bibitem [{\citenamefont {Guerra}\ \emph {et~al.}(2010)\citenamefont {Guerra},
  \citenamefont {Tartaglino}, \citenamefont {Vanossi},\ and\ \citenamefont
  {Tosatti}}]{Guerra.NatMater.2010}%
  \BibitemOpen
  \bibfield  {author} {\bibinfo {author} {\bibnamefont {Guerra}, \bibfnamefont
  {Roberto}}, \bibinfo {author} {\bibfnamefont {Ugo}\ \bibnamefont
  {Tartaglino}}, \bibinfo {author} {\bibfnamefont {Andrea}\ \bibnamefont
  {Vanossi}}, and\ \bibinfo {author} {\bibfnamefont {Erio}\ \bibnamefont
  {Tosatti}}} (\bibinfo {year} {2010}),\ \bibfield  {title} {\enquote {\bibinfo
  {title} {Ballistic nanofriction},}\ }\href {https://doi.org/10.1038/nmat2798}
  {\bibfield  {journal} {\bibinfo  {journal} {Nature Materials}\ }\textbf
  {\bibinfo {volume} {9}}~(\bibinfo {number} {8}),\ \bibinfo {pages}
  {634--637}}\BibitemShut {NoStop}%
\bibitem [{\citenamefont {Guerra}\ \emph {et~al.}(2017)\citenamefont {Guerra},
  \citenamefont {van Wijk}, \citenamefont {Vanossi}, \citenamefont {Fasolino},\
  and\ \citenamefont {Tosatti}}]{Guerra.Nanoscale.2017}%
  \BibitemOpen
  \bibfield  {author} {\bibinfo {author} {\bibnamefont {Guerra}, \bibfnamefont
  {Roberto}}, \bibinfo {author} {\bibfnamefont {Merel}\ \bibnamefont {van
  Wijk}}, \bibinfo {author} {\bibfnamefont {Andrea}\ \bibnamefont {Vanossi}},
  \bibinfo {author} {\bibfnamefont {Annalisa}\ \bibnamefont {Fasolino}}, and\
  \bibinfo {author} {\bibfnamefont {Erio}\ \bibnamefont {Tosatti}}} (\bibinfo
  {year} {2017}),\ \bibfield  {title} {\enquote {\bibinfo {title} {Graphene on
  h-bn: to align or not to align?}}\ }\href
  {https://doi.org/10.1039/C7NR02352A} {\bibfield  {journal} {\bibinfo
  {journal} {Nanoscale}\ }\textbf {\bibinfo {volume} {9}},\ \bibinfo {pages}
  {8799--8804}}\BibitemShut {NoStop}%
\bibitem [{\citenamefont {Hashimoto}\ \emph {et~al.}(2004)\citenamefont
  {Hashimoto}, \citenamefont {Suenaga}, \citenamefont {Gloter}, \citenamefont
  {Urita},\ and\ \citenamefont {Iijima}}]{Hashimoto.nature.2004}%
  \BibitemOpen
  \bibfield  {author} {\bibinfo {author} {\bibnamefont {Hashimoto},
  \bibfnamefont {Ayako}}, \bibinfo {author} {\bibfnamefont {Kazu}\ \bibnamefont
  {Suenaga}}, \bibinfo {author} {\bibfnamefont {Alexandre}\ \bibnamefont
  {Gloter}}, \bibinfo {author} {\bibfnamefont {Koki}\ \bibnamefont {Urita}},
  and\ \bibinfo {author} {\bibfnamefont {Sumio}\ \bibnamefont {Iijima}}}
  (\bibinfo {year} {2004}),\ \bibfield  {title} {\enquote {\bibinfo {title}
  {Direct evidence for atomic defects in graphene layers},}\ }\href
  {https://doi.org/10.1038/nature02817} {\bibfield  {journal} {\bibinfo
  {journal} {Nature}\ }\textbf {\bibinfo {volume} {430}}~(\bibinfo {number}
  {7002}),\ \bibinfo {pages} {870--873}}\BibitemShut {NoStop}%
\bibitem [{\citenamefont {He}\ \emph {et~al.}(1999)\citenamefont {He},
  \citenamefont {Müser},\ and\ \citenamefont {Robbins}}]{He.Science.1999}%
  \BibitemOpen
  \bibfield  {author} {\bibinfo {author} {\bibnamefont {He}, \bibfnamefont
  {Gang}}, \bibinfo {author} {\bibfnamefont {Martin~H.}\ \bibnamefont
  {Müser}}, and\ \bibinfo {author} {\bibfnamefont {Mark~O.}\ \bibnamefont
  {Robbins}}} (\bibinfo {year} {1999}),\ \bibfield  {title} {\enquote {\bibinfo
  {title} {Adsorbed layers and the origin of static friction},}\ }\href
  {https://doi.org/10.1126/science.284.5420.1650} {\bibfield  {journal}
  {\bibinfo  {journal} {Science}\ }\textbf {\bibinfo {volume} {284}}~(\bibinfo
  {number} {5420}),\ \bibinfo {pages} {1650--1652}}\BibitemShut {NoStop}%
\bibitem [{\citenamefont {Hermann}(2012)}]{Hermann.jpcm.2012}%
  \BibitemOpen
  \bibfield  {author} {\bibinfo {author} {\bibnamefont {Hermann}, \bibfnamefont
  {Klaus}}} (\bibinfo {year} {2012}),\ \bibfield  {title} {\enquote {\bibinfo
  {title} {Periodic overlayers and moir{\'{e}} patterns: theoretical studies of
  geometric properties},}\ }\href
  {https://doi.org/10.1088/0953-8984/24/31/314210} {\bibfield  {journal}
  {\bibinfo  {journal} {Journal of Physics: Condensed Matter}\ }\textbf
  {\bibinfo {volume} {24}}~(\bibinfo {number} {31}),\ \bibinfo {pages}
  {314210}}\BibitemShut {NoStop}%
\bibitem [{\citenamefont {Hirano}\ and\ \citenamefont
  {Shinjo}(1990)}]{Hirano.prb.1990}%
  \BibitemOpen
  \bibfield  {author} {\bibinfo {author} {\bibnamefont {Hirano}, \bibfnamefont
  {Motohisa}}, and\ \bibinfo {author} {\bibfnamefont {Kazumasa}\ \bibnamefont
  {Shinjo}}} (\bibinfo {year} {1990}),\ \bibfield  {title} {\enquote {\bibinfo
  {title} {Atomistic locking and friction},}\ }\href
  {https://doi.org/10.1103/PhysRevB.41.11837} {\bibfield  {journal} {\bibinfo
  {journal} {Phys. Rev. B}\ }\textbf {\bibinfo {volume} {41}},\ \bibinfo
  {pages} {11837--11851}}\BibitemShut {NoStop}%
\bibitem [{\citenamefont {Hirano}\ \emph {et~al.}(1997)\citenamefont {Hirano},
  \citenamefont {Shinjo}, \citenamefont {Kaneko},\ and\ \citenamefont
  {Murata}}]{Hirano.PhysRevLett.1997}%
  \BibitemOpen
  \bibfield  {author} {\bibinfo {author} {\bibnamefont {Hirano}, \bibfnamefont
  {Motohisa}}, \bibinfo {author} {\bibfnamefont {Kazumasa}\ \bibnamefont
  {Shinjo}}, \bibinfo {author} {\bibfnamefont {Reizo}\ \bibnamefont {Kaneko}},
  and\ \bibinfo {author} {\bibfnamefont {Yoshitada}\ \bibnamefont {Murata}}}
  (\bibinfo {year} {1997}),\ \bibfield  {title} {\enquote {\bibinfo {title}
  {Observation of superlubricity by scanning tunneling microscopy},}\ }\href
  {https://doi.org/10.1103/PhysRevLett.78.1448} {\bibfield  {journal} {\bibinfo
   {journal} {Phys. Rev. Lett.}\ }\textbf {\bibinfo {volume} {78}},\ \bibinfo
  {pages} {1448--1451}}\BibitemShut {NoStop}%
\bibitem [{\citenamefont {Hod}(2013)}]{Hod.chemphyschem.2013}%
  \BibitemOpen
  \bibfield  {author} {\bibinfo {author} {\bibnamefont {Hod}, \bibfnamefont
  {Oded}}} (\bibinfo {year} {2013}),\ \bibfield  {title} {\enquote {\bibinfo
  {title} {The registry index: A quantitative measure of materials' interfacial
  commensurability},}\ }\href
  {https://doi.org/https://doi.org/10.1002/cphc.201300259} {\bibfield
  {journal} {\bibinfo  {journal} {ChemPhysChem}\ }\textbf {\bibinfo {volume}
  {14}}~(\bibinfo {number} {11}),\ \bibinfo {pages} {2376--2391}}\BibitemShut
  {NoStop}%
\bibitem [{\citenamefont {Hod}\ \emph {et~al.}(2018)\citenamefont {Hod},
  \citenamefont {Meyer}, \citenamefont {Zheng},\ and\ \citenamefont
  {Urbakh}}]{Hod.nature.2018}%
  \BibitemOpen
  \bibfield  {author} {\bibinfo {author} {\bibnamefont {Hod}, \bibfnamefont
  {Oded}}, \bibinfo {author} {\bibfnamefont {Ernst}\ \bibnamefont {Meyer}},
  \bibinfo {author} {\bibfnamefont {Quanshui}\ \bibnamefont {Zheng}}, and\
  \bibinfo {author} {\bibfnamefont {Michael}\ \bibnamefont {Urbakh}}} (\bibinfo
  {year} {2018}),\ \bibfield  {title} {\enquote {\bibinfo {title} {Structural
  superlubricity and ultralow friction across the length scales},}\ }\href
  {https://doi.org/10.1038/s41586-018-0704-z} {\bibfield  {journal} {\bibinfo
  {journal} {Nature}\ }\textbf {\bibinfo {volume} {563}}~(\bibinfo {number}
  {7732}),\ \bibinfo {pages} {485--492}}\BibitemShut {NoStop}%
\bibitem [{\citenamefont {Huang}\ \emph {et~al.}(2020)\citenamefont {Huang},
  \citenamefont {Lin},\ and\ \citenamefont {Zheng}}]{Huang.nanoenergy.2020}%
  \BibitemOpen
  \bibfield  {author} {\bibinfo {author} {\bibnamefont {Huang}, \bibfnamefont
  {Xuanyu}}, \bibinfo {author} {\bibfnamefont {Li}~\bibnamefont {Lin}}, and\
  \bibinfo {author} {\bibfnamefont {Quanshui}\ \bibnamefont {Zheng}}} (\bibinfo
  {year} {2020}),\ \bibfield  {title} {\enquote {\bibinfo {title} {Theoretical
  study of superlubric nanogenerators with superb performances},}\ }\href
  {https://doi.org/https://doi.org/10.1016/j.nanoen.2020.104494} {\bibfield
  {journal} {\bibinfo  {journal} {Nano Energy}\ }\textbf {\bibinfo {volume}
  {70}},\ \bibinfo {pages} {104494}}\BibitemShut {NoStop}%
\bibitem [{\citenamefont {Huang}\ \emph {et~al.}(2021)\citenamefont {Huang},
  \citenamefont {Xiang}, \citenamefont {Nie}, \citenamefont {Peng},
  \citenamefont {Yang}, \citenamefont {Wu}, \citenamefont {Jiang},
  \citenamefont {Xu},\ and\ \citenamefont {Zheng}}]{Huang.nc.2021}%
  \BibitemOpen
  \bibfield  {author} {\bibinfo {author} {\bibnamefont {Huang}, \bibfnamefont
  {Xuanyu}}, \bibinfo {author} {\bibfnamefont {Xiaojian}\ \bibnamefont
  {Xiang}}, \bibinfo {author} {\bibfnamefont {Jinhui}\ \bibnamefont {Nie}},
  \bibinfo {author} {\bibfnamefont {Deli}\ \bibnamefont {Peng}}, \bibinfo
  {author} {\bibfnamefont {Fuwei}\ \bibnamefont {Yang}}, \bibinfo {author}
  {\bibfnamefont {Zhanghui}\ \bibnamefont {Wu}}, \bibinfo {author}
  {\bibfnamefont {Haiyang}\ \bibnamefont {Jiang}}, \bibinfo {author}
  {\bibfnamefont {Zhiping}\ \bibnamefont {Xu}}, and\ \bibinfo {author}
  {\bibfnamefont {Quanshui}\ \bibnamefont {Zheng}}} (\bibinfo {year} {2021}),\
  \bibfield  {title} {\enquote {\bibinfo {title} {Microscale schottky
  superlubric generator with high direct-current density and ultralong life},}\
  }\href {https://doi.org/10.1038/s41467-021-22371-1} {\bibfield  {journal}
  {\bibinfo  {journal} {Nature Communications}\ }\textbf {\bibinfo {volume}
  {12}}~(\bibinfo {number} {1}),\ \bibinfo {pages} {2268}}\BibitemShut
  {NoStop}%
\bibitem [{\citenamefont {Hurtado}\ and\ \citenamefont
  {Kim}(1999)}]{Hurtado.prsl.1999}%
  \BibitemOpen
  \bibfield  {author} {\bibinfo {author} {\bibnamefont {Hurtado}, \bibfnamefont
  {Juan~A}}, and\ \bibinfo {author} {\bibfnamefont {Kyung–Suk}\ \bibnamefont
  {Kim}}} (\bibinfo {year} {1999}),\ \bibfield  {title} {\enquote {\bibinfo
  {title} {Scale effects in friction of single\&\#x2013;asperity contacts. i.
  from concurrent slip to single\&\#x2013;dislocation\&\#x2013;assisted
  slip},}\ }\href {https://doi.org/10.1098/rspa.1999.0455} {\bibfield
  {journal} {\bibinfo  {journal} {Proceedings of the Royal Society of London.
  Series A: Mathematical, Physical and Engineering Sciences}\ }\textbf
  {\bibinfo {volume} {455}}~(\bibinfo {number} {1989}),\ \bibinfo {pages}
  {3363--3384}}\BibitemShut {NoStop}%
\bibitem [{\citenamefont {Jain}\ \emph {et~al.}(2016)\citenamefont {Jain},
  \citenamefont {Juri{\v{c}}i{\'{c}}},\ and\ \citenamefont
  {Barkema}}]{Jain.2dmater.2016}%
  \BibitemOpen
  \bibfield  {author} {\bibinfo {author} {\bibnamefont {Jain}, \bibfnamefont
  {Sandeep~K}}, \bibinfo {author} {\bibfnamefont {Vladimir}\ \bibnamefont
  {Juri{\v{c}}i{\'{c}}}}, and\ \bibinfo {author} {\bibfnamefont {Gerard~T}\
  \bibnamefont {Barkema}}} (\bibinfo {year} {2016}),\ \bibfield  {title}
  {\enquote {\bibinfo {title} {Structure of twisted and buckled bilayer
  graphene},}\ }\href {https://doi.org/10.1088/2053-1583/4/1/015018} {\bibfield
   {journal} {\bibinfo  {journal} {2D Materials}\ }\textbf {\bibinfo {volume}
  {4}}~(\bibinfo {number} {1}),\ \bibinfo {pages} {015018}}\BibitemShut
  {NoStop}%
\bibitem [{\citenamefont {Kawai}\ \emph {et~al.}(2016)\citenamefont {Kawai},
  \citenamefont {Benassi}, \citenamefont {Gnecco}, \citenamefont {Söde},
  \citenamefont {Pawlak}, \citenamefont {Feng}, \citenamefont {Müllen},
  \citenamefont {Passerone}, \citenamefont {Pignedoli}, \citenamefont
  {Ruffieux}, \citenamefont {Fasel},\ and\ \citenamefont
  {Meyer}}]{Kawai.Science.2016}%
  \BibitemOpen
  \bibfield  {author} {\bibinfo {author} {\bibnamefont {Kawai}, \bibfnamefont
  {Shigeki}}, \bibinfo {author} {\bibfnamefont {Andrea}\ \bibnamefont
  {Benassi}}, \bibinfo {author} {\bibfnamefont {Enrico}\ \bibnamefont
  {Gnecco}}, \bibinfo {author} {\bibfnamefont {Hajo}\ \bibnamefont {Söde}},
  \bibinfo {author} {\bibfnamefont {Rémy}\ \bibnamefont {Pawlak}}, \bibinfo
  {author} {\bibfnamefont {Xinliang}\ \bibnamefont {Feng}}, \bibinfo {author}
  {\bibfnamefont {Klaus}\ \bibnamefont {Müllen}}, \bibinfo {author}
  {\bibfnamefont {Daniele}\ \bibnamefont {Passerone}}, \bibinfo {author}
  {\bibfnamefont {Carlo~A.}\ \bibnamefont {Pignedoli}}, \bibinfo {author}
  {\bibfnamefont {Pascal}\ \bibnamefont {Ruffieux}}, \bibinfo {author}
  {\bibfnamefont {Roman}\ \bibnamefont {Fasel}}, and\ \bibinfo {author}
  {\bibfnamefont {Ernst}\ \bibnamefont {Meyer}}} (\bibinfo {year} {2016}),\
  \bibfield  {title} {\enquote {\bibinfo {title} {Superlubricity of graphene
  nanoribbons on gold surfaces},}\ }\href
  {https://doi.org/10.1126/science.aad3569} {\bibfield  {journal} {\bibinfo
  {journal} {Science}\ }\textbf {\bibinfo {volume} {351}}~(\bibinfo {number}
  {6276}),\ \bibinfo {pages} {957--961}}\BibitemShut {NoStop}%
\bibitem [{\citenamefont {Kazmierczak}\ \emph {et~al.}(2021)\citenamefont
  {Kazmierczak}, \citenamefont {Van~Winkle}, \citenamefont {Ophus},
  \citenamefont {Bustillo}, \citenamefont {Carr}, \citenamefont {Brown},
  \citenamefont {Ciston}, \citenamefont {Taniguchi}, \citenamefont {Watanabe},\
  and\ \citenamefont {Bediako}}]{Kazmierczak.NatMater.2021}%
  \BibitemOpen
  \bibfield  {author} {\bibinfo {author} {\bibnamefont {Kazmierczak},
  \bibfnamefont {Nathanael~P}}, \bibinfo {author} {\bibfnamefont {Madeline}\
  \bibnamefont {Van~Winkle}}, \bibinfo {author} {\bibfnamefont {Colin}\
  \bibnamefont {Ophus}}, \bibinfo {author} {\bibfnamefont {Karen~C.}\
  \bibnamefont {Bustillo}}, \bibinfo {author} {\bibfnamefont {Stephen}\
  \bibnamefont {Carr}}, \bibinfo {author} {\bibfnamefont {Hamish~G.}\
  \bibnamefont {Brown}}, \bibinfo {author} {\bibfnamefont {Jim}\ \bibnamefont
  {Ciston}}, \bibinfo {author} {\bibfnamefont {Takashi}\ \bibnamefont
  {Taniguchi}}, \bibinfo {author} {\bibfnamefont {Kenji}\ \bibnamefont
  {Watanabe}}, and\ \bibinfo {author} {\bibfnamefont {D.~Kwabena}\ \bibnamefont
  {Bediako}}} (\bibinfo {year} {2021}),\ \bibfield  {title} {\enquote {\bibinfo
  {title} {Strain fields in twisted bilayer graphene},}\ }\href
  {https://doi.org/10.1038/s41563-021-00973-w} {\bibfield  {journal} {\bibinfo
  {journal} {Nature Materials}\ }\textbf {\bibinfo {volume} {20}}~(\bibinfo
  {number} {7}),\ \bibinfo {pages} {956--963}}\BibitemShut {NoStop}%
\bibitem [{\citenamefont {Kerelsky}\ \emph {et~al.}(2019)\citenamefont
  {Kerelsky}, \citenamefont {McGilly}, \citenamefont {Kennes}, \citenamefont
  {Xian}, \citenamefont {Yankowitz}, \citenamefont {Chen}, \citenamefont
  {Watanabe}, \citenamefont {Taniguchi}, \citenamefont {Hone}, \citenamefont
  {Dean}, \citenamefont {Rubio},\ and\ \citenamefont
  {Pasupathy}}]{Kerelsky.nature.2019}%
  \BibitemOpen
  \bibfield  {author} {\bibinfo {author} {\bibnamefont {Kerelsky},
  \bibfnamefont {Alexander}}, \bibinfo {author} {\bibfnamefont {Leo~J.}\
  \bibnamefont {McGilly}}, \bibinfo {author} {\bibfnamefont {Dante~M.}\
  \bibnamefont {Kennes}}, \bibinfo {author} {\bibfnamefont {Lede}\ \bibnamefont
  {Xian}}, \bibinfo {author} {\bibfnamefont {Matthew}\ \bibnamefont
  {Yankowitz}}, \bibinfo {author} {\bibfnamefont {Shaowen}\ \bibnamefont
  {Chen}}, \bibinfo {author} {\bibfnamefont {K.}~\bibnamefont {Watanabe}},
  \bibinfo {author} {\bibfnamefont {T.}~\bibnamefont {Taniguchi}}, \bibinfo
  {author} {\bibfnamefont {James}\ \bibnamefont {Hone}}, \bibinfo {author}
  {\bibfnamefont {Cory}\ \bibnamefont {Dean}}, \bibinfo {author} {\bibfnamefont
  {Angel}\ \bibnamefont {Rubio}}, and\ \bibinfo {author} {\bibfnamefont
  {Abhay~N.}\ \bibnamefont {Pasupathy}}} (\bibinfo {year} {2019}),\ \bibfield
  {title} {\enquote {\bibinfo {title} {Maximized electron interactions at the
  magic angle in twisted bilayer graphene},}\ }\href
  {https://doi.org/10.1038/s41586-019-1431-9} {\bibfield  {journal} {\bibinfo
  {journal} {Nature}\ }\textbf {\bibinfo {volume} {572}}~(\bibinfo {number}
  {7767}),\ \bibinfo {pages} {95--100}}\BibitemShut {NoStop}%
\bibitem [{\citenamefont {Kietzig}\ \emph {et~al.}(2010)\citenamefont
  {Kietzig}, \citenamefont {Hatzikiriakos},\ and\ \citenamefont
  {Englezos}}]{Kietzig.jap.2010}%
  \BibitemOpen
  \bibfield  {author} {\bibinfo {author} {\bibnamefont {Kietzig}, \bibfnamefont
  {Anne-Marie}}, \bibinfo {author} {\bibfnamefont {Savvas~G.}\ \bibnamefont
  {Hatzikiriakos}}, and\ \bibinfo {author} {\bibfnamefont {Peter}\ \bibnamefont
  {Englezos}}} (\bibinfo {year} {2010}),\ \bibfield  {title} {\enquote
  {\bibinfo {title} {Physics of ice friction},}\ }\href
  {https://doi.org/10.1063/1.3340792} {\bibfield  {journal} {\bibinfo
  {journal} {Journal of Applied Physics}\ }\textbf {\bibinfo {volume}
  {107}}~(\bibinfo {number} {8}),\ \bibinfo {pages} {081101}}\BibitemShut
  {NoStop}%
\bibitem [{\citenamefont {Kim}\ \emph {et~al.}(2012)\citenamefont {Kim},
  \citenamefont {Coh}, \citenamefont {Tan}, \citenamefont {Regan},
  \citenamefont {Yuk}, \citenamefont {Chatterjee}, \citenamefont {Crommie},
  \citenamefont {Cohen}, \citenamefont {Louie},\ and\ \citenamefont
  {Zettl}}]{Kim.PRL.2012}%
  \BibitemOpen
  \bibfield  {author} {\bibinfo {author} {\bibnamefont {Kim}, \bibfnamefont
  {Kwanpyo}}, \bibinfo {author} {\bibfnamefont {Sinisa}\ \bibnamefont {Coh}},
  \bibinfo {author} {\bibfnamefont {Liang~Z.}\ \bibnamefont {Tan}}, \bibinfo
  {author} {\bibfnamefont {William}\ \bibnamefont {Regan}}, \bibinfo {author}
  {\bibfnamefont {Jong~Min}\ \bibnamefont {Yuk}}, \bibinfo {author}
  {\bibfnamefont {Eric}\ \bibnamefont {Chatterjee}}, \bibinfo {author}
  {\bibfnamefont {M.~F.}\ \bibnamefont {Crommie}}, \bibinfo {author}
  {\bibfnamefont {Marvin~L.}\ \bibnamefont {Cohen}}, \bibinfo {author}
  {\bibfnamefont {Steven~G.}\ \bibnamefont {Louie}}, and\ \bibinfo {author}
  {\bibfnamefont {A.}~\bibnamefont {Zettl}}} (\bibinfo {year} {2012}),\
  \bibfield  {title} {\enquote {\bibinfo {title} {Raman spectroscopy study of
  rotated double-layer graphene: Misorientation-angle dependence of electronic
  structure},}\ }\href {https://doi.org/10.1103/PhysRevLett.108.246103}
  {\bibfield  {journal} {\bibinfo  {journal} {Phys. Rev. Lett.}\ }\textbf
  {\bibinfo {volume} {108}},\ \bibinfo {pages} {246103}}\BibitemShut {NoStop}%
\bibitem [{\citenamefont {Koren}\ and\ \citenamefont
  {Duerig}(2016{\natexlab{a}})}]{Koren.prb.moire.2016}%
  \BibitemOpen
  \bibfield  {author} {\bibinfo {author} {\bibnamefont {Koren}, \bibfnamefont
  {E}}, and\ \bibinfo {author} {\bibfnamefont {U.}~\bibnamefont {Duerig}}}
  (\bibinfo {year} {2016}{\natexlab{a}}),\ \bibfield  {title} {\enquote
  {\bibinfo {title} {Moir\'e scaling of the sliding force in twisted bilayer
  graphene},}\ }\href {https://doi.org/10.1103/PhysRevB.94.045401} {\bibfield
  {journal} {\bibinfo  {journal} {Phys. Rev. B}\ }\textbf {\bibinfo {volume}
  {94}},\ \bibinfo {pages} {045401}}\BibitemShut {NoStop}%
\bibitem [{\citenamefont {Koren}\ and\ \citenamefont
  {Duerig}(2016{\natexlab{b}})}]{Koren.prb.quasi.2016}%
  \BibitemOpen
  \bibfield  {author} {\bibinfo {author} {\bibnamefont {Koren}, \bibfnamefont
  {Elad}}, and\ \bibinfo {author} {\bibfnamefont {Urs}\ \bibnamefont {Duerig}}}
  (\bibinfo {year} {2016}{\natexlab{b}}),\ \bibfield  {title} {\enquote
  {\bibinfo {title} {Superlubricity in quasicrystalline twisted bilayer
  graphene},}\ }\href {https://doi.org/10.1103/PhysRevB.93.201404} {\bibfield
  {journal} {\bibinfo  {journal} {Phys. Rev. B}\ }\textbf {\bibinfo {volume}
  {93}},\ \bibinfo {pages} {201404}}\BibitemShut {NoStop}%
\bibitem [{\citenamefont {Koren}\ \emph {et~al.}(2015)\citenamefont {Koren},
  \citenamefont {Lörtscher}, \citenamefont {Rawlings}, \citenamefont {Knoll},\
  and\ \citenamefont {Duerig}}]{Koren.science.2015}%
  \BibitemOpen
  \bibfield  {author} {\bibinfo {author} {\bibnamefont {Koren}, \bibfnamefont
  {Elad}}, \bibinfo {author} {\bibfnamefont {Emanuel}\ \bibnamefont
  {Lörtscher}}, \bibinfo {author} {\bibfnamefont {Colin}\ \bibnamefont
  {Rawlings}}, \bibinfo {author} {\bibfnamefont {Armin~W.}\ \bibnamefont
  {Knoll}}, and\ \bibinfo {author} {\bibfnamefont {Urs}\ \bibnamefont
  {Duerig}}} (\bibinfo {year} {2015}),\ \bibfield  {title} {\enquote {\bibinfo
  {title} {Adhesion and friction in mesoscopic graphite contacts},}\ }\href
  {https://doi.org/10.1126/science.aaa4157} {\bibfield  {journal} {\bibinfo
  {journal} {Science}\ }\textbf {\bibinfo {volume} {348}}~(\bibinfo {number}
  {6235}),\ \bibinfo {pages} {679--683}}\BibitemShut {NoStop}%
\bibitem [{\citenamefont {Krim}(2012)}]{Jacqueline.AdvPhys.2012}%
  \BibitemOpen
  \bibfield  {author} {\bibinfo {author} {\bibnamefont {Krim}, \bibfnamefont
  {Jacqueline}}} (\bibinfo {year} {2012}),\ \bibfield  {title} {\enquote
  {\bibinfo {title} {Friction and energy dissipation mechanisms in adsorbed
  molecules and molecularly thin films},}\ }\href@noop {} {\bibfield  {journal}
  {\bibinfo  {journal} {Advances in Physics}\ }\textbf {\bibinfo {volume}
  {61}}~(\bibinfo {number} {3}),\ \bibinfo {pages} {155--323}}\BibitemShut
  {NoStop}%
\bibitem [{\citenamefont {Krylov}\ \emph {et~al.}(2005)\citenamefont {Krylov},
  \citenamefont {Jinesh}, \citenamefont {Valk}, \citenamefont {Dienwiebel},\
  and\ \citenamefont {Frenken}}]{Krylov.pre.2005}%
  \BibitemOpen
  \bibfield  {author} {\bibinfo {author} {\bibnamefont {Krylov}, \bibfnamefont
  {S~Yu}}, \bibinfo {author} {\bibfnamefont {K.~B.}\ \bibnamefont {Jinesh}},
  \bibinfo {author} {\bibfnamefont {H.}~\bibnamefont {Valk}}, \bibinfo {author}
  {\bibfnamefont {M.}~\bibnamefont {Dienwiebel}}, and\ \bibinfo {author}
  {\bibfnamefont {J.~W.~M.}\ \bibnamefont {Frenken}}} (\bibinfo {year}
  {2005}),\ \bibfield  {title} {\enquote {\bibinfo {title} {Thermally induced
  suppression of friction at the atomic scale},}\ }\href
  {https://doi.org/10.1103/PhysRevE.71.065101} {\bibfield  {journal} {\bibinfo
  {journal} {Phys. Rev. E}\ }\textbf {\bibinfo {volume} {71}},\ \bibinfo
  {pages} {065101}}\BibitemShut {NoStop}%
\bibitem [{\citenamefont {Krylov}\ and\ \citenamefont
  {Frenken}(2014)}]{Krylov.physstatussolid.2014}%
  \BibitemOpen
  \bibfield  {author} {\bibinfo {author} {\bibnamefont {Krylov}, \bibfnamefont
  {Sergey~Yu}}, and\ \bibinfo {author} {\bibfnamefont {Joost W.~M.}\
  \bibnamefont {Frenken}}} (\bibinfo {year} {2014}),\ \bibfield  {title}
  {\enquote {\bibinfo {title} {The physics of atomic-scale friction: Basic
  considerations and open questions},}\ }\href
  {https://doi.org/https://doi.org/10.1002/pssb.201350154} {\bibfield
  {journal} {\bibinfo  {journal} {physica status solidi (b)}\ }\textbf
  {\bibinfo {volume} {251}}~(\bibinfo {number} {4}),\ \bibinfo {pages}
  {711--736}}\BibitemShut {NoStop}%
\bibitem [{\citenamefont {Trambly~de Laissardière}\ \emph
  {et~al.}(2010)\citenamefont {Trambly~de Laissardière}, \citenamefont
  {Mayou},\ and\ \citenamefont {Magaud}}]{Trambly.nanolett.2010}%
  \BibitemOpen
  \bibfield  {author} {\bibinfo {author} {\bibnamefont {Trambly~de
  Laissardière}, \bibfnamefont {G}}, \bibinfo {author} {\bibfnamefont
  {D.}~\bibnamefont {Mayou}}, and\ \bibinfo {author} {\bibfnamefont
  {L.}~\bibnamefont {Magaud}}} (\bibinfo {year} {2010}),\ \bibfield  {title}
  {\enquote {\bibinfo {title} {Localization of dirac electrons in rotated
  graphene bilayers},}\ }\href {https://doi.org/10.1021/nl902948m} {\bibfield
  {journal} {\bibinfo  {journal} {Nano Letters}\ }\textbf {\bibinfo {volume}
  {10}}~(\bibinfo {number} {3}),\ \bibinfo {pages} {804--808}}\BibitemShut
  {NoStop}%
\bibitem [{\citenamefont {Lang}\ \emph {et~al.}(2021)\citenamefont {Lang},
  \citenamefont {Xu}, \citenamefont {Zhu}, \citenamefont {Peng}, \citenamefont
  {Zou}, \citenamefont {Yu},\ and\ \citenamefont {Huang}}]{Lang.carbon.2021}%
  \BibitemOpen
  \bibfield  {author} {\bibinfo {author} {\bibnamefont {Lang}, \bibfnamefont
  {Haojie}}, \bibinfo {author} {\bibfnamefont {Yimeng}\ \bibnamefont {Xu}},
  \bibinfo {author} {\bibfnamefont {Pengzhe}\ \bibnamefont {Zhu}}, \bibinfo
  {author} {\bibfnamefont {Yitian}\ \bibnamefont {Peng}}, \bibinfo {author}
  {\bibfnamefont {Kun}\ \bibnamefont {Zou}}, \bibinfo {author} {\bibfnamefont
  {Kang}\ \bibnamefont {Yu}}, and\ \bibinfo {author} {\bibfnamefont {Yao}\
  \bibnamefont {Huang}}} (\bibinfo {year} {2021}),\ \bibfield  {title}
  {\enquote {\bibinfo {title} {Superior lubrication and electrical stability of
  graphene as highly effective solid lubricant at sliding electrical contact
  interface},}\ }\href
  {https://doi.org/https://doi.org/10.1016/j.carbon.2021.07.016} {\bibfield
  {journal} {\bibinfo  {journal} {Carbon}\ }\textbf {\bibinfo {volume} {183}},\
  \bibinfo {pages} {53--61}}\BibitemShut {NoStop}%
\bibitem [{\citenamefont {Larkin}\ and\ \citenamefont
  {Ovchinnikov}(1979)}]{Larkin.JLTP.1979}%
  \BibitemOpen
  \bibfield  {author} {\bibinfo {author} {\bibnamefont {Larkin}, \bibfnamefont
  {A~I}}, and\ \bibinfo {author} {\bibfnamefont {Yu.~N.}\ \bibnamefont
  {Ovchinnikov}}} (\bibinfo {year} {1979}),\ \bibfield  {title} {\enquote
  {\bibinfo {title} {Pinning in type ii superconductors},}\ }\href
  {https://doi.org/10.1007/BF00117160} {\bibfield  {journal} {\bibinfo
  {journal} {Journal of Low Temperature Physics}\ }\textbf {\bibinfo {volume}
  {34}}~(\bibinfo {number} {3}),\ \bibinfo {pages} {409--428}}\BibitemShut
  {NoStop}%
\bibitem [{\citenamefont {Lee}\ \emph {et~al.}(2010)\citenamefont {Lee},
  \citenamefont {Li}, \citenamefont {Kalb}, \citenamefont {Liu}, \citenamefont
  {Berger}, \citenamefont {Carpick},\ and\ \citenamefont
  {Hone}}]{Lee.science.2010}%
  \BibitemOpen
  \bibfield  {author} {\bibinfo {author} {\bibnamefont {Lee}, \bibfnamefont
  {Changgu}}, \bibinfo {author} {\bibfnamefont {Qunyang}\ \bibnamefont {Li}},
  \bibinfo {author} {\bibfnamefont {William}\ \bibnamefont {Kalb}}, \bibinfo
  {author} {\bibfnamefont {Xin-Zhou}\ \bibnamefont {Liu}}, \bibinfo {author}
  {\bibfnamefont {Helmuth}\ \bibnamefont {Berger}}, \bibinfo {author}
  {\bibfnamefont {Robert~W.}\ \bibnamefont {Carpick}}, and\ \bibinfo {author}
  {\bibfnamefont {James}\ \bibnamefont {Hone}}} (\bibinfo {year} {2010}),\
  \bibfield  {title} {\enquote {\bibinfo {title} {Frictional characteristics of
  atomically thin sheets},}\ }\href {https://doi.org/10.1126/science.1184167}
  {\bibfield  {journal} {\bibinfo  {journal} {Science}\ }\textbf {\bibinfo
  {volume} {328}}~(\bibinfo {number} {5974}),\ \bibinfo {pages}
  {76--80}}\BibitemShut {NoStop}%
\bibitem [{\citenamefont {Leven}\ \emph {et~al.}(2013)\citenamefont {Leven},
  \citenamefont {Krepel}, \citenamefont {Shemesh},\ and\ \citenamefont
  {Hod}}]{Leven.jpcl.2013}%
  \BibitemOpen
  \bibfield  {author} {\bibinfo {author} {\bibnamefont {Leven}, \bibfnamefont
  {Itai}}, \bibinfo {author} {\bibfnamefont {Dana}\ \bibnamefont {Krepel}},
  \bibinfo {author} {\bibfnamefont {Ortal}\ \bibnamefont {Shemesh}}, and\
  \bibinfo {author} {\bibfnamefont {Oded}\ \bibnamefont {Hod}}} (\bibinfo
  {year} {2013}),\ \bibfield  {title} {\enquote {\bibinfo {title} {Robust
  superlubricity in graphene/h-bn heterojunctions},}\ }\href
  {https://doi.org/10.1021/jz301758c} {\bibfield  {journal} {\bibinfo
  {journal} {The Journal of Physical Chemistry Letters}\ }\textbf {\bibinfo
  {volume} {4}}~(\bibinfo {number} {1}),\ \bibinfo {pages}
  {115--120}}\BibitemShut {NoStop}%
\bibitem [{\citenamefont {Leven}\ \emph {et~al.}(2016)\citenamefont {Leven},
  \citenamefont {Maaravi}, \citenamefont {Azuri}, \citenamefont {Kronik},\ and\
  \citenamefont {Hod}}]{Leven.jctc.2016}%
  \BibitemOpen
  \bibfield  {author} {\bibinfo {author} {\bibnamefont {Leven}, \bibfnamefont
  {Itai}}, \bibinfo {author} {\bibfnamefont {Tal}\ \bibnamefont {Maaravi}},
  \bibinfo {author} {\bibfnamefont {Ido}\ \bibnamefont {Azuri}}, \bibinfo
  {author} {\bibfnamefont {Leeor}\ \bibnamefont {Kronik}}, and\ \bibinfo
  {author} {\bibfnamefont {Oded}\ \bibnamefont {Hod}}} (\bibinfo {year}
  {2016}),\ \bibfield  {title} {\enquote {\bibinfo {title} {Interlayer
  potential for graphene/h-bn heterostructures},}\ }\href
  {https://doi.org/10.1021/acs.jctc.6b00147} {\bibfield  {journal} {\bibinfo
  {journal} {Journal of Chemical Theory and Computation}\ }\textbf {\bibinfo
  {volume} {12}}~(\bibinfo {number} {6}),\ \bibinfo {pages}
  {2896--2905}}\BibitemShut {NoStop}%
\bibitem [{\citenamefont {Li}\ \emph {et~al.}(2019)\citenamefont {Li},
  \citenamefont {Yin}, \citenamefont {Liu}, \citenamefont {Wu}, \citenamefont
  {Li}, \citenamefont {Li},\ and\ \citenamefont {Guo}}]{Li.NatNanotech.2019}%
  \BibitemOpen
  \bibfield  {author} {\bibinfo {author} {\bibnamefont {Li}, \bibfnamefont
  {Baowen}}, \bibinfo {author} {\bibfnamefont {Jun}\ \bibnamefont {Yin}},
  \bibinfo {author} {\bibfnamefont {Xiaofei}\ \bibnamefont {Liu}}, \bibinfo
  {author} {\bibfnamefont {Hongrong}\ \bibnamefont {Wu}}, \bibinfo {author}
  {\bibfnamefont {Jidong}\ \bibnamefont {Li}}, \bibinfo {author} {\bibfnamefont
  {Xuemei}\ \bibnamefont {Li}}, and\ \bibinfo {author} {\bibfnamefont {Wanlin}\
  \bibnamefont {Guo}}} (\bibinfo {year} {2019}),\ \bibfield  {title} {\enquote
  {\bibinfo {title} {Probing van der waals interactions at two-dimensional
  heterointerfaces},}\ }\href {https://doi.org/10.1038/s41565-019-0405-2}
  {\bibfield  {journal} {\bibinfo  {journal} {Nature Nanotechnology}\ }\textbf
  {\bibinfo {volume} {14}}~(\bibinfo {number} {6}),\ \bibinfo {pages}
  {567--572}}\BibitemShut {NoStop}%
\bibitem [{\citenamefont {Li}\ \emph {et~al.}(2017)\citenamefont {Li},
  \citenamefont {Wang}, \citenamefont {Gao}, \citenamefont {Chen},
  \citenamefont {Peng}, \citenamefont {Liu},\ and\ \citenamefont
  {Wei}}]{Li.Adv.Mat.2017}%
  \BibitemOpen
  \bibfield  {author} {\bibinfo {author} {\bibnamefont {Li}, \bibfnamefont
  {He}}, \bibinfo {author} {\bibfnamefont {Jinhuan}\ \bibnamefont {Wang}},
  \bibinfo {author} {\bibfnamefont {Song}\ \bibnamefont {Gao}}, \bibinfo
  {author} {\bibfnamefont {Qing}\ \bibnamefont {Chen}}, \bibinfo {author}
  {\bibfnamefont {Lianmao}\ \bibnamefont {Peng}}, \bibinfo {author}
  {\bibfnamefont {Kaihui}\ \bibnamefont {Liu}}, and\ \bibinfo {author}
  {\bibfnamefont {Xianlong}\ \bibnamefont {Wei}}} (\bibinfo {year} {2017}),\
  \bibfield  {title} {\enquote {\bibinfo {title} {Superlubricity between mos2
  monolayers},}\ }\href
  {https://doi.org/https://doi.org/10.1002/adma.201701474} {\bibfield
  {journal} {\bibinfo  {journal} {Advanced Materials}\ }\textbf {\bibinfo
  {volume} {29}}~(\bibinfo {number} {27}),\ \bibinfo {pages}
  {1701474}}\BibitemShut {NoStop}%
\bibitem [{\citenamefont {Li}\ \emph {et~al.}(2020{\natexlab{a}})\citenamefont
  {Li}, \citenamefont {Li}, \citenamefont {Jiang},\ and\ \citenamefont
  {Luo}}]{Li.nanoscale.2020}%
  \BibitemOpen
  \bibfield  {author} {\bibinfo {author} {\bibnamefont {Li}, \bibfnamefont
  {Jianfeng}}, \bibinfo {author} {\bibfnamefont {Jinjin}\ \bibnamefont {Li}},
  \bibinfo {author} {\bibfnamefont {Liang}\ \bibnamefont {Jiang}}, and\
  \bibinfo {author} {\bibfnamefont {Jianbin}\ \bibnamefont {Luo}}} (\bibinfo
  {year} {2020}{\natexlab{a}}),\ \bibfield  {title} {\enquote {\bibinfo {title}
  {Fabrication of a graphene layer probe to measure force interactions in
  layered heterojunctions},}\ }\href {https://doi.org/10.1039/C9NR09528D}
  {\bibfield  {journal} {\bibinfo  {journal} {Nanoscale}\ }\textbf {\bibinfo
  {volume} {12}},\ \bibinfo {pages} {5435--5443}}\BibitemShut {NoStop}%
\bibitem [{\citenamefont {Li}\ \emph {et~al.}(2018)\citenamefont {Li},
  \citenamefont {Gao},\ and\ \citenamefont {Luo}}]{Li.Adv.Science.2018}%
  \BibitemOpen
  \bibfield  {author} {\bibinfo {author} {\bibnamefont {Li}, \bibfnamefont
  {Jinjin}}, \bibinfo {author} {\bibfnamefont {Tianyang}\ \bibnamefont {Gao}},
  and\ \bibinfo {author} {\bibfnamefont {Jianbin}\ \bibnamefont {Luo}}}
  (\bibinfo {year} {2018}),\ \bibfield  {title} {\enquote {\bibinfo {title}
  {Superlubricity of graphite induced by multiple transferred graphene
  nanoflakes},}\ }\href
  {https://doi.org/https://doi.org/10.1002/advs.201700616} {\bibfield
  {journal} {\bibinfo  {journal} {Advanced Science}\ }\textbf {\bibinfo
  {volume} {5}}~(\bibinfo {number} {3}),\ \bibinfo {pages}
  {1700616}}\BibitemShut {NoStop}%
\bibitem [{\citenamefont {Li}\ \emph {et~al.}(2020{\natexlab{b}})\citenamefont
  {Li}, \citenamefont {Li}, \citenamefont {Chen}, \citenamefont {Liu},\ and\
  \citenamefont {Luo}}]{Li.carbon.2020}%
  \BibitemOpen
  \bibfield  {author} {\bibinfo {author} {\bibnamefont {Li}, \bibfnamefont
  {Jinjin}}, \bibinfo {author} {\bibfnamefont {Jianfeng}\ \bibnamefont {Li}},
  \bibinfo {author} {\bibfnamefont {Xinchun}\ \bibnamefont {Chen}}, \bibinfo
  {author} {\bibfnamefont {Yuhong}\ \bibnamefont {Liu}}, and\ \bibinfo {author}
  {\bibfnamefont {Jianbin}\ \bibnamefont {Luo}}} (\bibinfo {year}
  {2020}{\natexlab{b}}),\ \bibfield  {title} {\enquote {\bibinfo {title}
  {Microscale superlubricity at multiple gold–graphite heterointerfaces under
  ambient conditions},}\ }\href
  {https://doi.org/https://doi.org/10.1016/j.carbon.2020.01.070} {\bibfield
  {journal} {\bibinfo  {journal} {Carbon}\ }\textbf {\bibinfo {volume} {161}},\
  \bibinfo {pages} {827--833}}\BibitemShut {NoStop}%
\bibitem [{\citenamefont {Li}\ \emph {et~al.}(2015)\citenamefont {Li},
  \citenamefont {Song}, \citenamefont {Besenbacher},\ and\ \citenamefont
  {Dong}}]{Li.AccountChemRes.2015}%
  \BibitemOpen
  \bibfield  {author} {\bibinfo {author} {\bibnamefont {Li}, \bibfnamefont
  {Qiang}}, \bibinfo {author} {\bibfnamefont {Jie}\ \bibnamefont {Song}},
  \bibinfo {author} {\bibfnamefont {Flemming}\ \bibnamefont {Besenbacher}},
  and\ \bibinfo {author} {\bibfnamefont {Mingdong}\ \bibnamefont {Dong}}}
  (\bibinfo {year} {2015}),\ \bibfield  {title} {\enquote {\bibinfo {title}
  {Two-dimensional material confined water},}\ }\href
  {https://doi.org/10.1021/ar500306w} {\bibfield  {journal} {\bibinfo
  {journal} {Accounts of Chemical Research}\ }\textbf {\bibinfo {volume}
  {48}}~(\bibinfo {number} {1}),\ \bibinfo {pages} {119--127}}\BibitemShut
  {NoStop}%
\bibitem [{\citenamefont {Li}\ \emph {et~al.}(2013)\citenamefont {Li},
  \citenamefont {Wang}, \citenamefont {Kozbial}, \citenamefont {Shenoy},
  \citenamefont {Zhou}, \citenamefont {McGinley}, \citenamefont {Ireland},
  \citenamefont {Morganstein}, \citenamefont {Kunkel}, \citenamefont {Surwade},
  \citenamefont {Li},\ and\ \citenamefont {Liu}}]{Li.natmater.2013}%
  \BibitemOpen
  \bibfield  {author} {\bibinfo {author} {\bibnamefont {Li}, \bibfnamefont
  {Zhiting}}, \bibinfo {author} {\bibfnamefont {Yongjin}\ \bibnamefont {Wang}},
  \bibinfo {author} {\bibfnamefont {Andrew}\ \bibnamefont {Kozbial}}, \bibinfo
  {author} {\bibfnamefont {Ganesh}\ \bibnamefont {Shenoy}}, \bibinfo {author}
  {\bibfnamefont {Feng}\ \bibnamefont {Zhou}}, \bibinfo {author} {\bibfnamefont
  {Rebecca}\ \bibnamefont {McGinley}}, \bibinfo {author} {\bibfnamefont
  {Patrick}\ \bibnamefont {Ireland}}, \bibinfo {author} {\bibfnamefont
  {Brittni}\ \bibnamefont {Morganstein}}, \bibinfo {author} {\bibfnamefont
  {Alyssa}\ \bibnamefont {Kunkel}}, \bibinfo {author} {\bibfnamefont
  {Sumedh~P.}\ \bibnamefont {Surwade}}, \bibinfo {author} {\bibfnamefont {Lei}\
  \bibnamefont {Li}}, and\ \bibinfo {author} {\bibfnamefont {Haitao}\
  \bibnamefont {Liu}}} (\bibinfo {year} {2013}),\ \bibfield  {title} {\enquote
  {\bibinfo {title} {Effect of airborne contaminants on the wettability of
  supported graphene and graphite},}\ }\href {https://doi.org/10.1038/nmat3709}
  {\bibfield  {journal} {\bibinfo  {journal} {Nature Materials}\ }\textbf
  {\bibinfo {volume} {12}}~(\bibinfo {number} {10}),\ \bibinfo {pages}
  {925--931}}\BibitemShut {NoStop}%
\bibitem [{\citenamefont {Liao}\ \emph {et~al.}(2022)\citenamefont {Liao},
  \citenamefont {Nicolini}, \citenamefont {Du}, \citenamefont {Yuan},
  \citenamefont {Wang}, \citenamefont {Yu}, \citenamefont {Tang}, \citenamefont
  {Cheng}, \citenamefont {Watanabe}, \citenamefont {Taniguchi}, \citenamefont
  {Gu}, \citenamefont {Claerbout}, \citenamefont {Silva}, \citenamefont
  {Kramer}, \citenamefont {Polcar}, \citenamefont {Yang}, \citenamefont {Shi},\
  and\ \citenamefont {Zhang}}]{Liao.natmater.2022}%
  \BibitemOpen
  \bibfield  {author} {\bibinfo {author} {\bibnamefont {Liao}, \bibfnamefont
  {Mengzhou}}, \bibinfo {author} {\bibfnamefont {Paolo}\ \bibnamefont
  {Nicolini}}, \bibinfo {author} {\bibfnamefont {Luojun}\ \bibnamefont {Du}},
  \bibinfo {author} {\bibfnamefont {Jiahao}\ \bibnamefont {Yuan}}, \bibinfo
  {author} {\bibfnamefont {Shuopei}\ \bibnamefont {Wang}}, \bibinfo {author}
  {\bibfnamefont {Hua}\ \bibnamefont {Yu}}, \bibinfo {author} {\bibfnamefont
  {Jian}\ \bibnamefont {Tang}}, \bibinfo {author} {\bibfnamefont {Peng}\
  \bibnamefont {Cheng}}, \bibinfo {author} {\bibfnamefont {Kenji}\ \bibnamefont
  {Watanabe}}, \bibinfo {author} {\bibfnamefont {Takashi}\ \bibnamefont
  {Taniguchi}}, \bibinfo {author} {\bibfnamefont {Lin}\ \bibnamefont {Gu}},
  \bibinfo {author} {\bibfnamefont {Victor E.~P.}\ \bibnamefont {Claerbout}},
  \bibinfo {author} {\bibfnamefont {Andrea}\ \bibnamefont {Silva}}, \bibinfo
  {author} {\bibfnamefont {Denis}\ \bibnamefont {Kramer}}, \bibinfo {author}
  {\bibfnamefont {Tomas}\ \bibnamefont {Polcar}}, \bibinfo {author}
  {\bibfnamefont {Rong}\ \bibnamefont {Yang}}, \bibinfo {author} {\bibfnamefont
  {Dongxia}\ \bibnamefont {Shi}}, and\ \bibinfo {author} {\bibfnamefont
  {Guangyu}\ \bibnamefont {Zhang}}} (\bibinfo {year} {2022}),\ \bibfield
  {title} {\enquote {\bibinfo {title} {Uitra-low friction and edge-pinning
  effect in large-lattice-mismatch van der waals heterostructures},}\ }\href
  {https://doi.org/10.1038/s41563-021-01058-4} {\bibfield  {journal} {\bibinfo
  {journal} {Nature Materials}\ }\textbf {\bibinfo {volume} {21}}~(\bibinfo
  {number} {1}),\ \bibinfo {pages} {47--53}}\BibitemShut {NoStop}%
\bibitem [{\citenamefont {Liu}\ \emph {et~al.}(2017)\citenamefont {Liu},
  \citenamefont {Wang}, \citenamefont {Xu}, \citenamefont {Ma}, \citenamefont
  {Yu}, \citenamefont {Zhang}, \citenamefont {Geng}, \citenamefont {Yu},
  \citenamefont {Zhang}, \citenamefont {Wang}, \citenamefont {Hu},
  \citenamefont {Wang},\ and\ \citenamefont {Luo}}]{Liu.Nat.Comm.2017}%
  \BibitemOpen
  \bibfield  {author} {\bibinfo {author} {\bibnamefont {Liu}, \bibfnamefont
  {Shu-Wei}}, \bibinfo {author} {\bibfnamefont {Hua-Ping}\ \bibnamefont
  {Wang}}, \bibinfo {author} {\bibfnamefont {Qiang}\ \bibnamefont {Xu}},
  \bibinfo {author} {\bibfnamefont {Tian-Bao}\ \bibnamefont {Ma}}, \bibinfo
  {author} {\bibfnamefont {Gui}\ \bibnamefont {Yu}}, \bibinfo {author}
  {\bibfnamefont {Chenhui}\ \bibnamefont {Zhang}}, \bibinfo {author}
  {\bibfnamefont {Dechao}\ \bibnamefont {Geng}}, \bibinfo {author}
  {\bibfnamefont {Zhiwei}\ \bibnamefont {Yu}}, \bibinfo {author} {\bibfnamefont
  {Shengguang}\ \bibnamefont {Zhang}}, \bibinfo {author} {\bibfnamefont
  {Wenzhong}\ \bibnamefont {Wang}}, \bibinfo {author} {\bibfnamefont
  {Yuan-Zhong}\ \bibnamefont {Hu}}, \bibinfo {author} {\bibfnamefont {Hui}\
  \bibnamefont {Wang}}, and\ \bibinfo {author} {\bibfnamefont {Jianbin}\
  \bibnamefont {Luo}}} (\bibinfo {year} {2017}),\ \bibfield  {title} {\enquote
  {\bibinfo {title} {Robust microscale superlubricity under high contact
  pressure enabled by graphene-coated microsphere},}\ }\href
  {https://doi.org/10.1038/ncomms14029} {\bibfield  {journal} {\bibinfo
  {journal} {Nature Communications}\ }\textbf {\bibinfo {volume} {8}}~(\bibinfo
  {number} {1}),\ \bibinfo {pages} {14029}}\BibitemShut {NoStop}%
\bibitem [{\citenamefont {Liu}\ \emph {et~al.}(2012{\natexlab{a}})\citenamefont
  {Liu}, \citenamefont {Metcalf}, \citenamefont {Robinson}, \citenamefont
  {Houston},\ and\ \citenamefont {Scarpa}}]{Liu.Nanolett.2012}%
  \BibitemOpen
  \bibfield  {author} {\bibinfo {author} {\bibnamefont {Liu}, \bibfnamefont
  {Xiao}}, \bibinfo {author} {\bibfnamefont {Thomas~H.}\ \bibnamefont
  {Metcalf}}, \bibinfo {author} {\bibfnamefont {Jeremy~T.}\ \bibnamefont
  {Robinson}}, \bibinfo {author} {\bibfnamefont {Brian~H.}\ \bibnamefont
  {Houston}}, and\ \bibinfo {author} {\bibfnamefont {Fabrizio}\ \bibnamefont
  {Scarpa}}} (\bibinfo {year} {2012}{\natexlab{a}}),\ \bibfield  {title}
  {\enquote {\bibinfo {title} {Shear modulus of monolayer graphene prepared by
  chemical vapor deposition},}\ }\href {https://doi.org/10.1021/nl204196v}
  {\bibfield  {journal} {\bibinfo  {journal} {Nano Letters}\ }\textbf {\bibinfo
  {volume} {12}}~(\bibinfo {number} {2}),\ \bibinfo {pages}
  {1013--1017}}\BibitemShut {NoStop}%
\bibitem [{\citenamefont {Liu}\ \emph {et~al.}(1996)\citenamefont {Liu},
  \citenamefont {Erdemir},\ and\ \citenamefont {Meletis}}]{Liu.surfcoat.1996}%
  \BibitemOpen
  \bibfield  {author} {\bibinfo {author} {\bibnamefont {Liu}, \bibfnamefont
  {Y}}, \bibinfo {author} {\bibfnamefont {A.}~\bibnamefont {Erdemir}}, and\
  \bibinfo {author} {\bibfnamefont {E.I.}\ \bibnamefont {Meletis}}} (\bibinfo
  {year} {1996}),\ \bibfield  {title} {\enquote {\bibinfo {title} {A study of
  the wear mechanism of diamond-like carbon films},}\ }\href
  {https://doi.org/https://doi.org/10.1016/0257-8972(95)02623-1} {\bibfield
  {journal} {\bibinfo  {journal} {Surface and Coatings Technology}\ }\textbf
  {\bibinfo {volume} {82}}~(\bibinfo {number} {1}),\ \bibinfo {pages}
  {48--56}}\BibitemShut {NoStop}%
\bibitem [{\citenamefont {Liu}\ \emph {et~al.}(2018)\citenamefont {Liu},
  \citenamefont {Song}, \citenamefont {Xu}, \citenamefont {Zong}, \citenamefont
  {Zhang}, \citenamefont {Yang}, \citenamefont {Wang}, \citenamefont {Hu},
  \citenamefont {Luo},\ and\ \citenamefont {Ma}}]{Liu.acsnano.2018}%
  \BibitemOpen
  \bibfield  {author} {\bibinfo {author} {\bibnamefont {Liu}, \bibfnamefont
  {Yanmin}}, \bibinfo {author} {\bibfnamefont {Aisheng}\ \bibnamefont {Song}},
  \bibinfo {author} {\bibfnamefont {Zhi}\ \bibnamefont {Xu}}, \bibinfo {author}
  {\bibfnamefont {Ruilong}\ \bibnamefont {Zong}}, \bibinfo {author}
  {\bibfnamefont {Jie}\ \bibnamefont {Zhang}}, \bibinfo {author} {\bibfnamefont
  {Wenyan}\ \bibnamefont {Yang}}, \bibinfo {author} {\bibfnamefont {Rong}\
  \bibnamefont {Wang}}, \bibinfo {author} {\bibfnamefont {Yuanzhong}\
  \bibnamefont {Hu}}, \bibinfo {author} {\bibfnamefont {Jianbin}\ \bibnamefont
  {Luo}}, and\ \bibinfo {author} {\bibfnamefont {TianBao}\ \bibnamefont {Ma}}}
  (\bibinfo {year} {2018}),\ \bibfield  {title} {\enquote {\bibinfo {title}
  {Interlayer friction and superlubricity in single-crystalline contact enabled
  by two-dimensional flake-wrapped atomic force microscope tips},}\ }\href
  {https://doi.org/10.1021/acsnano.7b09083} {\bibfield  {journal} {\bibinfo
  {journal} {ACS Nano}\ }\textbf {\bibinfo {volume} {12}}~(\bibinfo {number}
  {8}),\ \bibinfo {pages} {7638--7646}}\BibitemShut {NoStop}%
\bibitem [{\citenamefont {Liu}\ \emph {et~al.}(2020)\citenamefont {Liu},
  \citenamefont {Wang}, \citenamefont {Xu}, \citenamefont {Zhang},
  \citenamefont {Hu}, \citenamefont {Ma}, \citenamefont {Zheng},\ and\
  \citenamefont {Luo}}]{Liu.App.Mat.2020}%
  \BibitemOpen
  \bibfield  {author} {\bibinfo {author} {\bibnamefont {Liu}, \bibfnamefont
  {Yanmin}}, \bibinfo {author} {\bibfnamefont {Kang}\ \bibnamefont {Wang}},
  \bibinfo {author} {\bibfnamefont {Qiang}\ \bibnamefont {Xu}}, \bibinfo
  {author} {\bibfnamefont {Jie}\ \bibnamefont {Zhang}}, \bibinfo {author}
  {\bibfnamefont {Yuanzhong}\ \bibnamefont {Hu}}, \bibinfo {author}
  {\bibfnamefont {Tianbao}\ \bibnamefont {Ma}}, \bibinfo {author}
  {\bibfnamefont {Quanshui}\ \bibnamefont {Zheng}}, and\ \bibinfo {author}
  {\bibfnamefont {Jianbin}\ \bibnamefont {Luo}}} (\bibinfo {year} {2020}),\
  \bibfield  {title} {\enquote {\bibinfo {title} {Superlubricity between
  graphite layers in ultrahigh vacuum},}\ }\href
  {https://doi.org/10.1021/acsami.0c05422} {\bibfield  {journal} {\bibinfo
  {journal} {ACS Applied Materials \& Interfaces}\ }\textbf {\bibinfo {volume}
  {12}}~(\bibinfo {number} {38}),\ \bibinfo {pages} {43167--43172}}\BibitemShut
  {NoStop}%
\bibitem [{\citenamefont {Liu}\ \emph {et~al.}(2016)\citenamefont {Liu},
  \citenamefont {Weiss}, \citenamefont {Duan}, \citenamefont {Cheng},
  \citenamefont {Huang},\ and\ \citenamefont
  {Duan}}]{Liu.NatureReviewsMaterials.2016}%
  \BibitemOpen
  \bibfield  {author} {\bibinfo {author} {\bibnamefont {Liu}, \bibfnamefont
  {Yuan}}, \bibinfo {author} {\bibfnamefont {Nathan~O.}\ \bibnamefont {Weiss}},
  \bibinfo {author} {\bibfnamefont {Xidong}\ \bibnamefont {Duan}}, \bibinfo
  {author} {\bibfnamefont {Hung-Chieh}\ \bibnamefont {Cheng}}, \bibinfo
  {author} {\bibfnamefont {Yu}~\bibnamefont {Huang}}, and\ \bibinfo {author}
  {\bibfnamefont {Xiangfeng}\ \bibnamefont {Duan}}} (\bibinfo {year} {2016}),\
  \bibfield  {title} {\enquote {\bibinfo {title} {Van der waals
  heterostructures and devices},}\ }\href
  {https://doi.org/10.1038/natrevmats.2016.42} {\bibfield  {journal} {\bibinfo
  {journal} {Nature Reviews Materials}\ }\textbf {\bibinfo {volume}
  {1}}~(\bibinfo {number} {9}),\ \bibinfo {pages} {16042}}\BibitemShut
  {NoStop}%
\bibitem [{\citenamefont {Liu}\ \emph {et~al.}(2011)\citenamefont {Liu},
  \citenamefont {B{\o}ggild}, \citenamefont {rui Yang}, \citenamefont {Cheng},
  \citenamefont {Grey}, \citenamefont {lun Liu}, \citenamefont {Wang},\ and\
  \citenamefont {shui Zheng}}]{Liu.nanotech.2011}%
  \BibitemOpen
  \bibfield  {author} {\bibinfo {author} {\bibnamefont {Liu}, \bibfnamefont
  {Ze}}, \bibinfo {author} {\bibfnamefont {Peter}\ \bibnamefont {B{\o}ggild}},
  \bibinfo {author} {\bibfnamefont {Jia}\ \bibnamefont {rui Yang}}, \bibinfo
  {author} {\bibfnamefont {Yao}\ \bibnamefont {Cheng}}, \bibinfo {author}
  {\bibfnamefont {Francois}\ \bibnamefont {Grey}}, \bibinfo {author}
  {\bibfnamefont {Yi}~\bibnamefont {lun Liu}}, \bibinfo {author} {\bibfnamefont
  {Li}~\bibnamefont {Wang}}, and\ \bibinfo {author} {\bibfnamefont {Quan}\
  \bibnamefont {shui Zheng}}} (\bibinfo {year} {2011}),\ \bibfield  {title}
  {\enquote {\bibinfo {title} {A graphite nanoeraser},}\ }\href
  {https://doi.org/10.1088/0957-4484/22/26/265706} {\bibfield  {journal}
  {\bibinfo  {journal} {Nanotechnology}\ }\textbf {\bibinfo {volume}
  {22}}~(\bibinfo {number} {26}),\ \bibinfo {pages} {265706}}\BibitemShut
  {NoStop}%
\bibitem [{\citenamefont {Liu}\ \emph {et~al.}(2012{\natexlab{b}})\citenamefont
  {Liu}, \citenamefont {Liu}, \citenamefont {Cheng}, \citenamefont {Li},
  \citenamefont {Wang},\ and\ \citenamefont {Zheng}}]{Liu.prb.2012}%
  \BibitemOpen
  \bibfield  {author} {\bibinfo {author} {\bibnamefont {Liu}, \bibfnamefont
  {Ze}}, \bibinfo {author} {\bibfnamefont {Jefferson~Zhe}\ \bibnamefont {Liu}},
  \bibinfo {author} {\bibfnamefont {Yao}\ \bibnamefont {Cheng}}, \bibinfo
  {author} {\bibfnamefont {Zhihong}\ \bibnamefont {Li}}, \bibinfo {author}
  {\bibfnamefont {Li}~\bibnamefont {Wang}}, and\ \bibinfo {author}
  {\bibfnamefont {Quanshui}\ \bibnamefont {Zheng}}} (\bibinfo {year}
  {2012}{\natexlab{b}}),\ \bibfield  {title} {\enquote {\bibinfo {title}
  {Interlayer binding energy of graphite: A mesoscopic determination from
  deformation},}\ }\href {https://doi.org/10.1103/PhysRevB.85.205418}
  {\bibfield  {journal} {\bibinfo  {journal} {Phys. Rev. B}\ }\textbf {\bibinfo
  {volume} {85}},\ \bibinfo {pages} {205418}}\BibitemShut {NoStop}%
\bibitem [{\citenamefont {Liu}\ \emph {et~al.}(2012{\natexlab{c}})\citenamefont
  {Liu}, \citenamefont {Yang}, \citenamefont {Grey}, \citenamefont {Liu},
  \citenamefont {Liu}, \citenamefont {Wang}, \citenamefont {Yang},
  \citenamefont {Cheng},\ and\ \citenamefont {Zheng}}]{Liu.prl.2012}%
  \BibitemOpen
  \bibfield  {author} {\bibinfo {author} {\bibnamefont {Liu}, \bibfnamefont
  {Ze}}, \bibinfo {author} {\bibfnamefont {Jiarui}\ \bibnamefont {Yang}},
  \bibinfo {author} {\bibfnamefont {Francois}\ \bibnamefont {Grey}}, \bibinfo
  {author} {\bibfnamefont {Jefferson~Zhe}\ \bibnamefont {Liu}}, \bibinfo
  {author} {\bibfnamefont {Yilun}\ \bibnamefont {Liu}}, \bibinfo {author}
  {\bibfnamefont {Yibing}\ \bibnamefont {Wang}}, \bibinfo {author}
  {\bibfnamefont {Yanlian}\ \bibnamefont {Yang}}, \bibinfo {author}
  {\bibfnamefont {Yao}\ \bibnamefont {Cheng}}, and\ \bibinfo {author}
  {\bibfnamefont {Quanshui}\ \bibnamefont {Zheng}}} (\bibinfo {year}
  {2012}{\natexlab{c}}),\ \bibfield  {title} {\enquote {\bibinfo {title}
  {Observation of microscale superlubricity in graphite},}\ }\href
  {https://doi.org/10.1103/PhysRevLett.108.205503} {\bibfield  {journal}
  {\bibinfo  {journal} {Phys. Rev. Lett.}\ }\textbf {\bibinfo {volume} {108}},\
  \bibinfo {pages} {205503}}\BibitemShut {NoStop}%
\bibitem [{\citenamefont {Luan}\ and\ \citenamefont
  {Robbins}(2005)}]{Luan.Nature.2005}%
  \BibitemOpen
  \bibfield  {author} {\bibinfo {author} {\bibnamefont {Luan}, \bibfnamefont
  {Binquan}}, and\ \bibinfo {author} {\bibfnamefont {Mark~O.}\ \bibnamefont
  {Robbins}}} (\bibinfo {year} {2005}),\ \bibfield  {title} {\enquote {\bibinfo
  {title} {The breakdown of continuum models for mechanical contacts},}\ }\href
  {https://doi.org/10.1038/nature03700} {\bibfield  {journal} {\bibinfo
  {journal} {Nature}\ }\textbf {\bibinfo {volume} {435}}~(\bibinfo {number}
  {7044}),\ \bibinfo {pages} {929--932}}\BibitemShut {NoStop}%
\bibitem [{\citenamefont {Luo}\ \emph {et~al.}(2011)\citenamefont {Luo},
  \citenamefont {Kim}, \citenamefont {Kawamoto}, \citenamefont {Rappe},\ and\
  \citenamefont {Johnson}}]{Luo.acsnano.2011}%
  \BibitemOpen
  \bibfield  {author} {\bibinfo {author} {\bibnamefont {Luo}, \bibfnamefont
  {Zhengtang}}, \bibinfo {author} {\bibfnamefont {Seungchul}\ \bibnamefont
  {Kim}}, \bibinfo {author} {\bibfnamefont {Nicole}\ \bibnamefont {Kawamoto}},
  \bibinfo {author} {\bibfnamefont {Andrew~M.}\ \bibnamefont {Rappe}}, and\
  \bibinfo {author} {\bibfnamefont {A.~T.~Charlie}\ \bibnamefont {Johnson}}}
  (\bibinfo {year} {2011}),\ \bibfield  {title} {\enquote {\bibinfo {title}
  {Growth mechanism of hexagonal-shape graphene flakes with zigzag edges},}\
  }\href {https://doi.org/10.1021/nn203381k} {\bibfield  {journal} {\bibinfo
  {journal} {ACS Nano}\ }\textbf {\bibinfo {volume} {5}}~(\bibinfo {number}
  {11}),\ \bibinfo {pages} {9154--9160}},\ \bibinfo {note} {pMID:
  21999584}\BibitemShut {NoStop}%
\bibitem [{\citenamefont {Ma}\ \emph {et~al.}(2015{\natexlab{a}})\citenamefont
  {Ma}, \citenamefont {Benassi}, \citenamefont {Vanossi},\ and\ \citenamefont
  {Urbakh}}]{Ma.prl.2015}%
  \BibitemOpen
  \bibfield  {author} {\bibinfo {author} {\bibnamefont {Ma}, \bibfnamefont
  {Ming}}, \bibinfo {author} {\bibfnamefont {Andrea}\ \bibnamefont {Benassi}},
  \bibinfo {author} {\bibfnamefont {Andrea}\ \bibnamefont {Vanossi}}, and\
  \bibinfo {author} {\bibfnamefont {Michael}\ \bibnamefont {Urbakh}}} (\bibinfo
  {year} {2015}{\natexlab{a}}),\ \bibfield  {title} {\enquote {\bibinfo {title}
  {Critical length limiting superlow friction},}\ }\href
  {https://doi.org/10.1103/PhysRevLett.114.055501} {\bibfield  {journal}
  {\bibinfo  {journal} {Phys. Rev. Lett.}\ }\textbf {\bibinfo {volume} {114}},\
  \bibinfo {pages} {055501}}\BibitemShut {NoStop}%
\bibitem [{\citenamefont {Ma}\ \emph {et~al.}(2015{\natexlab{b}})\citenamefont
  {Ma}, \citenamefont {Sokolov}, \citenamefont {Wang}, \citenamefont
  {Filippov}, \citenamefont {Zheng},\ and\ \citenamefont
  {Urbakh}}]{Ma.prx.2015}%
  \BibitemOpen
  \bibfield  {author} {\bibinfo {author} {\bibnamefont {Ma}, \bibfnamefont
  {Ming}}, \bibinfo {author} {\bibfnamefont {Igor~M.}\ \bibnamefont {Sokolov}},
  \bibinfo {author} {\bibfnamefont {Wen}\ \bibnamefont {Wang}}, \bibinfo
  {author} {\bibfnamefont {Alexander~E.}\ \bibnamefont {Filippov}}, \bibinfo
  {author} {\bibfnamefont {Quanshui}\ \bibnamefont {Zheng}}, and\ \bibinfo
  {author} {\bibfnamefont {Michael}\ \bibnamefont {Urbakh}}} (\bibinfo {year}
  {2015}{\natexlab{b}}),\ \bibfield  {title} {\enquote {\bibinfo {title}
  {Diffusion through bifurcations in oscillating nano- and microscale contacts:
  Fundamentals and applications},}\ }\href
  {https://doi.org/10.1103/PhysRevX.5.031020} {\bibfield  {journal} {\bibinfo
  {journal} {Phys. Rev. X}\ }\textbf {\bibinfo {volume} {5}},\ \bibinfo {pages}
  {031020}}\BibitemShut {NoStop}%
\bibitem [{\citenamefont {Maity}\ \emph {et~al.}(2020)\citenamefont {Maity},
  \citenamefont {Naik}, \citenamefont {Maiti}, \citenamefont {Krishnamurthy},\
  and\ \citenamefont {Jain}}]{Maity.prr.2020}%
  \BibitemOpen
  \bibfield  {author} {\bibinfo {author} {\bibnamefont {Maity}, \bibfnamefont
  {Indrajit}}, \bibinfo {author} {\bibfnamefont {Mit~H.}\ \bibnamefont {Naik}},
  \bibinfo {author} {\bibfnamefont {Prabal~K.}\ \bibnamefont {Maiti}}, \bibinfo
  {author} {\bibfnamefont {H.~R.}\ \bibnamefont {Krishnamurthy}}, and\ \bibinfo
  {author} {\bibfnamefont {Manish}\ \bibnamefont {Jain}}} (\bibinfo {year}
  {2020}),\ \bibfield  {title} {\enquote {\bibinfo {title} {Phonons in twisted
  transition-metal dichalcogenide bilayers: Ultrasoft phasons and a transition
  from a superlubric to a pinned phase},}\ }\href
  {https://doi.org/10.1103/PhysRevResearch.2.013335} {\bibfield  {journal}
  {\bibinfo  {journal} {Phys. Rev. Research}\ }\textbf {\bibinfo {volume}
  {2}},\ \bibinfo {pages} {013335}}\BibitemShut {NoStop}%
\bibitem [{\citenamefont {Mandelli}\ \emph {et~al.}(2017)\citenamefont
  {Mandelli}, \citenamefont {Leven}, \citenamefont {Hod},\ and\ \citenamefont
  {Urbakh}}]{Mandelli.scirep.2017}%
  \BibitemOpen
  \bibfield  {author} {\bibinfo {author} {\bibnamefont {Mandelli},
  \bibfnamefont {D}}, \bibinfo {author} {\bibfnamefont {I.}~\bibnamefont
  {Leven}}, \bibinfo {author} {\bibfnamefont {O.}~\bibnamefont {Hod}}, and\
  \bibinfo {author} {\bibfnamefont {M.}~\bibnamefont {Urbakh}}} (\bibinfo
  {year} {2017}),\ \bibfield  {title} {\enquote {\bibinfo {title} {Sliding
  friction of graphene/hexagonal --boron nitride heterojunctions: a route to
  robust superlubricity},}\ }\href {https://doi.org/10.1038/s41598-017-10522-8}
  {\bibfield  {journal} {\bibinfo  {journal} {Scientific Reports}\ }\textbf
  {\bibinfo {volume} {7}}~(\bibinfo {number} {1}),\ \bibinfo {pages}
  {10851}}\BibitemShut {NoStop}%
\bibitem [{\citenamefont {Mandelli}\ \emph {et~al.}(2018)\citenamefont
  {Mandelli}, \citenamefont {Guerra}, \citenamefont {Ouyang}, \citenamefont
  {Urbakh},\ and\ \citenamefont {Vanossi}}]{Mandelli.prm.2018}%
  \BibitemOpen
  \bibfield  {author} {\bibinfo {author} {\bibnamefont {Mandelli},
  \bibfnamefont {Davide}}, \bibinfo {author} {\bibfnamefont {Roberto}\
  \bibnamefont {Guerra}}, \bibinfo {author} {\bibfnamefont {Wengen}\
  \bibnamefont {Ouyang}}, \bibinfo {author} {\bibfnamefont {Michael}\
  \bibnamefont {Urbakh}}, and\ \bibinfo {author} {\bibfnamefont {Andrea}\
  \bibnamefont {Vanossi}}} (\bibinfo {year} {2018}),\ \bibfield  {title}
  {\enquote {\bibinfo {title} {Static friction boost in edge-driven
  incommensurate contacts},}\ }\href
  {https://doi.org/10.1103/PhysRevMaterials.2.046001} {\bibfield  {journal}
  {\bibinfo  {journal} {Phys. Rev. Materials}\ }\textbf {\bibinfo {volume}
  {2}},\ \bibinfo {pages} {046001}}\BibitemShut {NoStop}%
\bibitem [{\citenamefont {Mandelli}\ \emph {et~al.}(2019)\citenamefont
  {Mandelli}, \citenamefont {Ouyang}, \citenamefont {Hod},\ and\ \citenamefont
  {Urbakh}}]{Mandelli.PhysRevLett.2019}%
  \BibitemOpen
  \bibfield  {author} {\bibinfo {author} {\bibnamefont {Mandelli},
  \bibfnamefont {Davide}}, \bibinfo {author} {\bibfnamefont {Wengen}\
  \bibnamefont {Ouyang}}, \bibinfo {author} {\bibfnamefont {Oded}\ \bibnamefont
  {Hod}}, and\ \bibinfo {author} {\bibfnamefont {Michael}\ \bibnamefont
  {Urbakh}}} (\bibinfo {year} {2019}),\ \bibfield  {title} {\enquote {\bibinfo
  {title} {Negative friction coefficients in superlubric graphite--hexagonal
  boron nitride heterojunctions},}\ }\href
  {https://doi.org/10.1103/PhysRevLett.122.076102} {\bibfield  {journal}
  {\bibinfo  {journal} {Phys. Rev. Lett.}\ }\textbf {\bibinfo {volume} {122}},\
  \bibinfo {pages} {076102}}\BibitemShut {NoStop}%
\bibitem [{\citenamefont {Mandelli}\ \emph {et~al.}(2015)\citenamefont
  {Mandelli}, \citenamefont {Vanossi}, \citenamefont {Manini},\ and\
  \citenamefont {Tosatti}}]{Mandelli.prl.2015}%
  \BibitemOpen
  \bibfield  {author} {\bibinfo {author} {\bibnamefont {Mandelli},
  \bibfnamefont {Davide}}, \bibinfo {author} {\bibfnamefont {Andrea}\
  \bibnamefont {Vanossi}}, \bibinfo {author} {\bibfnamefont {Nicola}\
  \bibnamefont {Manini}}, and\ \bibinfo {author} {\bibfnamefont {Erio}\
  \bibnamefont {Tosatti}}} (\bibinfo {year} {2015}),\ \bibfield  {title}
  {\enquote {\bibinfo {title} {Friction boosted by equilibrium misalignment of
  incommensurate two-dimensional colloid monolayers},}\ }\href
  {https://doi.org/10.1103/PhysRevLett.114.108302} {\bibfield  {journal}
  {\bibinfo  {journal} {Phys. Rev. Lett.}\ }\textbf {\bibinfo {volume} {114}},\
  \bibinfo {pages} {108302}}\BibitemShut {NoStop}%
\bibitem [{\citenamefont {Mao}\ \emph {et~al.}(2003)\citenamefont {Mao},
  \citenamefont {kwang Mao}, \citenamefont {Eng}, \citenamefont {Trainor},
  \citenamefont {Newville}, \citenamefont {chang Kao}, \citenamefont {Heinz},
  \citenamefont {Shu}, \citenamefont {Meng},\ and\ \citenamefont
  {Hemley}}]{Wendy.science.2003}%
  \BibitemOpen
  \bibfield  {author} {\bibinfo {author} {\bibnamefont {Mao}, \bibfnamefont
  {Wendy~L}}, \bibinfo {author} {\bibfnamefont {Ho}~\bibnamefont {kwang Mao}},
  \bibinfo {author} {\bibfnamefont {Peter~J.}\ \bibnamefont {Eng}}, \bibinfo
  {author} {\bibfnamefont {Thomas~P.}\ \bibnamefont {Trainor}}, \bibinfo
  {author} {\bibfnamefont {Matthew}\ \bibnamefont {Newville}}, \bibinfo
  {author} {\bibfnamefont {Chi}\ \bibnamefont {chang Kao}}, \bibinfo {author}
  {\bibfnamefont {Dion~L.}\ \bibnamefont {Heinz}}, \bibinfo {author}
  {\bibfnamefont {Jinfu}\ \bibnamefont {Shu}}, \bibinfo {author} {\bibfnamefont
  {Yue}\ \bibnamefont {Meng}}, and\ \bibinfo {author} {\bibfnamefont
  {Russell~J.}\ \bibnamefont {Hemley}}} (\bibinfo {year} {2003}),\ \bibfield
  {title} {\enquote {\bibinfo {title} {Bonding changes in compressed superhard
  graphite},}\ }\href {https://doi.org/10.1126/science.1089713} {\bibfield
  {journal} {\bibinfo  {journal} {Science}\ }\textbf {\bibinfo {volume}
  {302}}~(\bibinfo {number} {5644}),\ \bibinfo {pages} {425--427}}\BibitemShut
  {NoStop}%
\bibitem [{\citenamefont {Martin}\ \emph {et~al.}(1993)\citenamefont {Martin},
  \citenamefont {Donnet}, \citenamefont {Le~Mogne},\ and\ \citenamefont
  {Epicier}}]{Martin.PhysRevB.1993}%
  \BibitemOpen
  \bibfield  {author} {\bibinfo {author} {\bibnamefont {Martin}, \bibfnamefont
  {J~M}}, \bibinfo {author} {\bibfnamefont {C.}~\bibnamefont {Donnet}},
  \bibinfo {author} {\bibfnamefont {Th.}\ \bibnamefont {Le~Mogne}}, and\
  \bibinfo {author} {\bibfnamefont {Th.}\ \bibnamefont {Epicier}}} (\bibinfo
  {year} {1993}),\ \bibfield  {title} {\enquote {\bibinfo {title}
  {Superlubricity of molybdenum disulphide},}\ }\href
  {https://doi.org/10.1103/PhysRevB.48.10583} {\bibfield  {journal} {\bibinfo
  {journal} {Phys. Rev. B}\ }\textbf {\bibinfo {volume} {48}},\ \bibinfo
  {pages} {10583--10586}}\BibitemShut {NoStop}%
\bibitem [{\citenamefont {Martin}\ and\ \citenamefont
  {Erdemir}(2018)}]{Martin.PhysicsToday.2018}%
  \BibitemOpen
  \bibfield  {author} {\bibinfo {author} {\bibnamefont {Martin}, \bibfnamefont
  {Jean~Michel}}, and\ \bibinfo {author} {\bibfnamefont {Ali}\ \bibnamefont
  {Erdemir}}} (\bibinfo {year} {2018}),\ \bibfield  {title} {\enquote {\bibinfo
  {title} {Superlubricity: Friction’s vanishing act},}\ }\href
  {https://doi.org/10.1063/PT.3.3897} {\bibfield  {journal} {\bibinfo
  {journal} {Physics Today}\ }\textbf {\bibinfo {volume} {71}}~(\bibinfo
  {number} {4}),\ \bibinfo {pages} {40--46}}\BibitemShut {NoStop}%
\bibitem [{\citenamefont {Martins}\ \emph {et~al.}(2017)\citenamefont
  {Martins}, \citenamefont {Matos}, \citenamefont {Paschoal}, \citenamefont
  {Freire}, \citenamefont {Andrade}, \citenamefont {Aguiar}, \citenamefont
  {Kong}, \citenamefont {Neves}, \citenamefont {de~Oliveira}, \citenamefont
  {Mazzoni}, \citenamefont {Filho},\ and\ \citenamefont
  {Can{\c{c}}ado}}]{Martins.nc.2017}%
  \BibitemOpen
  \bibfield  {author} {\bibinfo {author} {\bibnamefont {Martins}, \bibfnamefont
  {Luiz Gustavo~Pimenta}}, \bibinfo {author} {\bibfnamefont {Matheus J.~S.}\
  \bibnamefont {Matos}}, \bibinfo {author} {\bibfnamefont {Alexandre~R.}\
  \bibnamefont {Paschoal}}, \bibinfo {author} {\bibfnamefont {Paulo T.~C.}\
  \bibnamefont {Freire}}, \bibinfo {author} {\bibfnamefont {Nadia~F.}\
  \bibnamefont {Andrade}}, \bibinfo {author} {\bibfnamefont {Acr{\'i}sio~L.}\
  \bibnamefont {Aguiar}}, \bibinfo {author} {\bibfnamefont {Jing}\ \bibnamefont
  {Kong}}, \bibinfo {author} {\bibfnamefont {Bernardo R.~A.}\ \bibnamefont
  {Neves}}, \bibinfo {author} {\bibfnamefont {Alan~B.}\ \bibnamefont
  {de~Oliveira}}, \bibinfo {author} {\bibfnamefont {M{\'a}rio~S.C.}\
  \bibnamefont {Mazzoni}}, \bibinfo {author} {\bibfnamefont {Antonio G.~Souza}\
  \bibnamefont {Filho}}, and\ \bibinfo {author} {\bibfnamefont {Luiz~Gustavo}\
  \bibnamefont {Can{\c{c}}ado}}} (\bibinfo {year} {2017}),\ \bibfield  {title}
  {\enquote {\bibinfo {title} {Raman evidence for pressure-induced formation of
  diamondene},}\ }\href {https://doi.org/10.1038/s41467-017-00149-8} {\bibfield
   {journal} {\bibinfo  {journal} {Nature Communications}\ }\textbf {\bibinfo
  {volume} {8}}~(\bibinfo {number} {1}),\ \bibinfo {pages} {96}}\BibitemShut
  {NoStop}%
\bibitem [{\citenamefont {McMillan}(1976)}]{McMillan.PRB.1976}%
  \BibitemOpen
  \bibfield  {author} {\bibinfo {author} {\bibnamefont {McMillan},
  \bibfnamefont {W~L}}} (\bibinfo {year} {1976}),\ \bibfield  {title} {\enquote
  {\bibinfo {title} {Theory of discommensurations and the
  commensurate-incommensurate charge-density-wave phase transition},}\ }\href
  {https://doi.org/10.1103/PhysRevB.14.1496} {\bibfield  {journal} {\bibinfo
  {journal} {Phys. Rev. B}\ }\textbf {\bibinfo {volume} {14}},\ \bibinfo
  {pages} {1496--1502}}\BibitemShut {NoStop}%
\bibitem [{\citenamefont {Memarian}\ \emph {et~al.}(2015)\citenamefont
  {Memarian}, \citenamefont {Fereidoon},\ and\ \citenamefont {{Darvish
  Ganji}}}]{Memarian.supermicro.2015}%
  \BibitemOpen
  \bibfield  {author} {\bibinfo {author} {\bibnamefont {Memarian},
  \bibfnamefont {F}}, \bibinfo {author} {\bibfnamefont {A.}~\bibnamefont
  {Fereidoon}}, and\ \bibinfo {author} {\bibfnamefont {M.}~\bibnamefont
  {{Darvish Ganji}}}} (\bibinfo {year} {2015}),\ \bibfield  {title} {\enquote
  {\bibinfo {title} {Graphene young’s modulus: Molecular mechanics and dft
  treatments},}\ }\href
  {https://doi.org/https://doi.org/10.1016/j.spmi.2015.06.001} {\bibfield
  {journal} {\bibinfo  {journal} {Superlattices and Microstructures}\ }\textbf
  {\bibinfo {volume} {85}},\ \bibinfo {pages} {348--356}}\BibitemShut {NoStop}%
\bibitem [{\citenamefont {Mogera}\ and\ \citenamefont
  {Kulkarni}(2020)}]{Mogera.carbon.2020}%
  \BibitemOpen
  \bibfield  {author} {\bibinfo {author} {\bibnamefont {Mogera}, \bibfnamefont
  {Umesha}}, and\ \bibinfo {author} {\bibfnamefont {Giridhar~U.}\ \bibnamefont
  {Kulkarni}}} (\bibinfo {year} {2020}),\ \bibfield  {title} {\enquote
  {\bibinfo {title} {A new twist in graphene research: Twisted graphene},}\
  }\href {https://doi.org/https://doi.org/10.1016/j.carbon.2019.09.053}
  {\bibfield  {journal} {\bibinfo  {journal} {Carbon}\ }\textbf {\bibinfo
  {volume} {156}},\ \bibinfo {pages} {470--487}}\BibitemShut {NoStop}%
\bibitem [{\citenamefont {M\"user}(2011)}]{Muser.PhysRevB.2011}%
  \BibitemOpen
  \bibfield  {author} {\bibinfo {author} {\bibnamefont {M\"user}, \bibfnamefont
  {Martin~H}}} (\bibinfo {year} {2011}),\ \bibfield  {title} {\enquote
  {\bibinfo {title} {Velocity dependence of kinetic friction in the
  prandtl-tomlinson model},}\ }\href
  {https://doi.org/10.1103/PhysRevB.84.125419} {\bibfield  {journal} {\bibinfo
  {journal} {Phys. Rev. B}\ }\textbf {\bibinfo {volume} {84}},\ \bibinfo
  {pages} {125419}}\BibitemShut {NoStop}%
\bibitem [{\citenamefont {M\"user}\ \emph {et~al.}(2001)\citenamefont
  {M\"user}, \citenamefont {Wenning},\ and\ \citenamefont
  {Robbins}}]{Muser.PRL.2001}%
  \BibitemOpen
  \bibfield  {author} {\bibinfo {author} {\bibnamefont {M\"user}, \bibfnamefont
  {Martin~H}}, \bibinfo {author} {\bibfnamefont {Ludgar}\ \bibnamefont
  {Wenning}}, and\ \bibinfo {author} {\bibfnamefont {Mark~O.}\ \bibnamefont
  {Robbins}}} (\bibinfo {year} {2001}),\ \bibfield  {title} {\enquote {\bibinfo
  {title} {Simple microscopic theory of amontons's laws for static friction},}\
  }\href {https://doi.org/10.1103/PhysRevLett.86.1295} {\bibfield  {journal}
  {\bibinfo  {journal} {Phys. Rev. Lett.}\ }\textbf {\bibinfo {volume} {86}},\
  \bibinfo {pages} {1295--1298}}\BibitemShut {NoStop}%
\bibitem [{\citenamefont {Müser}(2004)}]{Muser.EPL.2004}%
  \BibitemOpen
  \bibfield  {author} {\bibinfo {author} {\bibnamefont {Müser}, \bibfnamefont
  {M~H}}} (\bibinfo {year} {2004}),\ \bibfield  {title} {\enquote {\bibinfo
  {title} {Structural lubricity: Role of dimension and symmetry},}\ }\href
  {https://doi.org/10.1209/epl/i2003-10139-6} {\bibfield  {journal} {\bibinfo
  {journal} {Europhysics Letters}\ }\textbf {\bibinfo {volume} {66}}~(\bibinfo
  {number} {1}),\ \bibinfo {pages} {97--103}}\BibitemShut {NoStop}%
\bibitem [{\citenamefont {Nigu{\`e}s}\ \emph {et~al.}(2014)\citenamefont
  {Nigu{\`e}s}, \citenamefont {Siria}, \citenamefont {Vincent}, \citenamefont
  {Poncharal},\ and\ \citenamefont {Bocquet}}]{Nigues.natmater.2014}%
  \BibitemOpen
  \bibfield  {author} {\bibinfo {author} {\bibnamefont {Nigu{\`e}s},
  \bibfnamefont {A}}, \bibinfo {author} {\bibfnamefont {A.}~\bibnamefont
  {Siria}}, \bibinfo {author} {\bibfnamefont {P.}~\bibnamefont {Vincent}},
  \bibinfo {author} {\bibfnamefont {P.}~\bibnamefont {Poncharal}}, and\
  \bibinfo {author} {\bibfnamefont {L.}~\bibnamefont {Bocquet}}} (\bibinfo
  {year} {2014}),\ \bibfield  {title} {\enquote {\bibinfo {title} {Ultrahigh
  interlayer friction in multiwalled boron nitride nanotubes},}\ }\href
  {https://doi.org/10.1038/nmat3985} {\bibfield  {journal} {\bibinfo  {journal}
  {Nature Materials}\ }\textbf {\bibinfo {volume} {13}}~(\bibinfo {number}
  {7}),\ \bibinfo {pages} {688--693}}\BibitemShut {NoStop}%
\bibitem [{\citenamefont {Nimbalkar}\ and\ \citenamefont
  {Kim}(2020)}]{Nimbalkar.nanomicro.2020}%
  \BibitemOpen
  \bibfield  {author} {\bibinfo {author} {\bibnamefont {Nimbalkar},
  \bibfnamefont {Amol}}, and\ \bibinfo {author} {\bibfnamefont {Hyunmin}\
  \bibnamefont {Kim}}} (\bibinfo {year} {2020}),\ \bibfield  {title} {\enquote
  {\bibinfo {title} {Opportunities and challenges in twisted bilayer graphene:
  A review},}\ }\href {https://doi.org/10.1007/s40820-020-00464-8} {\bibfield
  {journal} {\bibinfo  {journal} {Nano-Micro Letters}\ }\textbf {\bibinfo
  {volume} {12}}~(\bibinfo {number} {1}),\ \bibinfo {pages} {126}}\BibitemShut
  {NoStop}%
\bibitem [{\citenamefont {Novoselov}\ \emph {et~al.}(2016)\citenamefont
  {Novoselov}, \citenamefont {Mishchenko}, \citenamefont {Carvalho},\ and\
  \citenamefont {Neto}}]{Neto2016}%
  \BibitemOpen
  \bibfield  {author} {\bibinfo {author} {\bibnamefont {Novoselov},
  \bibfnamefont {K~S}}, \bibinfo {author} {\bibfnamefont {A.}~\bibnamefont
  {Mishchenko}}, \bibinfo {author} {\bibfnamefont {A.}~\bibnamefont
  {Carvalho}}, and\ \bibinfo {author} {\bibfnamefont {A.~H.~Castro}\
  \bibnamefont {Neto}}} (\bibinfo {year} {2016}),\ \bibfield  {title} {\enquote
  {\bibinfo {title} {2d materials and van der waals heterostructures},}\ }\href
  {https://doi.org/10.1126/science.aac9439} {\bibfield  {journal} {\bibinfo
  {journal} {Science}\ }\textbf {\bibinfo {volume} {353}}~(\bibinfo {number}
  {6298}),\ \bibinfo {pages} {aac9439}},\ \Eprint
  {https://arxiv.org/abs/https://www.science.org/doi/pdf/10.1126/science.aac9439}
  {https://www.science.org/doi/pdf/10.1126/science.aac9439} \BibitemShut
  {NoStop}%
\bibitem [{\citenamefont {Ouyang}\ \emph {et~al.}(2020)\citenamefont {Ouyang},
  \citenamefont {Azuri}, \citenamefont {Mandelli}, \citenamefont {Tkatchenko},
  \citenamefont {Kronik}, \citenamefont {Urbakh},\ and\ \citenamefont
  {Hod}}]{Ouyang.jctc.2020}%
  \BibitemOpen
  \bibfield  {author} {\bibinfo {author} {\bibnamefont {Ouyang}, \bibfnamefont
  {Wengen}}, \bibinfo {author} {\bibfnamefont {Ido}\ \bibnamefont {Azuri}},
  \bibinfo {author} {\bibfnamefont {Davide}\ \bibnamefont {Mandelli}}, \bibinfo
  {author} {\bibfnamefont {Alexandre}\ \bibnamefont {Tkatchenko}}, \bibinfo
  {author} {\bibfnamefont {Leeor}\ \bibnamefont {Kronik}}, \bibinfo {author}
  {\bibfnamefont {Michael}\ \bibnamefont {Urbakh}}, and\ \bibinfo {author}
  {\bibfnamefont {Oded}\ \bibnamefont {Hod}}} (\bibinfo {year} {2020}),\
  \bibfield  {title} {\enquote {\bibinfo {title} {Mechanical and tribological
  properties of layered materials under high pressure: Assessing the importance
  of many-body dispersion effects},}\ }\href
  {https://doi.org/10.1021/acs.jctc.9b00908} {\bibfield  {journal} {\bibinfo
  {journal} {Journal of Chemical Theory and Computation}\ }\textbf {\bibinfo
  {volume} {16}}~(\bibinfo {number} {1}),\ \bibinfo {pages} {666--676}},\
  \bibinfo {note} {pMID: 31815463}\BibitemShut {NoStop}%
\bibitem [{\citenamefont {Ouyang}\ \emph
  {et~al.}(2018{\natexlab{a}})\citenamefont {Ouyang}, \citenamefont {Mandelli},
  \citenamefont {Urbakh},\ and\ \citenamefont {Hod}}]{Ouyang.nanolett.2018}%
  \BibitemOpen
  \bibfield  {author} {\bibinfo {author} {\bibnamefont {Ouyang}, \bibfnamefont
  {Wengen}}, \bibinfo {author} {\bibfnamefont {Davide}\ \bibnamefont
  {Mandelli}}, \bibinfo {author} {\bibfnamefont {Michael}\ \bibnamefont
  {Urbakh}}, and\ \bibinfo {author} {\bibfnamefont {Oded}\ \bibnamefont {Hod}}}
  (\bibinfo {year} {2018}{\natexlab{a}}),\ \bibfield  {title} {\enquote
  {\bibinfo {title} {Nanoserpents: Graphene nanoribbon motion on
  two-dimensional hexagonal materials},}\ }\href
  {https://doi.org/10.1021/acs.nanolett.8b02848} {\bibfield  {journal}
  {\bibinfo  {journal} {Nano Letters}\ }\textbf {\bibinfo {volume}
  {18}}~(\bibinfo {number} {9}),\ \bibinfo {pages} {6009--6016}},\ \bibinfo
  {note} {pMID: 30109806}\BibitemShut {NoStop}%
\bibitem [{\citenamefont {Ouyang}\ \emph
  {et~al.}(2018{\natexlab{b}})\citenamefont {Ouyang}, \citenamefont {de~Wijn},\
  and\ \citenamefont {Urbakh}}]{Ouyang.nanoscale.2018}%
  \BibitemOpen
  \bibfield  {author} {\bibinfo {author} {\bibnamefont {Ouyang}, \bibfnamefont
  {Wengen}}, \bibinfo {author} {\bibfnamefont {Astrid~S.}\ \bibnamefont
  {de~Wijn}}, and\ \bibinfo {author} {\bibfnamefont {Michael}\ \bibnamefont
  {Urbakh}}} (\bibinfo {year} {2018}{\natexlab{b}}),\ \bibfield  {title}
  {\enquote {\bibinfo {title} {Atomic-scale sliding friction on a contaminated
  surface},}\ }\href {https://doi.org/10.1039/C7NR09530A} {\bibfield  {journal}
  {\bibinfo  {journal} {Nanoscale}\ }\textbf {\bibinfo {volume} {10}},\
  \bibinfo {pages} {6375--6381}}\BibitemShut {NoStop}%
\bibitem [{\citenamefont {Park}\ \emph {et~al.}(2021)\citenamefont {Park},
  \citenamefont {Cao}, \citenamefont {Watanabe}, \citenamefont {Taniguchi},\
  and\ \citenamefont {Jarillo-Herrero}}]{Park.nature.2021}%
  \BibitemOpen
  \bibfield  {author} {\bibinfo {author} {\bibnamefont {Park}, \bibfnamefont
  {Jeong~Min}}, \bibinfo {author} {\bibfnamefont {Yuan}\ \bibnamefont {Cao}},
  \bibinfo {author} {\bibfnamefont {Kenji}\ \bibnamefont {Watanabe}}, \bibinfo
  {author} {\bibfnamefont {Takashi}\ \bibnamefont {Taniguchi}}, and\ \bibinfo
  {author} {\bibfnamefont {Pablo}\ \bibnamefont {Jarillo-Herrero}}} (\bibinfo
  {year} {2021}),\ \bibfield  {title} {\enquote {\bibinfo {title} {Tunable
  strongly coupled superconductivity in magic-angle twisted trilayer
  graphene},}\ }\href {https://doi.org/10.1038/s41586-021-03192-0} {\bibfield
  {journal} {\bibinfo  {journal} {Nature}\ }\textbf {\bibinfo {volume}
  {590}}~(\bibinfo {number} {7845}),\ \bibinfo {pages} {249--255}}\BibitemShut
  {NoStop}%
\bibitem [{\citenamefont {Pellegrini}\ \emph {et~al.}(2019)\citenamefont
  {Pellegrini}, \citenamefont {Panizon}, \citenamefont {Santoro},\ and\
  \citenamefont {Tosatti}}]{Pellegrini.prb.2019}%
  \BibitemOpen
  \bibfield  {author} {\bibinfo {author} {\bibnamefont {Pellegrini},
  \bibfnamefont {Franco}}, \bibinfo {author} {\bibfnamefont {Emanuele}\
  \bibnamefont {Panizon}}, \bibinfo {author} {\bibfnamefont {Giuseppe~E.}\
  \bibnamefont {Santoro}}, and\ \bibinfo {author} {\bibfnamefont {Erio}\
  \bibnamefont {Tosatti}}} (\bibinfo {year} {2019}),\ \bibfield  {title}
  {\enquote {\bibinfo {title} {Thermally assisted lubricity and negative work
  tails in sliding friction},}\ }\href
  {https://doi.org/10.1103/PhysRevB.99.075428} {\bibfield  {journal} {\bibinfo
  {journal} {Phys. Rev. B}\ }\textbf {\bibinfo {volume} {99}},\ \bibinfo
  {pages} {075428}}\BibitemShut {NoStop}%
\bibitem [{\citenamefont {Peng}\ \emph {et~al.}(2021)\citenamefont {Peng},
  \citenamefont {Wang}, \citenamefont {Jiang}, \citenamefont {Zhao},
  \citenamefont {Wu}, \citenamefont {Tian}, \citenamefont {Ma},\ and\
  \citenamefont {Zheng}}]{Peng.nsr.2021}%
  \BibitemOpen
  \bibfield  {author} {\bibinfo {author} {\bibnamefont {Peng}, \bibfnamefont
  {Deli}}, \bibinfo {author} {\bibfnamefont {Jin}\ \bibnamefont {Wang}},
  \bibinfo {author} {\bibfnamefont {Haiyang}\ \bibnamefont {Jiang}}, \bibinfo
  {author} {\bibfnamefont {Shuji}\ \bibnamefont {Zhao}}, \bibinfo {author}
  {\bibfnamefont {Zhanghui}\ \bibnamefont {Wu}}, \bibinfo {author}
  {\bibfnamefont {Kaiwen}\ \bibnamefont {Tian}}, \bibinfo {author}
  {\bibfnamefont {Ming}\ \bibnamefont {Ma}}, and\ \bibinfo {author}
  {\bibfnamefont {Quanshui}\ \bibnamefont {Zheng}}} (\bibinfo {year} {2021}),\
  \bibfield  {title} {\enquote {\bibinfo {title} {{100 km wear-free sliding
  achieved by microscale superlubric graphite/DLC heterojunctions under ambient
  conditions}},}\ }\href {https://doi.org/10.1093/nsr/nwab109} {\bibfield
  {journal} {\bibinfo  {journal} {National Science Review}\ }\textbf {\bibinfo
  {volume} {9}}~(\bibinfo {number} {1}),\ 10.1093/nsr/nwab109}\BibitemShut
  {NoStop}%
\bibitem [{\citenamefont {Peng}\ \emph {et~al.}(2020)\citenamefont {Peng},
  \citenamefont {Wu}, \citenamefont {Shi}, \citenamefont {Qu}, \citenamefont
  {Jiang}, \citenamefont {Song}, \citenamefont {Ma}, \citenamefont {Aeppli},
  \citenamefont {Urbakh},\ and\ \citenamefont {Zheng}}]{Peng.pnas.2020}%
  \BibitemOpen
  \bibfield  {author} {\bibinfo {author} {\bibnamefont {Peng}, \bibfnamefont
  {Deli}}, \bibinfo {author} {\bibfnamefont {Zhanghui}\ \bibnamefont {Wu}},
  \bibinfo {author} {\bibfnamefont {Diwei}\ \bibnamefont {Shi}}, \bibinfo
  {author} {\bibfnamefont {Cangyu}\ \bibnamefont {Qu}}, \bibinfo {author}
  {\bibfnamefont {Haiyang}\ \bibnamefont {Jiang}}, \bibinfo {author}
  {\bibfnamefont {Yiming}\ \bibnamefont {Song}}, \bibinfo {author}
  {\bibfnamefont {Ming}\ \bibnamefont {Ma}}, \bibinfo {author} {\bibfnamefont
  {Gabriel}\ \bibnamefont {Aeppli}}, \bibinfo {author} {\bibfnamefont
  {Michael}\ \bibnamefont {Urbakh}}, and\ \bibinfo {author} {\bibfnamefont
  {Quanshui}\ \bibnamefont {Zheng}}} (\bibinfo {year} {2020}),\ \bibfield
  {title} {\enquote {\bibinfo {title} {Load-induced dynamical transitions at
  graphene interfaces},}\ }\href {https://doi.org/10.1073/pnas.1922681117}
  {\bibfield  {journal} {\bibinfo  {journal} {Proceedings of the National
  Academy of Sciences}\ }\textbf {\bibinfo {volume} {117}}~(\bibinfo {number}
  {23}),\ \bibinfo {pages} {12618--12623}}\BibitemShut {NoStop}%
\bibitem [{\citenamefont {Persson}\ and\ \citenamefont
  {Tosatti}(1999)}]{Persson.solidstatecommu.1999}%
  \BibitemOpen
  \bibfield  {author} {\bibinfo {author} {\bibnamefont {Persson}, \bibfnamefont
  {BNJ}}, and\ \bibinfo {author} {\bibfnamefont {E}~\bibnamefont {Tosatti}}}
  (\bibinfo {year} {1999}),\ \bibfield  {title} {\enquote {\bibinfo {title}
  {Theory of friction: elastic coherence length and earthquake dynamics},}\
  }\href {https://doi.org/https://doi.org/10.1016/S0038-1098(99)00034-4}
  {\bibfield  {journal} {\bibinfo  {journal} {Solid State Communications}\
  }\textbf {\bibinfo {volume} {109}}~(\bibinfo {number} {12}),\ \bibinfo
  {pages} {739--744}}\BibitemShut {NoStop}%
\bibitem [{\citenamefont {Peyrard}\ and\ \citenamefont
  {Aubry}(1983)}]{Peyrard.Aubry.1983}%
  \BibitemOpen
  \bibfield  {author} {\bibinfo {author} {\bibnamefont {Peyrard}, \bibfnamefont
  {M}}, and\ \bibinfo {author} {\bibfnamefont {S}~\bibnamefont {Aubry}}}
  (\bibinfo {year} {1983}),\ \bibfield  {title} {\enquote {\bibinfo {title}
  {Critical behaviour at the transition by breaking of analyticity in the
  discrete frenkel-kontorova model},}\ }\href
  {https://doi.org/10.1088/0022-3719/16/9/005} {\bibfield  {journal} {\bibinfo
  {journal} {Journal of Physics C: Solid State Physics}\ }\textbf {\bibinfo
  {volume} {16}}~(\bibinfo {number} {9}),\ \bibinfo {pages}
  {1593--1608}}\BibitemShut {NoStop}%
\bibitem [{\citenamefont {Pham}\ \emph {et~al.}(2022)\citenamefont {Pham},
  \citenamefont {Bodepudi}, \citenamefont {Shehzad}, \citenamefont {Liu},
  \citenamefont {Xu}, \citenamefont {Yu},\ and\ \citenamefont
  {Duan}}]{Pham.ChemRev.2022}%
  \BibitemOpen
  \bibfield  {author} {\bibinfo {author} {\bibnamefont {Pham}, \bibfnamefont
  {Phuong~V}}, \bibinfo {author} {\bibfnamefont {Srikrishna~Chanakya}\
  \bibnamefont {Bodepudi}}, \bibinfo {author} {\bibfnamefont {Khurram}\
  \bibnamefont {Shehzad}}, \bibinfo {author} {\bibfnamefont {Yuan}\
  \bibnamefont {Liu}}, \bibinfo {author} {\bibfnamefont {Yang}\ \bibnamefont
  {Xu}}, \bibinfo {author} {\bibfnamefont {Bin}\ \bibnamefont {Yu}}, and\
  \bibinfo {author} {\bibfnamefont {Xiangfeng}\ \bibnamefont {Duan}}} (\bibinfo
  {year} {2022}),\ \bibfield  {title} {\enquote {\bibinfo {title} {2d
  heterostructures for ubiquitous electronics and optoelectronics: Principles,
  opportunities, and challenges},}\ }\href
  {https://doi.org/10.1021/acs.chemrev.1c00735} {\bibfield  {journal} {\bibinfo
   {journal} {Chemical Reviews}\ }\textbf {\bibinfo {volume} {122}}~(\bibinfo
  {number} {6}),\ \bibinfo {pages} {6514--6613}}\BibitemShut {NoStop}%
\bibitem [{\citenamefont {Pierno}\ \emph {et~al.}(2015)\citenamefont {Pierno},
  \citenamefont {Bruschi}, \citenamefont {Mistura}, \citenamefont {Paolicelli},
  \citenamefont {di~Bona}, \citenamefont {Valeri}, \citenamefont {Guerra},
  \citenamefont {Vanossi},\ and\ \citenamefont
  {Tosatti}}]{Pierno.NatNano.2015}%
  \BibitemOpen
  \bibfield  {author} {\bibinfo {author} {\bibnamefont {Pierno}, \bibfnamefont
  {Matteo}}, \bibinfo {author} {\bibfnamefont {Lorenzo}\ \bibnamefont
  {Bruschi}}, \bibinfo {author} {\bibfnamefont {Giampaolo}\ \bibnamefont
  {Mistura}}, \bibinfo {author} {\bibfnamefont {Guido}\ \bibnamefont
  {Paolicelli}}, \bibinfo {author} {\bibfnamefont {Alessandro}\ \bibnamefont
  {di~Bona}}, \bibinfo {author} {\bibfnamefont {Sergio}\ \bibnamefont
  {Valeri}}, \bibinfo {author} {\bibfnamefont {Roberto}\ \bibnamefont
  {Guerra}}, \bibinfo {author} {\bibfnamefont {Andrea}\ \bibnamefont
  {Vanossi}}, and\ \bibinfo {author} {\bibfnamefont {Erio}\ \bibnamefont
  {Tosatti}}} (\bibinfo {year} {2015}),\ \bibfield  {title} {\enquote {\bibinfo
  {title} {Frictional transition from superlubric islands to pinned
  monolayers},}\ }\href {https://doi.org/10.1038/nnano.2015.106} {\bibfield
  {journal} {\bibinfo  {journal} {Nature Nanotechnology}\ }\textbf {\bibinfo
  {volume} {10}}~(\bibinfo {number} {8}),\ \bibinfo {pages}
  {714--718}}\BibitemShut {NoStop}%
\bibitem [{\citenamefont {Plimpton}(1995)}]{Plimpton.jcp.1995}%
  \BibitemOpen
  \bibfield  {author} {\bibinfo {author} {\bibnamefont {Plimpton},
  \bibfnamefont {Steve}}} (\bibinfo {year} {1995}),\ \bibfield  {title}
  {\enquote {\bibinfo {title} {Fast parallel algorithms for short-range
  molecular dynamics},}\ }\href
  {https://doi.org/https://doi.org/10.1006/jcph.1995.1039} {\bibfield
  {journal} {\bibinfo  {journal} {Journal of Computational Physics}\ }\textbf
  {\bibinfo {volume} {117}}~(\bibinfo {number} {1}),\ \bibinfo {pages}
  {1--19}}\BibitemShut {NoStop}%
\bibitem [{\citenamefont {Qu}\ \emph {et~al.}(2020)\citenamefont {Qu},
  \citenamefont {Wang}, \citenamefont {Wang}, \citenamefont {Gongyang},
  \citenamefont {Carpick}, \citenamefont {Urbakh},\ and\ \citenamefont
  {Zheng}}]{Qu.prl.2020}%
  \BibitemOpen
  \bibfield  {author} {\bibinfo {author} {\bibnamefont {Qu}, \bibfnamefont
  {Cangyu}}, \bibinfo {author} {\bibfnamefont {Kunqi}\ \bibnamefont {Wang}},
  \bibinfo {author} {\bibfnamefont {Jin}\ \bibnamefont {Wang}}, \bibinfo
  {author} {\bibfnamefont {Yujie}\ \bibnamefont {Gongyang}}, \bibinfo {author}
  {\bibfnamefont {Robert~W.}\ \bibnamefont {Carpick}}, \bibinfo {author}
  {\bibfnamefont {Michael}\ \bibnamefont {Urbakh}}, and\ \bibinfo {author}
  {\bibfnamefont {Quanshui}\ \bibnamefont {Zheng}}} (\bibinfo {year} {2020}),\
  \bibfield  {title} {\enquote {\bibinfo {title} {Origin of friction in
  superlubric graphite contacts},}\ }\href
  {https://doi.org/10.1103/PhysRevLett.125.126102} {\bibfield  {journal}
  {\bibinfo  {journal} {Phys. Rev. Lett.}\ }\textbf {\bibinfo {volume} {125}},\
  \bibinfo {pages} {126102}}\BibitemShut {NoStop}%
\bibitem [{\citenamefont {Rao}\ \emph {et~al.}(2009)\citenamefont {Rao},
  \citenamefont {Sood}, \citenamefont {Subrahmanyam},\ and\ \citenamefont
  {Govindaraj}}]{Rao.AngChemInter.2009}%
  \BibitemOpen
  \bibfield  {author} {\bibinfo {author} {\bibnamefont {Rao}, \bibfnamefont
  {C~N~R}}, \bibinfo {author} {\bibfnamefont {A.~K.}\ \bibnamefont {Sood}},
  \bibinfo {author} {\bibfnamefont {K.~S.}\ \bibnamefont {Subrahmanyam}}, and\
  \bibinfo {author} {\bibfnamefont {A.}~\bibnamefont {Govindaraj}}} (\bibinfo
  {year} {2009}),\ \bibfield  {title} {\enquote {\bibinfo {title} {Graphene:
  The new two-dimensional nanomaterial},}\ }\href
  {https://doi.org/https://doi.org/10.1002/anie.200901678} {\bibfield
  {journal} {\bibinfo  {journal} {Angewandte Chemie International Edition}\
  }\textbf {\bibinfo {volume} {48}}~(\bibinfo {number} {42}),\ \bibinfo {pages}
  {7752--7777}}\BibitemShut {NoStop}%
\bibitem [{\citenamefont {Reimann}\ and\ \citenamefont
  {Evstigneev}(2004)}]{Reimann.prl.2004}%
  \BibitemOpen
  \bibfield  {author} {\bibinfo {author} {\bibnamefont {Reimann}, \bibfnamefont
  {Peter}}, and\ \bibinfo {author} {\bibfnamefont {Mykhaylo}\ \bibnamefont
  {Evstigneev}}} (\bibinfo {year} {2004}),\ \bibfield  {title} {\enquote
  {\bibinfo {title} {Nonmonotonic velocity dependence of atomic friction},}\
  }\href {https://doi.org/10.1103/PhysRevLett.93.230802} {\bibfield  {journal}
  {\bibinfo  {journal} {Phys. Rev. Lett.}\ }\textbf {\bibinfo {volume} {93}},\
  \bibinfo {pages} {230802}}\BibitemShut {NoStop}%
\bibitem [{\citenamefont {Ribeiro-Palau}\ \emph {et~al.}(2018)\citenamefont
  {Ribeiro-Palau}, \citenamefont {Zhang}, \citenamefont {Watanabe},
  \citenamefont {Taniguchi}, \citenamefont {Hone},\ and\ \citenamefont
  {Dean}}]{Ribeiro.Science.2018}%
  \BibitemOpen
  \bibfield  {author} {\bibinfo {author} {\bibnamefont {Ribeiro-Palau},
  \bibfnamefont {Rebeca}}, \bibinfo {author} {\bibfnamefont {Changjian}\
  \bibnamefont {Zhang}}, \bibinfo {author} {\bibfnamefont {Kenji}\ \bibnamefont
  {Watanabe}}, \bibinfo {author} {\bibfnamefont {Takashi}\ \bibnamefont
  {Taniguchi}}, \bibinfo {author} {\bibfnamefont {James}\ \bibnamefont {Hone}},
  and\ \bibinfo {author} {\bibfnamefont {Cory~R.}\ \bibnamefont {Dean}}}
  (\bibinfo {year} {2018}),\ \bibfield  {title} {\enquote {\bibinfo {title}
  {Twistable electronics with dynamically rotatable heterostructures},}\ }\href
  {https://doi.org/10.1126/science.aat6981} {\bibfield  {journal} {\bibinfo
  {journal} {Science}\ }\textbf {\bibinfo {volume} {361}}~(\bibinfo {number}
  {6403}),\ \bibinfo {pages} {690--693}}\BibitemShut {NoStop}%
\bibitem [{\citenamefont {Riedl}\ \emph {et~al.}(2009)\citenamefont {Riedl},
  \citenamefont {Coletti}, \citenamefont {Iwasaki}, \citenamefont {Zakharov},\
  and\ \citenamefont {Starke}}]{Riedl.prl.2009}%
  \BibitemOpen
  \bibfield  {author} {\bibinfo {author} {\bibnamefont {Riedl}, \bibfnamefont
  {C}}, \bibinfo {author} {\bibfnamefont {C.}~\bibnamefont {Coletti}}, \bibinfo
  {author} {\bibfnamefont {T.}~\bibnamefont {Iwasaki}}, \bibinfo {author}
  {\bibfnamefont {A.~A.}\ \bibnamefont {Zakharov}}, and\ \bibinfo {author}
  {\bibfnamefont {U.}~\bibnamefont {Starke}}} (\bibinfo {year} {2009}),\
  \bibfield  {title} {\enquote {\bibinfo {title} {Quasi-free-standing epitaxial
  graphene on sic obtained by hydrogen intercalation},}\ }\href
  {https://doi.org/10.1103/PhysRevLett.103.246804} {\bibfield  {journal}
  {\bibinfo  {journal} {Phys. Rev. Lett.}\ }\textbf {\bibinfo {volume} {103}},\
  \bibinfo {pages} {246804}}\BibitemShut {NoStop}%
\bibitem [{\citenamefont {Riedo}\ \emph {et~al.}(2003)\citenamefont {Riedo},
  \citenamefont {Gnecco}, \citenamefont {Bennewitz}, \citenamefont {Meyer},\
  and\ \citenamefont {Brune}}]{Riedo.prl.2003}%
  \BibitemOpen
  \bibfield  {author} {\bibinfo {author} {\bibnamefont {Riedo}, \bibfnamefont
  {E}}, \bibinfo {author} {\bibfnamefont {E.}~\bibnamefont {Gnecco}}, \bibinfo
  {author} {\bibfnamefont {R.}~\bibnamefont {Bennewitz}}, \bibinfo {author}
  {\bibfnamefont {E.}~\bibnamefont {Meyer}}, and\ \bibinfo {author}
  {\bibfnamefont {H.}~\bibnamefont {Brune}}} (\bibinfo {year} {2003}),\
  \bibfield  {title} {\enquote {\bibinfo {title} {Interaction potential and
  hopping dynamics governing sliding friction},}\ }\href
  {https://doi.org/10.1103/PhysRevLett.91.084502} {\bibfield  {journal}
  {\bibinfo  {journal} {Phys. Rev. Lett.}\ }\textbf {\bibinfo {volume} {91}},\
  \bibinfo {pages} {084502}}\BibitemShut {NoStop}%
\bibitem [{\citenamefont {Rowe}\ \emph {et~al.}(2018)\citenamefont {Rowe},
  \citenamefont {Cs\'anyi}, \citenamefont {Alf\`e},\ and\ \citenamefont
  {Michaelides}}]{Rowe.prb.2018}%
  \BibitemOpen
  \bibfield  {author} {\bibinfo {author} {\bibnamefont {Rowe}, \bibfnamefont
  {Patrick}}, \bibinfo {author} {\bibfnamefont {G\'abor}\ \bibnamefont
  {Cs\'anyi}}, \bibinfo {author} {\bibfnamefont {Dario}\ \bibnamefont
  {Alf\`e}}, and\ \bibinfo {author} {\bibfnamefont {Angelos}\ \bibnamefont
  {Michaelides}}} (\bibinfo {year} {2018}),\ \bibfield  {title} {\enquote
  {\bibinfo {title} {Development of a machine learning potential for
  graphene},}\ }\href {https://doi.org/10.1103/PhysRevB.97.054303} {\bibfield
  {journal} {\bibinfo  {journal} {Phys. Rev. B}\ }\textbf {\bibinfo {volume}
  {97}},\ \bibinfo {pages} {054303}}\BibitemShut {NoStop}%
\bibitem [{\citenamefont {Sang}\ \emph {et~al.}(2001)\citenamefont {Sang},
  \citenamefont {Dub\'e},\ and\ \citenamefont {Grant}}]{Sang.prl.2001}%
  \BibitemOpen
  \bibfield  {author} {\bibinfo {author} {\bibnamefont {Sang}, \bibfnamefont
  {Yi}}, \bibinfo {author} {\bibfnamefont {Martin}\ \bibnamefont {Dub\'e}},
  and\ \bibinfo {author} {\bibfnamefont {Martin}\ \bibnamefont {Grant}}}
  (\bibinfo {year} {2001}),\ \bibfield  {title} {\enquote {\bibinfo {title}
  {Thermal effects on atomic friction},}\ }\href
  {https://doi.org/10.1103/PhysRevLett.87.174301} {\bibfield  {journal}
  {\bibinfo  {journal} {Phys. Rev. Lett.}\ }\textbf {\bibinfo {volume} {87}},\
  \bibinfo {pages} {174301}}\BibitemShut {NoStop}%
\bibitem [{\citenamefont {Sharp}\ \emph {et~al.}(2016)\citenamefont {Sharp},
  \citenamefont {Pastewka},\ and\ \citenamefont {Robbins}}]{Sharp.prb.2016}%
  \BibitemOpen
  \bibfield  {author} {\bibinfo {author} {\bibnamefont {Sharp}, \bibfnamefont
  {Tristan~A}}, \bibinfo {author} {\bibfnamefont {Lars}\ \bibnamefont
  {Pastewka}}, and\ \bibinfo {author} {\bibfnamefont {Mark~O.}\ \bibnamefont
  {Robbins}}} (\bibinfo {year} {2016}),\ \bibfield  {title} {\enquote {\bibinfo
  {title} {Elasticity limits structural superlubricity in large contacts},}\
  }\href {https://doi.org/10.1103/PhysRevB.93.121402} {\bibfield  {journal}
  {\bibinfo  {journal} {Phys. Rev. B}\ }\textbf {\bibinfo {volume} {93}},\
  \bibinfo {pages} {121402}}\BibitemShut {NoStop}%
\bibitem [{\citenamefont {Shinjo}\ and\ \citenamefont
  {Hirano}(1993)}]{Shinjo.surfsci.1993}%
  \BibitemOpen
  \bibfield  {author} {\bibinfo {author} {\bibnamefont {Shinjo}, \bibfnamefont
  {Kazumasa}}, and\ \bibinfo {author} {\bibfnamefont {Motohisa}\ \bibnamefont
  {Hirano}}} (\bibinfo {year} {1993}),\ \bibfield  {title} {\enquote {\bibinfo
  {title} {Dynamics of friction: superlubric state},}\ }\href
  {https://doi.org/https://doi.org/10.1016/0039-6028(93)91022-H} {\bibfield
  {journal} {\bibinfo  {journal} {Surface Science}\ }\textbf {\bibinfo {volume}
  {283}}~(\bibinfo {number} {1}),\ \bibinfo {pages} {473--478}}\BibitemShut
  {NoStop}%
\bibitem [{\citenamefont {Sokoloff}(1990)}]{Sokoloff.PhysRevB.1990}%
  \BibitemOpen
  \bibfield  {author} {\bibinfo {author} {\bibnamefont {Sokoloff},
  \bibfnamefont {J~B}}} (\bibinfo {year} {1990}),\ \bibfield  {title} {\enquote
  {\bibinfo {title} {Theory of energy dissipation in sliding crystal
  surfaces},}\ }\href {https://doi.org/10.1103/PhysRevB.42.760} {\bibfield
  {journal} {\bibinfo  {journal} {Phys. Rev. B}\ }\textbf {\bibinfo {volume}
  {42}},\ \bibinfo {pages} {760--765}}\BibitemShut {NoStop}%
\bibitem [{\citenamefont {Sokoloff}\ \emph {et~al.}(1978)\citenamefont
  {Sokoloff}, \citenamefont {Sacco},\ and\ \citenamefont
  {Weisz}}]{Sacco.PhysRevLett.1978}%
  \BibitemOpen
  \bibfield  {author} {\bibinfo {author} {\bibnamefont {Sokoloff},
  \bibfnamefont {J~B}}, \bibinfo {author} {\bibfnamefont {J.~E.}\ \bibnamefont
  {Sacco}}, and\ \bibinfo {author} {\bibfnamefont {J.~F.}\ \bibnamefont
  {Weisz}}} (\bibinfo {year} {1978}),\ \bibfield  {title} {\enquote {\bibinfo
  {title} {Undamped lattice vibrations in systems with two incommensurate
  periodicities},}\ }\href {https://doi.org/10.1103/PhysRevLett.41.1561}
  {\bibfield  {journal} {\bibinfo  {journal} {Phys. Rev. Lett.}\ }\textbf
  {\bibinfo {volume} {41}},\ \bibinfo {pages} {1561--1564}}\BibitemShut
  {NoStop}%
\bibitem [{\citenamefont {Song}\ \emph {et~al.}(2022)\citenamefont {Song},
  \citenamefont {Shi}, \citenamefont {Lu}, \citenamefont {Wang}, \citenamefont
  {Hu}, \citenamefont {Gao}, \citenamefont {Luo},\ and\ \citenamefont
  {Ma}}]{Song.nanolett.2022}%
  \BibitemOpen
  \bibfield  {author} {\bibinfo {author} {\bibnamefont {Song}, \bibfnamefont
  {Aisheng}}, \bibinfo {author} {\bibfnamefont {Ruoyu}\ \bibnamefont {Shi}},
  \bibinfo {author} {\bibfnamefont {Hongliang}\ \bibnamefont {Lu}}, \bibinfo
  {author} {\bibfnamefont {Xueyan}\ \bibnamefont {Wang}}, \bibinfo {author}
  {\bibfnamefont {Yuanzhong}\ \bibnamefont {Hu}}, \bibinfo {author}
  {\bibfnamefont {Hong-Jun}\ \bibnamefont {Gao}}, \bibinfo {author}
  {\bibfnamefont {Jianbin}\ \bibnamefont {Luo}}, and\ \bibinfo {author}
  {\bibfnamefont {Tianbao}\ \bibnamefont {Ma}}} (\bibinfo {year} {2022}),\
  \bibfield  {title} {\enquote {\bibinfo {title} {Fluctuation of interfacial
  electronic properties induces friction tuning under an electric field},}\
  }\href {https://doi.org/10.1021/acs.nanolett.1c04116} {\bibfield  {journal}
  {\bibinfo  {journal} {Nano Letters}\ }\textbf {\bibinfo {volume}
  {22}}~(\bibinfo {number} {5}),\ \bibinfo {pages} {1889--1896}}\BibitemShut
  {NoStop}%
\bibitem [{\citenamefont {Song}\ \emph {et~al.}(2018)\citenamefont {Song},
  \citenamefont {Mandelli}, \citenamefont {Hod}, \citenamefont {Urbakh},
  \citenamefont {Ma},\ and\ \citenamefont {Zheng}}]{Song.NatureMaterials.2018}%
  \BibitemOpen
  \bibfield  {author} {\bibinfo {author} {\bibnamefont {Song}, \bibfnamefont
  {Yiming}}, \bibinfo {author} {\bibfnamefont {Davide}\ \bibnamefont
  {Mandelli}}, \bibinfo {author} {\bibfnamefont {Oded}\ \bibnamefont {Hod}},
  \bibinfo {author} {\bibfnamefont {Michael}\ \bibnamefont {Urbakh}}, \bibinfo
  {author} {\bibfnamefont {Ming}\ \bibnamefont {Ma}}, and\ \bibinfo {author}
  {\bibfnamefont {Quanshui}\ \bibnamefont {Zheng}}} (\bibinfo {year} {2018}),\
  \bibfield  {title} {\enquote {\bibinfo {title} {Robust microscale
  superlubricity in graphite/hexagonal boron nitride layered
  heterojunctions},}\ }\href {https://doi.org/10.1038/s41563-018-0144-z}
  {\bibfield  {journal} {\bibinfo  {journal} {Nature Materials}\ }\textbf
  {\bibinfo {volume} {17}}~(\bibinfo {number} {10}),\ \bibinfo {pages}
  {894--899}}\BibitemShut {NoStop}%
\bibitem [{\citenamefont {Song}\ \emph {et~al.}(2021)\citenamefont {Song},
  \citenamefont {Wang}, \citenamefont {Wang}, \citenamefont {Urbakh},
  \citenamefont {Zheng},\ and\ \citenamefont {Ma}}]{Song.prm.2021}%
  \BibitemOpen
  \bibfield  {author} {\bibinfo {author} {\bibnamefont {Song}, \bibfnamefont
  {Yiming}}, \bibinfo {author} {\bibfnamefont {Jin}\ \bibnamefont {Wang}},
  \bibinfo {author} {\bibfnamefont {Yiran}\ \bibnamefont {Wang}}, \bibinfo
  {author} {\bibfnamefont {Michael}\ \bibnamefont {Urbakh}}, \bibinfo {author}
  {\bibfnamefont {Quanshui}\ \bibnamefont {Zheng}}, and\ \bibinfo {author}
  {\bibfnamefont {Ming}\ \bibnamefont {Ma}}} (\bibinfo {year} {2021}),\
  \bibfield  {title} {\enquote {\bibinfo {title} {Directional anisotropy of
  friction in microscale superlubric graphite/$h\mathrm{BN}$
  heterojunctions},}\ }\href
  {https://doi.org/10.1103/PhysRevMaterials.5.084002} {\bibfield  {journal}
  {\bibinfo  {journal} {Phys. Rev. Materials}\ }\textbf {\bibinfo {volume}
  {5}},\ \bibinfo {pages} {084002}}\BibitemShut {NoStop}%
\bibitem [{\citenamefont {Stampfli}(1986)}]{Stampfli.HPA.1986}%
  \BibitemOpen
  \bibfield  {author} {\bibinfo {author} {\bibnamefont {Stampfli},
  \bibfnamefont {Peter}}} (\bibinfo {year} {1986}),\ \bibfield  {title}
  {\enquote {\bibinfo {title} {A dodecagonal quasiperiodic lattice in two
  dimensions.}}\ }\href@noop {} {\bibfield  {journal} {\bibinfo  {journal}
  {Helvetica Physica Acta}\ }\textbf {\bibinfo {volume} {59}},\ \bibinfo
  {pages} {1260--1263}}\BibitemShut {NoStop}%
\bibitem [{\citenamefont {Stepanov}\ \emph {et~al.}(2020)\citenamefont
  {Stepanov}, \citenamefont {Das}, \citenamefont {Lu}, \citenamefont
  {Fahimniya}, \citenamefont {Watanabe}, \citenamefont {Taniguchi},
  \citenamefont {Koppens}, \citenamefont {Lischner}, \citenamefont {Levitov},\
  and\ \citenamefont {Efetov}}]{Stepanov.nature.2020}%
  \BibitemOpen
  \bibfield  {author} {\bibinfo {author} {\bibnamefont {Stepanov},
  \bibfnamefont {Petr}}, \bibinfo {author} {\bibfnamefont {Ipsita}\
  \bibnamefont {Das}}, \bibinfo {author} {\bibfnamefont {Xiaobo}\ \bibnamefont
  {Lu}}, \bibinfo {author} {\bibfnamefont {Ali}\ \bibnamefont {Fahimniya}},
  \bibinfo {author} {\bibfnamefont {Kenji}\ \bibnamefont {Watanabe}}, \bibinfo
  {author} {\bibfnamefont {Takashi}\ \bibnamefont {Taniguchi}}, \bibinfo
  {author} {\bibfnamefont {Frank H.~L.}\ \bibnamefont {Koppens}}, \bibinfo
  {author} {\bibfnamefont {Johannes}\ \bibnamefont {Lischner}}, \bibinfo
  {author} {\bibfnamefont {Leonid}\ \bibnamefont {Levitov}}, and\ \bibinfo
  {author} {\bibfnamefont {Dmitri~K.}\ \bibnamefont {Efetov}}} (\bibinfo {year}
  {2020}),\ \bibfield  {title} {\enquote {\bibinfo {title} {Untying the
  insulating and superconducting orders in magic-angle graphene},}\ }\href
  {https://doi.org/10.1038/s41586-020-2459-6} {\bibfield  {journal} {\bibinfo
  {journal} {Nature}\ }\textbf {\bibinfo {volume} {583}}~(\bibinfo {number}
  {7816}),\ \bibinfo {pages} {375--378}}\BibitemShut {NoStop}%
\bibitem [{\citenamefont {Stuart}\ \emph {et~al.}(2000)\citenamefont {Stuart},
  \citenamefont {Tutein},\ and\ \citenamefont {Harrison}}]{Stuart.jcp.2000}%
  \BibitemOpen
  \bibfield  {author} {\bibinfo {author} {\bibnamefont {Stuart}, \bibfnamefont
  {Steven~J}}, \bibinfo {author} {\bibfnamefont {Alan~B.}\ \bibnamefont
  {Tutein}}, and\ \bibinfo {author} {\bibfnamefont {Judith~A.}\ \bibnamefont
  {Harrison}}} (\bibinfo {year} {2000}),\ \bibfield  {title} {\enquote
  {\bibinfo {title} {A reactive potential for hydrocarbons with intermolecular
  interactions},}\ }\href {https://doi.org/10.1063/1.481208} {\bibfield
  {journal} {\bibinfo  {journal} {The Journal of Chemical Physics}\ }\textbf
  {\bibinfo {volume} {112}}~(\bibinfo {number} {14}),\ \bibinfo {pages}
  {6472--6486}}\BibitemShut {NoStop}%
\bibitem [{\citenamefont {Sun}\ \emph {et~al.}(2018)\citenamefont {Sun},
  \citenamefont {Zhang}, \citenamefont {Lu}, \citenamefont {Li}, \citenamefont
  {Xue}, \citenamefont {Du}, \citenamefont {Pu},\ and\ \citenamefont
  {Wang}}]{Sun.jpcl.2018}%
  \BibitemOpen
  \bibfield  {author} {\bibinfo {author} {\bibnamefont {Sun}, \bibfnamefont
  {Junhui}}, \bibinfo {author} {\bibfnamefont {Yanning}\ \bibnamefont {Zhang}},
  \bibinfo {author} {\bibfnamefont {Zhibin}\ \bibnamefont {Lu}}, \bibinfo
  {author} {\bibfnamefont {Qunyang}\ \bibnamefont {Li}}, \bibinfo {author}
  {\bibfnamefont {Qunji}\ \bibnamefont {Xue}}, \bibinfo {author} {\bibfnamefont
  {Shiyu}\ \bibnamefont {Du}}, \bibinfo {author} {\bibfnamefont {Jibin}\
  \bibnamefont {Pu}}, and\ \bibinfo {author} {\bibfnamefont {Liping}\
  \bibnamefont {Wang}}} (\bibinfo {year} {2018}),\ \bibfield  {title} {\enquote
  {\bibinfo {title} {Superlubricity enabled by pressure-induced friction
  collapse},}\ }\href {https://doi.org/10.1021/acs.jpclett.8b00877} {\bibfield
  {journal} {\bibinfo  {journal} {The Journal of Physical Chemistry Letters}\
  }\textbf {\bibinfo {volume} {9}}~(\bibinfo {number} {10}),\ \bibinfo {pages}
  {2554--2559}}\BibitemShut {NoStop}%
\bibitem [{\citenamefont {Theodorou}\ and\ \citenamefont
  {Rice}(1978)}]{Theodorou.PRB.1978}%
  \BibitemOpen
  \bibfield  {author} {\bibinfo {author} {\bibnamefont {Theodorou},
  \bibfnamefont {G}}, and\ \bibinfo {author} {\bibfnamefont {T.~M.}\
  \bibnamefont {Rice}}} (\bibinfo {year} {1978}),\ \bibfield  {title} {\enquote
  {\bibinfo {title} {Statics and dynamics of incommensurate lattices},}\ }\href
  {https://doi.org/10.1103/PhysRevB.18.2840} {\bibfield  {journal} {\bibinfo
  {journal} {Phys. Rev. B}\ }\textbf {\bibinfo {volume} {18}},\ \bibinfo
  {pages} {2840--2856}}\BibitemShut {NoStop}%
\bibitem [{\citenamefont {Thompson}\ \emph {et~al.}(2022)\citenamefont
  {Thompson}, \citenamefont {Aktulga}, \citenamefont {Berger}, \citenamefont
  {Bolintineanu}, \citenamefont {Brown}, \citenamefont {Crozier}, \citenamefont
  {{in 't Veld}}, \citenamefont {Kohlmeyer}, \citenamefont {Moore},
  \citenamefont {Nguyen}, \citenamefont {Shan}, \citenamefont {Stevens},
  \citenamefont {Tranchida}, \citenamefont {Trott},\ and\ \citenamefont
  {Plimpton}}]{Thompson.compphyscomm.2022}%
  \BibitemOpen
  \bibfield  {author} {\bibinfo {author} {\bibnamefont {Thompson},
  \bibfnamefont {Aidan~P}}, \bibinfo {author} {\bibfnamefont {H.~Metin}\
  \bibnamefont {Aktulga}}, \bibinfo {author} {\bibfnamefont {Richard}\
  \bibnamefont {Berger}}, \bibinfo {author} {\bibfnamefont {Dan~S.}\
  \bibnamefont {Bolintineanu}}, \bibinfo {author} {\bibfnamefont {W.~Michael}\
  \bibnamefont {Brown}}, \bibinfo {author} {\bibfnamefont {Paul~S.}\
  \bibnamefont {Crozier}}, \bibinfo {author} {\bibfnamefont {Pieter~J.}\
  \bibnamefont {{in 't Veld}}}, \bibinfo {author} {\bibfnamefont {Axel}\
  \bibnamefont {Kohlmeyer}}, \bibinfo {author} {\bibfnamefont {Stan~G.}\
  \bibnamefont {Moore}}, \bibinfo {author} {\bibfnamefont {Trung~Dac}\
  \bibnamefont {Nguyen}}, \bibinfo {author} {\bibfnamefont {Ray}\ \bibnamefont
  {Shan}}, \bibinfo {author} {\bibfnamefont {Mark~J.}\ \bibnamefont {Stevens}},
  \bibinfo {author} {\bibfnamefont {Julien}\ \bibnamefont {Tranchida}},
  \bibinfo {author} {\bibfnamefont {Christian}\ \bibnamefont {Trott}}, and\
  \bibinfo {author} {\bibfnamefont {Steven~J.}\ \bibnamefont {Plimpton}}}
  (\bibinfo {year} {2022}),\ \bibfield  {title} {\enquote {\bibinfo {title}
  {Lammps - a flexible simulation tool for particle-based materials modeling at
  the atomic, meso, and continuum scales},}\ }\href
  {https://doi.org/https://doi.org/10.1016/j.cpc.2021.108171} {\bibfield
  {journal} {\bibinfo  {journal} {Computer Physics Communications}\ }\textbf
  {\bibinfo {volume} {271}},\ \bibinfo {pages} {108171}}\BibitemShut {NoStop}%
\bibitem [{\citenamefont {Tison}\ \emph {et~al.}(2014)\citenamefont {Tison},
  \citenamefont {Lagoute}, \citenamefont {Repain}, \citenamefont {Chacon},
  \citenamefont {Girard}, \citenamefont {Joucken}, \citenamefont {Sporken},
  \citenamefont {Gargiulo}, \citenamefont {Yazyev},\ and\ \citenamefont
  {Rousset}}]{Tison.nanolett.2014}%
  \BibitemOpen
  \bibfield  {author} {\bibinfo {author} {\bibnamefont {Tison}, \bibfnamefont
  {Yann}}, \bibinfo {author} {\bibfnamefont {Jérôme}\ \bibnamefont
  {Lagoute}}, \bibinfo {author} {\bibfnamefont {Vincent}\ \bibnamefont
  {Repain}}, \bibinfo {author} {\bibfnamefont {Cyril}\ \bibnamefont {Chacon}},
  \bibinfo {author} {\bibfnamefont {Yann}\ \bibnamefont {Girard}}, \bibinfo
  {author} {\bibfnamefont {Frédéric}\ \bibnamefont {Joucken}}, \bibinfo
  {author} {\bibfnamefont {Robert}\ \bibnamefont {Sporken}}, \bibinfo {author}
  {\bibfnamefont {Fernando}\ \bibnamefont {Gargiulo}}, \bibinfo {author}
  {\bibfnamefont {Oleg~V.}\ \bibnamefont {Yazyev}}, and\ \bibinfo {author}
  {\bibfnamefont {Sylvie}\ \bibnamefont {Rousset}}} (\bibinfo {year} {2014}),\
  \bibfield  {title} {\enquote {\bibinfo {title} {Grain boundaries in graphene
  on sic(000$\bar{1}$) substrate},}\ }\href {https://doi.org/10.1021/nl502854w}
  {\bibfield  {journal} {\bibinfo  {journal} {Nano Letters}\ }\textbf {\bibinfo
  {volume} {14}}~(\bibinfo {number} {11}),\ \bibinfo {pages}
  {6382--6386}}\BibitemShut {NoStop}%
\bibitem [{\citenamefont {T{\"o}rm{\"a}}\ \emph {et~al.}(2022)\citenamefont
  {T{\"o}rm{\"a}}, \citenamefont {Peotta},\ and\ \citenamefont
  {Bernevig}}]{Torma.natrevphys.2022}%
  \BibitemOpen
  \bibfield  {author} {\bibinfo {author} {\bibnamefont {T{\"o}rm{\"a}},
  \bibfnamefont {P{\"a}ivi}}, \bibinfo {author} {\bibfnamefont {Sebastiano}\
  \bibnamefont {Peotta}}, and\ \bibinfo {author} {\bibfnamefont {Bogdan~A.}\
  \bibnamefont {Bernevig}}} (\bibinfo {year} {2022}),\ \bibfield  {title}
  {\enquote {\bibinfo {title} {Superconductivity, superfluidity and quantum
  geometry in twisted multilayer systems},}\ }\href
  {https://doi.org/10.1038/s42254-022-00466-y} {\bibfield  {journal} {\bibinfo
  {journal} {Nature Reviews Physics}\ }10.1038/s42254-022-00466-y}\BibitemShut
  {NoStop}%
\bibitem [{\citenamefont {Uchida}\ \emph {et~al.}(2014)\citenamefont {Uchida},
  \citenamefont {Furuya}, \citenamefont {Iwata},\ and\ \citenamefont
  {Oshiyama}}]{Uchida.prb.2014}%
  \BibitemOpen
  \bibfield  {author} {\bibinfo {author} {\bibnamefont {Uchida}, \bibfnamefont
  {Kazuyuki}}, \bibinfo {author} {\bibfnamefont {Shinnosuke}\ \bibnamefont
  {Furuya}}, \bibinfo {author} {\bibfnamefont {Jun-Ichi}\ \bibnamefont
  {Iwata}}, and\ \bibinfo {author} {\bibfnamefont {Atsushi}\ \bibnamefont
  {Oshiyama}}} (\bibinfo {year} {2014}),\ \bibfield  {title} {\enquote
  {\bibinfo {title} {Atomic corrugation and electron localization due to
  moir\'e patterns in twisted bilayer graphenes},}\ }\href
  {https://doi.org/10.1103/PhysRevB.90.155451} {\bibfield  {journal} {\bibinfo
  {journal} {Phys. Rev. B}\ }\textbf {\bibinfo {volume} {90}},\ \bibinfo
  {pages} {155451}}\BibitemShut {NoStop}%
\bibitem [{\citenamefont {Urbakh}(2013)}]{Urbakh.natnano.2013}%
  \BibitemOpen
  \bibfield  {author} {\bibinfo {author} {\bibnamefont {Urbakh}, \bibfnamefont
  {Michael}}} (\bibinfo {year} {2013}),\ \bibfield  {title} {\enquote {\bibinfo
  {title} {Towards macroscale superlubricity},}\ }\href
  {https://doi.org/10.1038/nnano.2013.244} {\bibfield  {journal} {\bibinfo
  {journal} {Nature Nanotechnology}\ }\textbf {\bibinfo {volume} {8}}~(\bibinfo
  {number} {12}),\ \bibinfo {pages} {893--894}}\BibitemShut {NoStop}%
\bibitem [{\citenamefont {Vanossi}\ \emph {et~al.}(2020)\citenamefont
  {Vanossi}, \citenamefont {Bechinger},\ and\ \citenamefont
  {Urbakh}}]{Vanossi.nc.2020}%
  \BibitemOpen
  \bibfield  {author} {\bibinfo {author} {\bibnamefont {Vanossi}, \bibfnamefont
  {Andrea}}, \bibinfo {author} {\bibfnamefont {Clemens}\ \bibnamefont
  {Bechinger}}, and\ \bibinfo {author} {\bibfnamefont {Michael}\ \bibnamefont
  {Urbakh}}} (\bibinfo {year} {2020}),\ \bibfield  {title} {\enquote {\bibinfo
  {title} {Structural lubricity in soft and hard matter systems},}\ }\href
  {https://doi.org/10.1038/s41467-020-18429-1} {\bibfield  {journal} {\bibinfo
  {journal} {Nature Communications}\ }\textbf {\bibinfo {volume}
  {11}}~(\bibinfo {number} {1}),\ \bibinfo {pages} {4657}}\BibitemShut
  {NoStop}%
\bibitem [{\citenamefont {Vanossi}\ \emph {et~al.}(2013)\citenamefont
  {Vanossi}, \citenamefont {Manini}, \citenamefont {Urbakh}, \citenamefont
  {Zapperi},\ and\ \citenamefont {Tosatti}}]{Vanossi.RevModPhys.2013}%
  \BibitemOpen
  \bibfield  {author} {\bibinfo {author} {\bibnamefont {Vanossi}, \bibfnamefont
  {Andrea}}, \bibinfo {author} {\bibfnamefont {Nicola}\ \bibnamefont {Manini}},
  \bibinfo {author} {\bibfnamefont {Michael}\ \bibnamefont {Urbakh}}, \bibinfo
  {author} {\bibfnamefont {Stefano}\ \bibnamefont {Zapperi}}, and\ \bibinfo
  {author} {\bibfnamefont {Erio}\ \bibnamefont {Tosatti}}} (\bibinfo {year}
  {2013}),\ \bibfield  {title} {\enquote {\bibinfo {title} {Colloquium:
  Modeling friction: From nanoscale to mesoscale},}\ }\href
  {https://doi.org/10.1103/RevModPhys.85.529} {\bibfield  {journal} {\bibinfo
  {journal} {Rev. Mod. Phys.}\ }\textbf {\bibinfo {volume} {85}},\ \bibinfo
  {pages} {529--552}}\BibitemShut {NoStop}%
\bibitem [{\citenamefont {Vanossi}\ and\ \citenamefont
  {Tosatti}(2012)}]{Vanossi.natmat.2012}%
  \BibitemOpen
  \bibfield  {author} {\bibinfo {author} {\bibnamefont {Vanossi}, \bibfnamefont
  {Andrea}}, and\ \bibinfo {author} {\bibfnamefont {Erio}\ \bibnamefont
  {Tosatti}}} (\bibinfo {year} {2012}),\ \bibfield  {title} {\enquote {\bibinfo
  {title} {Kinks in motion},}\ }\href {https://doi.org/10.1038/nmat3229}
  {\bibfield  {journal} {\bibinfo  {journal} {Nature Materials}\ }\textbf
  {\bibinfo {volume} {11}}~(\bibinfo {number} {2}),\ \bibinfo {pages}
  {97--98}}\BibitemShut {NoStop}%
\bibitem [{\citenamefont {Varini}\ \emph {et~al.}(2015)\citenamefont {Varini},
  \citenamefont {Vanossi}, \citenamefont {Guerra}, \citenamefont {Mandelli},
  \citenamefont {Capozza},\ and\ \citenamefont
  {Tosatti}}]{Varini.nanoscale.2015}%
  \BibitemOpen
  \bibfield  {author} {\bibinfo {author} {\bibnamefont {Varini}, \bibfnamefont
  {Nicola}}, \bibinfo {author} {\bibfnamefont {Andrea}\ \bibnamefont
  {Vanossi}}, \bibinfo {author} {\bibfnamefont {Roberto}\ \bibnamefont
  {Guerra}}, \bibinfo {author} {\bibfnamefont {Davide}\ \bibnamefont
  {Mandelli}}, \bibinfo {author} {\bibfnamefont {Rosario}\ \bibnamefont
  {Capozza}}, and\ \bibinfo {author} {\bibfnamefont {Erio}\ \bibnamefont
  {Tosatti}}} (\bibinfo {year} {2015}),\ \bibfield  {title} {\enquote {\bibinfo
  {title} {Static friction scaling of physisorbed islands: the key is in the
  edge},}\ }\href {https://doi.org/10.1039/C4NR06521B} {\bibfield  {journal}
  {\bibinfo  {journal} {Nanoscale}\ }\textbf {\bibinfo {volume} {7}},\ \bibinfo
  {pages} {2093--2101}}\BibitemShut {NoStop}%
\bibitem [{\citenamefont {\ifmmode~\check{C}\else \v{C}\fi{}ervenka}\ and\
  \citenamefont {Flipse}(2009)}]{Cervenka.prb.2009}%
  \BibitemOpen
  \bibfield  {author} {\bibinfo {author} {\bibnamefont {\ifmmode~\check{C}\else
  \v{C}\fi{}ervenka}, \bibfnamefont {J}}, and\ \bibinfo {author} {\bibfnamefont
  {C.~F.~J.}\ \bibnamefont {Flipse}}} (\bibinfo {year} {2009}),\ \bibfield
  {title} {\enquote {\bibinfo {title} {Structural and electronic properties of
  grain boundaries in graphite: Planes of periodically distributed point
  defects},}\ }\href {https://doi.org/10.1103/PhysRevB.79.195429} {\bibfield
  {journal} {\bibinfo  {journal} {Phys. Rev. B}\ }\textbf {\bibinfo {volume}
  {79}},\ \bibinfo {pages} {195429}}\BibitemShut {NoStop}%
\bibitem [{\citenamefont {Vu}\ \emph {et~al.}(2016)\citenamefont {Vu},
  \citenamefont {Zhang}, \citenamefont {Urbakh}, \citenamefont {Li},
  \citenamefont {He},\ and\ \citenamefont {Zheng}}]{Vu.prb.2016}%
  \BibitemOpen
  \bibfield  {author} {\bibinfo {author} {\bibnamefont {Vu}, \bibfnamefont
  {Cuong~Cao}}, \bibinfo {author} {\bibfnamefont {Shoumo}\ \bibnamefont
  {Zhang}}, \bibinfo {author} {\bibfnamefont {Michael}\ \bibnamefont {Urbakh}},
  \bibinfo {author} {\bibfnamefont {Qunyang}\ \bibnamefont {Li}}, \bibinfo
  {author} {\bibfnamefont {Q.-C.}\ \bibnamefont {He}}, and\ \bibinfo {author}
  {\bibfnamefont {Quanshui}\ \bibnamefont {Zheng}}} (\bibinfo {year} {2016}),\
  \bibfield  {title} {\enquote {\bibinfo {title} {Observation of
  normal-force-independent superlubricity in mesoscopic graphite contacts},}\
  }\href {https://doi.org/10.1103/PhysRevB.94.081405} {\bibfield  {journal}
  {\bibinfo  {journal} {Phys. Rev. B}\ }\textbf {\bibinfo {volume} {94}},\
  \bibinfo {pages} {081405}}\BibitemShut {NoStop}%
\bibitem [{\citenamefont {Wang}\ \emph {et~al.}(2016)\citenamefont {Wang},
  \citenamefont {Chen}, \citenamefont {Li}, \citenamefont {Cheng},
  \citenamefont {Yang}, \citenamefont {Wu}, \citenamefont {Xie}, \citenamefont
  {Zhang}, \citenamefont {Zhao}, \citenamefont {Lu}, \citenamefont {Chen},
  \citenamefont {Wang}, \citenamefont {Meng}, \citenamefont {Tang},
  \citenamefont {Yang}, \citenamefont {He}, \citenamefont {Liu}, \citenamefont
  {Shi}, \citenamefont {Watanabe}, \citenamefont {Taniguchi}, \citenamefont
  {Feng}, \citenamefont {Zhang},\ and\ \citenamefont {Zhang}}]{Wang.prl.2016}%
  \BibitemOpen
  \bibfield  {author} {\bibinfo {author} {\bibnamefont {Wang}, \bibfnamefont
  {Duoming}}, \bibinfo {author} {\bibfnamefont {Guorui}\ \bibnamefont {Chen}},
  \bibinfo {author} {\bibfnamefont {Chaokai}\ \bibnamefont {Li}}, \bibinfo
  {author} {\bibfnamefont {Meng}\ \bibnamefont {Cheng}}, \bibinfo {author}
  {\bibfnamefont {Wei}\ \bibnamefont {Yang}}, \bibinfo {author} {\bibfnamefont
  {Shuang}\ \bibnamefont {Wu}}, \bibinfo {author} {\bibfnamefont {Guibai}\
  \bibnamefont {Xie}}, \bibinfo {author} {\bibfnamefont {Jing}\ \bibnamefont
  {Zhang}}, \bibinfo {author} {\bibfnamefont {Jing}\ \bibnamefont {Zhao}},
  \bibinfo {author} {\bibfnamefont {Xiaobo}\ \bibnamefont {Lu}}, \bibinfo
  {author} {\bibfnamefont {Peng}\ \bibnamefont {Chen}}, \bibinfo {author}
  {\bibfnamefont {Guole}\ \bibnamefont {Wang}}, \bibinfo {author}
  {\bibfnamefont {Jianling}\ \bibnamefont {Meng}}, \bibinfo {author}
  {\bibfnamefont {Jian}\ \bibnamefont {Tang}}, \bibinfo {author} {\bibfnamefont
  {Rong}\ \bibnamefont {Yang}}, \bibinfo {author} {\bibfnamefont {Congli}\
  \bibnamefont {He}}, \bibinfo {author} {\bibfnamefont {Donghua}\ \bibnamefont
  {Liu}}, \bibinfo {author} {\bibfnamefont {Dongxia}\ \bibnamefont {Shi}},
  \bibinfo {author} {\bibfnamefont {Kenji}\ \bibnamefont {Watanabe}}, \bibinfo
  {author} {\bibfnamefont {Takashi}\ \bibnamefont {Taniguchi}}, \bibinfo
  {author} {\bibfnamefont {Ji}~\bibnamefont {Feng}}, \bibinfo {author}
  {\bibfnamefont {Yuanbo}\ \bibnamefont {Zhang}}, and\ \bibinfo {author}
  {\bibfnamefont {Guangyu}\ \bibnamefont {Zhang}}} (\bibinfo {year} {2016}),\
  \bibfield  {title} {\enquote {\bibinfo {title} {Thermally induced graphene
  rotation on hexagonal boron nitride},}\ }\href
  {https://doi.org/10.1103/PhysRevLett.116.126101} {\bibfield  {journal}
  {\bibinfo  {journal} {Phys. Rev. Lett.}\ }\textbf {\bibinfo {volume} {116}},\
  \bibinfo {pages} {126101}}\BibitemShut {NoStop}%
\bibitem [{\citenamefont {Wang}\ \emph
  {et~al.}(2015{\natexlab{a}})\citenamefont {Wang}, \citenamefont {Li},
  \citenamefont {Li}, \citenamefont {Cai}, \citenamefont {Zhu},\ and\
  \citenamefont {Jia}}]{Wang.friction.2015}%
  \BibitemOpen
  \bibfield  {author} {\bibinfo {author} {\bibnamefont {Wang}, \bibfnamefont
  {Jianjun}}, \bibinfo {author} {\bibfnamefont {Jinming}\ \bibnamefont {Li}},
  \bibinfo {author} {\bibfnamefont {Chong}\ \bibnamefont {Li}}, \bibinfo
  {author} {\bibfnamefont {Xiaolin}\ \bibnamefont {Cai}}, \bibinfo {author}
  {\bibfnamefont {Wenguang}\ \bibnamefont {Zhu}}, and\ \bibinfo {author}
  {\bibfnamefont {Yu}~\bibnamefont {Jia}}} (\bibinfo {year}
  {2015}{\natexlab{a}}),\ \bibfield  {title} {\enquote {\bibinfo {title}
  {Tuning the nanofriction between two graphene layers by external electric
  fields: A density functional theory study},}\ }\href
  {https://doi.org/10.1007/s11249-015-0624-0} {\bibfield  {journal} {\bibinfo
  {journal} {Tribology Letters}\ }\textbf {\bibinfo {volume} {61}}~(\bibinfo
  {number} {1}),\ \bibinfo {pages} {4}}\BibitemShut {NoStop}%
\bibitem [{\citenamefont {Wang}\ \emph
  {et~al.}(2019{\natexlab{a}})\citenamefont {Wang}, \citenamefont {Cao},
  \citenamefont {Song}, \citenamefont {Qu}, \citenamefont {Zheng},\ and\
  \citenamefont {Ma}}]{Wang.Nano.Letters.2019}%
  \BibitemOpen
  \bibfield  {author} {\bibinfo {author} {\bibnamefont {Wang}, \bibfnamefont
  {Jin}}, \bibinfo {author} {\bibfnamefont {Wei}\ \bibnamefont {Cao}}, \bibinfo
  {author} {\bibfnamefont {Yiming}\ \bibnamefont {Song}}, \bibinfo {author}
  {\bibfnamefont {Cangyu}\ \bibnamefont {Qu}}, \bibinfo {author} {\bibfnamefont
  {Quanshui}\ \bibnamefont {Zheng}}, and\ \bibinfo {author} {\bibfnamefont
  {Ming}\ \bibnamefont {Ma}}} (\bibinfo {year} {2019}{\natexlab{a}}),\
  \bibfield  {title} {\enquote {\bibinfo {title} {Generalized scaling law of
  structural superlubricity},}\ }\href
  {https://doi.org/10.1021/acs.nanolett.9b02656} {\bibfield  {journal}
  {\bibinfo  {journal} {Nano Letters}\ }\textbf {\bibinfo {volume}
  {19}}~(\bibinfo {number} {11}),\ \bibinfo {pages} {7735--7741}}\BibitemShut
  {NoStop}%
\bibitem [{\citenamefont {Wang}\ \emph
  {et~al.}(2023{\natexlab{a}})\citenamefont {Wang}, \citenamefont {Khosravi},
  \citenamefont {Silva}, \citenamefont {Fabrizio}, \citenamefont {Vanossi},\
  and\ \citenamefont {Tosatti}}]{Wang.buckling.2023}%
  \BibitemOpen
  \bibfield  {author} {\bibinfo {author} {\bibnamefont {Wang}, \bibfnamefont
  {Jin}}, \bibinfo {author} {\bibfnamefont {Ali}\ \bibnamefont {Khosravi}},
  \bibinfo {author} {\bibfnamefont {Andrea}\ \bibnamefont {Silva}}, \bibinfo
  {author} {\bibfnamefont {Michele}\ \bibnamefont {Fabrizio}}, \bibinfo
  {author} {\bibfnamefont {Andrea}\ \bibnamefont {Vanossi}}, and\ \bibinfo
  {author} {\bibfnamefont {Erio}\ \bibnamefont {Tosatti}}} (\bibinfo {year}
  {2023}{\natexlab{a}}),\ \href@noop {} {\enquote {\bibinfo {title} {Bending
  stiffness collapse, buckling, topological bands of freestanding twisted
  bilayer graphene},}\ }\Eprint {https://arxiv.org/abs/2305.07543}
  {arXiv:2305.07543 [cond-mat.mes-hall]} \BibitemShut {NoStop}%
\bibitem [{\citenamefont {Wang}\ \emph
  {et~al.}(2023{\natexlab{b}})\citenamefont {Wang}, \citenamefont {Ma},\ and\
  \citenamefont {Tosatti}}]{Wang.tobe.2023}%
  \BibitemOpen
  \bibfield  {author} {\bibinfo {author} {\bibnamefont {Wang}, \bibfnamefont
  {Jin}}, \bibinfo {author} {\bibfnamefont {Ming.}\ \bibnamefont {Ma}}, and\
  \bibinfo {author} {\bibfnamefont {Erio}\ \bibnamefont {Tosatti}}} (\bibinfo
  {year} {2023}{\natexlab{b}}),\ \bibfield  {title} {\enquote {\bibinfo {title}
  {Theoretical understanding of sliding friction in structural lubric 2d
  contacts},}\ }\href@noop {} {\bibinfo  {journal} {to be published}\
  }\BibitemShut {NoStop}%
\bibitem [{\citenamefont {Wang}\ \emph {et~al.}(2017)\citenamefont {Wang},
  \citenamefont {Ma},\ and\ \citenamefont {Sun}}]{Wang.royalSCh.2017}%
  \BibitemOpen
\bibfield  {journal} {  }\bibfield  {author} {\bibinfo {author} {\bibnamefont
  {Wang}, \bibfnamefont {Jingang}}, \bibinfo {author} {\bibfnamefont {Fengcai}\
  \bibnamefont {Ma}}, and\ \bibinfo {author} {\bibfnamefont {Mengtao}\
  \bibnamefont {Sun}}} (\bibinfo {year} {2017}),\ \bibfield  {title} {\enquote
  {\bibinfo {title} {Graphene{,} hexagonal boron nitride{,} and their
  heterostructures: properties and applications},}\ }\href
  {https://doi.org/10.1039/C7RA00260B} {\bibfield  {journal} {\bibinfo
  {journal} {RSC Adv.}\ }\textbf {\bibinfo {volume} {7}},\ \bibinfo {pages}
  {16801--16822}}\BibitemShut {NoStop}%
\bibitem [{\citenamefont {Wang}\ \emph {et~al.}(2021)\citenamefont {Wang},
  \citenamefont {Wang},\ and\ \citenamefont {Ma}}]{Wang.frontchem.2021}%
  \BibitemOpen
  \bibfield  {author} {\bibinfo {author} {\bibnamefont {Wang}, \bibfnamefont
  {Kehan}}, \bibinfo {author} {\bibfnamefont {Jin}\ \bibnamefont {Wang}}, and\
  \bibinfo {author} {\bibfnamefont {Ming}\ \bibnamefont {Ma}}} (\bibinfo {year}
  {2021}),\ \bibfield  {title} {\enquote {\bibinfo {title} {Negative or
  positive? loading area dependent correlation between friction and normal load
  in structural superlubricity},}\ }\href
  {https://doi.org/10.3389/fchem.2021.807630} {\bibfield  {journal} {\bibinfo
  {journal} {Frontiers in chemistry}\ }\textbf {\bibinfo {volume} {9}},\
  \bibinfo {pages} {807630}}\BibitemShut {NoStop}%
\bibitem [{\citenamefont {Wang}\ \emph {et~al.}(2022)\citenamefont {Wang},
  \citenamefont {He}, \citenamefont {Cao}, \citenamefont {Wang}, \citenamefont
  {Qu}, \citenamefont {Chai}, \citenamefont {Liu}, \citenamefont {Zheng},\ and\
  \citenamefont {Ma}}]{Wang.jmps.2022}%
  \BibitemOpen
  \bibfield  {author} {\bibinfo {author} {\bibnamefont {Wang}, \bibfnamefont
  {Kunqi}}, \bibinfo {author} {\bibfnamefont {Yuqing}\ \bibnamefont {He}},
  \bibinfo {author} {\bibfnamefont {Wei}\ \bibnamefont {Cao}}, \bibinfo
  {author} {\bibfnamefont {Jin}\ \bibnamefont {Wang}}, \bibinfo {author}
  {\bibfnamefont {Cangyu}\ \bibnamefont {Qu}}, \bibinfo {author} {\bibfnamefont
  {Maosheng}\ \bibnamefont {Chai}}, \bibinfo {author} {\bibfnamefont {Yuan}\
  \bibnamefont {Liu}}, \bibinfo {author} {\bibfnamefont {Quanshui}\
  \bibnamefont {Zheng}}, and\ \bibinfo {author} {\bibfnamefont {Ming}\
  \bibnamefont {Ma}}} (\bibinfo {year} {2022}),\ \bibfield  {title} {\enquote
  {\bibinfo {title} {Structural superlubricity with a contaminant-rich
  interface},}\ }\href
  {https://doi.org/https://doi.org/10.1016/j.jmps.2022.105063} {\bibinfo
  {journal} {Journal of the Mechanics and Physics of Solids}\ ,\ \bibinfo
  {pages} {105063}}\BibitemShut {NoStop}%
\bibitem [{\citenamefont {Wang}\ \emph
  {et~al.}(2019{\natexlab{b}})\citenamefont {Wang}, \citenamefont {Qu},
  \citenamefont {Wang}, \citenamefont {Ouyang}, \citenamefont {Ma},\ and\
  \citenamefont {Zheng}}]{Wang.acsami.2019}%
  \BibitemOpen
\bibfield  {journal} {  }\bibfield  {author} {\bibinfo {author} {\bibnamefont
  {Wang}, \bibfnamefont {Kunqi}}, \bibinfo {author} {\bibfnamefont {Cangyu}\
  \bibnamefont {Qu}}, \bibinfo {author} {\bibfnamefont {Jin}\ \bibnamefont
  {Wang}}, \bibinfo {author} {\bibfnamefont {Wengen}\ \bibnamefont {Ouyang}},
  \bibinfo {author} {\bibfnamefont {Ming}\ \bibnamefont {Ma}}, and\ \bibinfo
  {author} {\bibfnamefont {Quanshui}\ \bibnamefont {Zheng}}} (\bibinfo {year}
  {2019}{\natexlab{b}}),\ \bibfield  {title} {\enquote {\bibinfo {title}
  {Strain engineering modulates graphene interlayer friction by moiré pattern
  evolution},}\ }\href {https://doi.org/10.1021/acsami.9b09259} {\bibfield
  {journal} {\bibinfo  {journal} {ACS Applied Materials \& Interfaces}\
  }\textbf {\bibinfo {volume} {11}}~(\bibinfo {number} {39}),\ \bibinfo {pages}
  {36169--36176}}\BibitemShut {NoStop}%
\bibitem [{\citenamefont {Wang}\ \emph {et~al.}(2020)\citenamefont {Wang},
  \citenamefont {Qu}, \citenamefont {Wang}, \citenamefont {Quan},\ and\
  \citenamefont {Zheng}}]{Wang.prl.2020}%
  \BibitemOpen
  \bibfield  {author} {\bibinfo {author} {\bibnamefont {Wang}, \bibfnamefont
  {Kunqi}}, \bibinfo {author} {\bibfnamefont {Cangyu}\ \bibnamefont {Qu}},
  \bibinfo {author} {\bibfnamefont {Jin}\ \bibnamefont {Wang}}, \bibinfo
  {author} {\bibfnamefont {Baogang}\ \bibnamefont {Quan}}, and\ \bibinfo
  {author} {\bibfnamefont {Quanshui}\ \bibnamefont {Zheng}}} (\bibinfo {year}
  {2020}),\ \bibfield  {title} {\enquote {\bibinfo {title} {Characterization of
  a microscale superlubric graphite interface},}\ }\href
  {https://doi.org/10.1103/PhysRevLett.125.026101} {\bibfield  {journal}
  {\bibinfo  {journal} {Phys. Rev. Lett.}\ }\textbf {\bibinfo {volume} {125}},\
  \bibinfo {pages} {026101}}\BibitemShut {NoStop}%
\bibitem [{\citenamefont {Wang}\ \emph
  {et~al.}(2015{\natexlab{b}})\citenamefont {Wang}, \citenamefont {Dai},
  \citenamefont {Li}, \citenamefont {Yang}, \citenamefont {Srolovitz},\ and\
  \citenamefont {Zheng}}]{Wang.nc.2015}%
  \BibitemOpen
  \bibfield  {author} {\bibinfo {author} {\bibnamefont {Wang}, \bibfnamefont
  {Wen}}, \bibinfo {author} {\bibfnamefont {Shuyang}\ \bibnamefont {Dai}},
  \bibinfo {author} {\bibfnamefont {Xide}\ \bibnamefont {Li}}, \bibinfo
  {author} {\bibfnamefont {Jiarui}\ \bibnamefont {Yang}}, \bibinfo {author}
  {\bibfnamefont {David~J.}\ \bibnamefont {Srolovitz}}, and\ \bibinfo {author}
  {\bibfnamefont {Quanshui}\ \bibnamefont {Zheng}}} (\bibinfo {year}
  {2015}{\natexlab{b}}),\ \bibfield  {title} {\enquote {\bibinfo {title}
  {Measurement of the cleavage energy of graphite},}\ }\href
  {https://doi.org/10.1038/ncomms8853} {\bibfield  {journal} {\bibinfo
  {journal} {Nature Communications}\ }\textbf {\bibinfo {volume} {6}}~(\bibinfo
  {number} {1}),\ \bibinfo {pages} {7853}}\BibitemShut {NoStop}%
\bibitem [{\citenamefont {Wang}\ and\ \citenamefont
  {Li}(2019)}]{Wang.epl.2019}%
  \BibitemOpen
  \bibfield  {author} {\bibinfo {author} {\bibnamefont {Wang}, \bibfnamefont
  {Wen}}, and\ \bibinfo {author} {\bibfnamefont {Xide}\ \bibnamefont {Li}}}
  (\bibinfo {year} {2019}),\ \bibfield  {title} {\enquote {\bibinfo {title}
  {Interlayer motion and ultra-low sliding friction in microscale graphite
  flakes},}\ }\href {https://doi.org/10.1209/0295-5075/125/26003} {\bibfield
  {journal} {\bibinfo  {journal} {{EPL} (Europhysics Letters)}\ }\textbf
  {\bibinfo {volume} {125}}~(\bibinfo {number} {2}),\ \bibinfo {pages}
  {26003}}\BibitemShut {NoStop}%
\bibitem [{\citenamefont {Wen}\ \emph {et~al.}(2018)\citenamefont {Wen},
  \citenamefont {Carr}, \citenamefont {Fang}, \citenamefont {Kaxiras},\ and\
  \citenamefont {Tadmor}}]{Wen.prb.2018}%
  \BibitemOpen
  \bibfield  {author} {\bibinfo {author} {\bibnamefont {Wen}, \bibfnamefont
  {Mingjian}}, \bibinfo {author} {\bibfnamefont {Stephen}\ \bibnamefont
  {Carr}}, \bibinfo {author} {\bibfnamefont {Shiang}\ \bibnamefont {Fang}},
  \bibinfo {author} {\bibfnamefont {Efthimios}\ \bibnamefont {Kaxiras}}, and\
  \bibinfo {author} {\bibfnamefont {Ellad~B.}\ \bibnamefont {Tadmor}}}
  (\bibinfo {year} {2018}),\ \bibfield  {title} {\enquote {\bibinfo {title}
  {Dihedral-angle-corrected registry-dependent interlayer potential for
  multilayer graphene structures},}\ }\href
  {https://doi.org/10.1103/PhysRevB.98.235404} {\bibfield  {journal} {\bibinfo
  {journal} {Phys. Rev. B}\ }\textbf {\bibinfo {volume} {98}},\ \bibinfo
  {pages} {235404}}\BibitemShut {NoStop}%
\bibitem [{\citenamefont {Weston}\ \emph {et~al.}(2020)\citenamefont {Weston},
  \citenamefont {Zou}, \citenamefont {Enaldiev}, \citenamefont {Summerfield},
  \citenamefont {Clark}, \citenamefont {Z{\'o}lyomi}, \citenamefont {Graham},
  \citenamefont {Yelgel}, \citenamefont {Magorrian}, \citenamefont {Zhou},
  \citenamefont {Zultak}, \citenamefont {Hopkinson}, \citenamefont {Barinov},
  \citenamefont {Bointon}, \citenamefont {Kretinin}, \citenamefont {Wilson},
  \citenamefont {Beton}, \citenamefont {Fal'ko}, \citenamefont {Haigh},\ and\
  \citenamefont {Gorbachev}}]{Weston.NatNano.2020}%
  \BibitemOpen
  \bibfield  {author} {\bibinfo {author} {\bibnamefont {Weston}, \bibfnamefont
  {Astrid}}, \bibinfo {author} {\bibfnamefont {Yichao}\ \bibnamefont {Zou}},
  \bibinfo {author} {\bibfnamefont {Vladimir}\ \bibnamefont {Enaldiev}},
  \bibinfo {author} {\bibfnamefont {Alex}\ \bibnamefont {Summerfield}},
  \bibinfo {author} {\bibfnamefont {Nicholas}\ \bibnamefont {Clark}}, \bibinfo
  {author} {\bibfnamefont {Viktor}\ \bibnamefont {Z{\'o}lyomi}}, \bibinfo
  {author} {\bibfnamefont {Abigail}\ \bibnamefont {Graham}}, \bibinfo {author}
  {\bibfnamefont {Celal}\ \bibnamefont {Yelgel}}, \bibinfo {author}
  {\bibfnamefont {Samuel}\ \bibnamefont {Magorrian}}, \bibinfo {author}
  {\bibfnamefont {Mingwei}\ \bibnamefont {Zhou}}, \bibinfo {author}
  {\bibfnamefont {Johanna}\ \bibnamefont {Zultak}}, \bibinfo {author}
  {\bibfnamefont {David}\ \bibnamefont {Hopkinson}}, \bibinfo {author}
  {\bibfnamefont {Alexei}\ \bibnamefont {Barinov}}, \bibinfo {author}
  {\bibfnamefont {Thomas~H.}\ \bibnamefont {Bointon}}, \bibinfo {author}
  {\bibfnamefont {Andrey}\ \bibnamefont {Kretinin}}, \bibinfo {author}
  {\bibfnamefont {Neil~R.}\ \bibnamefont {Wilson}}, \bibinfo {author}
  {\bibfnamefont {Peter~H.}\ \bibnamefont {Beton}}, \bibinfo {author}
  {\bibfnamefont {Vladimir~I.}\ \bibnamefont {Fal'ko}}, \bibinfo {author}
  {\bibfnamefont {Sarah~J.}\ \bibnamefont {Haigh}}, and\ \bibinfo {author}
  {\bibfnamefont {Roman}\ \bibnamefont {Gorbachev}}} (\bibinfo {year} {2020}),\
  \bibfield  {title} {\enquote {\bibinfo {title} {Atomic reconstruction in
  twisted bilayers of transition metal dichalcogenides},}\ }\href
  {https://doi.org/10.1038/s41565-020-0682-9} {\bibfield  {journal} {\bibinfo
  {journal} {Nature Nanotechnology}\ }\textbf {\bibinfo {volume}
  {15}}~(\bibinfo {number} {7}),\ \bibinfo {pages} {592--597}}\BibitemShut
  {NoStop}%
\bibitem [{\citenamefont {van Wijk}\ \emph {et~al.}(2013)\citenamefont {van
  Wijk}, \citenamefont {Dienwiebel}, \citenamefont {Frenken},\ and\
  \citenamefont {Fasolino}}]{van.prb.2013}%
  \BibitemOpen
  \bibfield  {author} {\bibinfo {author} {\bibnamefont {van Wijk},
  \bibfnamefont {M~M}}, \bibinfo {author} {\bibfnamefont {M.}~\bibnamefont
  {Dienwiebel}}, \bibinfo {author} {\bibfnamefont {J.~W.~M.}\ \bibnamefont
  {Frenken}}, and\ \bibinfo {author} {\bibfnamefont {A.}~\bibnamefont
  {Fasolino}}} (\bibinfo {year} {2013}),\ \bibfield  {title} {\enquote
  {\bibinfo {title} {Superlubric to stick-slip sliding of incommensurate
  graphene flakes on graphite},}\ }\href
  {https://doi.org/10.1103/PhysRevB.88.235423} {\bibfield  {journal} {\bibinfo
  {journal} {Phys. Rev. B}\ }\textbf {\bibinfo {volume} {88}},\ \bibinfo
  {pages} {235423}}\BibitemShut {NoStop}%
\bibitem [{\citenamefont {de~Wijn}(2012)}]{deWijn.prb.2012}%
  \BibitemOpen
  \bibfield  {author} {\bibinfo {author} {\bibnamefont {de~Wijn}, \bibfnamefont
  {A~S}}} (\bibinfo {year} {2012}),\ \bibfield  {title} {\enquote {\bibinfo
  {title} {(in)commensurability, scaling, and multiplicity of friction in
  nanocrystals and application to gold nanocrystals on graphite},}\ }\href
  {https://doi.org/10.1103/PhysRevB.86.085429} {\bibfield  {journal} {\bibinfo
  {journal} {Phys. Rev. B}\ }\textbf {\bibinfo {volume} {86}},\ \bibinfo
  {pages} {085429}}\BibitemShut {NoStop}%
\bibitem [{\citenamefont {Woods}\ \emph {et~al.}(2014)\citenamefont {Woods},
  \citenamefont {Britnell}, \citenamefont {Eckmann}, \citenamefont {Ma},
  \citenamefont {Lu}, \citenamefont {Guo}, \citenamefont {Lin}, \citenamefont
  {Yu}, \citenamefont {Cao}, \citenamefont {Gorbachev}, \citenamefont
  {Kretinin}, \citenamefont {Park}, \citenamefont {Ponomarenko}, \citenamefont
  {Katsnelson}, \citenamefont {Gornostyrev}, \citenamefont {Watanabe},
  \citenamefont {Taniguchi}, \citenamefont {Casiraghi}, \citenamefont {Gao},
  \citenamefont {Geim},\ and\ \citenamefont {Novoselov}}]{Woods.NatPhys.2014}%
  \BibitemOpen
  \bibfield  {author} {\bibinfo {author} {\bibnamefont {Woods}, \bibfnamefont
  {C~R}}, \bibinfo {author} {\bibfnamefont {L.}~\bibnamefont {Britnell}},
  \bibinfo {author} {\bibfnamefont {A.}~\bibnamefont {Eckmann}}, \bibinfo
  {author} {\bibfnamefont {R.~S.}\ \bibnamefont {Ma}}, \bibinfo {author}
  {\bibfnamefont {J.~C.}\ \bibnamefont {Lu}}, \bibinfo {author} {\bibfnamefont
  {H.~M.}\ \bibnamefont {Guo}}, \bibinfo {author} {\bibfnamefont
  {X.}~\bibnamefont {Lin}}, \bibinfo {author} {\bibfnamefont {G.~L.}\
  \bibnamefont {Yu}}, \bibinfo {author} {\bibfnamefont {Y.}~\bibnamefont
  {Cao}}, \bibinfo {author} {\bibfnamefont {R.~ ~V.}\ \bibnamefont
  {Gorbachev}}, \bibinfo {author} {\bibfnamefont {A.~V.}\ \bibnamefont
  {Kretinin}}, \bibinfo {author} {\bibfnamefont {J.}~\bibnamefont {Park}},
  \bibinfo {author} {\bibfnamefont {L.~A.}\ \bibnamefont {Ponomarenko}},
  \bibinfo {author} {\bibfnamefont {M.~I.}\ \bibnamefont {Katsnelson}},
  \bibinfo {author} {\bibfnamefont {Yu. N.}\ \bibnamefont {Gornostyrev}},
  \bibinfo {author} {\bibfnamefont {K.}~\bibnamefont {Watanabe}}, \bibinfo
  {author} {\bibfnamefont {T.}~\bibnamefont {Taniguchi}}, \bibinfo {author}
  {\bibfnamefont {C.}~\bibnamefont {Casiraghi}}, \bibinfo {author}
  {\bibfnamefont {H.-J.}\ \bibnamefont {Gao}}, \bibinfo {author} {\bibfnamefont
  {A.~K.}\ \bibnamefont {Geim}}, and\ \bibinfo {author} {\bibfnamefont
  {K.~ ~S.}\ \bibnamefont {Novoselov}}} (\bibinfo {year} {2014}),\ \bibfield
  {title} {\enquote {\bibinfo {title} {Commensurate--incommensurate transition
  in graphene on hexagonal boron nitride},}\ }\href
  {https://doi.org/10.1038/nphys2954} {\bibfield  {journal} {\bibinfo
  {journal} {Nature Physics}\ }\textbf {\bibinfo {volume} {10}}~(\bibinfo
  {number} {6}),\ \bibinfo {pages} {451--456}}\BibitemShut {NoStop}%
\bibitem [{\citenamefont {Wu}\ \emph {et~al.}(2021)\citenamefont {Wu},
  \citenamefont {Huang}, \citenamefont {Xiang},\ and\ \citenamefont
  {Zheng}}]{Wu.commumater.2021}%
  \BibitemOpen
  \bibfield  {author} {\bibinfo {author} {\bibnamefont {Wu}, \bibfnamefont
  {Zhanghui}}, \bibinfo {author} {\bibfnamefont {Xuanyu}\ \bibnamefont
  {Huang}}, \bibinfo {author} {\bibfnamefont {Xiaojian}\ \bibnamefont {Xiang}},
  and\ \bibinfo {author} {\bibfnamefont {Quanshui}\ \bibnamefont {Zheng}}}
  (\bibinfo {year} {2021}),\ \bibfield  {title} {\enquote {\bibinfo {title}
  {Electro-superlubric springs for continuously tunable resonators and
  oscillators},}\ }\href {https://doi.org/10.1038/s43246-021-00207-1}
  {\bibfield  {journal} {\bibinfo  {journal} {Communications Materials}\
  }\textbf {\bibinfo {volume} {2}}~(\bibinfo {number} {1}),\ \bibinfo {pages}
  {104}}\BibitemShut {NoStop}%
\bibitem [{\citenamefont {Yang}\ and\ \citenamefont
  {Zhang}(2021)}]{Yang.acsami.2021}%
  \BibitemOpen
  \bibfield  {author} {\bibinfo {author} {\bibnamefont {Yang}, \bibfnamefont
  {Xing}}, and\ \bibinfo {author} {\bibfnamefont {Bin}\ \bibnamefont {Zhang}}}
  (\bibinfo {year} {2021}),\ \bibfield  {title} {\enquote {\bibinfo {title}
  {Rotational friction correlated with moiré patterns in strained bilayer
  graphene: Implications for nanoscale lubrication},}\ }\href
  {https://doi.org/10.1021/acsanm.1c01540} {\bibfield  {journal} {\bibinfo
  {journal} {ACS Applied Nano Materials}\ }\textbf {\bibinfo {volume}
  {4}}~(\bibinfo {number} {9}),\ \bibinfo {pages} {8880--8887}}\BibitemShut
  {NoStop}%
\bibitem [{\citenamefont {Yankowitz}\ \emph
  {et~al.}(2019{\natexlab{a}})\citenamefont {Yankowitz}, \citenamefont {Chen},
  \citenamefont {Polshyn}, \citenamefont {Zhang}, \citenamefont {Watanabe},
  \citenamefont {Taniguchi}, \citenamefont {Graf}, \citenamefont {Young},\ and\
  \citenamefont {Dean}}]{Yankowitz.science.2019}%
  \BibitemOpen
  \bibfield  {author} {\bibinfo {author} {\bibnamefont {Yankowitz},
  \bibfnamefont {Matthew}}, \bibinfo {author} {\bibfnamefont {Shaowen}\
  \bibnamefont {Chen}}, \bibinfo {author} {\bibfnamefont {Hryhoriy}\
  \bibnamefont {Polshyn}}, \bibinfo {author} {\bibfnamefont {Yuxuan}\
  \bibnamefont {Zhang}}, \bibinfo {author} {\bibfnamefont {K.}~\bibnamefont
  {Watanabe}}, \bibinfo {author} {\bibfnamefont {T.}~\bibnamefont {Taniguchi}},
  \bibinfo {author} {\bibfnamefont {David}\ \bibnamefont {Graf}}, \bibinfo
  {author} {\bibfnamefont {Andrea~F.}\ \bibnamefont {Young}}, and\ \bibinfo
  {author} {\bibfnamefont {Cory~R.}\ \bibnamefont {Dean}}} (\bibinfo {year}
  {2019}{\natexlab{a}}),\ \bibfield  {title} {\enquote {\bibinfo {title}
  {Tuning superconductivity in twisted bilayer graphene},}\ }\href
  {https://doi.org/10.1126/science.aav1910} {\bibfield  {journal} {\bibinfo
  {journal} {Science}\ }\textbf {\bibinfo {volume} {363}}~(\bibinfo {number}
  {6431}),\ \bibinfo {pages} {1059--1064}}\BibitemShut {NoStop}%
\bibitem [{\citenamefont {Yankowitz}\ \emph
  {et~al.}(2019{\natexlab{b}})\citenamefont {Yankowitz}, \citenamefont {Ma},
  \citenamefont {Jarillo-Herrero},\ and\ \citenamefont
  {LeRoy}}]{Yankowitz.NatRevPhys.2019}%
  \BibitemOpen
  \bibfield  {author} {\bibinfo {author} {\bibnamefont {Yankowitz},
  \bibfnamefont {Matthew}}, \bibinfo {author} {\bibfnamefont {Qiong}\
  \bibnamefont {Ma}}, \bibinfo {author} {\bibfnamefont {Pablo}\ \bibnamefont
  {Jarillo-Herrero}}, and\ \bibinfo {author} {\bibfnamefont {Brian~J.}\
  \bibnamefont {LeRoy}}} (\bibinfo {year} {2019}{\natexlab{b}}),\ \bibfield
  {title} {\enquote {\bibinfo {title} {van der waals heterostructures combining
  graphene and hexagonal boron nitride},}\ }\href
  {https://doi.org/10.1038/s42254-018-0016-0} {\bibfield  {journal} {\bibinfo
  {journal} {Nature Reviews Physics}\ }\textbf {\bibinfo {volume}
  {1}}~(\bibinfo {number} {2}),\ \bibinfo {pages} {112--125}}\BibitemShut
  {NoStop}%
\bibitem [{\citenamefont {Yazyev}\ and\ \citenamefont
  {Chen}(2014)}]{Yazyev.natnano.2014}%
  \BibitemOpen
  \bibfield  {author} {\bibinfo {author} {\bibnamefont {Yazyev}, \bibfnamefont
  {Oleg~V}}, and\ \bibinfo {author} {\bibfnamefont {Yong~P.}\ \bibnamefont
  {Chen}}} (\bibinfo {year} {2014}),\ \bibfield  {title} {\enquote {\bibinfo
  {title} {Polycrystalline graphene and other two-dimensional materials},}\
  }\href {https://doi.org/10.1038/nnano.2014.166} {\bibfield  {journal}
  {\bibinfo  {journal} {Nature Nanotechnology}\ }\textbf {\bibinfo {volume}
  {9}}~(\bibinfo {number} {10}),\ \bibinfo {pages} {755--767}}\BibitemShut
  {NoStop}%
\bibitem [{\citenamefont {Yazyev}\ and\ \citenamefont
  {Louie}(2010)}]{Yazyev.prb.2010}%
  \BibitemOpen
  \bibfield  {author} {\bibinfo {author} {\bibnamefont {Yazyev}, \bibfnamefont
  {Oleg~V}}, and\ \bibinfo {author} {\bibfnamefont {Steven~G.}\ \bibnamefont
  {Louie}}} (\bibinfo {year} {2010}),\ \bibfield  {title} {\enquote {\bibinfo
  {title} {Topological defects in graphene: Dislocations and grain
  boundaries},}\ }\href {https://doi.org/10.1103/PhysRevB.81.195420} {\bibfield
   {journal} {\bibinfo  {journal} {Phys. Rev. B}\ }\textbf {\bibinfo {volume}
  {81}},\ \bibinfo {pages} {195420}}\BibitemShut {NoStop}%
\bibitem [{\citenamefont {Yoo}\ \emph {et~al.}(2019)\citenamefont {Yoo},
  \citenamefont {Engelke}, \citenamefont {Carr}, \citenamefont {Fang},
  \citenamefont {Zhang}, \citenamefont {Cazeaux}, \citenamefont {Sung},
  \citenamefont {Hovden}, \citenamefont {Tsen}, \citenamefont {Taniguchi},
  \citenamefont {Watanabe}, \citenamefont {Yi}, \citenamefont {Kim},
  \citenamefont {Luskin}, \citenamefont {Tadmor}, \citenamefont {Kaxiras},\
  and\ \citenamefont {Kim}}]{Yoo.NatMater.2019}%
  \BibitemOpen
  \bibfield  {author} {\bibinfo {author} {\bibnamefont {Yoo}, \bibfnamefont
  {Hyobin}}, \bibinfo {author} {\bibfnamefont {Rebecca}\ \bibnamefont
  {Engelke}}, \bibinfo {author} {\bibfnamefont {Stephen}\ \bibnamefont {Carr}},
  \bibinfo {author} {\bibfnamefont {Shiang}\ \bibnamefont {Fang}}, \bibinfo
  {author} {\bibfnamefont {Kuan}\ \bibnamefont {Zhang}}, \bibinfo {author}
  {\bibfnamefont {Paul}\ \bibnamefont {Cazeaux}}, \bibinfo {author}
  {\bibfnamefont {Suk~Hyun}\ \bibnamefont {Sung}}, \bibinfo {author}
  {\bibfnamefont {Robert}\ \bibnamefont {Hovden}}, \bibinfo {author}
  {\bibfnamefont {Adam~W.}\ \bibnamefont {Tsen}}, \bibinfo {author}
  {\bibfnamefont {Takashi}\ \bibnamefont {Taniguchi}}, \bibinfo {author}
  {\bibfnamefont {Kenji}\ \bibnamefont {Watanabe}}, \bibinfo {author}
  {\bibfnamefont {Gyu-Chul}\ \bibnamefont {Yi}}, \bibinfo {author}
  {\bibfnamefont {Miyoung}\ \bibnamefont {Kim}}, \bibinfo {author}
  {\bibfnamefont {Mitchell}\ \bibnamefont {Luskin}}, \bibinfo {author}
  {\bibfnamefont {Ellad~B.}\ \bibnamefont {Tadmor}}, \bibinfo {author}
  {\bibfnamefont {Efthimios}\ \bibnamefont {Kaxiras}}, and\ \bibinfo {author}
  {\bibfnamefont {Philip}\ \bibnamefont {Kim}}} (\bibinfo {year} {2019}),\
  \bibfield  {title} {\enquote {\bibinfo {title} {Atomic and electronic
  reconstruction at the van der waals interface in twisted bilayer graphene},}\
  }\href {https://doi.org/10.1038/s41563-019-0346-z} {\bibfield  {journal}
  {\bibinfo  {journal} {Nature Materials}\ }\textbf {\bibinfo {volume}
  {18}}~(\bibinfo {number} {5}),\ \bibinfo {pages} {448--453}}\BibitemShut
  {NoStop}%
\bibitem [{\citenamefont {Zhang}\ and\ \citenamefont
  {Tadmor}(2018)}]{Zhang.Mech.Phys.Solids.2018}%
  \BibitemOpen
  \bibfield  {author} {\bibinfo {author} {\bibnamefont {Zhang}, \bibfnamefont
  {Kuan}}, and\ \bibinfo {author} {\bibfnamefont {Ellad~B.}\ \bibnamefont
  {Tadmor}}} (\bibinfo {year} {2018}),\ \bibfield  {title} {\enquote {\bibinfo
  {title} {Structural and electron diffraction scaling of twisted graphene
  bilayers},}\ }\href
  {https://doi.org/https://doi.org/10.1016/j.jmps.2017.12.005} {\bibfield
  {journal} {\bibinfo  {journal} {Journal of the Mechanics and Physics of
  Solids}\ }\textbf {\bibinfo {volume} {112}},\ \bibinfo {pages}
  {225--238}}\BibitemShut {NoStop}%
\bibitem [{\citenamefont {Zhang}\ \emph {et~al.}(2013)\citenamefont {Zhang},
  \citenamefont {Ning}, \citenamefont {Zhang}, \citenamefont {Zheng},
  \citenamefont {Chen}, \citenamefont {Xie}, \citenamefont {Zhang},
  \citenamefont {Qian},\ and\ \citenamefont {Wei}}]{Zhang.natnano.2013}%
  \BibitemOpen
  \bibfield  {author} {\bibinfo {author} {\bibnamefont {Zhang}, \bibfnamefont
  {Rufan}}, \bibinfo {author} {\bibfnamefont {Zhiyuan}\ \bibnamefont {Ning}},
  \bibinfo {author} {\bibfnamefont {Yingying}\ \bibnamefont {Zhang}}, \bibinfo
  {author} {\bibfnamefont {Quanshui}\ \bibnamefont {Zheng}}, \bibinfo {author}
  {\bibfnamefont {Qing}\ \bibnamefont {Chen}}, \bibinfo {author} {\bibfnamefont
  {Huanhuan}\ \bibnamefont {Xie}}, \bibinfo {author} {\bibfnamefont {Qiang}\
  \bibnamefont {Zhang}}, \bibinfo {author} {\bibfnamefont {Weizhong}\
  \bibnamefont {Qian}}, and\ \bibinfo {author} {\bibfnamefont {Fei}\
  \bibnamefont {Wei}}} (\bibinfo {year} {2013}),\ \bibfield  {title} {\enquote
  {\bibinfo {title} {Superlubricity in centimetres-long double-walled carbon
  nanotubes under ambient conditions},}\ }\href
  {https://doi.org/10.1038/nnano.2013.217} {\bibfield  {journal} {\bibinfo
  {journal} {Nature Nanotechnology}\ }\textbf {\bibinfo {volume} {8}}~(\bibinfo
  {number} {12}),\ \bibinfo {pages} {912--916}}\BibitemShut {NoStop}%
\bibitem [{\citenamefont {Zhang}\ \emph {et~al.}(2019)\citenamefont {Zhang},
  \citenamefont {Ma}, \citenamefont {Erdemir},\ and\ \citenamefont
  {Li}}]{Zhang.MaterToday.2019}%
  \BibitemOpen
  \bibfield  {author} {\bibinfo {author} {\bibnamefont {Zhang}, \bibfnamefont
  {Shuai}}, \bibinfo {author} {\bibfnamefont {Tianbao}\ \bibnamefont {Ma}},
  \bibinfo {author} {\bibfnamefont {Ali}\ \bibnamefont {Erdemir}}, and\
  \bibinfo {author} {\bibfnamefont {Qunyang}\ \bibnamefont {Li}}} (\bibinfo
  {year} {2019}),\ \bibfield  {title} {\enquote {\bibinfo {title} {Tribology of
  two-dimensional materials: From mechanisms to modulating strategies},}\
  }\href {https://doi.org/https://doi.org/10.1016/j.mattod.2018.12.002}
  {\bibfield  {journal} {\bibinfo  {journal} {Materials Today}\ }\textbf
  {\bibinfo {volume} {26}},\ \bibinfo {pages} {67--86}}\BibitemShut {NoStop}%
\bibitem [{\citenamefont {Zhang}\ \emph {et~al.}(2022)\citenamefont {Zhang},
  \citenamefont {Xu}, \citenamefont {Hou}, \citenamefont {Song}, \citenamefont
  {Ma}, \citenamefont {Gao}, \citenamefont {Zhu}, \citenamefont {Ma},
  \citenamefont {Liu}, \citenamefont {Feng},\ and\ \citenamefont
  {Li}}]{Zhang.NatMater.2022}%
  \BibitemOpen
  \bibfield  {author} {\bibinfo {author} {\bibnamefont {Zhang}, \bibfnamefont
  {Shuai}}, \bibinfo {author} {\bibfnamefont {Qiang}\ \bibnamefont {Xu}},
  \bibinfo {author} {\bibfnamefont {Yuan}\ \bibnamefont {Hou}}, \bibinfo
  {author} {\bibfnamefont {Aisheng}\ \bibnamefont {Song}}, \bibinfo {author}
  {\bibfnamefont {Yuan}\ \bibnamefont {Ma}}, \bibinfo {author} {\bibfnamefont
  {Lei}\ \bibnamefont {Gao}}, \bibinfo {author} {\bibfnamefont {Mengzhen}\
  \bibnamefont {Zhu}}, \bibinfo {author} {\bibfnamefont {Tianbao}\ \bibnamefont
  {Ma}}, \bibinfo {author} {\bibfnamefont {Luqi}\ \bibnamefont {Liu}}, \bibinfo
  {author} {\bibfnamefont {Xi-Qiao}\ \bibnamefont {Feng}}, and\ \bibinfo
  {author} {\bibfnamefont {Qunyang}\ \bibnamefont {Li}}} (\bibinfo {year}
  {2022}),\ \bibfield  {title} {\enquote {\bibinfo {title} {Domino-like
  stacking order switching in twisted monolayer--multilayer graphene},}\ }\href
  {https://doi.org/10.1038/s41563-022-01232-2} {\bibfield  {journal} {\bibinfo
  {journal} {Nature Materials}\ }\textbf {\bibinfo {volume} {21}}~(\bibinfo
  {number} {6}),\ \bibinfo {pages} {621--626}}\BibitemShut {NoStop}%
\bibitem [{\citenamefont {Zhao}\ \emph {et~al.}(2021)\citenamefont {Zhao},
  \citenamefont {Mei}, \citenamefont {Chang}, \citenamefont {Chen},
  \citenamefont {Cheng},\ and\ \citenamefont {Dassios}}]{Zhao.acsnano.2021}%
  \BibitemOpen
  \bibfield  {author} {\bibinfo {author} {\bibnamefont {Zhao}, \bibfnamefont
  {Yu}}, \bibinfo {author} {\bibfnamefont {Hui}\ \bibnamefont {Mei}}, \bibinfo
  {author} {\bibfnamefont {Peng}\ \bibnamefont {Chang}}, \bibinfo {author}
  {\bibfnamefont {Chao}\ \bibnamefont {Chen}}, \bibinfo {author} {\bibfnamefont
  {Laifei}\ \bibnamefont {Cheng}}, and\ \bibinfo {author} {\bibfnamefont
  {Konstantinos~G.}\ \bibnamefont {Dassios}}} (\bibinfo {year} {2021}),\
  \bibfield  {title} {\enquote {\bibinfo {title} {Infinite approaching
  superlubricity by three-dimensional printed structures},}\ }\href
  {https://doi.org/10.1021/acsnano.0c08713} {\bibfield  {journal} {\bibinfo
  {journal} {ACS Nano}\ }\textbf {\bibinfo {volume} {15}}~(\bibinfo {number}
  {1}),\ \bibinfo {pages} {240--257}}\BibitemShut {NoStop}%
\bibitem [{\citenamefont {Zheng}\ and\ \citenamefont
  {Jiang}(2002)}]{Zheng.prl.2002}%
  \BibitemOpen
  \bibfield  {author} {\bibinfo {author} {\bibnamefont {Zheng}, \bibfnamefont
  {Quanshui}}, and\ \bibinfo {author} {\bibfnamefont {Qing}\ \bibnamefont
  {Jiang}}} (\bibinfo {year} {2002}),\ \bibfield  {title} {\enquote {\bibinfo
  {title} {Multiwalled carbon nanotubes as gigahertz oscillators},}\ }\href
  {https://doi.org/10.1103/PhysRevLett.88.045503} {\bibfield  {journal}
  {\bibinfo  {journal} {Phys. Rev. Lett.}\ }\textbf {\bibinfo {volume} {88}},\
  \bibinfo {pages} {045503}}\BibitemShut {NoStop}%
\bibitem [{\citenamefont {Özoğul}\ \emph {et~al.}(2017)\citenamefont
  {Özoğul}, \citenamefont {İpek}, \citenamefont {Durgun},\ and\
  \citenamefont {Baykara}}]{Ozogul.apl.2017}%
  \BibitemOpen
  \bibfield  {author} {\bibinfo {author} {\bibnamefont {Özoğul},
  \bibfnamefont {Alper}}, \bibinfo {author} {\bibfnamefont {Semran}\
  \bibnamefont {İpek}}, \bibinfo {author} {\bibfnamefont {Engin}\ \bibnamefont
  {Durgun}}, and\ \bibinfo {author} {\bibfnamefont {Mehmet~Z.}\ \bibnamefont
  {Baykara}}} (\bibinfo {year} {2017}),\ \bibfield  {title} {\enquote {\bibinfo
  {title} {Structural superlubricity of platinum on graphite under ambient
  conditions: The effects of chemistry and geometry},}\ }\href
  {https://doi.org/10.1063/1.5008529} {\bibfield  {journal} {\bibinfo
  {journal} {Applied Physics Letters}\ }\textbf {\bibinfo {volume}
  {111}}~(\bibinfo {number} {21}),\ \bibinfo {pages} {211602}}\BibitemShut
  {NoStop}%
\end{thebibliography}%

\end{document}